\documentclass[11pt, a4paper,twoside]{mythesis}
\usepackage{graphicx}
\usepackage{amssymb}
\pdfoutput=1
\usepackage{epstopdf}
\usepackage{amsmath}
\usepackage{comment}
\usepackage{amsfonts}
\usepackage[applemac]{inputenc}
\usepackage{chapprelude}
\usepackage{fancyhdr}
\usepackage{color}
\usepackage{aas_macros}
\usepackage{multirow}
\usepackage{setspace}
\usepackage{moddeluxetable}
\usepackage[bookmarks,bookmarksnumbered,colorlinks=false]{hyperref}
\usepackage[round]{natbib}
\usepackage{hypernat}
\hypersetup{
	urlcolor=blue          
}

\fancypagestyle{plain}{
	\fancyhf{}
	\fancyhead{} 
	\renewcommand{\headrulewidth}{0pt} 
	\fancyfoot[LE,RO]{\thepage}
}

\begin{document}
\setcounter{secnumdepth}{2}

\textheight = 28.5 cm
\topmargin = 0.0 cm
\headheight = 0.0 cm
\headsep = 0.0 cm

\thispagestyle{empty}

\begin{center}
DISS. ETH NO. 20642\\
\vspace*{18truemm}
\Huge\textbf {The information content of galaxy surveys\\}
\vspace*{18truemm}
 
\small{A dissertation submitted to}\\
\vspace{5truemm}
\normalsize{ETH ZURICH}\\
\vspace{6truemm}
\small{for the degree of}\\
\vspace{6truemm}
\normalsize{Doctor of Sciences}\\
\vspace{18truemm}
\small{presented by}\\
\vspace{6truemm}
\normalsize{Julien Carron}\\
\vspace{4truemm}
\small{MSc ETH in Physics\\
\vspace{14truemm}
born on Dec 26th, 1985\\
\vspace{3truemm}
citizen of Sion}\\
\vspace*{14truemm}
\small{accepted on the recommendation of}\\                                                                                                                                                                                                                                  

\vspace{5truemm}
\normalsize{Prof. Dr. Simon Lilly, examiner\\
Dr. Adam Amara, co-examiner\\
Prof. Dr. Luca Amendola, co-examiner\\
\vspace{8truemm}
2012
}

\end{center}

\eject
 
\clearpage

\textheight = 22 cm
\topmargin = 0.0 cm
\headheight = 15 pt
\headsep = 1.0 cm
\footskip = 50pt
\pagenumbering{roman}
\setcounter{page}{2}

\clearpage{\pagestyle{empty}\cleardoublepage}

\begin{spacing}{0}
\tableofcontents
\addcontentsline{toc}{chapter}{Contents}
\cleardoublepage
\listoffigures
\addcontentsline{toc}{chapter}{List of Figures}
\cleardoublepage

\listoftables
\end{spacing}
\addcontentsline{toc}{chapter}{List of Tables}
\clearpage

\chapter*{Abstract}
\addcontentsline{toc}{chapter}{Abstract}
\markboth{ABSTRACT}{}

This research is a contribution to our understanding of the information content of the cosmological dark matter density field, and of the means to extract this information. These questions are of prime importance in order to reach closer for solutions to current fundamental issues in cosmology, such as the nature of dark matter and dark energy, that future large galaxy surveys are aiming at. The focus is on a traditional class of observables, the $N$-point functions, that we approach with known information theoretic tools, Fisher information and Shannon information entropy. It is built out of two main parts, the first presenting in details the mathematical methods we used and introduced, and the second the cosmological research that was performed with these tools.
\newline
\newline
A large fraction of this thesis is dedicated to the study of the information content of random fields with heavy tails, in particular the lognormal field, a model for the matter density fluctuation field. It is well known that in the nonlinear regime of structure formation, the matter fluctuation field develops such large tails. It has also been suggested that fields with large tails are not necessarily well described by the hierarchy of $N$-point functions. In this thesis, we are able to make this last statement precise and with the help of the lognormal model to quantify precisely its implications for inference on cosmological parameters : we find as our main result that only a tiny fraction of the total Fisher information of the field is still contained in the hierarchy of $N$-point moments in the nonlinear regime, rendering parameter inference from such moments very inefficient. We show that the hierarchy fails to capture the information that is contained in the underdense regions, which at the same time are found to be the most rich in information.  We find further our results to be very consistent with numerical analysis using $N$-body simulations. We also discuss these issues with the help of explicit families of fields with the same hierarchy of $N$-point moments defined in this work. A similar analysis is then applied to the convergence field, the weighted projection of the matter density fluctuation field along the line of sight, with similar conclusions. We also show how simple mappings can correct for this inadequacy, consistently with previous findings in the literature.
\newline
\newline
These results were made possible using an expansion of the Fisher information matrix in uncorrelated components associated to $N$-points moments of successive orders. An entire chapter is dedicated to this expansion, investigating its  properties and making a connection to the moment problem in the field of mathematics. Some simple models exactly solvable at all orders are also presented.
\newline
\newline
Beside these investigations of the statistical power of the hierarchy of $N$-point moments, we also study the combination of various probes of the convergence field, including the magnification, shear and flexion fields, in particular at the two-point level. We use Shannon information entropy to discuss the simple structure of the information within these tracers of the lensing potential field. We then evaluate the prospects for such a combination according to current understanding of the relevant dispersion parameters. Finally, we revisit known derivations of the Fisher information matrix for Gaussian variables, commenting in this light on the use of Gaussian likelihoods for power spectra or two-point correlation function estimators in cosmology. We point towards the fact that despite their motivation from the central limit theorem, care must be taken in the case of a large number of fields, as this assumption assigns too much information to the observables. 
\chapter*{R\'esum\'e}
\addcontentsline{toc}{chapter}{R\'esum\'e}
\markboth{R\'esum\'e}{}

Cette thèse est une contribution à notre compréhension de l'information utile à des fins cosmologiques contenue dans le champ de matière noire de notre Univers, ainsi que des procédés pour extraire cette information. Ces questions sont essentielles dans l'optique de s'approcher d'éléments de réponses à des questions fondamentales de la cosmologie moderne, telles que la nature de la matière noire et de l'énergie sombre. Dans ce travail, nous concentrons principalement nos efforts sur une classe traditionnelle d'observables, c'est à dire la hiérarchie des fonctions à $N$ points, que nous approchons en utilisant deux outils empruntés à la théorie de l'information : l'information de Fisher et l'entropie de Shannon. Cette thèse est constituée de deux parties principales, la première présentant et développant ces méthodes mathématiques pour notre fin, la deuxième la recherche en cosmologie proprement dite qui a été effectuée avec ces outils.
\newline
\newline
Une portion importante de ce travail est dédiée à l'étude de l'information de Fisher contenue dans les champs aléatoires avec forte asymétrie droite, et en particulier dans les champs log-normaux. Il est bien connu que dans le régime non linéaire de la formation des structures, le champ de densité de matière noire développe une telle asymétrie, et il a été également suggéré que cette asymétrie pénalise les fonctions à $N$ points dans leur capacité à décrire ces champs. Dans cette thèse, nous formulons ce dernier aspect plus précisément, et calculons dans plusieurs modèles son impact sur notre capacité à extraire les paramètres cosmologiques de ces observables. Nous trouvons qu'une petite fraction seulement d'information reste accessible à la hiérarchie des fonctions à $N$ points dans le régime non linéaire. Nous montrons que ces observables sont inadéquates à capturer l'information du champ dans les régions sous-denses, qui elles-mêmes contiennent la plus grande part de l'information dans ce régime, et confrontons avec succès ces résultats à des simulations numériques à $N$ corps. Nous discutons également plusieurs de ces aspects avec l'aide de quelques exemples explicites de champs aléatoires qui possèdent les mêmes fonctions à $N$ points à tous les ordres. Nous effectuons une analyse similaire pour le champ de convergence avec des résultats inchangés. Nous montrons comment de simples transformations non linéaire permettent de corriger ces problèmes, de manière comparable à d'autres résultats déjà présents dans la littérature. 
\newline
\newline
Ces résultats sont rendus possibles par une expansion de la matrice d'information de Fisher en composantes associées de manière univoque aux membres successifs de la hiérarchie des moments à $N$ points que nous introduisons dans ce travail. Un chapitre entier est consacré à cette expansion, où nous discutons ses propriétés principales, ainsi que les liens avec le problème des moments en mathématique. Nous résolvons aussi quelques modèles suffisamment simples pour permettre la dérivation analytique d'une solution exacte à tous les ordres.
\newline
\newline
Nous présentons également une étude sur la combinaison de différents traceurs du champ de convergence. Nous considérons les champs de grandissement, de cisaillement faible et de flexion, tous directement reliés à la convergence, notamment leurs fonctions à deux points, et utilisons l'entropie de Shannon pour discuter leur information jointe. Nous évaluons les avantages d'une analyse jointe de ces traceurs selon notre compréhension actuelle des paramètres de dispersion.
Finalement, nous revoyons des dérivations connues de la matrices de Fisher pour des champs Gaussiens, ce qui nous permet de commenter l'usage fréquent de statistiques Gaussiennes pour des estimateurs de spectres de puissance ou de fonctions à deux points en cosmologie. Bien que la forme Gaussienne soit motivée par le théorème de la limite centrale, nous notons que cette hypothèse assigne trop d'information aux spectres dans le cas d'une analyse jointe de plusieurs champs aléatoires.

\cleardoublepage

\pagenumbering{arabic}
\setcounter{page}{1}

\newcommand{\ca}[1]{s_{#1}}
\newcommand{\mlnrho}{\bar{\ln \rho}}
\newcommand{\mrho}{\bar{\rho}}
\newcommand{\vecun}{\mathbf 1}

\newcommand{\beq}{\begin{equation}}
\newcommand{\enq}{\end{equation}}
\newcommand{\beqa}{\begin{eqnarray}}
\newcommand{\enqa}{\end{eqnarray}}
\newcommand{\beit}{\begin{itemize}}
\newcommand{\enit}{\end{itemize}}
\newcommand{\bem}{\begin{pmatrix}}
\newcommand{\enm}{\end{pmatrix}}
\newcommand{\vect}{\mathbf t} 	
\newcommand{\vecT}{\mathbf T}
\newcommand{\vecF}{\mathbf F}
\newcommand{\vecC}{\mathbf C}
\newcommand{\vecalpha}{\boldsymbol{\alpha}}	
\newcommand{\veca}{\mathbf a}
\newcommand{\vecx}{\mathbf{x} }
\newcommand{\vecy}{\mathbf{y} }
\newcommand{\vectheta}{\bm{\theta}}
\newcommand{\veck}{\mathbf{k}}
\newcommand{\vecv}{\mathbf v}
\newcommand{\vecr}{\mathbf{r}}
\newcommand{\vecp}{\mathbf p}
\newcommand{\vecq}{\mathbf{q}}
\newcommand{\vecu}{\mathbf u}
\newcommand{\vecw}{\mathbf w}
\newcommand{\beps}{\boldsymbol{\epsilon}}
\newcommand{\vecnorm}[1]{\left\|#1\right\|}
\newcommand{\bi}[1]{\mathbf i_{#1}}
\newcommand{\Tr}{\mathrm{Tr}}
\newcommand{\pa}{\partial_\alpha}
\newcommand{\pb}{\partial_\beta}
\newcommand{\bmu}{\boldsymbol {\mu}}
\newcommand{\bphi}{\boldsymbol {\phi}}
\newcommand{\vecl}{\boldsymbol {l}}
 
\newcommand{\lmin}{\mathrm{l_{\min}}}
\newcommand{\lmax}{\mathrm{l_{\max}}}

\newcommand{\intz}{\int_{0}^{\infty}dz}
\newcommand{\intk}{\int_{0}^{\infty}dk}
\newcommand{\intkkk}{\int\frac{d^3k}{(2\pi)^3}}
\newcommand{\intV}{\int_{V}\frac{d^3x}{V}}
\newcommand{\lat}{\left\langle}
\newcommand{\rat}{\right\rangle}
\newcommand{\av}[1]{\lat #1 \rat}
\newcommand{\Var}[1]{\textrm{Var}\left(#1\right)}
\newcommand{\MSE}[1]{\textrm{MSE}\left(#1\right)}
\newcommand{\binq}[3]{\bem #1 \\ #2 \enm_{#3}}
\newcommand{\bin}[2]{\bem #1 \\ #2 \enm}

\newcommand{\pochh}[3]{\lp #1 : #2 \rp_{#3}}
\newcommand{\pdf}{\textit{pdf }}
\newcommand{\Stir}[2]{\left\{ \begin{matrix} #1 \\ #2 \end{matrix} \right\}}
\newcommand{\PI}[1]{\int \mathcal D #1}
\newcommand{\FD}[2]{\frac{\delta #1}{\delta #2 }}
\newcommand{\ft}[2]{\tilde {#1} \left( #2 \right)}
\newcommand{\obs}{\textrm{obs}}
\newcommand{\old}{\textrm{old}}
\newcommand{\new}{\textrm{new}}
\newcommand{\true}{\textrm{true}}
\newcommand{\cst}{\textrm{cst}}
\newcommand{\eff}{\textrm{eff}}
\newcommand{\lb}{\left [}
\newcommand{\rb}{\right ]}
\newcommand{\lp}{\left (}
\newcommand{\rp}{\right )}
\newcommand{\fsky}{f_{\mathrm{sky}}}
\renewcommand{\max}{\mathrm{max}}
\renewcommand{\min}{\mathrm{min}}
\newcommand{\inv}{^{-1}}
\newcommand{\norm}[1]{\left\|#1\right\|}
\newcommand{\Fab}{F_{\alpha\beta}}\newcommand{\xref}{\mathbf x_{\textrm{ref}} }
\newcommand{\xquant}{\mathbf x_{\textrm{quant}} }
\newcommand{\vecK}{\mathbf K}

\newcommand{\btheta}{\boldsymbol{\theta}}
\chapter{Introduction}
After the cosmic microwave background radiation, the study of the formation and evolution of the structures on the large scales of our Universe forms one of the pillars of modern cosmology. These structures can be mapped by galaxy surveys, and cosmological observables derived from these surveys such as the galaxy two-point correlation function or its Fourier transform the power spectrum are central to the field, and are used to contrast predictions of cosmological models to observations \citep{2004ApJ...607..655P,2006PhRvD..74l3507T,2010MNRAS.401.2148P}.  Despite essential successes in the last two decades with the emergence an observationally very successful concordance cosmology, this description of our Universe, the $\Lambda$CDM model, is still very mysterious, with only about 5\% of the energy budget of the Universe being made of the matter which we have daily experience of \citep{2011ApJS..192...18K}. The real nature of the two dominating components, the dark energy and the dark matter, remains unclear to this day and is the heart of a large scientific effort. Both the dark matter density field as well as the impact of dark energy on the geometry of the Universe can now in principle both be observed with the help of weak lensing \citep{2001PhR...340..291B,2006glsw.book.....S}. It is thus believed that large galaxy surveys able to reach for the lensing signal are going to play an increasingly important role towards these fundamental issues in cosmology \citep{2006astro.ph..9591A}. 
\newline
\newline
Of course, in order to assess some set of observables as valuable for cosmology, and to design an experiment towards its extraction, it is essential to understand both our capabilities to extract it, as well as the robustness and pertinence of the predictions of our model. These aspects, often of statistical nature, are the very backbones of this thesis. Our main aim in the present research was to try and quantify these aspects in several situations relevant to cosmology, focussing on the dark matter density field, or its weighted projection along the line of sight, the weak lensing convergence field, contributing in this way to our understanding of the information content of galaxy surveys. We review in this introductory chapter the known tools that we have built upon as well as the class of observables we have focused on, putting thus our work in context.

 \section{Stochasticity in cosmological observables}
All major predictions and measurements that are used to test our understanding of our cosmological model are meaningful only in a statistical sense. Indeed, our inability to observe initial conditions, which we may tentatively evolve, as well as the complexity of some of the physical processes involved render in general a statistical description unavoidable. For this reason, a key element that determine to an often decisive extent what observable will be of interest for the purpose of the analyst is the probability density for the realisation of the fields from which the observables are derived. Typically a CMB temperature map, or a galaxy density field, from which one measures for instance the two-point correlation function. This element of stochasticity is sometimes referred to as \textit{cosmic variance}, a denomination that we adopt in the following.  One must generically include other sort of stochasticity on top of the cosmic variance, that we refer to as \textit{noise}, for instance due to the specificities of the instrumentation, filling another gap between model predictions and actual data outputs.
\newline
\newline
We need to introduce some notation :
\newline
\newline
We always write a probability density with $p$, at times adding  a subscript indicating to which random variable it refers to for clarity. In the case of cosmological fields, these probability densities are generically high dimensional, describing the joint occurrence of fields values at different points. Typically, when the random variable are the values $(\phi_1,\cdots,\phi_d)$ of a a field $\phi$ at points $(x_1,\cdots,x_d)$, then $p$ is a function of $d$ variables,  a $d$-point probability density function.
 The position label $x$ itself can have various meanings in diverse cosmologically relevant situation. It can have for instance dimension $n = 1$  (Lyman-$\alpha$ forest), $n = 2$ (weak lensing tomography, projected density fields, CMB), or $n = 3$ (redshift surveys).
 \newline
 \newline
The joint density for the realisation of the field $\phi$ at all points  can be written conveniently as the functional $p[\phi]$.  Expectation values of observables $f$ are given formally as
\beq \label{average}
\av{f}  = \int \mathcal D \phi\:p[\phi] f[\phi],
\enq
an infinite dimensional integral. It should be kept in mind that such probability densities $p[\phi]$ are however not always very well defined and intrinsically difficult to handle, except in some cases. Expectation values \ref{average} can nevertheless be understood as the limit of a finite dimensional, well defined average
\beq
\av{f} = \int d^d\phi\:p\lp \phi_1,\cdots,\phi_d\rp f \lp \phi_1,\cdots,\phi_d \rp
\enq
over a finite sample of the field, with large $d$. In a harmless abuse of terminology we may identify at times in this work such finite samples of the field with the field itself, especially when dealing with $N$-body simulations, that have of course only a finite number of spatial resolution elements.
\subsubsection{Homogeneity, isotropy, ergodicity}
Cosmic variance in the sense defined above is the stochasticity of the data due to the fact that we observe one particular realisation of a random field, namely that of our own Universe (or of the observed part of the Universe, in which case one can also refer to a component of sample variance). It is a fundamental limitation in the sense that this variability can never be beaten down, as this would ultimately require the observation of several universes governed by the same density functions, which is a mathematical construct useless to our purposes.
\newline
\newline
Within this framework, one relies on several assumptions, namely that of statistical homogeneity, isotropy and ergodicity.
The first two express the absence of preferred locations and directions in the Universe. Mathematically speaking, all density functions are required to be invariant under spatial translations and rotations,
\beq
p \lp \phi(x_1),\cdots,\phi(x_d)\rp = p\lp \phi(x_1 + r),\cdots,\phi(x_d + r)\rp  =  p \lp \phi(R\cdot x_1),\cdots,\phi(R\cdot x_d)\rp,
\enq
for any translation vector $r$ and rotation matrix $R$. These two important assumptions can be tested and are confronted to observations. Of course,  homogeneity and isotropy do not apply to fields in redshift space coordinates. The third, ergodicity, states that we can reinterpret the ensemble averages in equation \ref{average} to be spatial averages. We expect this assumption to be correct as long as the spatial averages can be made over sufficiently large volumes, or using widely separated samples, assuming that correlations at large distances decays quickly enough to zero. Under these conditions, so called ergodic theorems can indeed be proven. However, this assumption cannot be fundamentally tested and we have no choice but to take it as granted in order to obtain useful results out of this mathematical approach. 
\newline
\newline
Very often of primary interest are the zero mean, dimensionless fluctuations $\delta$ of $\phi$, defined as
\beq
\delta(x) = \frac{\phi(x) - \bar \phi}{\bar \phi},
\enq
where $\bar \phi = \av{\phi(x)}$ is the mean of the field, independent of position $x$ by homogeneity.


\section{Fisher information for cosmology : a first look}
Inference on model parameters might appear extremely simple in principle.  For a set of model parameters $\btheta = (\alpha,\beta,\cdots)$ of interest, and the observed field $\phi$, probability theory tells us that we must update our knowledge of $\btheta$ with the simple rule,
\beq
p(\btheta | \phi) =\frac{ p(\phi | \btheta ) p(\btheta) }{\int d\btheta\: p(\phi |\btheta)p(\btheta) }.
\enq
In this equation, the density $p(\btheta)$ describes our prior state of knowledge on $\btheta$, and $p(\phi|\btheta)$, viewed as function of the parameter is called the likelihood. On the lefthand side, $p(\btheta|\phi)$ is called the posterior. Of course, the simplicity of this formula should not hide the very high complexity of its implementation for typical cosmological instances. In particular, the likelihoods $p(\phi|\btheta)$ are in general only poorly known, and the very high dimensionality of this object requires the compression of $\phi$ to some smaller subset of observables, whose statistics are set by the likelihood, all carrying some of the amount of the information that the likelihood carried originally. It is thus clearly of the uttermost interest to be able to quantify more precisely this information, both that of the original likelihood as well as that of the different observables. This is where Fisher information comes into play.
\newline
\newline
It seems fair to say that the use of Fisher information in cosmology begins, though indirectly, with
\cite{1996PhRvD..54.1332J,1996PhRvL..76.1007J}, two works in the context of CMB experiments aiming at measuring the temperature fluctuation spectrum $C_l$. In these works, it is argued that the posterior for the parameters will be approximately Gaussian. Let us consider for simplicity the case of a single parameter of interest, $\alpha$, as the discussion or an arbitrary number of parameters holds essentially unchanged. With the true value, or best fit value of the parameter defined as $\alpha_0$, the posterior is assumed to have the form
\beq \label{assumeGaussian}
p(\alpha | \left \{ C_l \right\}) \propto \exp \lp - \frac 12 \lp \alpha-\alpha_0\rp^2 i_\alpha(\alpha_0) \rp,
\enq
with the number $i_\alpha$ is defined as
\beq \label{Amatrix}
i_{\alpha} (\alpha_0)= \sum_{l}\frac{1}{\sigma^2_l} \lp \frac{\partial C_l(\alpha_0)}{\partial \alpha} \rp^2.
\enq
In this equation, $\sigma^2_l$ is the variance of the estimates of $C_l$, including cosmic variance, incomplete sky coverage and detector noise.
\newline
\newline
Under the assumption \eqref{assumeGaussian}, it is clear that $1/i_\alpha$ is the variance $\sigma^2_\alpha$ of the parameter. Very interestingly, from its definition \eqref{Amatrix} we see that this variance can be evaluated prior obtaining data, if a reasonable fiducial point $\alpha_0$ can be chosen and the model predictions of the spectrum are given. Provided the assumptions made there are correct, this is making the approach of \cite{1996PhRvD..54.1332J,1996PhRvL..76.1007J} quite powerful, providing us with a rather great understanding of the capabilities of the experiment.
\newline
\newline
It is worthwhile spending a bit more thoughts on $i_\alpha$ defined in \eqref{Amatrix}. It is a special case of an expression that weights the derivatives of some set of observables $O_i$ according to their covariance matrix $\Sigma_{ij}$,
\beq \label{Amatrix2}
i_{\alpha}(\alpha_0) = \sum_{i,j}\frac{\partial O_i(\alpha_0)}{\partial \alpha} \lb \Sigma^{-1} \rb_{ij} \frac{\partial O_j(\alpha_0)}{\partial \alpha}.
\enq
We recover \eqref{Amatrix} by setting the observables to be the spectrum, and the covariance matrix to be diagonal, as required for a perfectly Gaussian CMB map. The number \eqref{Amatrix2} has
an array of fundamental properties, none of them being difficult to show :
\beit
\item It is a non negative number, that becomes larger for a smaller covariance matrix or a larger impact of $\alpha$ on the observable, and vice versa. 
\item $i_\alpha$ corresponding to independent  observables (i.e. with no covariance) is simply the sum of their respective $i_\alpha$. 
\item Adding an observable $O_{n+1}$ to a set $\lp O_1,\cdots O_n \rp$ can only increase $i_\alpha$, and not decrease it.
\item $i_\alpha$ is identical to the expected curvature of a least squares fit to the observables with the given covariance matrix\footnote{It should be noted that this identification to a curvature in a least square fitting procedure holds only if the covariance matrix is treated as parameter independent.}, provided $\alpha_0$ is the parameter value that gives the least squared residuals.
\enit
These properties are very consistent with what we would expect from a measure of information on the parameter $\alpha$ and are making the number \eqref{Amatrix2} a promising candidate for such a measure. However, it is clearly not the end of the story, since it depends only on the chosen set of observables and their covariance matrix, but neglects all other aspects of the probability density $p(\phi | \alpha)$.
The link with Fisher information, a well known tool in statistics, was then exposed and extended to other areas of cosmology in works such as  
\cite{1997ApJ...480...22T,Tegmark97b}. It was noted that the number \eqref{Amatrix} is identical to
\beq \label{FIfirstlook}
I(\alpha):= -\av{\frac{\partial^2 \ln p}{\partial \alpha^2}},
\enq
where $p$ is a Gaussian likelihood for the noisy CMB temperature fluctuation field. The connection between equation \eqref{FIfirstlook} and covariances on parameters was also used in an astrophysical context earlier in \cite{1996ApL&C..33...63A}. Equation \ref{FIfirstlook} is the Fisher information in $p$ on $\alpha$, a most sensible measure  of information on parameters, whose properties and link to \eqref{Amatrix2} we will have the occasion to discuss extensively. For several parameters, it becomes the Fisher information matrix
\beq \label{FII}
F_{\alpha\beta}:= -\av{\frac{\partial^2 \ln p}{\partial \alpha\partial\beta}},
\enq
Since then, such Fisher information matrices for Gaussian variables and the assumption \eqref{assumeGaussian} have been used routinely in cosmology in order to assess the capabilities of some future experiments. 
\newline
\newline
Two comments are in order at this point :
\newline
\newline
First, the definition \eqref{FII} of the Fisher information matrix is the most common in cosmology. However, in this thesis, we will rather use the alternative
\beq \label{FIII}
F_{\alpha\beta} := \av{\frac{\partial \ln p}{\partial \alpha}\frac{\partial \ln p}{\partial \beta}}
\enq
as the definition of the information matrix.
These two definitions can be shown to be equivalent for any probability density function using the fact that probability densities are normalised to unity, $0 = \frac{\partial}{\partial \alpha}\av{1} = \av{\partial_\alpha \ln p}$.
While \eqref{FII} conveniently presents  the information matrix as a curvature matrix, it will become clear in chapter \ref{ch2} that \eqref{FIII}, making a reference to the \textit{score function} $\partial_\alpha \ln p$, is in fact much more fundamental for our purposes.  This form generalises more easily to non normalised density functions as well.
\newline
\newline
Second, as discussed above, Fisher information is often interpreted in cosmology as an approximation to the parameter posterior, approximated as a Gaussian with covariance matrix $F^{-1}$. The Gausisan approximation, as well as the identification of the Fisher information matrix with the inverse covariance matrix are of course only assumptions that can fail, at times severely. This is especially true when marginalising within this approach over poorly constrained parameters, whose distribution often cannot be approximated by a Gaussian shape (see for example \cite{2012arXiv1205.3984W}), giving rise to results that are difficult to interpret. In this thesis the focus is on the more orthodox interpretation of the Fisher information matrix as a very meaningful and well defined measure of information, and not as an approximation to a posterior. In particular, we are not going to inverse the Fisher matrix or marginalise over a set of parameters, except in some instances making connections to results in the literature. 
\section{$N$-point functions}
A very common class of observables, at the heart of this thesis, are the $N$-point functions, that we review briefly in this section. They are very convenient at least for two reasons. First, for Gaussian fields the mean and two-point function do contain the entire information in the field : in the language of orthodox statistics, they form a set of sufficient statistics. Second, the measurement of a (connected) three-point or higher order point function directly tests for non Gaussianity of the field.\subsection{$N$-point functions from the density : characteristic functional}
In cosmology, the $N$-point function $\xi_N$ is defined as the connected part of the $N$-point moment of the fluctuation field. These are most easily defined using the generating function technology, ubiquitous in any field theory. Consider first an arbitrary $N$-point moment of the field,
\beq
 \av{\delta(x_1)\cdots\delta(x_N)}.
\enq
We can write it, at least formally, as a derivative of the generating functional $Z$, or \textit{characteristic functional}, essentially the Fourier transform of the density :
\beq \label{disconnected}
 \av{\delta(x_1)\cdots\delta(x_N)} = \frac{1}{i^N}\left. \frac{\partial^N}{\partial J(x_1)\cdots\partial J(x_N)}Z[J]\right|_{J = 0}, 
\enq
with
\beq \label{generatingfunctional}
Z[J] := \av{\exp\lp i \int d^nx \:J(x) \delta(x)\rp}. 
\enq
In other words, the $N$-point moments can be considered as the successive terms in an expansion of the generating functional in a power series in $J$. Note that there are cases, such as for instance the lognormal field, where the generating functional cannot be written as a power series, even for $J$ very close to zero. In this case, the series should be considered as a formal power series regardless of convergence. The connected $N$-point correlation functions are then defined as the successive terms in the formal expansion of $\ln Z$ :   
\beq \label{connected}
\xi_N(x_1,\cdots,x_N) =  \frac{1}{i^N}\left. \frac{\partial^N}{\partial J(x_1)\cdots\partial J(x_N)} \ln Z\right|_{J = 0}.
\enq
If all arguments $x_1$ to $x_N$ are identical, these connected point functions become the familiar cumulants of the one dimensional density function $p(\phi(x))$.
\newline
\newline
The connected point functions are convenient since they are additive for uncorrelated fields. Indeed, if two fields are uncorrelated, then one finds directly from its very definition \eqref{generatingfunctional} that the characteristic function $Z$ of the joint density is the product of the characteristic functions of each of the densities. Taking the logarithm and using the definition \eqref{connected} shows that the connected functions just add up. From these relations \eqref{disconnected} and \eqref{connected} one can infer recursion relations for the connected point functions, as well as convenient diagrammatic representations, Feynman diagrams alike, where connected point functions are represented by connected graphs \citep[e.g.]{2002PhR...367....1B,2005astro.ph..5391S}.
\newline
\newline
It holds that the very first connected point functions are identical to the first moments of the delta field,
\beq\begin{split}
\xi_2(x_1,x_2) \equiv \xi(x_1,x_2)  &= \av{\delta(x_1)\delta(x_2)} \\
\xi_3(x_1,x_2,x_3) &= \av{\delta(x_1)\delta(x_2)\delta(x_3)},
\end{split}
\enq
but this is not the case anymore for higher $N$.
\newline
\newline
From homogeneity and isotropy, these functions are invariant under translations and rotations. In particular the two-point function  is a function of a single argument,
\beq
\xi(x_1,x_2) = \xi\lp |x_1-x_2|\rp.
\enq
Translation invariance allows conveniently the use of a description in terms of harmonics.  In Cartesian space, with Fourier transform
\beq
\tilde \delta(\veck) =  \int d^nx\: \delta(x) e^{ -i \veck\cdot x },
\enq
we have for any statistically homogeneous field the simple relation
\beq \label{spectrumP}
\av{\delta(\veck)\delta^*(\veck')} = \lp 2\pi \rp^n \delta^D(\veck -\veck')P(\veck),
\enq
where $\delta^D$ is the Dirac $\delta$ function and $P(\veck)$, the power spectrum, is the Fourier transform of the two-point function
\beq
P(\veck)  = \int d^nx \:\xi(x) e^{-i \veck\cdot x}.
\enq
Is the field further statistically isotropic, the spectrum is only a function of the modulus $k$ of the wavenumber. Similarly, one can define higher order spectra, the polyspectra, prominently the bispectrum for $N = 3$ and trispectrum  for $N = 4$ through the Fourier transforms of the connected $N$-point functions, or equivalently the expectation of products of the Fourier modes of the field. 
\newline
\newline
In this thesis, the distinction between connected and disconnected point functions,  or the use of polyspectra rather than the $N$-point functions, are of no fundamental relevance, as they provide equivalent descriptions of the same source of information. We will not make a difference between a connected or disconnected point function. We regard a generic $N$-point moment
\beq
\av{\phi(x_1)\cdots\phi(x_N)}
\enq
as a $N$-point function.
\newline
\newline
The prime example of a homogeneous isotropic random fluctuation field is of course the Gaussian field. Gaussian fields are very convenient for many reasons. They are stable under any linear transformations, such as smoothing, and also under convolutions.  The celebrated central limit theorem states that sums of a large number of independent variables tend to have a Gaussian distribution under fairly generic conditions.
Besides, they also arise as fields of  maximum information entropy for a given two-point function. The Gaussian field is defined through
\beq \label{Gfield}
\ln p[\delta] = - \frac 12 \int d^nx \int d^ny\: \delta(x) \xi^{-1}(x-y) \delta(y)+ \cst,  
\enq
or, in Fourier space,
\beq
\ln p[\delta] = - \frac 12 \int \frac{d^nk}{(2\pi)^n} \frac{|\tilde \delta(\veck)|^2}{P(\veck)}+ \cst.  
\enq
The second representation shows that the Fourier modes of such a field are independent complex Gaussian variables with the correlations as given in \eqref{spectrumP}. All finite $d$-dimensional joint densities are $d$-dimensional multivariate Gaussian distributions.
\newline
\newline
The characteristic functional can be evaluated in closed form. It is a standard result called the Gaussian integral.
\beq
Z[J] = \exp \lp -\frac 12  \int d^nx \int d^ny J(x) \xi(x-y) \:J(y) \rp =  \exp \lp -\frac 12  \int \frac{d^nk}{(2\pi)^n} P(\veck) |\tilde J(\veck)|^2 \rp.
\enq
 It follows that $\ln Z$ is a polynomial second order in $J$. It is then immediate that the connected point functions of the Gaussian field vanish for $N > 2$, since the derivatives of that order do vanish.
\subsection{The density from $N$-point functions : determinacy of the moment problem}
A key to several results of this thesis is the so-called \textit{moment problem} and its \textit{determinacy}. These are respectively the problem of finding a density given the hierarchy of $N$-point moments, and the question of whether a solution is unique or not. While not part of the usual cosmological literature, this topic is a well known area of research of mathematics, in particular for one dimensional densities \citep{Akhiezer65,Simon97theclassical}. For such one dimensional densities examples of different distributions with the same moment series have been known for more than a century \citep{Stieltjes,Heyde63}.
\newline
\newline
It is not uncommonly argued in the cosmological literature that the relation between the moments, the characteristic functional and the density function can be inverted, suggesting that the density is always uniquely set by the $N$-point moments \citep[e.g.]{1985ApJ...289...10F,mo2010galaxy}. It is important to keep in mind that this holds only when the characteristic functional can be written as a convergent power series in the moments in a region around $J = 0$. As already mentioned this is not always true, in which case the characteristic functional cannot be expressed in terms of $N$-point moments. However, it is true that the mapping between the characteristic functional and the density is one to one. The indeterminacy of the moment problem was touched upon in a cosmological context in \cite{1991MNRAS.248....1C}, though it did not attract much attention in the cosmological literature since then. 
\newline
\newline
Obviously, an indeterminate moment problem is relevant for our purposes as it implies that the entire $N$-point function hierarchy contains less information than the density itself. In that case, the entire $N$-point hierarchy is an inefficient set of observables. Namely, it is impossible to reconstruct uniquely the density from the hierarchy. The implications for cosmological parameter inference are discussed in several chapters of this thesis, notably in chapter \ref{ch4}, where to the best of our knowledge first explicit examples of densities of any dimensionality with identical $N$-point moments at all orders are presented. 
\newline
\newline
 The Gaussian field is an example of a density that can be uniquely recovered from the $N$-point moment hierarchy. On the other hand, the lognormal field, first introduced later in \ref{GNG} is an example where this is not possible.
\subsection{$N$-point functions from discrete populations, poisson samples}
In galaxy or weak lensing surveys, the fields that are observed are rather discrete than continuous. Namely positions of galaxies are recorded, and additional information such as the distortion of galaxy images can be effectively measured only on these positions. 
Discreteness adds some complexity. We are not able to predict galaxy positions in the sky, and this discrete field may not trace in an obvious manner the underlying, interesting field, typically the dark matter density field or its projection. Rather, they are only tracers that can be biased in several ways, and the measured $N$-point functions need not always be representative of that of the underlying field.
\newline
\newline
There is no unique manner to create a point process from a continuous random field, but for density fields the infinitesimal Poisson model is rather natural and gives a direct interpretation of the point functions. Within this prescription, one divides the total volume in infinitesimal cells and simply set the probability for a point in a cell to be proportional to the value of the continuous field $\phi$ at that point, and this independently from cells to cells. In that case, given a total number of $N$ objects in a total volume $V$, the probability density to find these at $x_1\cdots,x_N$ conditional on $\phi$ is by definition
\beq
p(x_1,\cdots,x_N|\phi) = \phi(x_1)\cdots \phi(x_N)  \frac{1}{\lp \bar \phi V \rp^N}.
\enq
Note that the normalisation $(\bar\phi V)^N$ must require ergodicity, in order to be able to identify $\int d^nx_i \: \phi(x_i)$ with $\bar \phi V$.
This condition implies that for any $k$ the following link between $N$-point functions must hold
\beq
\frac 1 {\bar \phi V}\int_V d^nx \av{\phi(x) \phi(x_1)\cdots \phi(x_k)} = \av{\phi(x_1) \cdots \phi(x_k)},
\enq
which is a very non trivial condition on a density function, requiring in fact the zeroth mode of the field, $\int_V d^nx \:\phi(x) = \tilde \phi(0)$ to be actually no random variable but a usual number. For a Gaussian field, this is equivalent to require the condition on the spectrum $P(0) = 0$ and thus presents no difficulty for any volume $V$. On the other hand, as we will discuss in chapter \ref{ch3}, a lognormal field never has this property fulfilled exactly in a finite volume, since its power at zero must be strictly positive.
\newline
\newline
We find the probability density $p(x_1,\cdots,x_N)$, unconditional to $\phi$, to find these objects at these positions by marginalising over the unseen underlying field $\phi$, 
\beq \label{ppoints}
\begin{split}
p(x_1,\cdots,x_N)&= \int \mathcal D \phi \: p[\phi]\: p(x_1,\cdots,x_N|\phi) \\
&=  \frac{1}{\lp \bar \phi V \rp^N} \av{\phi(x_1) \cdots\phi(x_N)}.
\end{split}
\enq
From \eqref{ppoints} we find a direct interpretation of the two-point function. From the rule of probability theory we find that the probability density of observing an object at $x_2$ given that there is one at $x_1$ is given by
\beq
p(x_2|x_1)d^nx_2 = \frac{1}{ \bar\phi^2 V} \av{\phi(x_1)\phi(x_2)}d^nx_2 = \lp 1 + \xi\lp x_2 - x_1\rp \rp \frac{d^nx_2}{V}. 
\enq
Thus for such processes the connected two-point function describes directly the clustering of the points, by enhancing or reducing this conditional probability to find particles separated by some distance \cite[e.g.]{1980lssu.book.....P,2002PhR...367....1B}.
\section{Gaussian and non-Gaussian matter density fields \label{GNG}}
In cosmology, the Gaussian field \eqref{Gfield} is fundamental. It is used routinely in order to describe the statistics of the small fluctuations present in the early Universe that we observe in the CMB radiation, or more importantly for us that of the density we observe on the largest scales. While it is possible to treat this working hypothesis as an ad-hoc assumption adopted for convenience or lack of a better prescription, it has now some theoretical support as well, in that the simplest model of inflation predict initial conditions that are extremely close to Gaussian \citep{2000cils.conf.....L}. It is fortunately possible to see where this comes about without entering any details : in such models a nearly free scalar field is responsible for the rapid expansion of the Universe, and its fluctuations give rise to the primordial deviations from homogeneity. Since the action of a free field is quadratic in the field, and since the action plays the same role in a quantum field theory as $\ln p[\phi]$ in a statistical field theory, we see that it corresponds to a Gaussian field.
\newline
\newline
Nevertheless, it should be noted that in the particular case of the matter density or fluctuation field relevant for galaxy surveys, the assumption of Gaussianity is in fact flawed from the very beginning. This is because the matter density is positive, while the Gaussian assumption assigns non-zero probability density to negative values. As long as the variance is small, this is however not an essential shortcoming of the model. 
\newline
\newline
Of course, there are many situations where non-Gaussian statistics play a major role, even in the noise-free fields. For instance, signatures from  non-Gaussianities in the primordial fluctuations can be used to try and constrain more sophisticated inflationary models.  This is often dubbed as primordial non-Gaussianity. 
In this thesis we are going to deal with the non-Gaussianity sourced from the nonlinear evolution of the density field. We already mentioned that Gaussian statistics cannot provide a perfect description of a density field, and this is even more true as nonlinear evolution take place. This is illustrated in figure \ref{figmatter}, from   A. Pillepich \citep{2010MNRAS.402..191P}. The four panels on the left show the evolution of the matter density field in a $N$-body simulation, from redshift 50 to redshift 0 downwards. The inbox inside each panel shows the one-point probability density $p(1 + \delta)$ of the matter fluctuation field as the dark line, together with that of the previous panel in grey. As the                                                                                                                                                                                                                                                                                                      fluctuations grows, one observes the field to develop large tails in the overdense regions, and a cutoff in the underdense regions. The right panels are the same simulations where an amount of primordial non Gaussianity of the local type roughly 10 times larger than current observational constraints \citep{2011ApJS..192...18K} was added, represented as the blue line in the inboxes. It is obvious that the non Gaussianity induced by the formation of structures is both very strong and completely dominant over the primordial non-Gaussianity in the late Universe.
\newline
\newline
\begin{figure}
  \includegraphics[width = 0.8\textwidth]{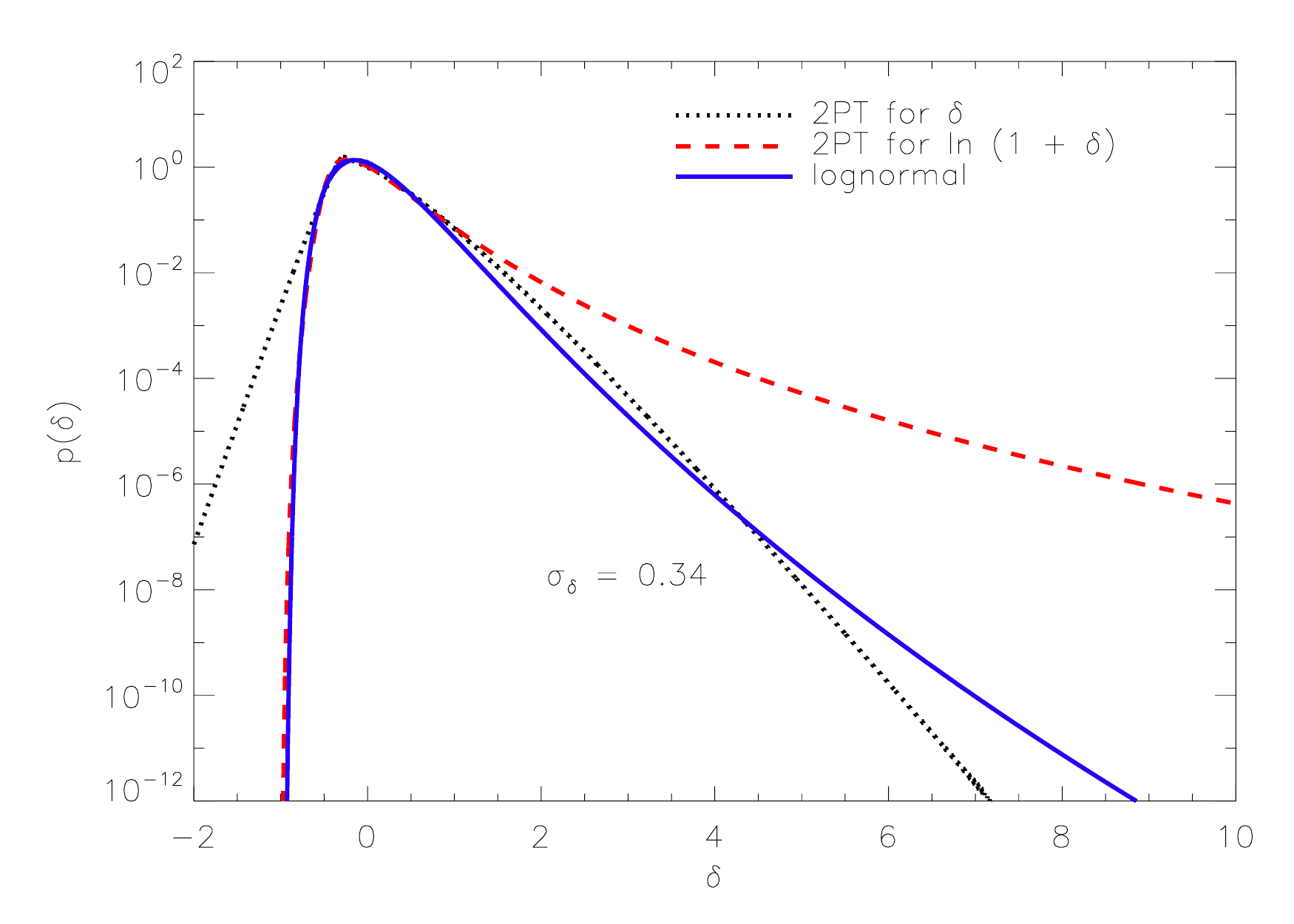}  
  \centering
  \caption[The shape of the one-point distribution function of the matter density fluctuation field according to second order perturbation theory.]{The shape of the one-point distribution function of the fluctuation field according to second order perturbation theory for $\delta$, dotted, and for $\ln (1 + \delta)$, dashed, calculated with methods exposed in \cite{2000MNRAS.314...92T}. The solid line is the lognormal distribution. \label{fig2PT}}
\end{figure}
\begin{figure}[htbp]
  \includegraphics[width = 0.38\textwidth]{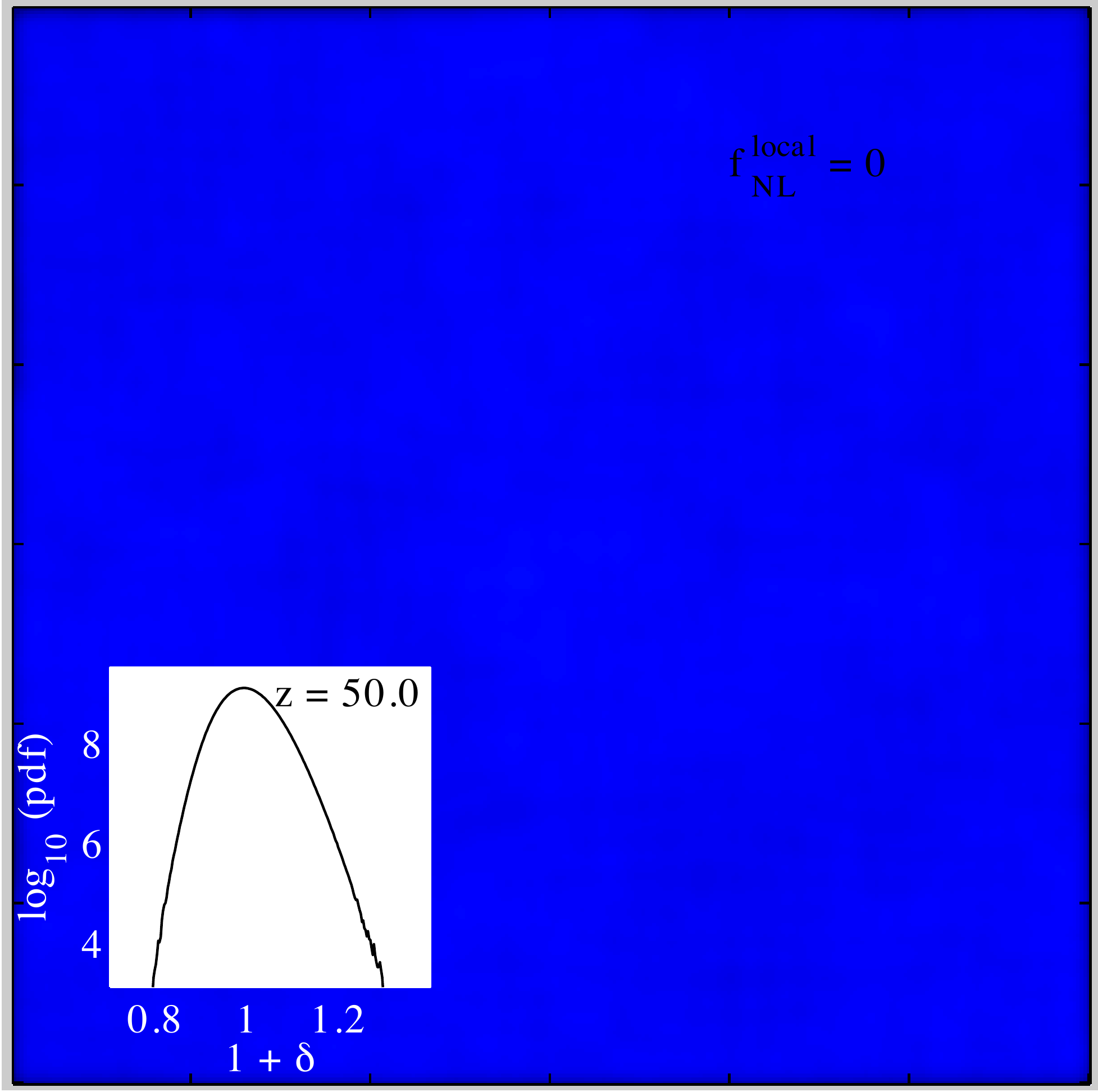}  
  \includegraphics[width = 0.38\textwidth]{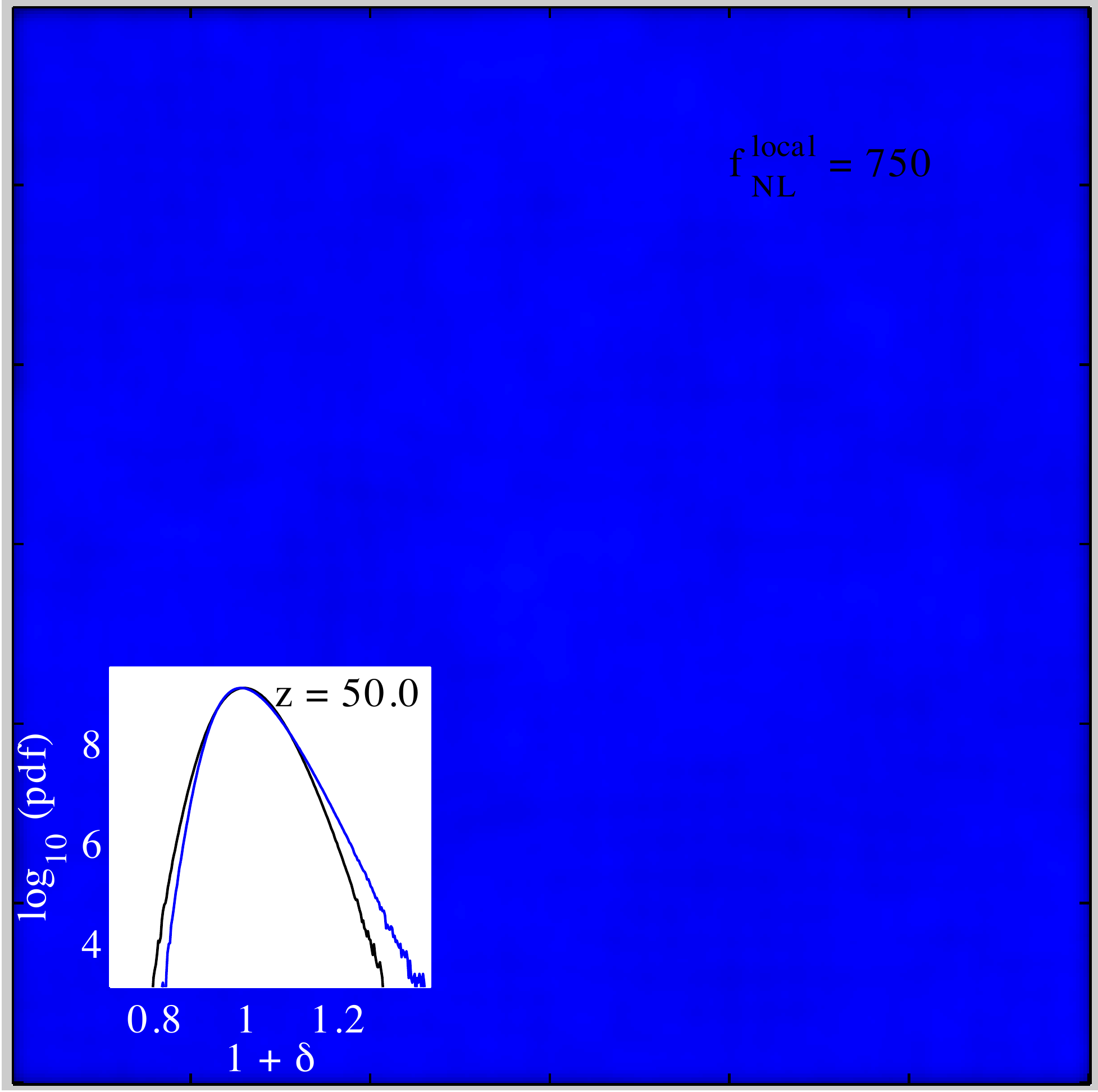}
  \includegraphics[width = 0.38\textwidth]{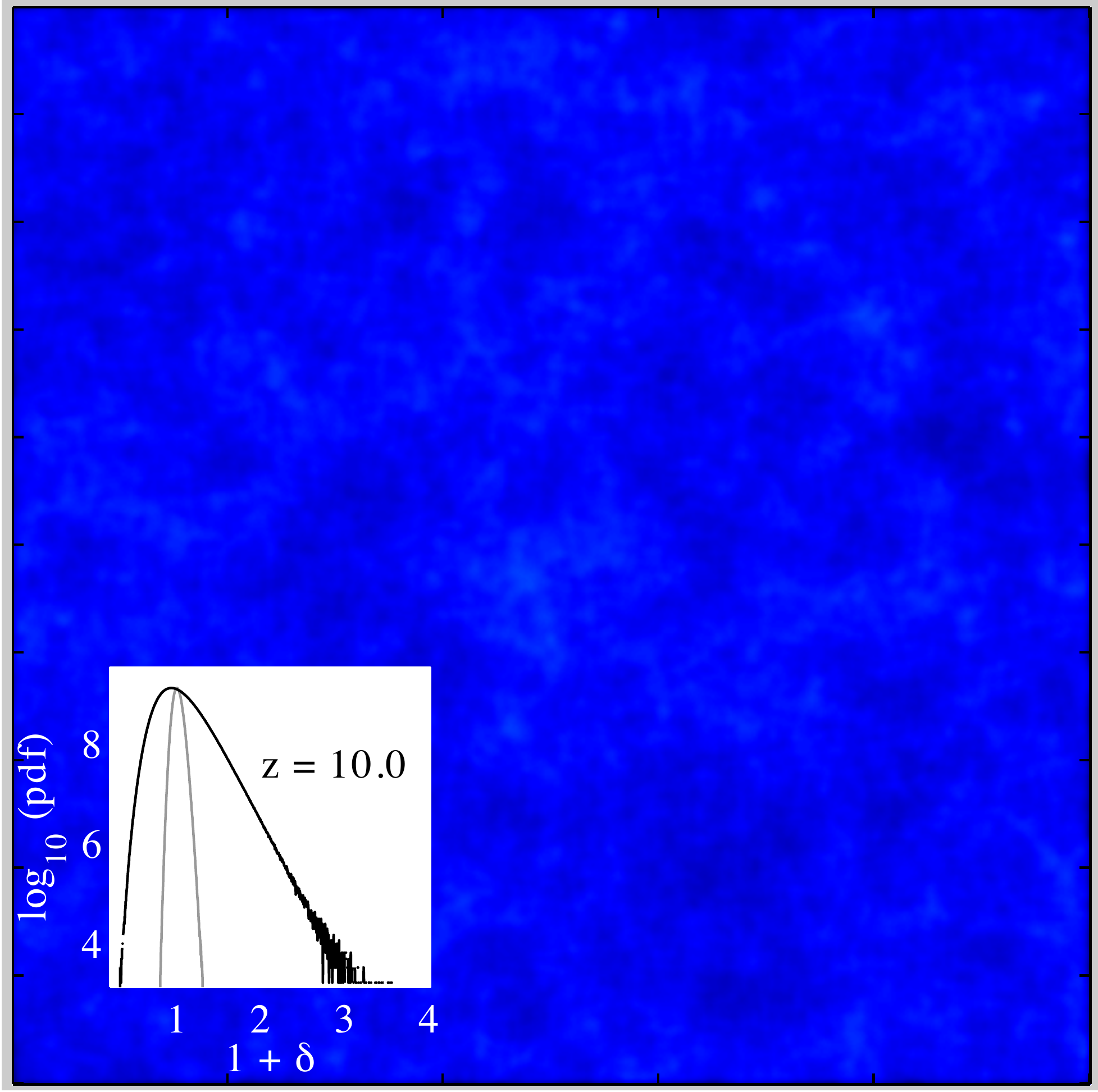}  
  \includegraphics[width = 0.38\textwidth]{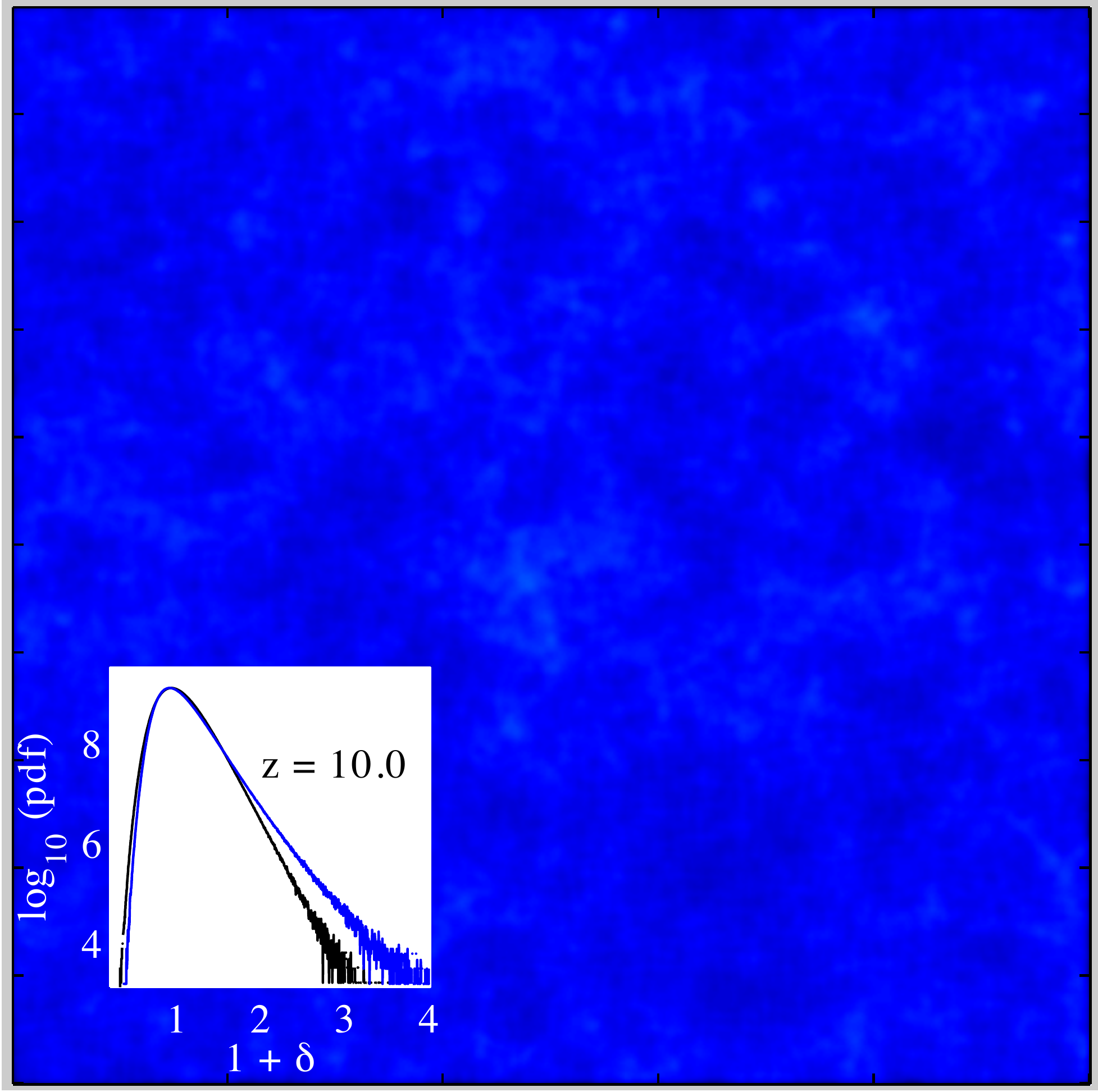}
   \includegraphics[width = 0.38\textwidth]{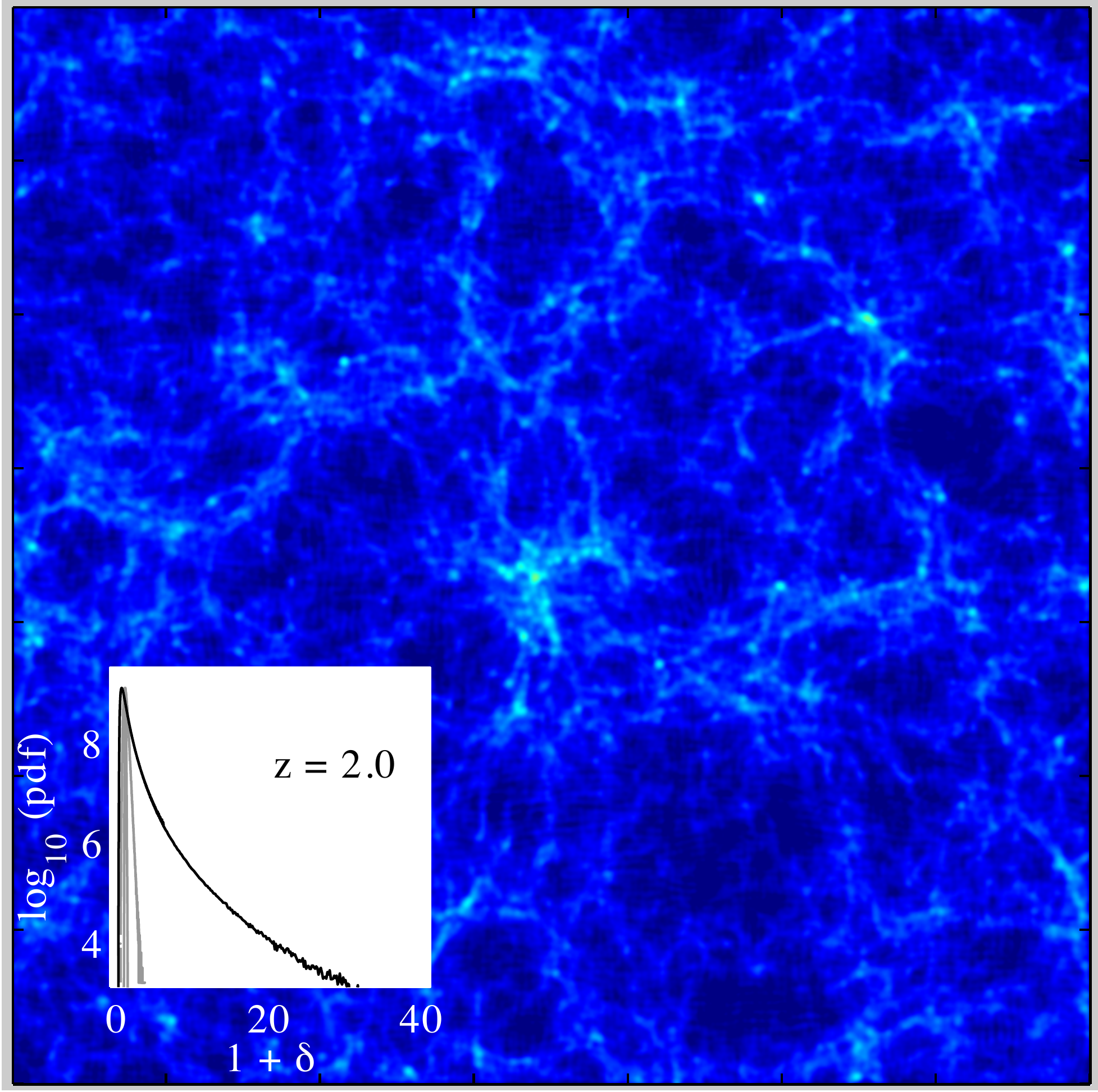}  
 \includegraphics[width = 0.38\textwidth]{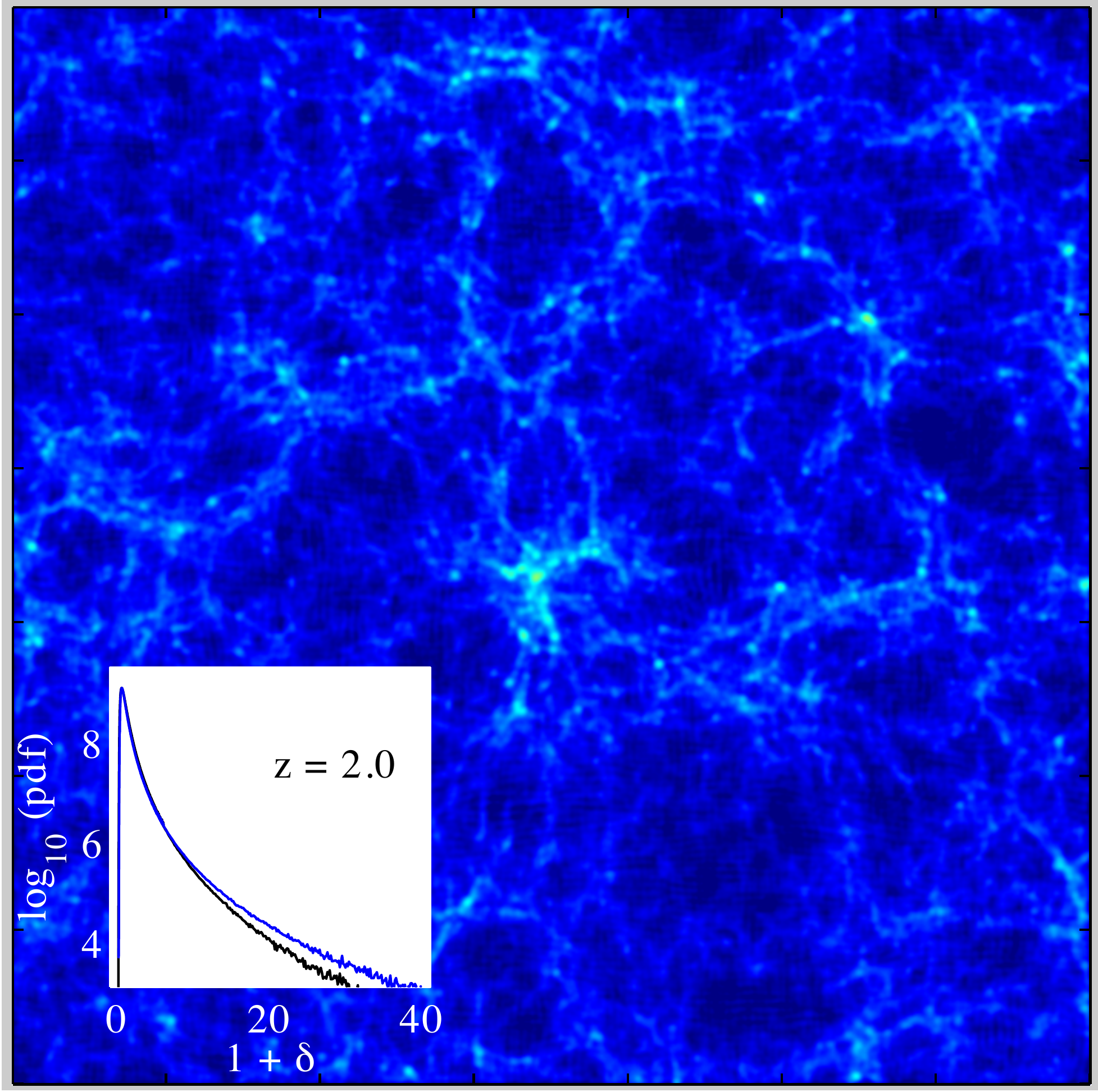}
  \includegraphics[width = 0.38\textwidth]{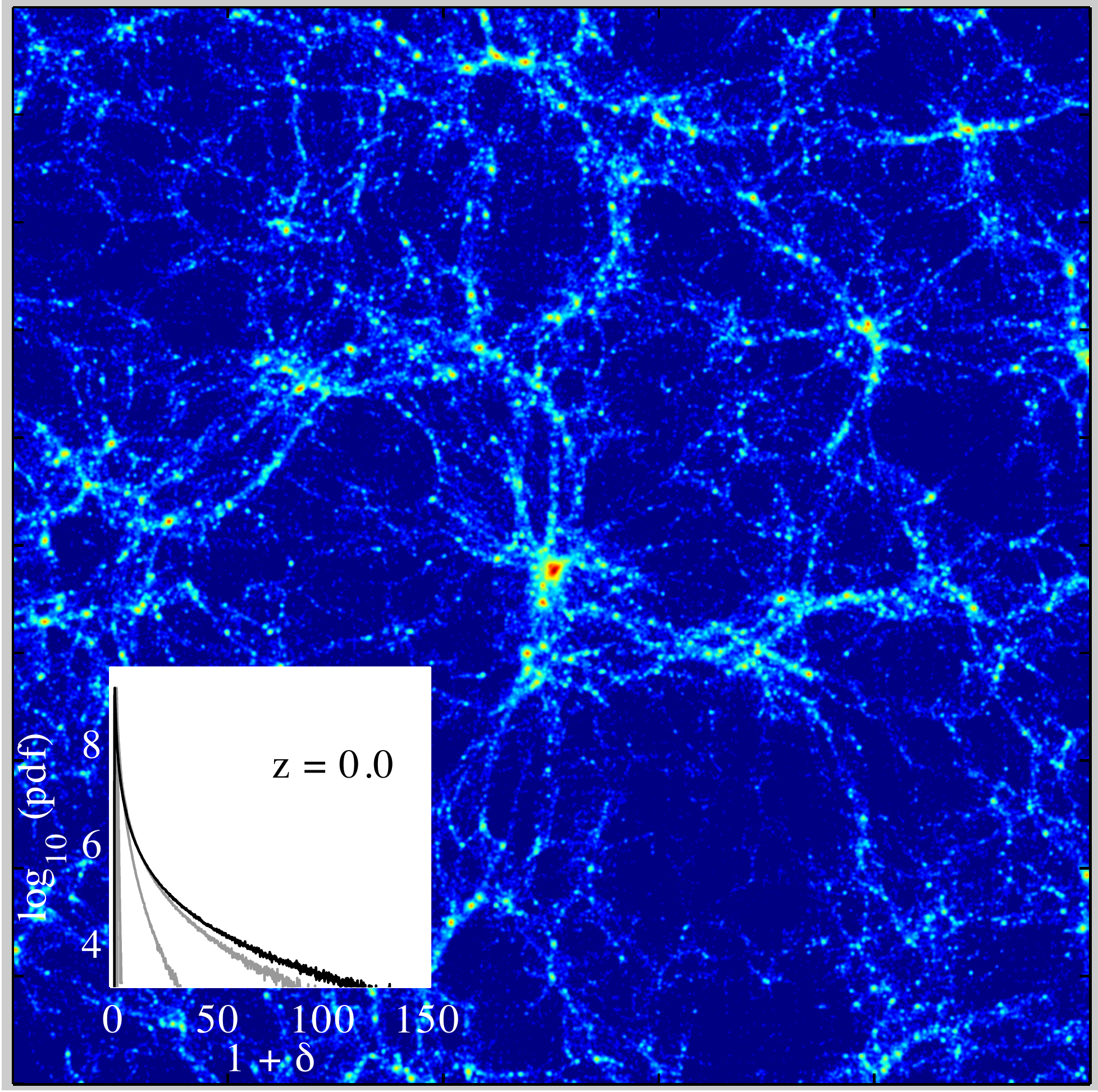}  
  \centering
 \includegraphics[width = 0.38\textwidth]{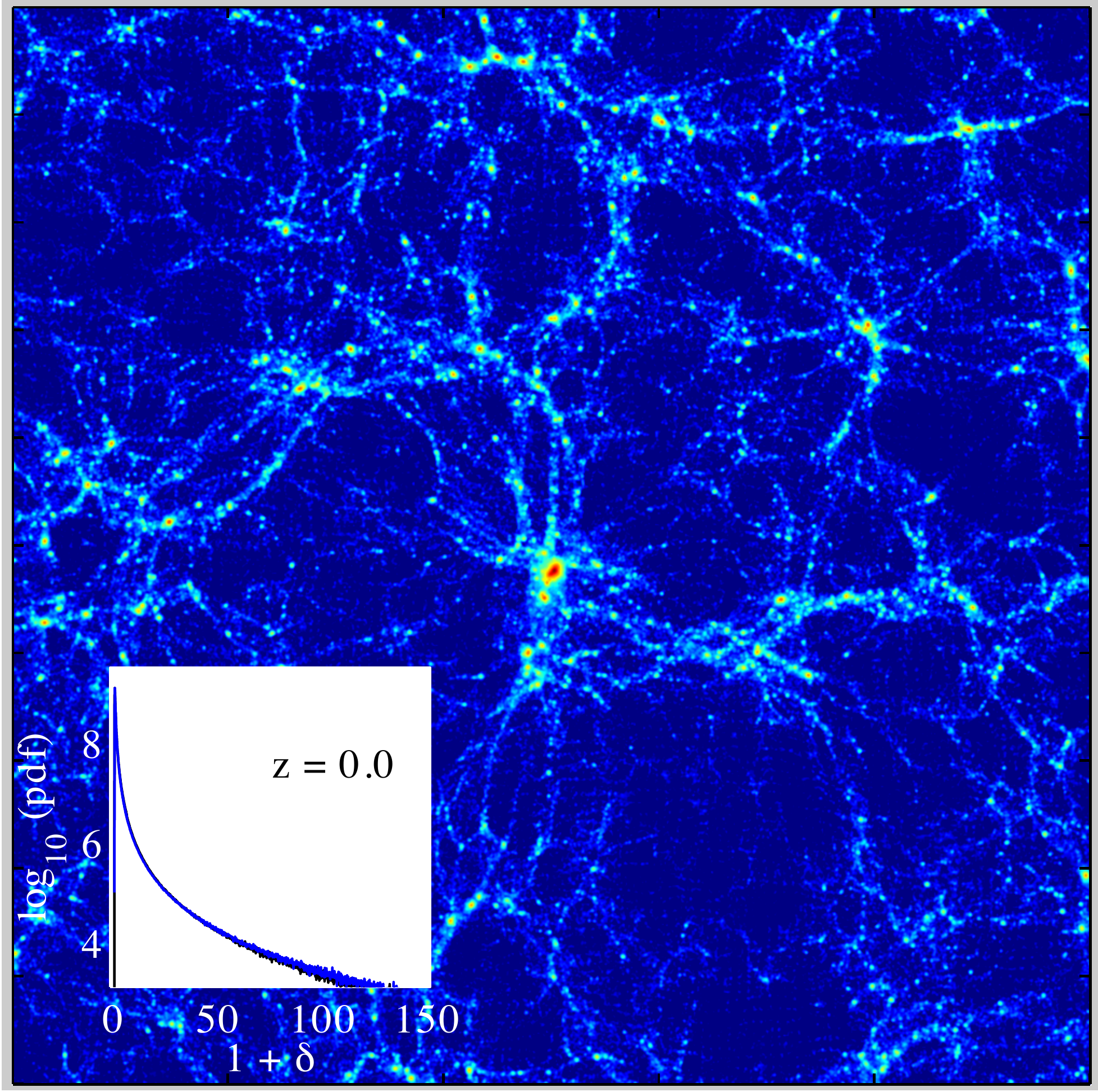}
  \caption[Evolution of the matter density field distribution with cosmic time in a N-body simulation. Image courtesy of A. Pillepich]{Evolution of the matter density field distribution with cosmic time in a N-body simulation. See text for more details. Image courtesy of A. Pillepich \citep{2010MNRAS.402..191P}. \label{figmatter}}
\end{figure}
\noindent More to the point in the case of the density field is the assumption of a lognormal field \citep{1991MNRAS.248....1C}, where essentially the logarithm $A = \ln \lp 1 + \delta \rp$ is set to be a Gaussian field.  Since $\ln \lp 1 + \delta\rp$ is very close to $\delta$ for small fluctuations, these two fields are indistinguishable for any practical purposes as long as the variance is small. The lognormal is however always positive definite, correcting for the defect of the Gaussian prescription. In the nonlinear regime, it also shows large tails and a cutoff in the underdense regions, reproducing the qualitative features the one-point distribution of the fluctuation field remarkably well given the simplicity of the prescription.
\newline
\newline
Let us illustrate these aspects  in figure \ref{fig2PT} with the help of perturbation theory. As the dotted line is shown the prediction of second order perturbation theory for the distribution function, calculated with the same methods as \cite{2000MNRAS.314...92T}, assuming Gaussian initial conditions for $\delta$, at a variance of $\sigma_\delta = 0.34$. We see that the assumption of Gaussian initial conditions gives rise to a nonsensical non zero probability density for negative matter density. On the other hand, as expected, it predicts a large tail in the overdense regions. The dashed line shows the probability density obtained from the same type of perturbative calculations for $A$, and the solid line the lognormal distribution. It is rather remarkable how both the tails and the underdense regions of the lognormal are reproduced with these two perturbative calculations. The agreement between the lognormal prescription and the matter fluctuations measured in the $N$-body simulations is also seen to be very good, see \cite{2000MNRAS.314...92T}.
\newline
\newline
We can see from the apparition of a long tail and a sharp cutoff in the density fluctuations that the statistics of the nonlinear regime are not going to obey the same rules as that of the linear Gaussian field. For instance, it is worth mentioning at this point that the traditional observable the spectrum of the field, that contains the entire information in the field in the Gaussian regime, seems to lose most of its advantages leaving the nonlinear regime. In particular, heavy correlations appear between the Fourier modes, and little information can apparently be extracted from the spectrum on these scales \citep{2005MNRAS.360L..82R,2006MNRAS.370L..66N,2008ApJ...686L...1L}. The study of the information within the lognormal field will form a large part of this thesis, chapter \ref{ch4}.
\clearpage
\section{Structure of the thesis}
The thesis is built out of two parts. The first part, that includes the first and second chapter, describes and builds upon the mathematical tools that are used in later chapters. The second part, chapter 3 to chapter 7, contains the cosmological research properly speaking.
Each chapter begins with a more detailed description of its content, as well as the references to our corresponding publications when appropriate.
\newline
\newline
In chapter 2, we discuss two measures of information, Fisher's information matrix and Shannon's information entropy, and the duality between them. We comment extensively on the information inequality, fundamental for much of this thesis. We discuss the information content of maximal entropy distributions, and identify these distributions as those for which the information inequality is an equality.
\newline
\newline
In chapter 3 we deal with the information content of $N$-point moments for a given density function. We present how to decompose the Fisher information matrix in uncorrelated components, associated to the $N$-point moments of a given order, with the help of orthogonal polynomials. The properties of this expansion are discussed. In particular, it is shown that the hierarchy of $N$-point moments does not necessarily carry the entire Fisher information content of the distribution for indeterminate moment problems, while the entire information is recovered for determinate moment problems. We also present several models for which we could obtain this expansion explicitly at all orders.
\newline
\newline
Chapter 4 uses the tools introduced in chapter 2 in the context of weak lensing. We show in this chapter how the information from different probes of the weak lensing convergence field, the magnification, shears and flexion fields, do combine in a very simple way. We then evaluate their information content using current values for the dispersion parameters and discuss the benefits of their combination.
\newline
\newline
Chapter 5 is a little note on the use of Gaussian distributions for the statistics of estimators of second order statistics in cosmology. We show that we can use these information-theoretic concepts to clarify some issues in the literature and the signification of the parameter dependence of the covariance matrices for two-point correlation functions or power spectra.
\newline
\newline
Chapter 6 discusses the information content of $N$-point moments in the lognormal density field. It is the largest chapter in this thesis, and contains our main results as well. After discussing some fundamental limitations of the lognormal field to describe the matter density field of the $\Lambda$CDM universe, we present families of different fields all having the same hierarchy of $N$-point moments than the lognormal, and discuss the implications for parameter inference.
The expansion of the Fisher information matrix introduced in chapter 2 is then performed exactly at all orders in two simplified but tractable situations. This then allows us to make successful connections with $N$-body simulation results on the extraction of power spectra
\newline
\newline
Finally, in chapter 7 we evaluate the information content of the moment hierarchy of the one-point distribution of the weak lensing convergence field, demonstrating that the nonlinearities generically lead to distributions that are very poorly described by their moments. On the other hand, it is shown that simple mappings are able to correct for this deficiency.

\part{Quantifying information \label{part1}}
\cleardoublepage

\chapter{Shannon entropy and Fisher information}
\label{ch1}


The primary aim of this first chapter is to introduce in a rather detailed and comprehensive manner the tools that form the building blocks of this thesis, which are Fisher's matrix valued measure of information, as well as Shannon's measure of entropy.
As such, unlike the subsequent chapters, it does not contain exclusively original material. Namely, the sections \ref{Setting the scene} and \ref{Jaynes}, while important parts of our publication \cite{2011MNRAS.417.1938C} can be considered to some extent a review of our perspective on the properties of Fisher information and Shannon entropy that are then built upon in the later parts of this thesis. 
\newline
\newline
This chapter is built as follows :
\newline
\newline
In section \ref{Setting the scene}, we introduce and discuss the main properties of the Fisher information, with an emphasis on its information theoretic properties. Essential to most of this thesis is the information inequality, equation \eqref{MultiCramer}, and its consequences. In section \ref{Jaynes}, we discuss maximal entropy distributions associated to a prescribed set of observables, and that these distributions are precisely those for which the information inequality is an equality.  We find with equation (\ref{fishermaxent}) a measure of information content that depends only on the constraints put on the data and the physical model, written in terms of the curvature of Shannon's entropy surface. We recover the Fisher information matrices for Gaussian fields of common use in cosmology as the special case of fields with prescribed two-point functions.
\newline
\newline
The text in these two sections is based to a large extent on the first part of  \cite{2011MNRAS.417.1938C}, with the exception of appendix \ref{appendix_unique}.



\section{Fisher information and the information inequality}
\label{Setting the scene}
We first review here a few simple points of interest that justify the interpretation of the Fisher matrix as a measure of the information content of an experiment. Let us begin by considering the case of a single measurement $X$, with different possible outcomes, or realisations, $x$, and our model has a single parameter $\alpha$. We also assume that we have knowledge, prior to the given experiment, of the probability density function $p_X(x,\alpha)$, which depends on our parameter $\alpha$, that gives the probability of observing particular realisations for each value of the model parameter.
The Fisher information, $F$, in $X$ on $\alpha$, is a non-negative scalar in this one parameter case. It is defined  in a fully general way as a sum over all realisations of the data \citep{fisher25}:
\beq \label{FisherI}
I_X(\alpha) = \av{\left( \frac{\partial \ln p_X(x,\alpha)}{\partial\alpha} \right)^2}.
\enq
Three simple but important properties of Fisher information are worth highlighting at this point.
\newline
\beit
\item The first is that $I_X(\alpha)$ is positive definite, and it vanishes if and only if the parameter $\alpha$ does not impact the data, i.e. if the derivative of $p_X(x,\alpha)$ with respect to $\alpha$ is zero for every realisation $x$.

\item The second point is that it is invariant to invertible manipulations of the observed data. This can be seen by considering an invertible change of variable $y = f(x)$, which, due to the rules of probability theory can be expressed as 
\beq
p_Y(y,\alpha) = p_X(x,\alpha)\left |\frac {dx}{dy}\right|.
\enq
Thus
\beq
\frac{\partial \ln p_Y(y,\alpha)}{\partial \alpha} = \frac{\partial \ln p_X(x,\alpha)}{\partial \alpha},
\enq 
leading to the simple equivalence that $I_X(\alpha) = I_Y(\alpha)$. On the other hand, information may be lost when the transformation is not unique in both directions. For instance, if the data is combined to produce a new variable that could arise from different sets of data points.
This is only the statement that manipulations of the data leads, at best, only to conservation of the information.\newline

\item The third point is that information from independent experiments add together. Indeed, if two experiments with data $X$ and $Y$ are independent, then the joint probability density factorises,
\beq
p_{XY}(x,y) = p_X(x)p_Y(y),
\enq
and it is easy to show that the joint information in the observations decouples,
\beq
I_{XY}(\alpha) = I_X(\alpha) + I_Y(\alpha).
\enq
\enit 
These three properties satisfy what we might intuitively expect from a mathematical implementation of an abstract concept such as information. However, we can ask the reverse question and try to find an alternative measure of information that may be better suited, for a particular purpose, than the Fisher information.  In appendix \ref{appendix_unique}, we discuss to what extent the information measure in \eqref{FisherI} is in fact uniquely set by the these requirements above.
\newline
\newline
These properties are making the Fisher information a meaningful measure of information. This is independent of its interpretation as providing error bars on parameters. It further implies that once a physical model is specified with a given set of parameters, a given experiment has a definite information content that can only decrease with data processing.
\subsection{The case of a single observable}\label{single observable}
To quantify the last point above, and in order to get an understanding of the structure of the information in a data set, we first discuss a simple situation, common in cosmology, where the extraction of the model parameter $\alpha$ from the data goes through the intermediate step of estimating a particular observable, $D$, from the data, $x$, with the help of which $\alpha$  will be inferred. A typical example could be, from the temperature map of the  CMB ($x$), the measurement of the power spectra of the fluctuations ($D$), from which a cosmological parameter ($\alpha$) is extracted.
The observable $D$ is measured from $x$ with the help of an estimator, that we call $\hat D$, and that we will take as unbiased. This means that its mean value, as would be obtained for instance if many realizations of the data were available, converges to the actual value that we want to compare with the model prediction,
\beq
\lat \hat D\rat = D(\alpha).
\enq
A measure for its deviations from sample to sample, or the uncertainty in the actual measurement, is then given by the variance of $\hat D$, defined as
\beq \label{variance}
\Var{\hat D} = \av{\hat D^2} - \av{\hat D}^2.
\enq
In such a situation, a major role is played by the so-called Cram\'{e}r-Rao 
 inequality (\cite{Rao}),  that links the Fisher information content of the data to the variance of the estimator, stating that
\beq \label{CR}
\textrm{Var}(\hat D) I_X(\alpha) \ge \left(\frac{\partial D(\alpha)}{\partial \alpha}\right)^2.
\enq
This equation holds for any such estimator $\hat D$ and any model parameter $\alpha$. Two different interpretations of this equation are possible: \newline

\indent The first bounds the variance of $\hat D$ by the inverse of the Fisher information. To see this, we consider the special case of the model parameter $\alpha$ being $D$ itself. Although we are making in general a conceptual distinction between the observable $D$ and the model parameter $\alpha$, nothing requires us from doing so. Since $\alpha$ is now equal to $D$, the derivative on the right hand side becomes unity, and one obtains
\beq \label{CRB}
\textrm{Var}(\hat D) \ge \frac 1 {I_X(D)}.
\enq
The variance of any unbiased estimator $\hat D$ of  $D$ is therefore bounded by the inverse of the amount of information $I_X(D)$ the data possess on $D$. If $I_X(D)$ is known it gives a useful lower limit on the error bars that the analysis of the data can put on this observable. We emphasise at this point that this bound only holds in the case of unbiased estimators. There are cosmologically relevant situations where biased estimators can go beyond this level, and thus perform better according to the minimal squared error criterion than any unbiased one.
\newline
\newline
The second reading of the Cram\'{e}r-Rao 
 inequality, closer in spirit to the present thesis, is to look at how information is lost by constructing the observable $D$, and discarding the rest of the data set. For this, we rewrite trivially equation (\ref{CR}) as
\beq \label{CramerRao}
I_X(\alpha) \ge  \left(\frac{\partial D}{\partial \alpha}\right)^{2} \frac 1 {\textrm{Var}(\hat D)}.
\enq
The expression on the right hand side is the ratio of the sensitivity of the observable to the model parameter $\lp \frac{ \partial D}{\partial \alpha}\rp^2 $, to the accuracy with which the observable can be extracted from the data, $\textrm{Var}(\hat D)$. One of the conceivable approaches in order to estimate the true value of the parameter $\alpha$, is to perform a $\chi^2$ fit to the measured value of $D$ . It is simple to show that this ratio, evaluated at the best fit value, is in fact proportional to the expected value of the curvature of $\chi^2(\alpha)$ at this value. Since the curvature of the $\chi^2$ surface describes how fast the value of the $\chi^2$ is increasing when moving away from the best fit value, its inverse may be interpreted as an approximation to the error estimate that the analysis with the help of  $\hat D$ will put on $\alpha$. 
\newline
\newline
Thus, equation (\ref{CramerRao}) shows that by only considering $D$ and not the full data set, we may have lost information on $\alpha$, a loss given by the difference between the left and right hand side of that equation. While the latter may be interpreted as the information on $\alpha$ contained in the part of the data represented by $D$, we may have lost trace of any other source of information.
\newline
\newline
It should be noted that while we have just chosen to interpret the right hand side of \eqref{CramerRao} as the information in $\hat D$, this is a slight abuse of terminology. More rigorously, the Fisher information in $\hat D$ is not the right hand side of \eqref{CramerRao} but the Fisher information of its density function
\beq
I_{\hat D}(\alpha) = \av{\lp \frac{\partial \ln p_{\hat D}}{\partial \alpha} \rp^2},\quad  p_{\hat D}(\hat D) = \int dx\: p_X(x) \delta^D(\hat D - \hat D(x)),
\enq
but it will always be clear in this thesis from the context which one is meant. Anticipating the nomenclature of chapter \ref{ch2}, the right hand side of \eqref{CramerRao} is actually the information in the mean of $\hat D$. From equation \eqref{CramerRao} we infer
\beq
I_{\hat D}(\alpha) \ge \left(\frac{\partial D}{\partial \alpha}\right)^{2} \frac 1 {\textrm{Var}(\hat D)}.
\enq

\subsection{The general case} \label{CramerRaoMulti}
These considerations on the Cram\'{e}r-Rao 
 bound can be easily generalised to the case of many parameters and many estimators of as many observables. Still dealing with a measurement $X$ with outcomes $x$, we want to estimate a set of parameters
\beq
\boldsymbol \theta = (\alpha, \beta,\cdots)
\enq
 with the help of some vector of observables,
\beq
\mathbf D = (D_1,\cdots,D_n)
\enq
that are extracted from $x$ with the help of an array of unbiased estimators,
\beq
\mathbf {\hat D} = \bem \hat D_1,\cdots, \hat D_n \enm,\quad \av{\mathbf{\hat D}} = \mathbf D
\enq
In this multidimensional setting, all the three scalar quantities that played a role in our discussion in section \ref{single observable}, i.e. the variance of the estimator, the derivative of the observable with respect to the parameter, and the Fisher information, are now matrices.
\newline
\newline
The Fisher information $F$ in $X$ on the parameters $\boldsymbol \theta$ is defined as the square matrix
\beq \label{FisherII}
\lb F^X\lp \boldsymbol \theta \rp \rb_{\alpha\beta} = \av{ \ \frac {\partial \ln p_X}{\partial \alpha}\frac{\partial\ln p_X}{\partial \beta}}.
\enq
While the diagonal elements $F^X_{\alpha \alpha}$ are the information scalars $I_X(\alpha)$ in equation (\ref{FisherI}), the off diagonal ones describe correlated information.
The Fisher information matrix still carries the three properties we discussed in section \ref{Setting the scene}.
\newline
\newline
The variance of the estimator in equation (\ref{variance}) now becomes the covariance matrix $\textrm{cov}(\mathbf {\hat D})$ of the estimators $\mathbf {\hat D}$, defined as
\beq
\textrm{cov} \lp \mathbf {\hat D} \rp _{ij} = \av{\hat D_i \hat D_j} - D_iD_j.
\enq
Finally, the derivative of the observable with respect to the parameter, in the right hand side of (\ref{CR}), becomes a matrix $\Delta$, in general rectangular, defined as
\beq
\Delta_{\alpha \:i} = \frac{\partial D_i}{\partial \alpha},
\enq
where $\alpha$ runs over all elements of the set $\btheta$ of model parameters.
Again, the Cram\'{e}r-Rao 
 inequality provides a useful link between these three matrices, and again there are two approaches to that equation : first, as usually presented in the literature \citep{Rao}, in the form of a lower bound to the covariance matrix of the estimators,
\beq \label{CRmulti}
\textrm{cov} \left( \mathbf {\hat D} \right) \ge \Delta^T \lb  F^X\lp \boldsymbol \theta \rp \rb ^{-1}  \Delta.
\enq
The inequality between two symmetric matrices $A \ge B$ having the meaning that the matrix $A - B$ is positive definite. \footnote{A symmetric matrix $A$ is called positive definite when for any vector $x$ holds that $x^TAx \ge 0$. Further we say for such matrices that $A$  is larger than $B$, or $A> B$, whenever $A-B >  0$. A concrete implication for our purposes is e.g. that the diagonal entries of the left hand side of (\ref{CRmulti}) or (\ref{covF}), which are the individual variances of each estimator $\hat D_i$, are greater than those of the right hand side.   For many more properties of positive definite matrices, see for instance \citep{Bhatia}}. If, as above, we consider the special case of identifying the parameters with the observables themselves, the matrix $\Delta$ is the identity matrix, and so we obtain that the covariance of the vector of the estimators is bounded by the inverse of the amount of Fisher information that there is on the observables in the data,
\beq \label{covF}
\textrm{cov}(\mathbf {\hat D}) \ge \lb F^X(\mathbf D)\rb^{-1}.
\enq
\newline
 Second, we can turn this lower bound on the covariance to a lower bound on the amount of information in the data set as well. By rearranging equation (\ref{CRmulti}), we obtain the multidimensional analogue of equation (\ref{CramerRao}), the information inequality,  which describes the loss of information that occurs when the data is reduced to a set of estimators,
\beq \label{MultiCramer}
 F^X\lp \boldsymbol \theta \rp \ge \Delta \lb \textrm {cov} \lp \mathbf {\hat D }\rp\rb^{-1} \Delta^{T}.
\enq
This information inequality is a central piece to much of this thesis. A proof can be found in the appendix. A maybe simpler proof follows also for instance from the discussion in the appendix of chapter \ref{ch2}.
\newline
\newline
Instead of giving a useful lower bound to the covariance of the estimator as in the Cram\'{e}r-Rao inequality, equation (\ref{CRmulti}), the information inequality makes clear how information is in general lost when reducing the data to any particular set of estimators. The right hand side may be seen, as before, as the expected curvature of a $\chi^2$ fit to the estimates produced by the estimators $\mathbf {\hat D}$, when evaluated at the best fit value, with all correlations fully and consistently taken into account. Note that as before the right hand side of the information inequality is not the Fisher information content of the joint probability density function of the estimators, but only that of their means.
 \newline
 \newline
 In section \ref{Jaynes}, we discuss how Jaynes' Maximal Entropy Principle allow us to understand the total information content of a data set, once a model is specified, in very similar terms.
  \newcommand{\cov}[1]{\textrm{cov}\lp #1\rp}

 \subsection{Resolving the density function: Fisher information density}
 Due to its generality, the information inequality \eqref{MultiCramer} is very powerful. We now have a deeper look at a special case that sheds some light on the definition of the Fisher information matrix \eqref{FisherII}, and that we will use in part \ref{part2}.
 \newline\newline
 Assuming that the variable $x$ is continuous and one dimensional, pick a set of points $x_i$ with separation $d x$ covering some range $A$ of the variable, such that in the limit of a large number of points we can write
 \beq
 \sum_i p_X(x_i) dx \rightarrow \int_A dx\: p_X(x),
 \enq
Generalisation to discrete variables or multidimensional cases will be obvious.
 \newline
 Define then a set of estimator $\hat D_i$ as follows
 \beq
 \hat D_i(x) := \begin{cases} 1, \quad x &\in (x_i,x_i+dx) \\ 0 ,\quad x &\notin (x_i,x_i+dx)\ \end{cases}.
 \enq  
 These estimators simply build an histogram of the variable over $A$. In other words, our set of estimators are defined such the entire density function is resolved over $A$.
 \newline
 \newline
 We want to evaluate the information inequality for this set of estimators. We have
 \beq
 \av{\hat D_i}  =p_X(x_i)dx,
 \enq
 with covariance matrix
 \beq
 \cov {\mathbf{\hat D}}_{ij}=  \delta_{ij} p_X(x_i)dx - dx\:p_X(x_i)\:dx\:p_X(x_j).
 \enq
 Define as $p_X^-$ the probability that a realisation of the variable does not belong to $A$ :
 \beq
 p_X^- := \int_{\mathbb R /A}dx\:p_X(x). 
 \enq
 It is then easily seen that the inverse covariance matrix is
 \beq
 \lb \cov {\mathbf{\hat D}}^{-1} \rb_{ij} = \frac{\delta_{ij}}{p_X(x_i) dx} + \frac{1}{p_X^{-}}.
 \enq
 It follows that the right hand side of the information inequality becomes
 \beq
 \sum_{i,j}\lp dx \rp^2 \frac{\partial p_X(x_i)}{\partial \alpha}\lb \cov {\mathbf{\hat D}}^{-1} \rb_{ij} \frac{\partial p_X(x_j)}{\partial \beta} = \int_A dx\:\frac{1}{p_X(x)} \frac{\partial p_X(x)}{\partial \alpha}\frac{\partial p_X(x)}{\partial \beta} + \frac 1 {p_X^{-}}\frac{\partial p_X^{-}}{\partial \alpha}\frac{\partial p_X^{-}}{\partial \beta}.
 \enq
 This is nothing else than
 \beq \label{density}
  \int_A dx\:p_X(x) \frac{\partial \ln p_X(x)}{\partial \alpha}\frac{\partial \ln p_X(x)}{\partial \beta} + p_X^{-}\frac{\partial \ln p_X^{-}}{\partial \alpha}\frac{\partial \ln p_X^{-}}{\partial \beta}.
 \enq
 If $A$ covers the full range of the variable, then the first term is precisely the total Fisher information matrix, and the second vanishes, since $p^-_X = 0$. It is clear from \eqref{density} that we can interpret
 \beq
 p_X(x) \frac{\partial \ln p_X(x)}{\partial \alpha}\frac{\partial \ln p_X(x)}{\partial \beta} = \frac{1}{p_X(x)} \frac{\partial p_X(x)}{\partial \alpha}\frac{\partial p_X(x)}{\partial \beta}
 \enq
 as a Fisher information density, representing information from observations of the variables around $x$. The additional term in \eqref{density} involving $p^-_X$ originates from the fact the the density is normalised to unity : observations of the density over the range $A$ provides some information on the density in the complement to $A$,
 \beq
 p^-_X = \int_{\mathbb R /A}dx\:p_X(x) = 1 - \int_A dx \: p_X(x).
 \enq
 However, it is not possible to resolve the individual contributions of $p_X(x)$ to $p^{-}_X$ for each $x$ on the complement of $A$, and thus the derivatives act in this case outside of the integrals, unlike the first term in  \eqref{density}.
\section{Jaynes Maximal Entropy Principle \label{Jaynes}}
\label{The structure of the information in a physical model}
In cosmology,  the knowledge of the probability distribution  of the data as function of the parameters, $p_X(x,\btheta)$,  which is compulsory in order to evaluate its Fisher information content, is usually very limited. In a galaxy survey, a data outcome $x$ would be typically the full set of angular positions of the galaxies, together with some redshift estimation if available, to which we may add any other kind of information, such as luminosities, shapes, etc.  Our ignorance of both initial conditions and of many relevant physical processes does not allow us to predict either galaxy positions in the sky, or all interconnections with all this additional information.  Our predictions of the shape of $p_X$ is thus limited to some statistical properties, that are sensitive to the model parameters $\btheta$, such as the mean density over some large volume, or certain types of correlation functions.\newline\newline
In fact, even if it were possible to devise some procedure in order to get the exact form of $p_X$, it may eventually turn out to be useless, or even undesirable, to do so. The incredibly large number of degrees of freedom of such a function is very likely to overwhelm the analyst with a mass of irrelevant details,  which may have no relevant significance on their own, or improve the analysis in any meaningful way.
\newline 
\newline
These arguments call for a kind a thermodynamical approach, which would try and capture those aspects of the data which are relevant to our purposes, reducing the number of degrees of freedom in a drastic way. Such an approach already exists in the field of probability theory \citep{jaynes57}. It is based on Shannon's concept of entropy of a probability distribution \citep{shannon48} and did shed new light on the connection between probability theory and statistical mechanics.
\newline
\newline
As we have just argued, our predictive knowledge of $p_X(x,\btheta)$ is limited to some statistical properties. Let us formalise this mathematically, in a similar way as in section \ref{CramerRaoMulti}.  Astrophysical theory gives us a set of constraints on the shape of $p_X$, in the form of averages of some functions $o_i$,
\beq \label{constraints} 
O_i(\btheta) = \lat o_i(x) \rat (\btheta), \quad i = 1,\cdots,n.
\enq
where $p_X$ enters through the angle brackets. As an example, suppose the data outcome $x$ is a map of the matter density field as a function of position. In this case, one of these constraints  $O_i$  could be the mean of the field or its power spectrum, as given by some cosmological model.
\newline
\newline
The role of this array $\mathbf O = (O_1,\cdots,O_n)$ is to represent faithfully the physical understanding we have of $p_X$, according to the model, as a function of the model parameters $\btheta$.  In the ideal case, some way can be devised to extract each one of these quantities $O_i$ from the data and to confront them to theory. The set of observables $\mathbf D$, that we used in section \ref{CramerRaoMulti},  would be a subset of these predictions $\mathbf O$, and we henceforth refer to $\mathbf O $ as the 'constraints'.
\newline
\newline
Although $p_X$ must satisfy the constraints (\ref{constraints}), there may still be a very large number of different distributions compatible with these. However, a very special status among these distributions has the one which maximises the value of Shannon's entropy\footnote{Formally, for continuous distributions the reference to another distribution is needed to render S invariant with respect to invertible transformations, leading to the concept of the entropy of $p_X$ relative to another distribution $q_X$, $S = \int dx \:p_X(x)\ln \frac {p_X(x)}{q_X(x)}$, also called Kullback-Leibler divergence. The quantity defined in the text is more precisely the entropy of $p_X(x)$ relative to a uniform probability density function. For an recent account on this, close in spirit to this work, see \cite{caticha08}.}, defined as
\beq\label{Entropy}
S = - \int dx \:p_X(x,\btheta)\ln p_X(x,\btheta).
\enq
First introduced by Shannon \citep{shannon48} as a  measure of the uncertainty in a distribution on the actual outcome, Shannon's entropy is now the cornerstone of information theory. Jaynes' Maximal Entropy Principle states that the $p_X$  for which this measure $S$ is maximal is the one that best deals with our insufficient knowledge of the distribution, and should be therefore preferred. We refer the reader to Jaynes' work \citep{jaynes83,jaynes2003} and to \cite{caticha08} for detailed discussions of the role of entropy in probability theory and for the conceptual basis of maximal entropy methods. Astronomical applications related to some extent to Jaynes's ideas include image reconstruction from noisy data, (see e.g. \cite{1984MNRAS.211..111S,1996VA.....40..563S,2004MNRAS.347..339M} and references therein) , mass profiles reconstruction from shear estimates \citep{1998MNRAS.299..895B,2002MNRAS.335.1037M}, as well as model comparison when very few data is available \citep{2007MNRAS.380..865Z}. We will see that for our purposes as well it provides us a powerful tool, and that the Maximal Entropy Principle is the ideal complement to Fisher information, fitting very well within our discussions in section \ref{Setting the scene} on the information inequality.\newline
\newline
Intuitively, the entropy $S$ of $p_X$ tells us how sharply constrained the possible outcomes $x$ are, and Jaynes' Maximal Entropy Principle selects the $p_X$ which is as wide as possible, but at the same time consistent with the constraints (\ref{constraints}) that we put on it. 
The actual maximal value attained by the entropy $S$, among all the possible distributions which satisfy (\ref{constraints}), is a function of the constraints $\mathbf O$, which we denote by
\beq \label{EntropyII}
S(O_1,\cdots,O_n).
\enq
Of course it is a function of the model parameters $\btheta$ as well, since they enter the constraints.  As we will see, the shape of that surface as a function of $\mathbf O$, and thus implicitly as a function of $\btheta$, is the key point in understanding the Fisher information content of the data. In the following, in order to keep the notation simple, we will omit the dependency on $\btheta$ of most of our expressions, though it will always be implicit.
\newline
\newline
The problem of finding the distribution $p_X$ that maximises the entropy (\ref{Entropy}), while satisfying the set of constraints (\ref{constraints}), is an optimization exercise. We can quote the end result \cite[chap. 11]{jaynes83},\cite[chap. 4]{caticha08}: \newline
The probability density function $p_X$, when it exists, has the following exponential form,
\beq \label{MaxEnt}
p_X(x) = \frac 1 {Z} \exp\left(-\sum_{i = 1}^{n}\lambda_i o_i(x)\right),
\enq
in which to each constraint $O_i$ is associated a conjugate quantity $\lambda_i$, that arises formally as a Lagrange multiplier in this optimization problem with constraints. The conjugate variables $\lambda$'s are also called  'potentials', terminology that we will adopt in the following. We will see below in equation (\ref{potentials}) that the potentials have a clear interpretation, in the sense that the each potential $\lambda_i$ quantifies how sensitive is the entropy function $S$ in (\ref{EntropyII}) to its associated constraint $O_i$. The quantity $Z$, that plays the role of the normalisation factor, is called the partition function. Since equation (\ref{MaxEnt}) must integrate to unity, the explicit form of the partition function is 
\beq \label{partitionfunction}
Z(\lambda_1,\cdots,\lambda_n) = \int dx \: \exp\left(-\sum_{i=1}^{n}\lambda_i o_i(x)\right).
\enq
The actual values of the potentials are set by the constraints (\ref{constraints}). They reduce namely, in terms of the partition function, to a system of equations to solve for the potentials,
\beq \label{lambda}
 O_i = -\frac{\partial}{\partial \lambda_i}\ln Z,\quad i = 1,\cdots,n.
\enq
The partition function $Z$ is closely related to the entropy $S$ of $p_X$. It is simple to show that the following relation holds,
\beq \label{legendre}
S = \ln Z + \sum_{i = 1}^n \lambda_iO_i,
\enq
and the values of the potentials can be explicitly written as function of the entropy, in a relation mirroring equation (\ref{lambda}),
\beq \label{potentials}
 \lambda_i =  \frac{\partial S}{\partial O_i}, \quad i = 1,\cdots,n
\enq
Given the nomenclature, it is of no surprise that a deep analogy between this formalism and statistical physics does exist. Just as the entropy, or partition function, of a physical system determines the physics of the system, the statistical properties of these maximal entropy distributions follow from the functional form of the Shannon entropy or its partition function as a function of the constraints. For instance, the covariance matrix of the constraints is given by
\beq \label{fluctuations}
\lat \left(o_i(x) - O_i\right)\left(o_j(x) - O_j\right)\rat = \frac{\partial^2\ln Z}{\partial \lambda_i\partial\lambda_j} 
\enq
In statistical physics the constraints can be the mean energy, the volume or the mean particle number, with potentials being the temperature, the pressure and the chemical potential. We refer to \cite{jaynes57} for the connection to the physical concept of entropy in thermodynamics and statistical physics.
\subsection{Information in maximal entropy distributions} \label{combinations}
With our choice of probabilities $p_X$ given by equation (\ref{MaxEnt}), the amount of Fisher information on the parameters $\btheta = (\alpha,\beta,\cdots)$ of the model can be evaluated in a straightforward way. The dependence on the model goes through the constraints, or, equivalently, through their associated potentials. It holds therefore that
\beq \begin{split}
\frac{ \partial \ln p_X(x)}{\partial \alpha}& =-\frac {\partial \ln Z}{\partial \alpha} -\sum_{i= 1}^n\frac{\partial\lambda_i}{\partial \alpha} o_i(x)
\\ &= \sum_{i= 1}^n \frac{\partial\lambda_i}{\partial \alpha}\left[O_i - o_i(x)\right],
\end{split}
\enq
where the second line follows from the first after application of the chain rule and equation (\ref{lambda}).
Using the covariance matrix of the constraints given in (\ref{fluctuations}), the Fisher information matrix, defined in (\ref{FisherII}), can then be written as a double sum over the potentials,
\beq
\begin{split}\label{Zform}
F^X_{\alpha\beta} &= \sum_{i,j = 1}^n\frac{\partial\lambda_i}{\partial \alpha}\frac{\partial^2\ln Z}{\partial \lambda_i\partial\lambda_j} \frac{\partial\lambda_j}{\partial \beta}.
\end{split}
\enq
There are several ways to rewrite this expression as a function of the constraints and/or their potentials. First, it can be written as a single sum by using equation (\ref{lambda}) as
\beq \label{sumform}
F^X_{\alpha\beta} = -\sum_{i = 1}^n\frac{\partial\lambda_i}{\partial\alpha}\frac{\partial O_i}{\partial \beta}.
\enq
Alternatively, since we will be more interested in using the constraints as the main variables, and not the potentials,  we can show, using equation (\ref{potentials}), that it also takes the form \footnote{We note that this result is valid only for maximal entropy distributions and is not equivalent to the second derivative of the entropy with respect to the parameters themselves. However it is formally identical to the corresponding expression for the information content of distributions within the exponential family \citep{Jennrich75}, or \citep[chapter 4]{vandenbos07}, once the curvature of the entropy surface is identified with the generalized inverse of the covariance matrix.}
\beq \label{fishermaxent}
F^X_{\alpha\beta} = -\sum_{i,j = 1}^n\frac{\partial O_i}{\partial \alpha} \frac{\partial^2 S}{\partial O_i O_j}\frac{\partial O_j}{\partial \beta}.
\enq
We will use both of these last expressions in chapter \ref{ch3} of this thesis.
\newline
\newline
Equation (\ref{fishermaxent}) presents the total amount of information on the model parameters $\btheta$ in the data $X$, when the model predicts the set of constraints $O_i$. The amount of information is in the form of a sum of the information contained in each constraint, with correlations taken into account, as in the right hand side in equation (\ref{MultiCramer}). In particular, it is a property of the maximal entropy distributions, that if the constraints $O_i$ are not redundant, then it follows that the curvature matrix of the entropy surface $- \partial^2S$ is invertible and is the inverse of the covariance matrix $\partial^2 \ln Z $  between the observables. To see this explicitly,  consider the derivative of equation (\ref{lambda}) with respect to the potentials,
\beq
-\frac{\partial O_i }{\partial \lambda_j}  = \frac{\partial^2\ln Z}{\partial \lambda_i\partial\lambda_j}.
\enq
The inverse of the matrix on the left hand side, if it can be inverted, is $-\frac{\partial \lambda_i }{\partial O_j}$, which can be obtained taking the derivative of equation (\ref{potentials}), with the result
\beq
-\frac{\partial \lambda_i }{\partial O_j} =- \frac{\partial^2 S}{\partial O_i\partial O_j}.
\enq
We have thus obtained in equation (\ref{fishermaxent}), combining Jaynes' Maximal Entropy Principle together with Fisher's information, the exact expression of the information 
 inequality (\ref{MultiCramer}) for our full set of constraints,  but with an equality sign.
 \newline
 \newline
We see that the choice of maximal entropy probabilities is fair, in the sense that all the Fisher information comes from what was forced upon the probability density function, i.e. the constraints. No additional Fisher information is added when these probabilities are chosen. In fact, as shown in the appendix this requirement alone is enough to single out the maximal entropy distributions, as being precisely those for which the information 
 inequality is an equality. This can be understood in terms of sufficient statistics and goes back to \cite{1936PCPS...32..567P} and \cite{koopman36}. For a discussion in the language of the exponential family of distribution see \cite{RePEc:spr:metrik:v:41:y:1994:i:1:p:109-119}.
 \newline
 \newline
In the special case that the model parameters are the constraints themselves, we have
\beq \label{entropysurface}
F^X_{O_iO_j}= - \frac{\partial^2 S}{\partial O_i O_j} = -\frac{\partial \lambda_i }{\partial O_j},
\enq
which means that the Fisher information on the model predictions contained in the expected future data is directly given by the sensitivity of their corresponding potential. Also, the application of the Cram\'{e}r-Rao 
 inequality, in the form given in equation (\ref{covF}), to any set of unbiased estimators of $\mathbf O$, shows that the best joint, unbiased, reconstruction of $\mathbf O$ is given by the inverse curvature of the entropy surface $-\partial^2S$, which is, as we have shown, $\partial^2 \ln Z$.\newline\newline
We emphasise at this point that although the amount of information is seen to be identical to the Fisher information in a Gaussian distribution of the observables with the above correlations, nowhere in our approach do we assume Gaussian properties. The distribution of the constraints $o_i(x)$ themselves is set by the  maximal entropy distribution of the data.
\subsection{Redundant observables}\label{redundant}
We have just seen that in the case of independent constraints, the entropy of $p_X$ provides through equation (\ref{fishermaxent}) both the joint information content of the data, as well as the inverse covariance matrix between the observables. However, if the constraints put on the distribution are redundant, the covariance matrix is not invertible, and the curvature of the entropy surface cannot be inverted either. We show however that in these cases, our equations for the Fisher information content (\ref{Zform}, \ref{sumform}, \ref{fishermaxent}) are still fully consistent, dealing automatically with redundant information to provide the correct answer.

\noindent An example of redundant information occurs trivially if one of the functions $o_i(x)$ can be written in terms of the others. For instance, for galaxy survey data, the specification of the galaxy power spectrum as an constraint, together with the mean number of galaxy pairs as function of distance, and/or the two-points correlation function, which are three equivalent descriptions of the same statistical property of the data. Although the number of observables $\mathbf O$, and thus the number of potentials, describing the maximal entropy distribution greatly increases by doing so, it is clear that we should expect the Fisher matrix to be unchanged, by adding such superfluous pieces of information. A small calculation shows that the potentials adjust themselves so that it is actually the case, meaning that this type of redundant information is automatically discarded within this approach.  Therefore, we need not worry about the independency of the constraints when evaluating the information content of the data, which will prove convenient in some cases.

\noindent There is another, more relevant type of redundant information, that allow us to understand better the role of the potentials. Consider that we have some set of constraints $\{O_i\}_{i =1}^n$, and that we obtain the corresponding $p_X$ that maximises the entropy. This $p_X$ could then be used to predict the value $O_{n+1}$ of the average some other function $o_{n+1}(x)$, that is not contained in our set of predictions, 
\beq
\av{o_{n +1}(x)}  = : O_{n +1}.
\enq
For instance, the maximal entropy distribution built with constraints on the first $n$ moments of $p_X$, will predict some particular value for the $n+1$-th moment, $O_{n+1}$, that the model was unable to predict by itself.\newline
Suppose now some new theoretical work provides the shape of $O_{n+1}$ as a function of the model parameters. This new constraint can thus now be added to the previous set, and a new, updated $p_X$ is obtained by maximising the entropy. There are two possibilities at this point :
\beit
\item It may occur that the value of $O_{n+1}$ as provided by the model is identical to the prediction by the maximal entropy distribution that was built without that constraint. Since the new constraint was automatically satisfied, the maximal entropy distribution satisfying the full set of $n +1$ constraints must be equal to the one satisfying the original set. From the equality of the two distributions, which are both of the form (\ref{MaxEnt}), it follows that the additional constraint must have vanishing associated potential,
\beq
\lambda_{n+1} = 0,
\enq
while the other potentials are pairwise identical. It follows immediately that the total information, as seen from equation (\ref{sumform}) is unaffected, and no information on the model parameters was gained by this additional prediction.
A cosmological example would be to enforce on the distribution of some field, together with the two-points correlation function, fully disconnected higher order correlation functions. It is well known that the maximal entropy distribution with constrained two-points correlation function has a Gaussian shape, and that Gaussian distributions have disconnected points function at any order. No information is thus provided by these field moments of higher order in this case.\newline
This argument shows that, for a given set of original constraints and associated maximal entropy distribution, any function $f(x)$, which was not contained in this set, with average $F$, can be seen as being set to zero potential. Such $F$'s therefore do not contribute to the information.
\item More interesting is, of course, the case where this additional constraint differs from the predictions obtained from the original set $\left\{O_i\right\}_{i = 1}^n$. Suppose that there is a mismatch $\delta O_{n+1}$ between the predictions of the maximal entropy distribution and the model. In this case, when updating $p_X$ to include this constraint, the potentials are changed by this new information, a change given to first order by
\beq
\delta\lambda_i = \frac{\partial^2S}{\partial O_i \partial O_{n+1}}\delta O_{n+1}, \quad  i = 1,\cdots, n+1,
\enq
and the amount of Fisher information changes accordingly.
It is interesting to note that the entropy itself is invariant at this order. From equation (\ref{potentials}) we have namely
\beq
\delta S = \sum_{i = 1}^{n+1} \lambda_{i}\:\delta O_i =  \lambda_{n + 1}\delta O_{n+1} = 0,
\enq
since the new constraint was originally at zero potential.
The entropy is, therefore, stationary not only with respect to changes in the probability distribution function, but also with respect to the predictions its associated maximal entropy distribution makes on any other quantities.
\enit
Of course, although the formulae of this section are valid for any model, it requires numerical work in order to get the partition function and/or the entropy surface in a general situation.
\subsection{The entropy and Fisher information content of Gaussian homogeneous fields} \label{Examples}
We now obtain the Shannon entropy of a family of fields when only the two-point correlation function is the relevant constraint, that we will use later in this thesis. It is easily obtained by a straightforward generalisation of the finite dimensional multivariate case, where the means and covariance matrix of the variables are known. It is well known  \citep{shannon48}  that the maximal entropy distribution is in this case the multivariate Gaussian distribution. Denoting the constraints on $p_X$ with the matrix $D$ and vector $\boldsymbol \mu$
\beq \label{cov} \begin{split}
D_{ij} &=  \av{x_{i}x_{j}} \\
\mu_i &= \av{x_i}, \quad i,j = 1,\cdots,N
\end{split}
\enq
the associated potentials are given explicitly by the relations
\beq \begin{split} \label{Dm}
 \lambda &= \frac 12 C^{-1}
\\\eta &= - C^{-1}\bmu,
\end{split}
\enq
where the matrix $C$ is the covariance matrix
\beq
C := D - \bmu \bmu^T.
\enq
The Shannon entropy is given by, up to some irrelevant additive constant,
\beq \label{entropyG}
S(D,\bmu) = \frac 12 \ln \det (D -  \bmu \bmu^T).
\enq
The fact that about half of the constraints are redundant, due to the symmetry of the $D$ and $C$ matrices, is reflected by the fact that the corresponding inverse correlation matrix in equation (\ref{fishermaxent}),
\beq
-\frac{\partial^2 S}{\partial D_{ij}\partial D_{kl}} = - \frac{\partial \lambda_{ij}}{\partial D_{kl}} =  \frac 12  C\inv_{ik}C\inv_{jl},
\enq
is not invertible as such if we considers all entries of the matrix $D$ as constraints. Of course,  this is not the case anymore if only the independent entries of $D$ form the constraints.
\newline 
\newcommand{\bbphi}{\bar\bphi}
Using  the handy formalism of functional calculus, we can straightforwardly extend the above relations to systems with infinite degrees of freedom, i.e. fields, where means as well as the two-point correlation functions are constrained. A realisation of the variable $X$ is now a field, or a family of fields $\bphi = (\phi_1,\cdots,\phi_N)$, taking values on some $n$-dimensional space.  The expressions above in the multivariate case all stays valid, with the understanding that operations such as matrix multiplications have to be taken with respect to the discrete indices as well as the continuous ones.
\newline
\newline
With the two-point correlation function and means
\beq
\begin{split}
\rho_{ij}(\vecx,\vecy) &= \av{\phi_i(\vecx)\phi_j(\vecy)} \\
\bar\phi_i(\vecx) &= \av{\phi_i(\vecx)}
\end{split}
\enq
we still have, up to an unimportant constant,
\beq \label{entropyfield}
S = \frac 12 \ln \det (\rho - \bphi\bphi^T). 
\enq
\newline\newline
In n-dimensional Euclidean space, within a box of volume $V$ for a family of homogeneous fields, it is simplest to work with the spectral matrices. These are defined as
\beq \label{spectralmatrix}
\frac 1 V \lat \tilde\phi_i(\veck) \tilde\phi_j^*(\veck')\rat =  P_{ij}(\veck)\:\delta_{\veck\veck'},
\enq
where the Fourier transforms of the fields are defined through
\beq
\tilde\phi_i(\veck) = \int_V d^nx \: \phi_i(x)\:e^{-i\veck\cdot\vecx}.
\enq
It is well known that these matrices provide an equivalent description of the correlations, since the they form Fourier pairs with the correlation functions
\beq
\rho_{ij}(\vecx,\vecy) = \frac 1 V \sum_{\veck} P_{ij}(\veck) e^{i\veck \cdot (\vecx - \vecy)} = \rho_{ij}(\vecx -\vecy).
\enq
In this case, the entropy in equation (\ref{entropyfield}) reduces, again discarding irrelevant constants, to an uncorrelated sum over the modes,
\beq \label{entropy field}
S = \frac 12 \ln\det\left[ \frac {P(0)}{V} - \bar\phi\bar\phi^T \right]+ \frac 12\sum_{\veck}\ln \det \frac{ P(\veck) }{V},
\enq
which is the straightforward mutlidimensional version of \citep[eq. 39]{2001MNRAS.328.1027T}.
Comparison with equation (\ref{entropyG}) shows the well-known fact that the modes can be seen as Gaussian, uncorrelated and complex variables with correlation matrices proportional to $P(\veck)$.  All modes have zero mean, except for the zero-mode, which, as seen from its definition, is proportional to the mean of the field itself. Accordingly, taking the appropriate derivatives, the potentials $\lambda(\veck)$ associated to $P(\veck) $ read
\beq
\begin{split} 
\lambda(\veck) &= \frac V{2} P(\veck) ^{-1},\quad \veck \ne 0 \\
\lambda(0) &=  \frac 1{2} \lb  \frac{P(0)}{V} - \bphi\bphi^T \rb^{-1} . 
\end{split}
\enq
and those associated to the means $\bphi$,
\beq
\boldsymbol \eta = - \lb  \frac{P(0)}{V} - \bphi\bphi^T \rb^{-1}\bphi
\enq
Note that although the spectral matrices are, in general, complex, they are hermitian, so that the determinants are real. The amount of Fisher information in the family of fields is easily obtained with the help of equation (\ref{sumform}) , with the familiar result
\beq \label{FPk} \begin{split}
\Fab &= \frac 1 2\sum_{\veck } \Tr\left[P_c^{-1}(\veck)\frac{\partial P_c(\veck)}{\partial \alpha}P_c^{-1}(\veck)\frac{\partial P_c(\veck)}{\partial \beta}\right]\\
&\quad+ \frac{\partial \bbphi^T}{\partial \alpha}\lb \frac{P_c(0)}{V}\rb^{-1} \frac{\partial\bbphi}{\partial \beta},
\end{split}
\enq
with $P_c(\veck)$ being the connected part of the spectral matrices,
\beq
\begin{split}
P_c(\veck) &= P(\veck) - \delta_{\veck0}V\bphi\bphi^T.
\end{split}
\enq
These expressions are of course also valid for isotropic fields on the sphere. With a decomposition in spherical harmonics, the sum runs over the multipoles.
\newline
\newline
The Fisher matrices in common use in weak lensing or clustering can thus all be seen as special cases of this approach, namely equation (\ref{FPk}), when knowledge of the statistical properties of the future data does not go beyond the two-point statistics. Indeed, in the case that the model does not predict the means, and knowing that for discrete fields the spectral matrices, equation (\ref{spectralmatrix}), carry a noise term due to the finite number of galaxies, or, in the case of weak lensing, also due to the intrinsic ellipticities of galaxies, the amount of information in (\ref{FPk})  is essentially identical to the standard expressions used to predict the accuracy with which parameters will be extracted from power-spectra analysis. 
\newline
\newline
Of course, the maximal entropy approach, which tries to capture the relevant properties of $p_X$ through a sophisticated guess, gives no guaranties that its predictions are actually correct. Nevertheless, as discussed in section \ref{redundant}, it provides a systematic approach with which to update the probability density function in case of improved knowledge of the relevant physics.
\clearpage
\newpage
\section{Appendix} In \ref{appendix_unique}, we look at what possible 'information densities' do in fact satisfy those conditions that we would like any measure of information about the true value of a model parameter to possess. We then provide in \ref{CramerRao bound} a unified derivation of the Cram\'{e}r-Rao and information inequality in the multidimensional case (following a similar argumentation than in \cite{Rao}) and then show its relation to maximal entropy distributions. 
\subsection{Measures of information on a parameter}\label{appendix_unique} 
Denoting with $p_X(x,\alpha)$ the probability density function of some variable, we look for functionals $i$
\beq
i[p_X(x,\alpha)]
\enq
such that our candidate of the measure of information on $\alpha$ is given by 
\beq
I_X(\alpha) = \int dx \: i[p_X(x,\alpha)],
\enq
and has the following properties :
\newline
\newline
\textbf{C1: } We would want our functional to be a regular function of two arguments, one being the value of the probability function at  $\alpha$ itself, and the second the value of its derivative with respect to the same $\alpha$,
\beq i[p_X(x,\alpha)] := i(p_X(x,\alpha),\partial_\alpha p_X(x,\alpha)).
\enq
While the main reason for this very strong requirement is simplicity, it appears nonetheless reasonable to us. This condition simply reflects the fact that the values of $p_X(x,\alpha)$ in regions far away from the true value of $\alpha$ should not provide much information, or that the information is provided by the probabilities and the linear impact of $\alpha$ around the true value.
\newline
\newline
\textbf{C2:}
The choice of coordinates must not carry information, i.e. we want
\beq
I_X(\alpha) = I_{g(X)}(\alpha) 
\enq
whenever $g$ is an invertible function. While this is automatic for discrete probabilities, this is not the case for continuous distributions.\newline
\newline
\textbf{C3:} The information on $\alpha$ from independent experiments should add
\beq
I_{XY}(\alpha) = I_X(\alpha) + I_Y(\alpha).
\enq
The last two are the key requirements for the interpretation of $I$ as information. In combination with the first requirement, it  leads to the following result: up to a multiplicative constant, there is a unique positive definite density satisfying these conditions. it reads
\beq
i(p_X,\partial_\alpha p_X)  = \frac{\lp\partial_\alpha p_X\rp^2}{p_X},
\enq
which is easily seen to be precisely the Fisher information measure, as defined in equation (\ref{FisherI}).
\newline
\newline
From now on, in order to simplify the notation, we write the two arguments $p_X$ and $\partial_\alpha p_X$ of $i$ as $p$ and $\beta$.
If we drop the positive definite condition, the general solution is
\beq
i(p,\beta) = c_1 \frac{\beta^2}{p} + c_2 \frac{\beta^3}{p^2} + c_3 \beta,
\enq
for arbitrary constants $c_{1-3}$.
This result follows from the claim that the most general smooth function satisfying $C1$ and $C3$ should be of the form
\beq \label{claim}
i(p,\beta) = c_1 \frac{\beta^2}{p} + c_2 \frac{\beta^3}{p^2} + c_3\beta + c_4 \beta \ln(p) + c_5 p\ln(p),
\enq
for a set of arbitrary constants $c_{1-5}$. The two last terms do not meet $C2$ and are thus to be discarded, while the first one is the only positive definite among the three others.
In order to see why this holds, we first note, as can be checked by direct calculation, that in this form $i(p,\beta)$ fulfill the requirements $C1$ and $C3$. We then use extensively $C3$ for particular instances of variables with different distributions and derivatives, sketching the proof that these criteria do not allow other functional forms.
\newline
\newline
We first notice that, since any probability density function must be normalised to unity, for any value of the model parameter, the following relations must hold for any variable $X$,
\beq
\int dx \: p_X(x,\alpha) = 1\label{first},
\enq
as well as
\beq
\int dx \: \partial_\alpha p_X(x,\alpha) = 0.\label{second}
\enq
We now consider, beside an arbitrary distribution $p_X(x,\alpha)$, the distribution of some independent variable $Y$, given by  $q_Y(y,\alpha)$, and denote the derivatives with respect to $\alpha$ associated to $p$  and $q$, as $f$ and $g$,
\beq \begin{split}
f(x) &= \partial_\alpha p_X(x,\alpha) \\
g(x) &= \partial_\alpha q_Y(x,\alpha),
\end{split}
\enq
where the dependency on $\alpha$ is omitted.
The joint probability density function of the independent variables $X$ and $Y$ is given by the product of $p_X$ with $q_Y$, with associated derivative
\beq
\frac{\partial\:  p_Xq_Y}{\partial \alpha } = fq_Y + p_Xg.
\enq
Condition $C3$ states in these terms explicitly that for any such $p,q$ such that (\ref{first}) holds, and $f,g$ such that (\ref{second}) holds, the following relation, describing the additivity of information, must be true,
\beq \begin{split}
&\int dx \int dy \: i\lp p_X(x)q_Y(y),f(x)q_Y(y) + p_X(x)g(y) \rp  \\
&= \int dx \: i \lp p_X(x),f(x) \rp+ \int dy \: i\lp q_Y(y),g(y)  \rp.
\end{split}
\enq
With the help of this relation we can constrain the functional form of $i(p,\beta)$.
\newline
\newline
Let us first pick a uniform probability density for  the variable $Y$,
\beq
q_Y(y) \equiv 1,\quad \textrm{with }  y \in [0,1].
\enq
It must hold, using $C3$, that
\beq \begin{split}
&\int dx \int dy \:i\left(p_X(x),p_X(x)g(y) + f(x)\right)\\
 = & \int dx\: i(p_X(x),f(x)) + \int dy \:i(1,g(y)),
\end{split}
\enq
for any allowed $g(y)$. It can be shown, for instance by performing variations with respect to  $g(y)$, that any solution to this integral equation must obey the relation
\beq
\frac{\partial^2i }{\partial \beta^2}(p,\beta) = \frac 1 p j\left(\frac \beta p\right)
\enq
for some function $j$, that satisfies
\beq
j(\beta) = \int dx\:p_X(x)j\left(\beta + \frac{f(x)}{p_X(x)}\right).
\enq
This must still hold as above for any $p_X$ and associated derivative function $f$, satisfying (\ref{first}) and (\ref{second}) respectively.
Taking a variation with respect to $f$ shows that the only solutions for $j(\beta)$ of this equation are
\beq
j(\beta) = c_1 \beta + c_2
\enq
for some arbitrary constants $c_1$ and $c_2$.
Therefore, we have constrained the full function $i(p,\beta)$ to be of the form
\beq \label{fullhouse}
i(p,\beta) = c_1 \frac{\beta^2}{p} + c_2 \frac{\beta^3}{p^2} + \beta\: r(p) + s(p)
\enq
for some unknown functions $r(p)$ and $s(p)$.
The first two terms are full solutions of $C2$.
With very similar methods, the two other terms can be reduced to
\beq
r(p) = c_3\ln(p) + c_4
\enq
and
\beq
s(p) = c_5\: p\ln(p).
\enq
In this form, all terms in equation (\ref{fullhouse}) are consistent with $C3$ and we have thus proved our claim (\ref{claim}).
\subsection{Cram\'er-Rao bound and maximal entropy distributions}\label{CramerRao bound}
We denote the vector of model parameters of dimension $n$ with
\beq
\vecalpha = \bem\alpha_1,\cdots,\alpha_n\enm
\enq
and a vector of functions of dimension $m$ the estimators
\beq
\hat {\mathbf D} = \lp \hat D_1,\cdots \hat D_m \rp,
\enq
with expectation values $D_i(\vecalpha)=  \av {\hat D_i(x)} $. 
In the following, we rely on Gram matrices, whose elements are defined by scalar products. Namely, for a set of vectors $\vecy_i$, the Gram matrix $Y$ generated by this set of vectors is defined as
\beq
Y_{ij} = \vecy_i \cdot \vecy_j.
\enq
Gram matrices are positive definite and have the same rank as the set of vectors that generate them. Especially, if the vectors are linearly independent, the Gram matrix is strictly positive definite and invertible. \newline
We adopt a vectorial notation for functions, writing scalar products between vectors as
\beq
f\cdot g \equiv \int dx\:p_X(x,\vecalpha) f(x) g(x),
\enq
with $p_X(x,\vecalpha)$ being the probability density function of the variable $X$ of interest.
In this notation, both the Fisher information matrix and covariance matrix are seen to be Gram matrices. We have namely that the Fisher information matrix reads
\beq
 F_{ {\alpha_i}{\alpha_j}} = f_{\alpha_i} \cdot f_{\alpha_j},\quad f_{\alpha_i}(x) = \frac{\partial \ln p_X(x,\vecalpha)}{\partial \alpha_i},
\enq
while the covariance matrix of the estimators is
\beq
C_{ij} = g_i\cdot g_j,\quad g_i(x,\vecalpha) = \hat D_i(x) - D_i(\vecalpha). 
\enq
For simplicity and since it is sufficently generic for our purpose, we will assume that both sets of vectors $f$ and $g$ are lineary independent, so that both matrices can be inverted. Note that we also have
\beq
\frac{\partial D_i}{\partial {\alpha_j}} = \int dx \:p_X(x,\vecalpha) \:\hat D_i(x) \frac{\partial \ln p_X(x,\vecalpha)}{\partial {\alpha_j}} = g_i\cdot f_{\alpha_j}.
\enq
\newline
The  Gram matrix $ G$ of dimension $\lp (m + n) \times (m + n) \rp$ generated by the set of vectors $\lp g_1,\cdots, g_m ,f_{\alpha_1},\cdots, f_{\alpha_n}\rp $ takes the form
\beq
G = \bem  C & \Delta \\  \Delta^T & F \enm , \quad \Delta_{i {\alpha_j}} = g_i \cdot f_{\alpha_j}
\enq
and is also positive definite due to its very definition. It is congruent to the matrix
\beq
 Y G Y^T = \bem  C - \Delta F \inv \Delta^T & 0 \\ 0 & F \enm,
\enq
with
\beq
Y = \bem 1_{m\times m} & - \Delta F^{-1} \\ 0 & 1_{n\times n} \enm.
\enq
Since two congruent matrices have the same number of positive, zero and negative eigenvalues respectively and since both $ F$ and $G$  are positive, we can conclude that
\beq
C \ge \Delta F \inv \Delta^T,
\enq 
which is the Cram\'{e}r-Rao 
 inequality. The lower bound on the amount of information is seen from the fact that for any matrix written in block form holds
\beq
\bem  C & \Delta \\  \Delta^T & F \enm\ge 0 \Leftrightarrow \bem F & \Delta^T \\  \Delta & C \enm \ge 0
\enq
and using the same congruence argument leads to the lower bound on information
\beq
F \ge \Delta^T C \inv \Delta.
\enq
Assume now that we have a probability density function such that this inequality
is in fact an equality, i.e.
\beq
F = \Delta^T C \inv \Delta.
\enq
By the above argument, the Gram matrix generated by
\beq
\lp f_{\alpha_1},\cdots, f_{\alpha_n}, g_1,\cdots, g_m\rp
\enq
is congruent to the matrix
\beq
\bem 0_{n \times n} & 0 \\ 0& C \enm
\enq
and has rank $m$. By assumption, the covariance matrix is invertible, such that the set $\lp g_1,\cdots, g_m \rp$ alone has rank $m$. It implies that each of the $f$ vector can be written as linear combination of the $g$ vectors,
\beq
f_{\alpha_i} = \sum_{j = 1}^{m} A_{j} g_j,
\enq
or, more explicitly,
\beq
\frac{\partial \ln p_X(x,\vecalpha)}{\partial {\alpha_i}} =  \sum_{j = 1}^{m} A_{ j}(\vecalpha) \left[\hat{ D_j}(x) - D_j(\vecalpha)\right],
\enq
where the key point is that the coefficients $A_j$ are independent of $x$. Integrating this equation, we obtain
\beq
\ln p_X(x,\vecalpha) = -\sum_{i = 1}^m \lambda_i(\vecalpha)\hat {D_i}(x) - \ln Z(\vecalpha) + \ln q_X(x)
\enq
for some functions $\lambda$ and $Z$ of the model parameters only, and a function $q_X$ of $x$ only. We obtain thus
\beq
p_X(x,\vecalpha) = \frac {q_X(x)} {Z(\vecalpha)} \exp\left(-\sum^m_{i=1} \lambda_i (\vecalpha)\hat {D}_i(x) \right).
\enq
This is precisely the distribution that we obtain by maximising the entropy relative to $q_X(x)$, while satisfying the constraints
\beq
D_i(\vecalpha) = \av{\hat D_i(x)},\quad i = 1,\cdots, m.
\enq 
Taking $q_X$ as the uniform distribution makes it identical with the formula in equation (\ref{MaxEnt}).

\clearpage
\newpage

\chapter{Information within N-point moments}
\label{ch2}
In this chapter we demonstrate how to decompose the Fisher information matrix into components unambigously associated to independent information from $N$-point moments of each order. The general approach to decompose the Fisher information matrix in uncorrelated components according to an orthogonal system was briefly discussed in a statistical journal in theorem 3.1 in \cite{Jarrett84}. It seems however that this procedure was not given further attention. In this chapter, similar ideas are taken further, dealing mostly with the system of moments, where the associated orthogonal system are orthogonal polynomials.
\newline
\newline
We start in section \ref{generaltheory} with one dimensional variables. We will in a first step define for a probability distribution $p(x,\btheta)$, coefficients $s_n(\alpha)$ which unambiguously represent the independent information content of the moment of order $n$ on $\alpha$. These coefficients can then be used to reconstruct the Fisher information matrix order by order.
The straight forward generalisation to any number of variables is performed in \ref{svariables}, or to any hierarchical system other than the $N$-point moments in the appendix. Properties of this expansion under the presence of noise are discussed in \ref{noiseN}.
\newline
\newline
In section \ref{determ}, we then present this exact decomposition for a few determinate probability density functions. We give in closed form these coefficients for several common classes of distributions, spanning a wide range of different situations. We solve these decomposition for the normal, the gamma and the beta families of distributions, as well as for an extended Poisson model. 
\newline
\newline
We use this decomposition in \ref{indeterm} to approach the indeterminate moment problem. Namely, the Fisher information content of a moments series can be less than the information content of the probability density function it originates from, if that function cannot be uniquely recovered from the moment series. Therefore, this loss of information is also a useful measure of the indeterminacy of the moment problem. After dealing with the lognormal distribution, whose expansion we will present in greater details in chapter \ref{ch4}, we treat numerically the Weibull distribution and strectched exponential. For each of those we compare the information lost to the moment hierarchy to Stoyanov's dissimilarity index of associated Stieltjes classes \citep{Stoyanov04}.
\section{One variable}
\label{generaltheory}
Our starting point is the information inequality of chapter \ref{ch1} : for any set of unbiased estimators $\hat{ \mathbf O}$ aimed at extracting the vector of observables $\mathbf O = \av{\hat {\mathbf O}}$, the following inequality between positive definite matrices holds,
\beq \label{infoinequality}
F  \ge \Delta \Sigma^{-1}  \Delta^T,
\enq
where $\Sigma$ is the covariance matrix of the estimators,
\beq
\Sigma_{ij} = \av{\hat O_i \hat O_j} - O_iO_j,
\enq
and $\Delta$ is the matrix of derivatives,
\beq
\Delta_{\alpha i} =\frac{\partial O_i}{\partial \alpha}.
\enq
When the vector $\mathbf O$ are the moments themselves, $O_i = m_i = \av{x^i},$for $\quad i = 1,\cdots N$,
it is possible to rewrite the information inequality in a more insightful form. Consider a set of polynomials $\left\{Q_n\right\}_{n = 1}^\infty$, $n$ the degree of the polynomial, orthogonal with respect to $p(x,\btheta)$,
\beq \label{orthopol}
\begin{split}
Q_n(x) &= \sum_{k = 0}^nC_{nk}x^k \\
\av{Q_n(x)Q_m(x)}  &= h_n\delta_{mn}, \quad n,m = 0,\cdots,N
\end{split}
\enq
where $h_n$ is some strictly positive number, equal to unity if the polynomials are orthonormal. Since the normalisation of many common families of orthogonal polynomials is not unity, we will at the expense of extra notation keep track of the terms $h_n$ in the following. Of course, the polynomials
\beq
P_n(x) :=Q_n(x) / \sqrt{h_n}
\enq
are then orthonormal. We refer to \cite{Freud71} or \cite{Szego39} for the theory of orthogonal polynomials. The polynomial $P_n$ is always set by the first $2N$ moments, and is unique up to an overall sign, which we set by requiring the coefficient of $x^n$ in $P_n(x)$ (the leading coefficient) to be positive. A simple way to build formally the orthonormal polynomials with this sign convention is for instance to apply the Gram-Schmidt orthonormalisation process to the set $(1,x,\cdots,x^n)$ with respect to the scalar product $(f,g) := \av{f(x)g(x)}$.
\newline
\newline
The key point  for our purposes is to realise that the inverse covariance matrix between the moments can be written \footnote{A proof can be found in a slightly more general setting later in section \ref{hierarchy}.}
\beq \label{sigmaP}
\lb \Sigma^{-1} \rb_{ij} = \sum_{n= 0}^N\frac{C_{in}C_{jn}}{h_n},\quad i,j = 1,\cdots N.
\enq
This identity allows us to express the right hand side of the information inequality in the form of a sum of uncorrelated pieces : using (\ref{sigmaP}), it holds
\beq \label{chi2}
\sum_{i,j = 1}^N \frac{\partial m_i}{\partial\alpha}\Sigma^{-1}_{ij}\frac{\partial m_j}{\partial \beta} = \sum_{n = 1}^Ns_n(\alpha)s_n(\beta).
\enq
where
\beq \label{def}
s_n(\alpha) = \frac{1}{\sqrt{h_n}}\sum_{k = 1}^nC_{nk}\frac{\partial m_k}{\partial \alpha}.
\enq
The matrix $s_N(\alpha)s_N(\beta)$
is therefore the part of the Fisher information in the $N$th moment
that was not contained in the moments of lower order.\newline
These coefficients have a straightforward interpretation. They can be namely written as
\beq\label{sn}
s_n(\alpha) := \av{s(x,\alpha) P_n(x)},\quad s(x,\alpha) = \frac{\partial \ln p(x,\btheta)}{\partial \alpha},
\enq
which we will take as the definition of $s_n(\alpha)$. This can be seen from expanding $P_n(x)$ in terms of the matrix $C$, and noting that $s(x,\alpha)p(x,\btheta) = \partial_\alpha p(x,\btheta)$, recovering \eqref{def}. In other words, $s_n(\alpha)$ is nothing else than the component of the corresponding function $\partial_\alpha \ln p$ (the score function) parallel to the orthonormal polynomial of order $n$.
\newline
\newline
It follows immediately from equation (\ref{sn}) that the Fisher information content of the moments depends on how well the score functions can be approximated through polynomials. With increasing $N$, one expects the score function to be better and better reproduced by the series
\beq
s_{\le N}(x,\alpha) := \sum_{n  =1}^Ns_n(\alpha)P_n(x).
\enq
We see that the following inequality between positive matrices holds
\beq
\begin{split}
0 &\le \av{ \lp s(x,\alpha) - s_{\le N}(x,\alpha)\rp\lp s(x,\beta) - s_{\le N}(x,\beta)\rp } 
= F_{\alpha\beta} -  \sum_{i,j = 1}^N \frac{\partial m_i}{\partial\alpha}\lb \Sigma^{-1}\rb_{ij}\frac{\partial m_j}{\partial \beta}.
\end{split}
\enq
In particular, for any parameter $\alpha$ holds that the residual to the best approximation of the score function with polynomials is given by
\beq \label{residuals}
 \av{\lp s(x,\alpha) - s_{\le N}(x,\alpha)\rp^2} = I(\alpha) -  \sum_{i,j = 1}^N \frac{\partial m_i}{\partial\alpha}\lb \Sigma^{-1}\rb_{ij}\frac{\partial m_j}{\partial \alpha},
\enq
where $I(\alpha) = F_{\alpha\alpha}$ is the Fisher information on $\alpha$. The bits of Fisher information that are absent from the set of moments $m_1$ to $m_N$ are thus precisely the mean squared error of the fit of the score function through polynomials throughout the range of $p(x,\btheta)$.
\newline
\newline
We define for further reference the matrices
\beq \label{Fn1}
\lb F_n \rb_{\alpha \beta}  := s_n(\alpha)s_n(\beta) 
\enq
 as well as
 \beq
 \label{Fn}
\lb F_{\le N} \rb_{\alpha\beta} := \sum_{n= 1}^Ns_n(\alpha) s_n(\beta) \enq
These are the matrices representing the independent information content of the $n$th moment and of the first $N$ moments respectively. By construction holds
\beq
\lb F_{\le N} \rb_{\alpha\beta}=   \sum_{i,j = 1}^N \frac{\partial m_i}{\partial\alpha}\lb \Sigma^{-1} \rb_{ij}\frac{\partial m_j}{\partial \beta}.
\enq
We stress that these matrices are strictly speaking associated to a moment series rather than a density function. These matrices are namely identical for different densities having the same moment series.
\newline
\newline
We see that if the score function is itself a polynomial, of degree $N$, for each values of the parameters, then only the first $N$ of these coefficients are possibly non-zero. A finite number of moments do catch all the Fisher information content of $p(x,\btheta)$ in this case. In fact, the reverse statement is also true. This can be understood in the framework of orthodox statistics : in this case $p(x,\btheta)$ is proportional to the exponential of a polynomial with parameter dependent coefficients, where a finite number of sufficient statistics exist \cite{1936PCPS...32..567P,koopman36}. In the light of chapter \ref{ch1}, these distributions are precisely those that maximise Shannon entropy for fixed values of the first $N$ moments. We thus have
\beq
F_{\le N} = F \quad \textrm{Max. entropy distributions with constrained first $N$ moments}.
\enq
The ubiquitous example of this family being of course the Gaussian distribution, for which $N = 2$.
\newline
\newline
More generally, a sufficient criterium for the moment hierarchy to possess the same amount of Fisher information as the density function itself, 
is that the moment problem associated to the moment series is \textit{determinate}, that is to say that the density can be uniquely recovered from the moment series. In this case, indeed, by a well known theorem due to M. Riesz, \citep{Riesz23}, the orthogonal polynomials associated to the density form a complete set of basis functions for square integrable functions. Therefore, the square residual in equation \eqref{residuals} must go to zero for $N \rightarrow \infty$. We have in this case
\beq
\lim_{N \rightarrow \infty} F_{\le N} = F  \quad \textrm{Moment determinate density functions}.
\enq
Finally, for moment \textit{indeterminate} density functions, for which different density functions exist with the same moment series, we have in general an inequality (again, an inequality between positive matrices, not matrix element to matrix element)
\beq
\lim_{N \rightarrow \infty} F_{\le N} \le F  \quad \textrm{Moment indeterminate density functions}.
\enq
The amplitude of the mismatch can vary very substantially from case to case. In chapter \ref{ch4}, we study extensively the lognormal field, for which the inequality is always a strict inequality, but is a very strong function of the variance of the field.  

\subsection{The Christoffel-Darboux kernel as information on the density itself}
To conclude this section, it is interesting to understand the interpretation of the orthogonal polynomials themselves, in this information theoretic framework. To this aim, consider that the model parameter of interests are the values of the density function themselves, allowing thus complete freedom. In the following, we do not require the density to be normalised to unity for convenience. Take for simplicity discrete values $x_i, i =0,1,\cdots $, with associated probability density
$p(x_i)$
and the set of model parameters being precisely the density, i.e. $\alpha_i = p(x_i)$. Straight calculation leads to the following expression for the Fisher information matrix elements on the parameters $p(x_i)$, $p(x_j)$,
\beq
F_{ij}=  \frac{ \delta_{ij}} {p(x_i)}.
\enq
On the other hand, we have $\partial m_k/ \partial p(x_i) = x_i^k$, leading to
\beq
s_n\lp p(x_i) \rp = P_n(x_i).
\enq
Therefore, the $n$th orthogonal polynomial $P_n(x_i)$ is the information content of the $n$th moment on the density function  $p(x_i)$ itself.
The matrices $F_{\le N}$ become the celebrated Christoffel-Darboux kernel $K_N$ \citep{2008arXiv0806.1528S},
\beq
K_N(x_i,x_j) = \sum_{n = 0}^NP_n(x_i)P_n(x_j) = \lb F_{\le N} \rb_{ij}
\enq
We have from these relations that the information escaping the first $N$ moments is given by
\beq \label{CDK}
F_{ij} - \lb F_{\le N} \rb_{ij} =  \frac{\delta_{ij}}{p(x_i)} -  K_N(x_i,x_j).
\enq
In accordance with our discussions above, this right hand side of \eqref{CDK} is known to tend to zero as $N \rightarrow \infty$ precisely for determinate moment problem.  This is  can be seen from the fact that in this case, from Riesz theorem, the reproducing property holds
\beq
\sum_i p(x_i)\lb f(x_i) - \sum_{n = 0}^N \lp \sum_{k} p(x_k) f(x_k) P_n(x_k)\rp P_n(x_i) \rb^2  \stackrel{N\rightarrow \infty}{\rightarrow} 0,
\enq
for any function $f(x_i)$ that is square summable with respect to $p(x_i)$.
This implies
\beq
\sum_{k}p(x_k)f(x_k) K_N(x_k,x_i)  \stackrel{N\rightarrow \infty}{\rightarrow}f(x_i)
\enq
for any such function and therefore
\beq
K_N(x_i,x_j)  \stackrel{N\rightarrow \infty}{\rightarrow}\frac{\delta_{ij}}{p(x_j)}.
\enq
\def\apj{The Astrophysical Journal}                 

\section{Several variables\label{svariables} }

\newcommand{\vecn}{\mathbf n}
\newcommand{\vecm}{\mathbf m}
\newcommand{\veci}{\mathbf i}
\newcommand{\vecj}{\mathbf j}
The general theory on the statistical power of moments exposed in section \ref{generaltheory} extends is a straightforward way to density functions of any number of variables and $N$-point moments.
We first need a little bit of notation. For a $d$-dimensional variable $X$ taking values $x = \lp x_1,\cdots x_d\rp$, a multiindex
\beq
\vecn = (n_1,\cdots,n_d),\quad n_i = 0,1,2,\cdots,
\enq
is a $d$-dimensional vector of non negative integers.
The order of the multiindex is defined as
\beq
|\vecn| := \sum_{i = 1}^dn_i.
\enq
For a given order $|\vecn| = n$, there are  $\begin{pmatrix} n+ d -1\\  n \end{pmatrix}$ different such multiindices $\vecn$.
We define further the notation $x^\vecn$ as
\beq
x^\vecn = x_1^{n_1}\cdots x_d^{n_d}.
\enq
With this notation in place, a moment of order $n$ is given by
\beq
m_\vecn := \av{x^\vecn} = \av{x_1^{n_1}\cdots x_d^{n_d}}, \quad |\vecn| = n,
\enq
and the covariance matrix between the moments is
\beq
\av{m_{\vecn + \vecm}} - m_\vecn m_\vecm =: \Sigma_{\vecn\vecm}.
\enq
In this notation, the decomposition of the information in independent bits of order $n$ proceeds by strict analogy with the one dimensional case. We refer to \cite{Dunkl01} for the general theory of orthogonal polynomials in several variables. A main difference being that at a fixed order $N$ there are not one but $\begin{pmatrix} n+ d -1\\  n \end{pmatrix}$ independent orthogonal polynomials. This number is the same as the number of the above multiindices of that order $N$. Each multiindex defines namely an independent monomial $x^\vecn$ of that order. These polynomials are not defined in an unique way. The orthogonality of the polynomials of same order is not essential for our purposes, but requiring the following condition is enough,
\beq \label{conditions}
\begin{split}
\av{P_\vecn(x) P_\vecm(x)} &= 0 ,\quad |\vecn| \ne |\vecm| \\
\av{P_\vecn(x) P_\vecm(x)} &= \lb H_n\rb_{\vecn\vecm} ,\quad |\vecn| = |\vecm| = n\\
\end{split}
\enq
for some positive matrices $H_n$, which replace the normalisation $h_n$ in \ref{generaltheory}.
The component of the score function $\partial_\alpha \ln p$ parallel to the polynomial $P_\vecn$ is
\beq
s_\vecn(\alpha) := \av{ \frac{\partial \ln p(x,\btheta)}{\partial \alpha} P_\vecn(x)}, 
\enq
and the expansion of the score function  in terms of these polynomials reads
\beq \label{sxa}
s_{\le N}(x,\alpha) := \sum_{n = 0}^N\sum_{|\vecn|,|\vecm| = n}s_\vecn(\alpha) \lb H_n^{-1}\rb_{\vecn\vecm}P_\vecm(x).
\enq
It will converge to the score function for $N \rightarrow \infty$ if the set of polynomials is complete, whereas it may not if not. We note that there is some freedom in the definition \eqref{conditions}. This freedom is that of the choice of a basis in the vector space of polynomials of order $n$ orthogonal to all polynomials of lower order. For this reason, $s_\vecn(\alpha)$ depends on the particular basis. However, the expansion \eqref{sxa} does not, and so will not the information matrices at fixed order.
\newline
\newline
Writing now the orthogonal polynomials in terms of a triangular transition matrix
\beq
P_\vecn(x) = \sum_{|\vecm|\le |\vecn|}C_{\vecn \vecm}x^\vecm
\enq
we see that the information matrix of order $n$ \eqref{Fn1} becomes
\beq
\begin{split} \label{smulti}
\lb F_n \rb_{\alpha \beta} &= \sum_{|\vecn|,|\vecm| = n} s_\vecn(\alpha) \lb H^{-1}_n \rb_{\vecn\vecm}s_\vecm(\beta) \\
&=    \sum_{|\veci|,|\vecj| \le n} \lb C^TH_n^{-1}C\rb_{\veci \vecj} \frac{\partial m_\veci}{\partial \alpha}\frac{\partial m_\vecj}{\partial \alpha},
\end{split}
\enq
The strict analog of equation (\ref{Fn}) holds for each $N$,
\beq
\lb F_{\le N} \rb_{\alpha \beta} =  \sum_{n= 1}^N \lb F_n \rb_{\alpha \beta} = \sum_{|\veci|,|\vecj| = 1 }^N \frac{\partial m_\veci}{\partial \alpha}\lb \Sigma^{-1} \rb_{\veci\vecj} \frac{\partial m_\vecj}{\partial \beta},
\enq
recovering the right hand side of the information inequality for all moments of order up to $N$. Just as before, the missing piece between $F$ and $F_{\le N}$ is the least squared residual the approximation of the score function through polynomials of order up to $N$.
\newline
The matrices $F_n$ and $F_{\le N}$ are easily seen to be invariant under mappings $y = Ax + b$, where $A$ is an invertible square matrix of size $d\times d$ and $b$ a $d$ dimensional vector, provided both are parameter independent.    
\newline\newline
As a simple illustration, for the multivariate Gaussian distribution with mean vector $\bmu$ and covariance matrix $C$  we have
\beq \label{FGm}
\lb F_{1} \rb_{\alpha \beta} =  \frac{\partial \bmu}{\partial \alpha}C^{-1}\frac{\partial \bmu}{\partial \beta}
\enq
and
\beq \label{FGs}
\lb F_{2} \rb_{\alpha \beta}=  \frac 12 \Tr \lb C^{-1} \frac{\partial C}{\partial \alpha}C^{-1} \frac{\partial C}{\partial \beta}\rb,
\enq
summing up to the total information, $F_{n} = 0, n > 2$.
\subsection{Independent variables and uncorrelated fiducial} \label{independent}
In general, it is a rather difficult problem to obtain explicit expressions for the orthogonal polynomials or the matrices $F_N$ and $F_{\le N}$ , especially in the case of several variables. Using exact though formal expressions as starting point requires the evaluation of determinants of moment matrices, and cases are rare when this is tractable.
For independent variables $(x_1,\cdots,x_d)$, however, a canonical choice of $P_\vecn$ can be written down in terms of the ones of associated to one dimensional problem. More specifically, if the variables are independent, then
\beq \label{pindependent}
p(x,\btheta) =  \prod_{i = 1}^d p_i(x_i,\btheta),
\enq
where $p_i$ denotes the one dimensional probability density function of the $i$th variable.
Define then the polynomial in $d$ variables of order $|\vecn|$ through
\beq \label{productform}
P_{\vecn}(x) := \prod_{i = 1}^dP_{n_i}(x_i),
\enq
where $P_{n_i}(x_i)$ is the orthonormal polynomial in one variable of order $i$ with respect to $p_i$.
It is not difficult to see that for any multiindices $\vecn$ and $\vecm$ the average over $x$ factorizes in averages with respect to each variable $x_i$. We have namely
\beq
\begin{split}
\av{P_\vecn (x)P_\vecm(x)} = \prod_{i = 1}^d \av{P_{n_i}(x_i) P_{m_i}(x_i)} = \prod_{i = 1}^d \delta_{n_im_i} =  \delta_{\vecn\vecm}.
\end{split}
\enq
The so defined polynomials are therefore orthogonal with matrices $H_n$ being unit matrices.
From equation \eqref{pindependent} follows that the function $\partial_\alpha \ln p$ is the sum of the functions $\partial_\alpha \ln p_i$ of the individual variables. Therefore, using the above polynomials defined in \eqref{productform}, it is not difficult to see that all the coefficients $s_\vecn(\alpha)$ that couples different variables vanishes, i.e. $s_\vecn(\alpha) = 0$ if $\vecn$ has two or more non zero indices. The information content of order $N$ becomes then simply the sum of the information of order $N$ within each variable, as expected. This is an manifestation of the additivity of Fisher information for independent variables, which is thus seen to hold order by order.
\newline
\newline
Note that polynomials also take this product form \eqref{productform} if as above the probability density function factorizes at the fiducial values of the model parameters, but with the difference that there is no splitting of the derivative functions $\partial_\alpha \ln p$ as a uncorrelated sum. In this case, the fiducial model is uncorrelated, but model parameters create correlations away from their fiducial values. $N$-point moments at non zero lag may carry genuine information in this case. In mathematical terms, all $s_\vecn(\alpha)$ can possibly be non zero, depending on how the score functions couple the different variables.

\subsection{On the impact of noise \label{noiseN}}
Very often the observed random variable is the  sum of two random variables,
\beq
z = x + y,
\enq
where $x$ is a $d$-dimensional variable carrying the information on the parameters $\btheta$, and $y$ is some additive noise. We set
\beq
p_X(x,\btheta) 
\enq
to be the probability density function for $x$, and
\beq
p_Y(y)
\enq
being that for $y$, which is parameter independent. From the rules of probability theory, we have
\beq
p_Z(z,\btheta) = \int d^dx \: p_X(x,\btheta)\: p_Y(z-x).
\enq
In this section, we prove the following, maybe rather intuitive fact, valid for any functional form of $p_X $ and of $p_Y$, any model parameters and any dimensionality $d$:
\newline
For any $N$, the following relation between positive matrices holds
\beq \label{Fnless}
F_{\le N}^{X +Y} \le F_{\le N}^X.
\enq
On the other hand, this relation does not hold for the matrices $F_n$. For instance, a non Gaussian noise $Y$ on a Gaussian signal $X$ will create in general third and higher order terms in the score functions. However, their amplitude is constrained by \eqref{Fnless}.
\newline
\newline
We prove this inequality here only for one dimensional variables, $d = 1$. The general proof consists in replacing indices such as $n$ with multindices $\vecn$ and so on.
\newline
\newline
Our proof is based on the following basic fact concerning blockwise positive matrices (for a proof of this particular fact, see \cite[e.g.]{Bhatia} :
Whenever $A$ and $D$ are strictly positive matrices, then
\beq \label{fact}
\bem A & C \\ C^T & D \enm \ge 0 \Leftrightarrow A \ge CD^{-1}C^T.
\enq
We proceed as follows : in general we can write the moments of $Z$ as a linear combination of those of $X$
\beq
m^Z_n  = \av{\lp x  + y \rp^n} = \sum_{k = 0}^n\bin n k m_n^Xm_{n-k}^Y = : \sum_{k = 0}^n B_{nk}m_k^X.
\enq
Note that $B$ is a triangular matrix with non vanishing diagonal elements, and is therefore invertible.
Writing $\mathbf m = (m_0,\cdots m_n)^T$, it holds therefore
\beq
\lb F_{\le N}^Z \rb_{\alpha \beta} = \lp \frac{\partial\mathbf m^Z}{\partial \alpha}\rp^TM_Z^{-1}\lp \frac{\partial\mathbf m^Z}{\partial \beta}\rp = \lp \frac{\partial\mathbf m^X}{\partial \alpha}\rp^TB^TM_Z^{-1}B\lp \frac{\partial\mathbf m^X}{\partial \beta}\rp,
\enq
where $M_X$, respectively $M_Z$ is the moment matrix $M_{ij} = m_{i+j}$ of $X$, respectively $Z$.
The claim \eqref{Fnless} is proven provided
\beq
\lp \frac{\partial\mathbf m^X}{\partial \alpha}\rp^TB^TM_Z^{-1}B\lp \frac{\partial\mathbf m^X}{\partial \beta}\rp \le \lp \frac{\partial\mathbf m^X}{\partial \alpha}\rp^TM_X^{-1}\lp \frac{\partial\mathbf m^X}{\partial \beta}\rp,
\enq
which is equivalent to
\beq
 B^TM_Z^{-1}B \le M_X^{-1}.
\enq
We now prove this last relation.
\newline
\newline
Consider the following vector
\beq
\mathbf v := (z^0,\cdots,z^N,x^0,\cdots,x^N)
\enq
 and the matrix $G$ defined with the help of this vector,
\beq
G_{ij} := \av{v_iv_j} := \int dx \int dy\:p_X(x)p_Y(y) v_iv_j.
\enq
Such a matrix of scalar products is called a Gram matrix. $G$ has by definition the block form
\beq
G = \bem M_Z & BM_X \\ M_XB^T & M_X \enm.
\enq
It is well known that any Gram matrix is positive : from the definition of $G$, we have that $u^TGu = \av{(u\cdot v)^2} \ge 0 $ for any vector $u$.
Using the above fact \eqref{fact} we have
\beq
M_Z \ge BM_XB^T.
\enq
Equivalently
\beq
M_X \le B^{-1}M_Z B^{-T}.
\enq
Taking the inverse gives
\beq
M_X^{-1} \ge B^TM_Z^{-1}B,
\enq
which concludes the proof.
\subsection{Poissonian discreteness effects \label{poissoneffects}}
The same relations hold for another  source of noise relevant in cosmological surveys, i.e. discreteness effects due to a finite number of tracers of the underlying fields. A common parametrisation is the Poisson model, where the observed number of tracers in a cell is given by a Poisson variable with intensity the value of the underlying continuous field $x$ at that point. Explicitly, with $N = (N_1,\cdots,N_d)$ the number of tracers in $d$ cells, we set their joint probability to be
\beq
p_N(N_1,\cdots,N_d) = \int d^dx\: p_X(x,\btheta)\prod_{i = 1}^d e^{- x_i} \frac{x_i^{N_i}}{N_i !}.
\enq
This model has the peculiar property of transforming the moments of $X$ to factorial moments : the falling factorial in $d$ variables\footnote{see \eqref{fallingfactorial} for the falling factorial in one dimension} becomes
\beq
(N)_{\vecm} = \prod_{i = 1}^d (N_i)_{m_i}
\enq
and we have indeed the known curious relation
\beq
\av{N_{(\vecm)}} = \av{x^\vecm} = m^X_\vecm.
\enq
Using this relation, it is easy to prove with the same methods than above that
\beq
F^N_{\le M} \le F_{\le M}^X
\enq
for any $M$.
This does not appear to be necessarily the case for more generic functional form of $p_N(N | x)$, since there are no obvious relations between the first $M$ moments of $N$ and those of $X$.

\section{Some exactly solvable models for moment determinate densities\label{determ}}
We derive in this section the exact analytical expressions  for the information coefficients at all orders of well known families of moment determinate probability density functions. We will deal with the normal distribution, the beta and gamma families as well as an extended Poisson model. For all these instances, the matrices $F_{\le N}$ converges to $F$ as $N\rightarrow \infty$.
\paragraph{Normal distribution}
The normal distribution provides us with a very simple illustration of the approach. Its probability density function is
\beq
p(x,\mu,\sigma^2) := \frac{1}{\sqrt{2\pi}\sigma}\exp\lp -\frac 12 \frac{\lp x - \mu\rp^2}{\sigma^2} \rp.
\enq
Its Fisher information matrix is well known,
\beq\label{Fgauss}
F = \bem \frac{1}{\sigma^2} &  0 \\  0 & \frac{1}{2\sigma^4} \enm.
\enq
The information coefficients take the form
\beq \label{smuG}
s_1(\mu) = \frac 1 \sigma, \quad s_n(\mu) = 0, \quad n \ne 1,
\enq
and
\beq\label{ssigG}
s_2(\sigma^2) =  \frac{1}{\sqrt{2}\sigma^2},\quad s_n(\sigma^2) = 0, \quad n \ne 2.
\enq
We recover already the full matrix with the first two moments, $F_{\le 2} = F$. The matrix is diagonal 
because there is no order $n$ for which both coefficients $s_n(\mu)$ and $s_n(\sigma^2)$ do not vanish.
\paragraph{Derivation :}
It is easily seen that the score function $\partial_\mu \ln p$ is a polynomial of first order in $x$. This implies immediately that only $s_1(\mu)$ is non-zero. For any probability density function and parameter $\alpha$, it holds from \eqref{def} that
\beq
s_1(\alpha) = \frac{\partial m_1}{\partial \alpha}\frac 1 \sigma,
\enq
which proves \eqref{smuG}. Very similarly, the score function associated to $\sigma^2$ is a polynomial of second order, such that only the first two coefficients can possibly be non-zero. However, $s_1(\sigma^2)$ vanishes since $\sigma^2$ does not impact the mean. Finally, $s_2(\sigma^2)$ can be gained by noting that the polynomials orthogonal to the normal 
distribution are the Hermite polynomials \citep{Szego39},
\beq
P_n(x) = H_n\lp \frac{x-\mu}{\sigma}\rp,
\enq
with normalisation $h_n = 1 / n! $. From $H_2(x) = x^2 - 1$
and equation (\ref{def}) follows $C_{22} = 1 / \sigma^2$ and $s_2(\sigma^2) = \frac{1}{\sqrt{2}\sigma^2}.$
\paragraph{Beta distribution}
The beta distribution is defined as
\beq
p(x,\alpha,\beta) := \frac{x^{\alpha - 1}\lp 1 - x\rp ^{\beta-1}}{B(\alpha,\beta)},\quad 0 < x < 1,\:\:\alpha,\beta > 0
\enq
where $B(\alpha,\beta)$ is the beta integral, which will also enter the following section on the gamma distribution. It has the well known representation in terms of the gamma function $\Gamma$,  $B(\alpha,\beta) = \Gamma(\alpha)\Gamma(\beta) / \Gamma(\alpha + \beta)$.
It plays a fundamental role in order statistics.
\newline
\newline
The full Fisher information matrix can be conveniently expressed in terms of the second derivative of the logarithm of the gamma function, called the trigamma function $\psi_1$ \cite[p. 258-260]{abramowitz70a},
\beq \label{Fbeta}
F = 
\bem 
\psi_1(\alpha) & 0 \\ 0 & \psi_1(\beta)
\enm
- \psi_1(\alpha + \beta)
\bem 1 & 1 \\1 & 1
\enm,
\enq
Our derivation of the information coefficients associated to $\alpha$ and $\beta$, based on the explicit expressions of the orthogonal polynomials, in this case the Jacobi polynomials, is rather lengthy. We defer it to the appendix.
\newline
\newline
The result is
\beq
s_n(\alpha) = - \frac {(-1)^{n}}{n} \lp \frac{(\beta)^{(n)}n!}{(\alpha)^{(n)}(\alpha + \beta)^{(n)}} \rp^{1/2} \lp \frac{2n + \alpha + \beta -1}{n + \alpha + \beta -1} \rp^{1/2}.
\enq
and
\beq
s_n(\beta) = - \frac {1}{n} \lp \frac{(\alpha)^{(n)}n!}{(\beta)^{(n)}(\alpha + \beta)^{(n)}} \rp^{1/2} \lp \frac{2n + \alpha + \beta -1}{n + \alpha + \beta -1} \rp^{1/2},
\enq
where $\lp x \rp^{(n)} := x(x+1)\cdots (x + n -1)$ is the rising factorial. For the symmetric case $\alpha = \beta = 1/2$, these expressions simplify considerably. We have namely
\beq
s_n(\alpha) = (-1)^{n-1}\frac{\sqrt 2}{n},\quad s_n(\beta) = -\frac{\sqrt 2}{n}.
\enq
Since $\psi_1(1/2) = \pi^2 / 2$ and $\psi_1(1) = \pi^2 / 6$, a short calculation shows that we recover indeed, summing these coefficients, the full matrix given  in \eqref{Fbeta}
\beq
\left. F \right|_{\alpha = \beta = 1/2} = \frac{\pi^2}{6} \bem 2 & - 1 \\ -1  & 2 \enm.
\enq  
\paragraph{Gamma distribution}
The gamma distribution is a two parameter family,
\beq
p(x,\ln \theta,k) := \frac{e^{-x/\theta}}{\theta^k\Gamma(k)}x^{k-1} \quad k,\theta > 0,\quad x > 0.
\enq
where $\Gamma(k)$ is the gamma function. Special cases include the exponential distribution ($k = 1$), or the chi squared distribution with $n$ degrees of freedom ($k = n/2, \theta = 2$).
The calculation of the Fisher information matrix associated to $\ln \theta$ and $k$ is not difficult. Again, the trigamma function shows up,
\beq \label{Fgamma}
F = \bem k  & 1    \\ & \\ 1   & \psi_1(k) \enm. 
\enq
\newcommand{\lag}[2]{L_{#1}^{(#2)}}
The information coefficients are evaluated below, with the result
\beq
s_1(\ln\theta) = \sqrt{k},\quad s_n(\ln \theta) =0 ,\quad n \ne 1,
\enq
and
\beq
s_n( k) = (-1)^{n-1}\sqrt{ \frac{ B(k,n)}{n}}.
\enq
As a consistency check, we see that since $s_1( k) = \frac{1}{\sqrt k}$, we recover trivially the $\ln \theta \ln \theta$ and $\ln \theta k$ elements of the matrix  (\ref{Fgamma}).
Its $kk$ element implies that the following identity must hold,
\beq
\sum_{n = 1}^{\infty} \frac{B(k,n)}{n} =  \psi_1(k),\quad k > 0,
\enq
which reduces for $k = 1$ to Euler's famous formula $\sum_{n = 1}^\infty \frac{1}{n^2} = \frac{\pi^2}{6}$. 
\paragraph{Derivation :}
The polynomials associated to that distribution are the generalized Laguerre polynomials $L_n^{(k-1)}$\citep[p.775]{abramowitz70a} . More precisely, we have
\beq
\int_0^\infty t^{k-1} e^{-t} \lag n {k-1} (t)\lag m {k-1}(t) = \frac{\Gamma(n + k)}{n!}\delta_{mn}.
\enq
For this reason, the polynomials orthogonal to the gamma distribution are
\beq
P_n(x) := \lag n {k-1}\lp \frac{x}{\theta} \rp,
\enq
with normalisation
\beq \label{norm}
h_n = \frac{\Gamma(n + k)}{\Gamma(k)}\frac{1}{n!} = \frac 1 n \frac 1 {B(k,n)}. 
\enq
They have the explicit matrix elements $C_{ni}$\footnote{We added the factor $(-1)^n$ to the conventions of \cite{abramowitz70a}, in accordance with our own conventions of having a positive leading coefficient $C_{nn}$.},
\beq 
C_{ni} = (-1)^{n-i} \bin {n+k-1}{n-i} \frac{\theta^{-i}}{i!}.
\enq
The moments of the gamma function are given by
\beq
m_n = \theta^n\frac{\Gamma(n + k)}{\Gamma(k)}.
\enq
The Fisher information on $\theta$ is the simplest. It holds namely that the score function associated to $\theta$ is a polynomial first order in $x$, and therefore that only $s_1(\ln \theta)$ is non-zero. It follows
\beq
s_1(\ln\theta) = \sqrt{k},\quad s_n(\ln \theta) =0 ,\quad n \ne 1.
\enq
The calculation of $s_n( k)$ requires a little bit more work, but we can make use of previous results we derived when dealing with the beta family. The derivatives of the moments with respect to $k$ are given by
\beq
\frac{\partial \ln m_n}{\partial k} = \psi_0(k + n) - \psi_0(k).
\enq
Using the representation (\ref{def}), together with 
\beq
\begin{split}
\bin {n + k -1}{n -i}\frac{1}{i!}\frac 1 {\Gamma(k + i)} = \frac{\Gamma(n + k)}{\Gamma(n + 1)} \bin n i.
\end{split}
\enq
one obtains the following expression
\beq
\begin{split}
s_n(k)  &=  \frac {1} { \sqrt {n\:B(k,n)}}\sum_{i = 0}^n (-1)^{n-i} \bin ni \lp  \psi_0(k + i) - \psi_0(k) \rp.
\end{split}
\enq
The sum is precisely the function $f_q(t)$ defined in (\ref{defft}), evaluated a $t = 1$ and $k = q$, and that we proved in (\ref{ftbeta}) to be a representation of the beta function,
\beq
f_k(1) = (-1)^{n-1} B(k,n).
\enq
We conclude therefore
\beq
s_n( k) = (-1)^{n-1}\sqrt{ \frac{ B(k,n)}{n}}.
\enq
\paragraph{Poisson model}
Consider the probability density function
\beq
p(x,\mu,\ln A) = \sum_{n = 0}^\infty e^{-\mu} \frac{\mu^n}{n!} \delta^D(x - An),\quad \mu, A > 0
\enq
with $\delta^D$ the Dirac delta function. This is the usual Poisson law, together with some amplitude $A$ that is left free. If $A$ is set to unity, we recover the Poisson law.
This is a peculiar situation in terms of Fisher information. Unlike the previous situations, the parameter $A$ impacts indeed the range of $x$. The Fisher information on $A$ is not well defined, formally infinite.
\newline
\newline
Even though the Fisher information matrix is not well defined, the information coefficients as given in equations (\ref{def}) or (\ref{sn}) are still meaningful, and so are their interpretation (\ref{chi2}) in terms of the expected curvature of a $\chi^2$ fit. 
The result is
\beq
s_1(\ln \mu) = \sqrt{\mu},\quad s_n(\ln \mu) = 0,\quad n \ne 1.
\enq
and
\beq
\label{snlnA}
\begin{split}
s_1(\ln A) &= \sqrt{\mu}, \quad s_n(\ln A) = (-1)^n \frac{\sqrt {n!}}{n(n-1)}\mu^{1-n/2},\quad n \ge 2 .
\end{split}
\enq
Note that it is compatible with the fact that the Fisher information on $A$ is infinite. For any $\mu$ the sum
\beq
\sum_{n = 1}^\infty s_n^2(\ln A)
\enq
is divergent. The turnover of the coefficients occurs around $n \approx \mu$ where they start to increase. Atypically, the higher order the moment the most interesting it becomes for inference on $A$ in this model.
\paragraph{Derivation :}
Consider the Charlier polynomials \cite[pp.788]{abramowitz70a}, defined by
\beq \label{Charlier}
c_n(x) := \sum_{k = 0}^{n}(-1)^{n-k}\bin nk \mu^{-k}(x)_k,
\enq
where
\beq \label{fallingfactorial}
x(x-1)\cdots(x-m+1) =: (x)_m
\enq
is the falling factorial. These are the polynomials orthogonal with respect to the Poisson distribution with intensity $\mu$. Clearly, the polynomials
\beq \label{Ppoisson}
P_n(x) := c_n\lp \frac x A\rp
\enq
are then the polynomials orthogonal to $p(x,\mu,\ln A)$. The normalization is
\beq \label{normalisation}
h_n = \frac{n!}{\mu^n}.
\enq
On the other hand, the moments are given by $m_n =  A^nm^P_n$,
where $m^P_n$ is the $n$th moment of the Poisson distribution.
For this reason, it holds that $\frac{\partial m_n}{\partial \ln A} = n \:m_n$.
This equation, together with (\ref{Ppoisson}) implies that the information coefficients on $\ln A$ are independent of the actual value of $A$, since $C_{nk} \propto A^{-k}$ and $m_k \propto A^k$. So are the coefficients $s_n(\ln \mu)$. We can therefore from now on safely chose $A = 1$ as the fiducial value, and work with the usual Poisson distribution and associated Charlier polynomials.
\newline
\newline
The coefficients associated to $\ln \mu$ are the simplest to obtain. It is not difficult to see that the score function associated to $\ln \mu$ of the Poisson distribution is a polynomial of first order. We obtain $s_1(\ln \mu) = \sqrt{\mu}$
and $s_n(\ln \mu) = 0,\quad n \ne 1$.
Turning to $\ln A$, we need to evaluate
\beq
s_n(\ln A) =\frac{\mu^{n/2}}{\sqrt{n!}} \sum_{k = 0}^n C_{nk}\:k \:m_k.
\enq 
The derivation we propose uses extensively the technique of generating functions \citep{Wilf_1994}, and is a variation on the following theme : consider the known expression for the factorial moments of the Poisson distribution,
\beq\label{facmoment}
\av{n(n-1)\cdots (n-m +1)}  = \mu^m.
\enq
Even though it may appear mysterious at first sight, a simple way to prove this identity is to notice
\beq
\av{n(n-1)\cdots (n-m +1)}  =  \left.\frac{d^m}{dt^m}G(t)\right|_{t = 1},\quad G(t) := \sum_{n = 0}^{\infty}p(n)t^n,
\enq
where $G(t)$ is called the probability generating function, and to try and evaluate the right hand side. This is indeed very convenient, since the probability generating function of the poisson law takes a very simple form,
\beq
G(t) =  e^{-\mu}\sum_{n = 0}^{\infty}\frac{\mu^n}{n!}t^n  = e^{\mu \lp t -1\rp }.
\enq
The derivatives are trivial to evaluate, proving (\ref{facmoment}). We will need the following result, that comes out directly from an essentially identical argument,
\beq \label{avfallingfactorial}
\av{(an)_m}  = \left.\frac{d^m}{dt^m}e^{\mu(t^a-1)}\right|_{t = 1},
\enq
where $a$ is some number.
\newline\newline
We now turn to the evaluation of
\beq
\sum_{k = 0}^nC_{nk}k\:m_k.
\enq
This sum can be written as the first derivative with respect to $a$ evaluated at $a = 1$ of the following expression
\beq
\av{c_n(ax)} = \sum_{k = 0}^\infty  \frac{e^{-\mu}\mu^k}{k!}c_n(ak),
\enq
as can be seen directly from an expansion of the Charlier polynomials $c_n$ in terms of the coefficients $C_{nk}$. Using equation (\ref{Charlier}) and our result (\ref{avfallingfactorial}), we obtain
\beq \label{eq}
 \sum_{k = 0}^nC_{nk}k\: m_k = e^{-\mu} \sum_{k = 0}^n(-1)^{n-k}\mu^{-k}\bin nk \left.\frac{d}{da}\frac{d^k}{dt^k}\exp\lp \mu t^a \rp\right|_{t,a= 1}.
\enq
The derivative with respect to $a$ is easily performed :
\beq
\left.\frac{d}{da}\exp\lp \mu t^a \rp\right|_{a= 1} =\mu\:  e^{\mu t} \:t \ln t.
\enq
Also, using Leibniz rule for derivatives of products, we can write
\beq
\left. \frac{d^k}{dt^k}\exp\lp \mu t \rp \:t \ln t \right |_{t= 1} = e^\mu\sum_{i = 0}^k \bin  k i \mu^{k-i} \left.\frac{d^i}{dt^i} \:t \ln t\right|_{t = 1}.
\enq
The factor of $\mu^k$ cancels in equation (\ref{eq}), so that
\beq
 \sum_{k = 0}^nC_{nk}k\: m_k  = \mu \sum_{k =0}^{n} \sum_{i =0}^{k}\bin nk \bin ki (-1)^{n-k}\mu^{-i}\left.\frac{d^i}{dt^i} \:t \ln t\right|_{t = 1}.
\enq
From the properties of the binomial coefficients follows that the sum over $k$ is just $(-1)^n\delta_{ni}$. Therefore,
\beq
s_n(\ln A) =\mu \frac{\mu^{-n/2}}{\sqrt {n!}} \left.\frac{d^n}{dt^n} \:t \ln t\right|_{t = 1}.
\enq
The derivatives are easily computed :
\beq
\left.\frac{d^n}{dt^n} \:t \ln t\right|_{t = 1} = \begin{cases} 0 & n = 0 \\ 1 & n = 1 \\ (-1)^n(n-2)! & n \ge 2 \end{cases}
\enq
with the final result given in equation \eqref{snlnA}.
\subsection{Moment indeterminate densities\label{indeterm}}
We now turn to indeterminate distributions, i.e. those that cannot be uniquely recovered from their moments.
In this case, the Fisher information content of the moments will generally not converge to the total amount. In this respect, the limit
\beq
\lim_{N \rightarrow \infty} \frac{\sum_{n = 1}^{N}s_n^2(\alpha) }{I(\alpha)},
\enq
a number in $(0,1]$, may be thought of as an indirect measure of the amount of indeterminacy of the moment problem. A value of unity means that the moments do carry the same information on the parameter $\alpha$ as the probability density function, while for values close to zero a large amount of information on $\alpha$ is not present in the hierarchy.  The quantity $\sqrt{\epsilon(\alpha)}$ has via the information inequality the interpretation of the ratio of the expected constraints on $\alpha$ by extraction of the full set of moments to the best constraints achievable with the help of unbiased estimators of $\alpha$.
When more than one parameter are of interest, this definition is not really satisfactory anymore, since it is tied  to $\alpha$ exclusively. One way to get around this is by defining a ratio of determinants,
\beq
\epsilon := \lim_{N \rightarrow \infty} \epsilon_N = \lim_{N \rightarrow \infty}  \det F_{\le N}  / { \det  F   },
\enq
where $F_{\le N}$ was defined in equation \eqref{Fn}. Clearly, $\epsilon_N$ is the cumulative efficiency to catch the information. Since the matrices in the numerator and denominator do transform in the same way under any smooth reparametrisation of the parameters, or coordinates, of the family, this ratio is left unchanged under such transformations. Thus, $\epsilon$  and $\epsilon_N$ are well defined quantities associated to that family of distributions, independently of the chosen coordinates. 
\newline
\newline
Naively, one might expect this lost information to reflect in some way the freedom there is in the choice of a distribution with the very same series of moments. We will therefore compare our results to a measure of this freedom, $D_S$, that was proposed recently \citep{Stoyanov04}
 as the index of dissimilarity of a Stieltjes class. A Stieltjes class is a family of  probability density functions of the form
 \beq
 p_\delta(x) = p(x)\lp1 + \delta h(x)\rp,\quad |h(x)| \le 1,\:\: |\delta| \le 1,
 \enq
 all having the same moment series as the central probability density function $p$.
 The index $D_S$, defined as
 \beq
 \int dx \:p(x) |h(x)|,
 \enq
 in $[0,1]$ is the maximal distance between two members of the class, being zero for determinate moment problems, where $h$ must identically vanish.
 \newline
 \newline
 We will consider three types of probability density functions, the lognormal, which is indeterminate for any values of its parameter space, as well as the Weibull distribution and the stretched exponential, which are indeterminate solely for parts of their parameter range.
For these last two distributions we could not find in all cases exact analytical expressions for the coefficients $s_n$. We generated them therefore numerically, using a fine discretization procedure to obtain the orthogonal polynomials, following the exposition in \cite{Gautschi04}.
\paragraph{Lognormal distribution}
The lognormal distribution has the following functional form,
\beq
p(x,\mu_Y,\sigma_Y) = \frac{1}{\sqrt{2\pi}\sigma_Y}\frac 1 x \exp\lp -\frac1 2 \frac{\lp\ln x - \mu_Y\rp^2 }{\sigma_Y^2} \rp, \quad x > 0
\enq
The parameters $\mu_Y$ and $\sigma_Y^2$ are the mean and the variance of $ y = \ln x$ which is normally distributed. They are related to the mean $\mu$ and variance $\sigma^2$ of $x$ by the relations
\beq
\begin{split}
\mu_Y &= \ln \mu - \frac 12 \ln \lp 1 + \frac{ \sigma^2}{ \mu^2} \rp \\
\sigma^2_Y &= \ln \lp 1 + \frac{\sigma^2}{\mu^2} \rp.
\end{split}
\enq
The only relevant parameter for our purposes is the reduced variance $\sigma^2 / \mu^2$, in terms of which our results can be expressed. The information coefficients associated to the parameters $\ln \mu$ and $\sigma^2 / \mu^2$ we will derive in details in chapter \ref{ch4}, (see also \cite{2011ApJ...738...86C}), with the help of $q$-series. These are given by 
\beq\label{snln}
\begin{split}
s_n(\ln \mu)& = (-1)^{n-1} \lp  \frac{q^n}{1-q^n} \rp^{1/2} \sqrt{\pochh q q {n-1}} \\
s_n(\sigma^2 / \mu^2) &= q(-1)^n\lp  \frac{q^n}{1-q^n} \rp^{1/2} \sqrt{\pochh q q {n-1}} \lp\sum_{k = 1}^{n-1} \frac{q^k}{1-q^k} \rp.
\end{split}
\enq
In these equations,
\beq
q := \frac{1}{1 + \sigma^2/\mu^2}
\enq
and
\beq
\pochh q q n := \prod_{k = 1}^n\lp 1 -q^k \rp
\enq
is the q-Pochammer symbol \cite{Andrews99}.
According to the chain rule of derivation, we obtain $s_n(\mu_Y) = s_n(\ln \mu)$ and $s_n(\sigma_Y^2) = \frac 12(s_n(\ln \mu) + s_n(\sigma^2/\mu^2))$. On the other hand, since $\ln x$ is normally distributed, the total Fisher information matrix $F$ associated to $\mu_Y$ and $\sigma^2_Y$ is diagonal with $1/\sigma^2_Y$ and $1/ 2\sigma^4_Y$ as diagonal elements. It follows that in these coordinates  $\det F = 1/(2\sigma^6_Y)$.
\newline
\newline
The Stieltjes class we consider is set by
\beq \label{Stln}
h(x) = \sin\lb \frac{2\pi}{\sigma_Y^2} \lp \ln x - \mu_Y \rp \rb.
\enq
Others are given in \cite{Heyde63} but are equivalent to a uninteresting rescaling of $\sigma_Y$.
After an obvious variable substitution, the dissimilarity index becomes
\beq \label{DSln}
D_S = \frac{1}{\sqrt{2\pi}}\int_{-\infty}^\infty \exp \lp-\frac{x^2}{2} \rp\left | \sin\lp \frac{2\pi}{\sigma_Y}x \rp\right|\: dx. 
\enq
\begin{figure}[htbp]
\begin{center}
\includegraphics[width= 0.8\textwidth]{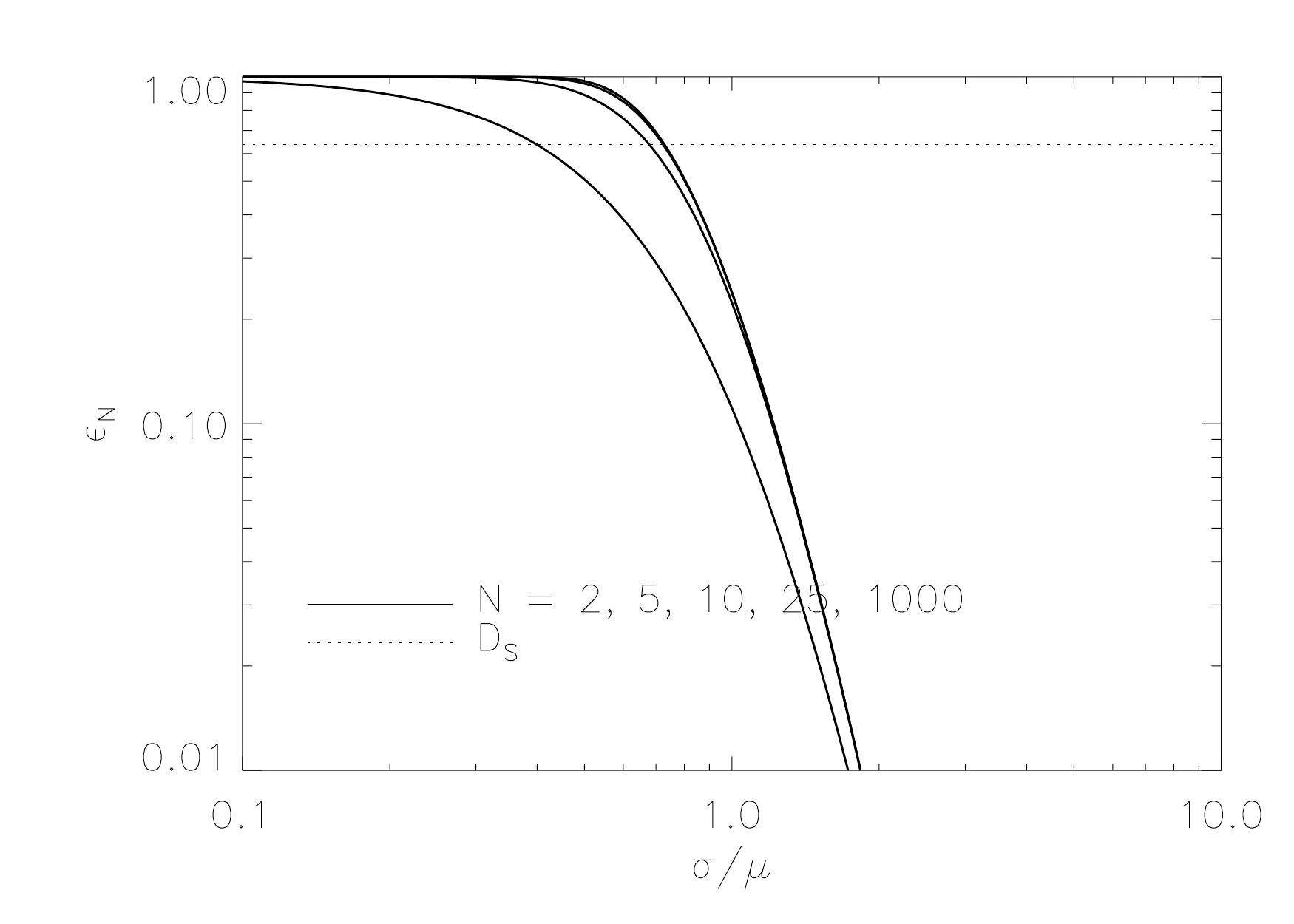}
\caption[The cumulative efficiencies of the moments of the lognormal distribution to capture its information content ]{Solid lines : the cumulative efficiencies $\epsilon_N$, for $N = 2,5,10,25$ and $1000$, from bottom to top, as function of the reduced variance of the lognormal distribution. Dashed : The dissimilarity index for the Stieltjes class given in  equation \eqref{Stln}.}
\label{fig:ln}
\end{center}
\end{figure}
In figure \ref{fig:ln}, we show the cumulative efficiency $\epsilon_N$, for $N = 2,5,10,15$ and $1000$ for the lognormal family, evaluated with the exact expressions given in \eqref{snln}, solid lines from bottom to top, together with the dissimilarity index $D_S$, shown as the dashed line, evaluated by numerical integration of equation \eqref{DSln}.
It is not difficult to see from the above expressions that the coefficients $s_n$ decay exponentially for large $N$, such that the convergence is rather quick over the full range.
We observe a very sharp transition in $\epsilon$ as  soon as the reduced variance approaches unity, from a regime where the entire information content is within the second moment, as for the normal distribution (and thus where the indeterminacy of the moment problem is irrelevant for parameter estimation),  to the opposite regime where all moments completely fail to capture the information. This cutoff is discussed at length in chapter \ref{ch4}.  On the other hand. the index $D_S$ is seen to be remarkably constant over the range shown, equal to its limiting values for $\sigma_Y \rightarrow 0$, which can be evaluated from \eqref{DSln} to be $2/\pi$.
\paragraph{Weibull distribution}
We consider now the Weibull distribution with shape parameter $k$ and scale parameter $\lambda$,
\beq
p(x,\lambda,k)= \frac k \lambda  \lp\frac{x}{\lambda}\rp^{k-1}e^{- (x/\lambda) ^k}, \quad x,\lambda,k > 0.
\enq
The variable $x$ can be seen as a power of an exponentially distributed variable with intensity unity, $
p(t) = e^{-t}$, for $ t =  \lp \frac x \lambda \rp^{k}$. It is known that the moment problem  associated to the moments of $x$ is determinate for $k \ge 1/2 $ and indeterminate  for $k < 1/2$ \cite[section 11.3 e.g.]{Stojanov87}. A Stieltjes class in the latter regime is provided in the same reference, which after some algebraic manipulations reduces to
\beq \label{StW}
h(x) = \sin \lp c_k (x/\lambda)^k - k\pi \rp,\quad c_k= \tan\lp  k \pi \rp,
\enq
The index $D_S$ becomes
\beq
D_S = \int_0^\infty e^{-t} \left| \sin \lp c_k \: t - k\pi\rp  \right|\:dt.
\enq
It is interesting to note that this integral can be performed analytically, for instance with the help of partial integration, with the result
\beq
D_S = \sin\lp k\:2\pi\rp \frac{e^{-k\pi / c_k}}{1 - e^{-\pi/c_k}}.
\enq
\begin{figure}[htbp]
\begin{center}
\includegraphics[width = 0.8\textwidth]{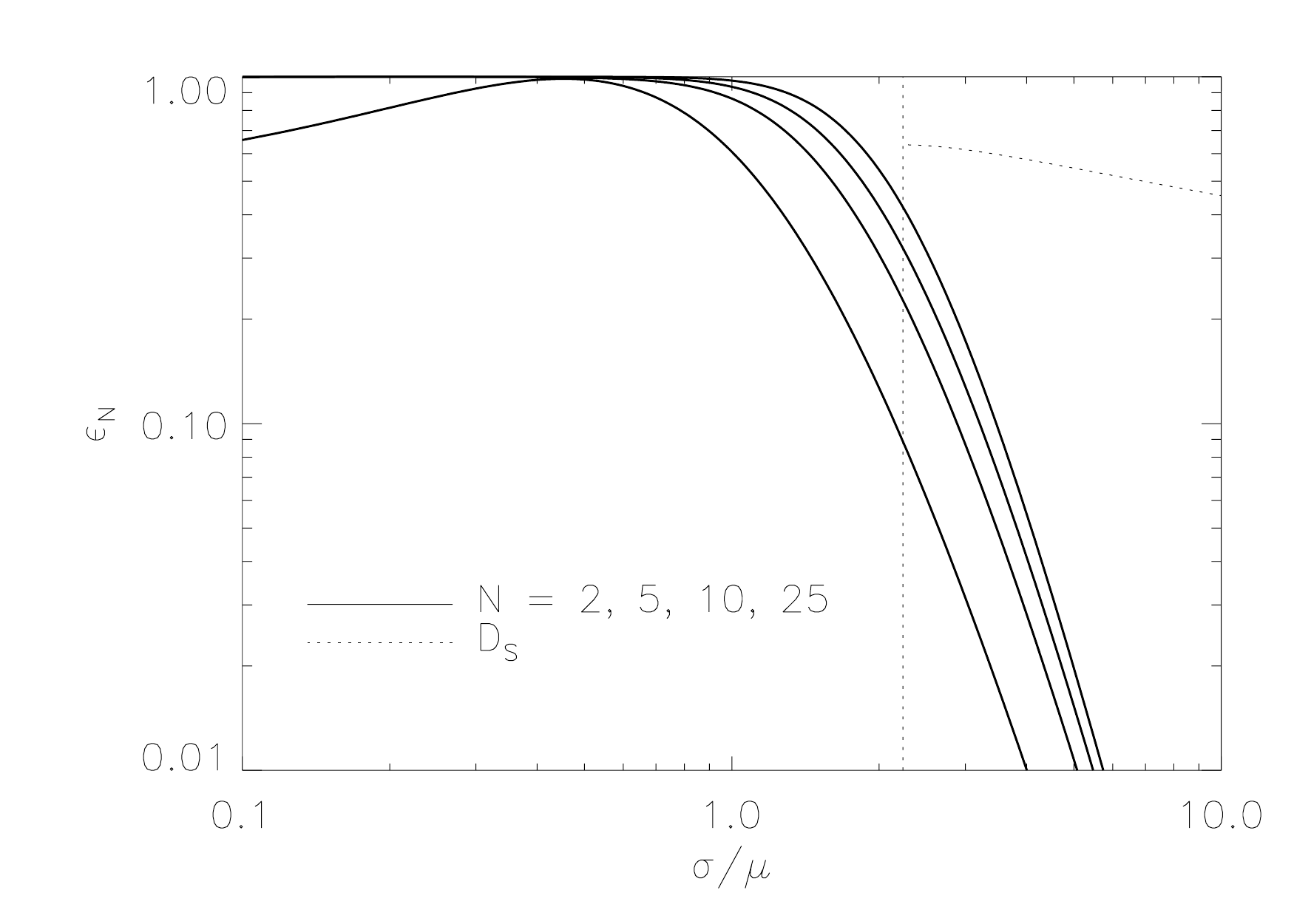}
\caption[The cumulative efficiencies of the moments of the Weibull distribution to capture its information content]{Solid lines : the cumulative efficiencies $\epsilon_N$, for $N = 2,5,10,25$, from bottom to top, as function of $\sigma/\mu$ for the Weibull distribution. Dashed : The dissimilarity index for the Stieltjes class given in  equation \eqref{StW}. The vertical dashed line, corresponding to $k = 1/2$, separates the regimes of determinacy and indeterminacy of the associated moment problem.}
\label{fig:W}
\end{center}
\end{figure}
We evaluated $\epsilon_N$ and the associated $F_{\le N}$ numerically, using the convenient coordinates  $\ln \lambda$ and $\ln k$. It is not difficult to see that $\epsilon_N$ is independent of $\lambda$. The full information matrix for these parameters can be evaluated analytically, with the result $\det F = k\pi^2/6$.  The results for $\epsilon_N$ are shown in figure \ref{fig:W}, that we present as for the lognormal as function of the reduced variance
\beq
\frac \sigma \mu = \lp \frac{\Gamma \lp 1 + 2/k \rp}{\Gamma^2 \lp 1 + 1/k \rp} - 1 \rp^{1/2}.
\enq
The dashed vertical line corresponding to $k = 1/2$ separates the two regimes where, on the left, $\epsilon$ is unity and $D_S$ zero since the moment problem is determinate, and on the right, where the moment problem is indeterminate. The scale $\sigma/ \mu = 1$ corresponds to $k = 1$.\newline
While the decay of $\epsilon_N$ is also very sharp in the indeterminate regime, very slow convergence of $\epsilon_N$ is seen to occur in the large reduced variance regime unlike for the lognormal distribution. This also in the region $1 < k <2 $, which corresponds to the phase where the Weibull distribution goes from a unimodal to a monotonically decreasing distribution, but $\epsilon$ still is unity, since the moment problem still is determinate. For instance, for $k = 1$, the following exact result can be gained with the same methods as exposed in this section,
\beq
\left. \epsilon_N \right|_{k = 1} = \lp \sum_{n = 1}^{N-1}\frac 1 {n^2}  \rp\frac 6 {\pi^2}.
\enq
\paragraph{Stretched exponential function}
Finally, we treat the case of the stretched exponential,
\beq
p(x,\lambda,k) =  \frac{k}{\lambda \Gamma(1/k)} e^{-(x/\lambda)^k}, \quad x \ge 0.
\enq
Just as for the Weibull distribution, the moment problem  associated to the moments of $x$ is determinate for $k \ge 1/2 $ and indeterminate  for $k < 1/2$ \cite[section 11.4]{Stojanov87}
A Stieltjes class is given by
\beq \label{StS}
h(x) = \sin\lp c_k (x/\lambda)^k\rp,\quad x \ge 0,\:k > 0
\enq
where $c_k$ is as above given by $c_k = \tan \lp k \:\pi\rp$.
\begin{figure}[htbp]
\begin{center}
\includegraphics[width= 0.8\textwidth]{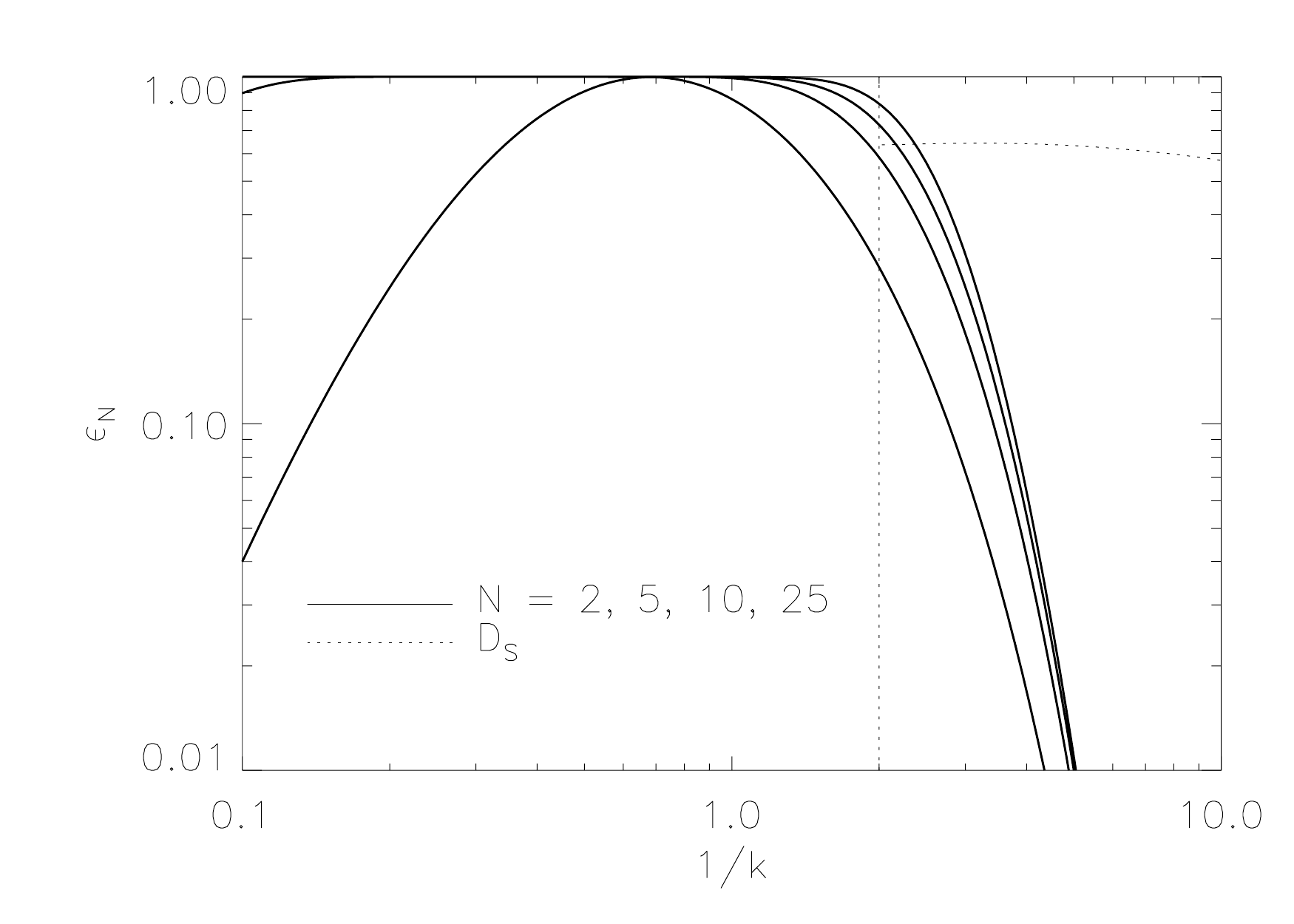}
\caption[The cumulative efficiencies of the moments of the stretched exponential function to capture its information content]{Solid lines : the cumulative efficiencies $\epsilon_N$, for $N = 2,5,10,25$, from bottom to top, as function of the inverse shape parameter $1/k$ of the stretched exponential. Dashed : The dissimilarity index for the Stieltjes class given in  equation \eqref{StS}. The vertical dashed line at $k = 1/2$ separates the regimes of determinacy and indeterminacy of the associated moment problem.}
\label{fig:S}
\end{center}
\end{figure}
Numerical evaluation of $\epsilon_N$, and of the dissimilarity index is shown in fig \ref{fig:S}. As for the Weibull distribution, these results are independent of the scale parameter $\lambda$. We can conclude that in none of the situations we investigated is $D_S$ a good tracer of the importance of the indeterminacy of the moment problem for parameter inference. 

\clearpage
\newpage
\section{Appendix}
\subsection{Derivation for the beta family}\label{appendixbeta}
In the following we will need the first two derivatives of the logarithm of the gamma function. These are called the digamma $\psi_0$ and trigamma $\psi_1$ functions respectively \cite[p. 258-260]{abramowitz70a},
\beq
\begin{split} \label{ditrigamma}
\psi_0(x) &= \frac{d}{dx}\ln \Gamma(x) \\
\psi_1(x) &= \frac{d^2}{dx^2}\ln \Gamma(x).
\end{split}
\enq
The Fisher information matrix can be gained by noting that by differentiation under the integral sign, we have
\beq
\begin{split}
\Fab = \frac{\partial^2\ln B(\alpha,\beta)}{\partial\alpha\partial\beta} .
\end{split}
\enq
Using the representation of the beta integral in terms of the Gamma function, we conclude that
\beq
F = 
\bem 
\psi_1(\alpha) & 0 \\ 0 & \psi_1(\beta)
\enm
- \psi_1(\alpha + \beta)
\bem 1 & 1 \\1 & 1
\enm.
\enq
In order to obtain the information coefficients $s_n(\alpha)$ and $s_n(\beta)$, it is more convenient to start with $s_n(\beta)$. $s_n(\alpha)$ will then be gained effortlessly by looking at the symmetry of the problem.
\newline\newline
From the definition of the beta integral, the moments of the beta distribution are given by
\beq
m_k = \frac{B(\alpha + k, \beta)}{B(\alpha,\beta)}.
\enq
The derivatives of the moments with respect to $\beta$ are given by
\beq \label{dermoments}
\begin{split}
\partial_\beta \ln m_k &= - \psi_0(\alpha + \beta + k) +  \psi_0(\alpha + \beta).
\end{split}
\enq
\newcommand{\Jac}[2]{P_n^{(#1-1,#2 - 1)}}
\newcommand{\Gj}[2]{G_n^{(#1,#2)}}
\noindent
The orthogonal polynomials are the Jacobi polynomials $\Gj \alpha \beta (x)$ \citep[page 774]{abramowitz70a}. Instead of the parameters $p,q$ used in \cite{abramowitz70a}, we stick to $\alpha$ and $\beta$, which are more appropriate for our purposes \footnote{We have in the notation of  \cite{abramowitz70a} $q = \alpha, p = \alpha + \beta -1$}. These polynomials are proportional to the Jacobi polynomials $\Jac \alpha \beta (2x -1)$ \cite[chap. IV]{Szego39}, orthogonal on the interval $(-1,1)$.
Their matrix elements of $G_n$ are given explicitly by
\beq
C_{nk} = \frac{\Gamma(\alpha + n)}{\Gamma(\alpha + \beta + 2n - 1)} (-1)^k\bin nk \frac{\Gamma(\alpha + \beta + k +n -1 )}{\Gamma(\alpha + k)}.
\enq
We first note the following relation,
\beq \label{cnkmk}
C_{nk}m_k = \frac{\Gamma(\alpha + n)\Gamma(\beta)}{\Gamma(2n + \alpha + \beta -1)B(\alpha,\beta)} (-1)^{n-k}\bin nk  \lp \alpha + \beta + k\rp^{(n-1)},
\enq
where
\beq
(x)^{(n)} = x(x+1)\cdots(x + n - 1)
\enq
is the rising factorial. Since $s_n(\beta)$ is given by
\beq\label{s_nb}
s_n(\beta) = \sum_{k = 0}^n C_{nk}m_k \partial_\beta \ln m_k,
\enq
the evaluation of the following sum is necessary, 
\beq
A_n(q) :=\sum_{k = 0}^n\bin nk (-1)^{n-k}\lp q + k\rp^{(n-1)} \lp \psi_0(q + k) - \psi_0(q) \rp,\quad q := \alpha + \beta > 0.
\enq
The following paragraphs are dedicated to the lengthy but straightforward proof of the following result,
\beq\label{anq}
A_n(q) = \frac{(n -1) !}{q + n -1}.
\enq
The proof consists of a number of steps.
\paragraph{Step 1 :} 
Using an always useful trick, we turn the rising factorial in the sum into a power and differentiate : from
\beq
(q + k)^{(n-1)} = (q + k)\cdots (q + k + n - 2) = \left.\frac{d^{n-1}}{dt^{n-1}} \right|_{t = 1} t^{k + q  + n - 2},
\enq
follows
\beq
\begin{split}
A_n(q) &= \left.\frac{d^{n-1}}{dt^{n-1}}\right|_{t = 1}  t^{q + n-2} \sum_{k = 0}^n\bin nk (-1)^{n-k}t^k (\psi(q + k) - \psi(q)) \\
&=:\left.\frac{d^{n-1}}{dt^{n-1}}\right|_{t = 1}  t^{q + n-2} f_q(t). \label{defft}
\end{split}
\enq
We then perform the derivatives using the Leibniz rule of derivation. Since
\beq
\left.\frac{d^{n - 1 - m}}{dt^{n-1-m}}\right|_{t = 1}  t^{q + n-2} = \frac{\Gamma(q + n - 1)}{\Gamma(q + m)},
\enq
one obtains
\beq \label{Anq}
A_n(q) = \Gamma(q + n -1)\sum_{m= 0}^{n-1}\bin {n-1}{m} \frac{ f_q^{(m)}(t = 1) }{\Gamma(q + m)}.
\enq
\paragraph{Step 2}
To obtain $f_q^{(m)}(1)$, we first construct an integral representation of the function $f_q(t)$. From the relation \cite[theorem 1.2.7]{Andrews99}
\beq
\psi_0(q + 1) = \frac 1 q + \psi_0(q)
\enq
of the digamma function, we note that
\beq
\psi_0(q + k) - \psi_0(q) = \sum_{j = 0}^{k-1}\frac{1}{q + j}.
\enq
Writing
\beq
\frac 1 {q + j } = \int_0^{1} dx\: x^{q + j -1},
\enq
the sum over $j$ becomes a geometric series.
We obtain therefore,
\beq
\psi_0(q + k) - \psi_0(q) = \int_0^1dx\:x^{q-1}\frac{1 - x^k}{1-x}.
\enq
We can now perform the sum over $k$ in $f_q(t)$. We have indeed
\beq
\sum_{k = 1}^{n}\bin nk (-1)^{n-k} t^k (1-x^k) = (t-1)^n + (-1)^n (1 - tx)^n.
\enq
It follows immediately
\beq \label{ftbeta}
f_q(t) = \int_0^1dx\: x^{q-1}\frac{(t-1)^n -(-1)^n(1-tx)^n}{ 1- x}.
\enq
We only need derivatives of that function evaluated at $t = 1$. The order of each derivative is $< n$. For this reason the term $(t-1)^n$ in this expression actually plays no role. Each such derivative is thus a beta integral :
\beq\label{bb}
f_q^{(m)}(1) = (-1)^{n + m - 1}n\cdots (n - m + 1)B(q + m, n - m),\quad m < n.
\enq
\paragraph{Step 3 :}
To go further, we use the following property of the beta integral,
\beq
B(x,y) = \frac{(x + y)^{(n)}}{(y)^{(n)}}B(x,y +n), 
\enq
which can be seen from its representation in terms of the gamma function, or from its integral representation \cite[page 5]{Andrews99}. Two applications of this rule leads to
\beq
B(q + m,n - m) =  B(q,n)\frac{(q)^{(m)}}{m !}\frac{1}{\bin {n-1} m}.
\enq
By combining this relation with (\ref{bb}), and using
\beq
(q)^{(m)} = \frac{\Gamma(q + m)}{\Gamma(q)},
\enq
we have shown
\beq
f^{(m)}(1) = (-1)^{n + m -1}B(q,n) \frac{\Gamma(q+m)}{\Gamma(q)} \bin n m \frac{1}{\bin {n-1}m }.
\enq
This form is very convenient, since many terms now cancel in equation (\ref{Anq}). We obtain
\beq
A_n(q) = -\underbrace{B(q,n)\frac{\Gamma( n + q -1)}{\Gamma(q)}}_{\frac{(n-1)!}{n + q -1}} \sum_{m = 0}^{n-1}\bin nm (-1)^{n-m}.
\enq
The sum over $m$ would vanish would it run up to $n$. Its value is therefore minus the $m = n$ term, which is unity. It follows
\beq
A_n(q) = \frac{(n-1)!}{n + q -1},
\enq
which was to be proved.
\newline\newline
Getting $s_n(\beta)$ requires now only to keep track of the normalization.
The normalization of the Jacobi polynomials $G_n$ \cite{abramowitz70a} is 
\beq
h_n  = \frac{n!}{B(\alpha,\beta)} \frac{\Gamma(\alpha + n)\Gamma(\alpha + \beta -1 + n)\Gamma(\beta + n)}{(\alpha + \beta - 1 + 2n)\Gamma^2(\alpha + \beta  -1 + 2n)}.
\enq
(Note that there is a additional factor of $1/B$ with respect to \cite{abramowitz70a} since there the measure is not normalized to unit integral).
We have from equations (\ref{dermoments}), (\ref{cnkmk}), (\ref{s_nb}) and (\ref{anq}) together with some algebra
\beq
s_n(\beta) = - \frac {1}{n} \lp \frac{(\alpha)^{(n)}n!}{(\beta)^{(n)}(\alpha + \beta)^{(n)}} \rp^{1/2} \lp \frac{2n + \alpha + \beta -1}{n + \alpha + \beta -1} \rp^{1/2}.
\enq
Tedious calculations are not needed to get $s_n(\alpha)$, but symmetry considerations are enough.
The Jacobi polynomials obeys the symmetry relation,
\beq
\Jac \alpha \beta(-x) = (-1)^n\Jac \beta \alpha (x),
\enq
and therefore
\beq
\Gj \alpha\beta(1-x) = (-1)^n\Gj \beta \alpha (x).
\enq
It is then not difficult\footnote{for instance from the representation (\ref{sn}) } to see that $s_n(\alpha)$ is proportional to $s_n(\beta)$ with $\alpha$ and $\beta$ exchanged.
We conclude
\beq
s_n(\alpha) = - \frac {(-1)^{n}}{n} \lp \frac{(\beta)^{(n)}n!}{(\alpha)^{(n)}(\alpha + \beta)^{(n)}} \rp^{1/2} \lp \frac{2n + \alpha + \beta -1}{n + \alpha + \beta -1} \rp^{1/2}.
\enq
\subsection{General hierarchical systems, recursion relations.\label{hierarchy}}
This chapter was focussed on the hierarchy of $N$-point moments, with associated orthogonal system the orthogonal polynomials. One of course expects the approach of this chapter to extend in some way to any system of observables. It is the aim of this section to discuss briefly the case of other hierarchical systems.
\newline
\newline
Given a density function $p(x,\btheta)$, where $x = (x_1,\cdots,x_d)$,
consider a set of functions,  (a \textit{hierarchy} $\lb f \rb$ ),
\beq
\lb f \rb = f_0(x),f_1(x),f_2(x),\cdots
\enq
which can be finite or infinite. The restrictions are $f_0(x) = 1$, the functions to be linearly independent with respect to the density $p(x,\btheta)$, as well as $\av{f_if_j}$ to be well defined.
\newline
\newline
The density provides us with a scalar product,
\beq
f \cdot g :=  \av{f(x)g(x)} = \int dx\: p(x,\btheta) f(x) g(x).
\enq
Consider now the orthonormal system built from $\lb f \rb$, following the Gram-Schmidt orthogonalisation procedure. 
Explicitly, they can be built recursively from
\beq \label{GS}
P_n(x) = \frac{f_n(x) - \sum_{k = 0}^n \av{f_nP_k}P_k(x)}{\sqrt{\av{ \lp f_n(x) - \sum_{k = 0}^n \av{  f_nP_k}P_k(x)  \rp^2}}}
\enq
By construction, we have indeed
\beq
\av{P_n(x)P_m(x)} = \delta_{mn}.
\enq
$P_n$ is a polynomial in the hierarchy members, in the sense that
\beq
P_n(x) = \sum_{k = 0}^nC_{nk}f_k(x),
\enq
for some matrix elements $C_{nk}$, with $C_{nn} > 0$, and $C_{nk} = 0, \quad k > n$.
\newline
\newline
Very useful is the following identity between matrices of size  $N + 1 \times N + 1$ :
\beq \label{cholesky}
C^TC = \lb M_N \rb^{-1},\quad \lb M_N \rb_{ij} = \av{f_i(x)f_j(x)},\quad i,j = 0,\cdots N.
\enq
\paragraph{Derivation} Consider the expansion of $f_k$ in terms of $P_n$ :
\beq
f_k(x) = \sum_{n = 0}^kB_{kn}P_n(x).
\enq
Such an expansion is always possible since the set of $P_n$ and $f_n$ span the same space by construction.
It follows $B_{kn} = \av{f_k P_n}$. Expanding in this relation $P_n$ with the help of the matrix $C$, we obtain
\beq
f_k(x) = \sum_{n = 0}^k\sum_{i,j = 0}^nC_{n i}C_{n j}\av{f_i f_k}f_j(x),\quad k = 0,1,\cdots, \forall x
\enq
Relation \eqref{cholesky} follows.
\newline\newline
We are writing now the moments of the hierarchy as $m_k := \av{f_k}$.
The inverse of the covariance matrix $ \lb \Sigma_{N}\rb_{ij} = \av{f_i f_j} - m_im_j$ between the members of  $\lb h \rb$ is given by
\beq
\lb \Sigma_N^{-1} \rb_{ij} = \lb C^TC \rb_{ij} = \sum_{n = 0}^NC_{ni}C_{nj},\quad i,j= 1,\cdots,N
\enq
This is indeed the $N \times N$ lower right block of $M_N^{-1}$.
\newline
\newline From these considerations follows :
Define the (positive) Fisher information matrix within the first $N$ hierarchy members, $\lb F_N\rb$ as 
\beq
\lb F_{\le N} \rb_{\alpha \beta} :=  \sum_{i,j = 0}^N\frac{\partial m_i}{\partial \alpha}\lb M_N^{-1} \rb_{ij} \frac{\partial m_j}{\partial \beta}.
\enq
If the density is a probability density, then $\partial_\alpha m_0 = 0$, and therefore
\beq
\lb F_{\le N} \rb_{\alpha \beta} =  \sum_{i,j = 1}^N\frac{\partial m_i}{\partial \alpha}\lb \Sigma_N^{-1} \rb_{ij} \frac{\partial m_j}{\partial \beta}.
\enq
Writing $M_N^{-1}$ as $C^TC$ we can expand the matrix as
\beq
\lb F_{\le N} \rb_{\alpha \beta} = \sum_{n = 1}^N s_n(\alpha)s_n(\beta),\textrm{   with   }s_n(\alpha) = \av{\frac{\partial \ln p}{\partial_\alpha }\: P_n} = \sum_{k = 0}^n C_{nk} \frac{\partial m_k}{\partial \alpha}.
\enq
The coefficients $s_n(\alpha)$ is therefore the information on $\alpha$ in $f_n$ that was not contained already in the previous members of the hierarchy.
The bits of information absent from the hierarchy are given by
\beq
\av{ \lp \frac{\partial \ln p}{\partial \alpha} - \sum_{k = 1}^ns_n(\alpha)P_n(x) \rp\lp \frac{\partial \ln p}{\partial \beta} - \sum_{k = 1}^ns_n(\beta)P_n(x) \rp} = F_{\alpha\beta} - \lb F_{\le N} \rb_{\alpha \beta}.
\enq
This matrix on the right hand side of that relation is positive (it is a Gram matrix), and thus this relation states that the information within the hierarchy is always less than the total information. From this inequality follows thus the information inequality for any set of probes. Convergence to the total information occurs when
\beq
s_{\le N}(x,\alpha) := \sum_{n = 0}^Ns_n(\alpha) P_n(x) \stackrel{N \rightarrow \infty}{\rightarrow} \frac{\partial \ln p(x,\btheta)}{\partial \alpha} ,
\enq
in the mean square error sense with respect to $p(x,\btheta)$. This implies that the hierarchy is efficient precisely when $\partial_\alpha \ln p$  is sparse in $\lb f \rb$. In particular, maximal entropy distributions with the prescribed moments of $\lb f\rb$ have a finite non zero numbers of coefficients. Note that $s_{\le N}(x,\alpha)$ is the best approximation of $\partial_\alpha \ln p$ with the given hierarchy up to $N$ according to this mean square criterium.
\paragraph{Recursion relations}
Remember the triangular matrix $B$ of size $N$ defined above, $B_{nk} = \av{f_nP_k}$. The following relations are easily seen to hold,
\beq
BB^T = M_N, \quad B = C^{-1}.
\enq
It follows that the matrix $B$ is nothing else than the Cholesky decomposition of the moment matrix. Multiplying \eqref{GS} with the score function and integrating, one obtains
\beq
s_n(\alpha)  = \frac{1}{B_{nn}}\lp \frac{\partial m_n}{\partial \alpha} - \sum_{k =0}^nB_{nk}\:s_k(\alpha)\rp.
\enq
This can provide a way to evaluate numerically these coefficient, after a Cholesky decomposition of the moment matrix. However, it is necessary to keep in mind that large moment matrices are infamously known for being generically badly conditioned \citep{Tyrtyshnikov:1994fk}.

\cleardoublepage
\part{Applications for cosmology\label{part2}}

\chapter{On the combination of shears, magnification and flexion fields\label{ch3}}

At the heart of weak lensing as a cosmological probe lies the convergence field, the weighted projection along the line of sight of of the density fluctuation field, with weights sensitive to the geometry of the Universe. In this chapter we use the duality of the Shannon entropy and Fisher information introduced in chapter \ref{ch1} to discuss quantitatively the combination of different probes of the convergence field, both for mass reconstruction as well as for cosmological purposes. The probes we include are the galaxy shears, the magnification as well as the flexion fields, i.e. all modifications to the galaxy images up to second order. Section \ref{probess} describes the observables and the approach, and section \ref{results} the results.
\newline
\newline
We find that flexion alone outperforms the well established shears on the arcsecond scale, making flexion well suited for mass reconstruction on small scales. At the same time, it complements powerfully the shears on the scale of the arcminute.  We find size information to carry some modest, scale independent amount of information. Besides, the results of this chapter show how standard cosmological Fisher matrix analysis based on Gaussian statistics can incorporate these other probes in the most simple way. From (\ref{multis}) follows namely that the inclusion of all the two-point correlations of these additional weak lensing probes can be accounted for by adapting the noise term.
\newline
\newline
The text in this section follows the second part of \cite{2011MNRAS.417.1938C}.
\section{Introduction}
Gravitational lensing, which can be used to measure the distribution of mass along the line of sight, has been recognized as powerful probe of the dark components of the Universe \citep{1992grle.book.....S,2001PhR...340..291B,2003ARA&A..41..645R,2006astro.ph.12667M,2006glsw.book.....S} since it is  sensitive to both the geometry of the Universe, and to the growth of structure. Weak lensing data is typically used in two ways. The first, which is deployed for cosmological parameter fitting, relies on measuring the correlated distortions in galaxy images \citep{2006astro.ph..9591A}. The second approach uses each galaxy to make a noisy measurement of the lensing signal at that position. These point estimates are then used to reconstruct the dark matter density distribution \citep[e.g.][]{1993ApJ...404..441K,2001A&A...374..740S}. Most of the measurements of weak lensing to date have focused on the shearing that galaxy images experience. However, gravitational lensing causes a number of other distortions of galaxy images. These include change in size, which is related to the magnification, and higher order image distortions known as the flexion \citep{2006MNRAS.365..414B}. A number of techniques have been developed for measuring these higher order images distortions, such as HOLICS \citep{2007ApJ...660..995O} and shapelets methods \citep{2007MNRAS.380..229M}. Since all of the image distortions originate from the same source, the lensing potential field, the information content of any two lensing measurements must be degenerate. At the same time, since each method has different systematics and specific noise properties, combining multiple measurement may bring substantial benefits. Some recent works have looked at the impact of combining shear and flexion measurements for mass reconstruction \citep{2010arXiv1008.3088E,2010ApJ...723.1507P,2010arXiv1011.3041V}, as well as the benefits for breaking multiplicative bias of including galaxy size measurements  \citep{2010arXiv1009.5590V}.
\newline
\newline
We will focus on tracers of galaxy image distortions up to second order in the distortions. These include, to first order, the change in apparent size of the galaxies, due to the magnification source by the convergence field, the two components of the shear, and, to second order, the four components of the two flexion fields. We will limit ourselves to the case where the noise contaminating each probe can be effectively treated as independent of the model parameters. The common point of the probes cited above is that they all trace, in some noisy way, the same central field on a discrete set of points, represented by the  positions of the galaxies on which the tracers are measured. The framework we presented in chapter \ref{ch1} is ideal to deal with the special structure of the situation. We will first show how we can understand the total information content of such degenerate probes of a central field in a general situation, and then make quantitative evaluations at the two-point level. 
\section{Linear tracers \label{probess}}
The situation that we consider here is one where a broad number of observables are linked in some way to a central field. We will limit ourselves to the case where the noise contaminating each probe can be effectively treated as independent of the model parameters.  
Imagine one plans to perform different measurements, of observables that are all linked in some way or another to the same central field, which is the actually interesting quantity from the point of view of the analyst. 
The most straightforward example of such a situation, occurs when the two point correlations of the central field, say the the power spectrum of the convergence in weak lensing, is predicted from theory, while we try to extract it looking at the correlations of the two ellipticity components of galaxies. In this case, predictive power of the power spectrum of the convergence turns into predictive power of (parts of) the three correlation functions there are between the two ellipticity components.
\newline\newline
The predictive power of some observable  $O_c$ of a central field (for instance its power spectrum at some mode)  translates into an array of constraints $O_i,\:\: i = 1,\cdots,n$ in the noisy probes, that we could try and extract and confront to theory :
\beq
\begin{split}\label{assum}
O_i(\btheta) = f_i(O_c(\btheta)), \quad i = 1,\cdots, n
\end{split}
\enq
for some functions $f_i$.\newline
For the purpose of this work,  the case of functions linear with respect to $O_c$ is generic enough, i.e. we will consider that
\beq
\frac{\partial^2 f_i}{\partial O_c^2} = 0,\quad i = 1,\cdots,n.
\enq
The entropy $S$ of the data is a function of the $n$ constraints $\mathbf O$. It is however fundamentally a function of $O_c$ since it does enter all of these observables. It is therefore very natural to associate a potential $\lambda_c$ to $O_c$, although it is not itself a constraint on the probability density function. In analogy with
\beq
\lambda_i =  \frac{\partial S}{\partial O_i}, \quad i = 1,\cdots,n
\enq
we define
\beq
\lambda_c := \frac{d S}{dO_c}(O_1,\cdots,O_m),
\enq
with the result, given by application of the chain rule, of
\beq  \label{un}
\lambda_c = \boldsymbol \lambda \cdot \frac{\partial \mathbf f}{\partial O_c}.
\enq
On the other hand, the impact of a model parameter on each observables can be similarly written in terms of the central observable $O_c$,
\beq \label{deux}
\frac{\partial \mathbf O}{\partial \alpha} = \frac{\partial O_c}{\partial \alpha} \frac{\partial \mathbf f}{\partial O_c} .
\enq
It follows directly from these relations (\ref{un}) and (\ref{deux}), and the linearity of $f_i$, that the joint information in the full set of constraints $\mathbf O$, given in equation (\ref{sumform}) as a sum over all $n$ constraints, reduces to a formally identical expression with the only difference that only $O_c$ enters :
\beq
\Fab^X = \frac{\partial \mathbf O}{\partial \alpha }\cdot \frac{\partial \boldsymbol \lambda}{ \partial \beta} = \frac{\partial \lambda_c}{\partial \alpha}\frac{\partial O_c}{\partial \beta},
\enq
which can also be written in the form analog to (\ref{fishermaxent}),
\beq  \label{infreduction}
\Fab^X =  -\frac{\partial O_c}{\partial \alpha}\frac{d^2 S}{dO_c^2} \frac{\partial O_c}{\partial \beta}.
\enq
This last equation shows that all the effect of combining this set of constraints have been absorbed into the second total derivative of the entropy. This second total derivative is the total amount of information there is on the central quantity $O_c$ in the data. Indeed, taking as a special case of model parameter to the central quantity itself, i.e. 
\beq
\alpha =  \beta =  O_c, 
\enq
one obtains now that the full amount of information in $X$ on $O_c$ is 
\beq \label{Infcentr}
F^X_{O_cO_c} = -\frac{d^2 S}{d O_c^2}\lp O_1,\cdots,O_n \rp \equiv \frac{1}{\sigma^2_{\eff}}.
\enq
A simple application of the Cramer Rao inequality presented in equation (\ref{CramerRao}) shows that this effective variance $\sigma_{\eff}$ is the lower bound to an unbiased reconstruction of the central observable from the noisy probes. \newline\newline
These considerations on the effect of probe combination in the case of a single central field observable $O_c$ generalize easily to the case where there are many, $(O_c^1,\cdots, O_c^m)$. In this case, each central field quantity leads to an array of constraints in the form of equation (\ref{assum}), it is simple to show that the amount of Fisher information can again be written in terms of the information associated to the central field, with an effective covariance matrix between the $O_c's$. The result is
\beq
\Fab^X = -\sum_{i,j = 1}^{m}\frac{\partial O_c^i}{\partial \alpha}\frac{d^2S}{dO_c^iO_c^j}\frac{\partial O_c^j}{\partial \beta}.
\enq
All the effects of probe combination are thus encompassed in  an effective covariance matrix $\Sigma_\eff$ of the central field observables,
\beq
-\frac{d^2S}{dO_c^iO_c^j} \equiv \lb \Sigma^{-1}_\eff\rb_{ij}.
\enq
Again, an application of the Cramer Rao inequality, in the multi-dimensional case, shows that this effective covariance matrix is the best achievable unbiased joint reconstruction of  $(O_c^1,\cdots, O_c^m)$.
\newline\newline
We now explore further the case of linear probes of homogeneous Gaussian fields, which is cosmologically relevant and can be solved analytically to full extent. We will focus on zero mean fields,  for which according to our previous section the entropy can be written in terms of the spectral matrices, up to a constant,
\beq \label{EntropyField}
S = \frac 12 \sum_{\veck} \ln \det P(\veck).
\enq
\color{black}
\subsection{Linear tracers at the two-point level}
The standard instance of a linear tracer $\phi_i$ of some central field $\kappa$ in weak lensing is provided by a relation in Fourier space of the form
\beq \label{example}
\tilde \phi_i(\veck) = v_i \tilde \kappa(\veck) +\tilde \epsilon_i(\veck),
\enq
for some noise term $\tilde\epsilon_i$, uncorrelated with $\kappa$, and coefficient $v_i$. Typically, if one observes a tracer of the derivative of the field $\kappa$, then the vector $\vecv$ would be proportional to $-i\veck$. We are ignoring here any observational effect, such as incomplete sky coverage, that would require corrections to this relation. It is clear from this relation that the spectral matrices of this family take a special form of equation (\ref{assum}): defining the spectrum of the $\kappa$ field by $P^\kappa$, we obtain by putting this relation (\ref{example}) into (\ref{spectralmatrix}), that the spectral matrices can be written at each mode in the form
\beq \label{specmatrix}
P =  P^\kappa \vecv\vecv^\dagger + N,
\enq
where $\vecv^\dagger$ is the hermitian conjugate of $\vecv = \lp v_1,\cdots,v_n\rp$. The matrix $N$ is the spectrum of the noise components $\epsilon$,
\beq
N_{ij}(\veck) = \frac 1 V\av{\tilde \epsilon_i(\veck)\tilde \epsilon^*(\veck)}.
\enq 
Our subsequent results hold for any family of tracers that obey this relation. While the special case in (\ref{example}) enter this category, this need not be the only instance. All the weak lensing observables we deal with in this work will satisfy equation (\ref{specmatrix}).
\newline\newline
Both the n-dimensional vector $\vecv$ and the noise matrix $N$ can depend on the wave vector $\veck$, but they are independent from the model parameters. The  matrix $N$ of dimension $n\times n$ is the noise component of the spectra of the fields, typically built from two parts. The first is due to the discrete nature of the fields, since such data consist in quantities measured where galaxies sits, and the second to the intrinsic dispersion of the measured values.
\subsection{Joint entropy and information content}
\indent Information on the model parameters enters through $P^\kappa$ only.
To evaluate the full information content, we need only evaluate eq. (\ref{EntropyField}) with the spectral matrix given in (\ref{specmatrix}), keeping in mind the result from last section, that we need only the total derivative with respect to $P^\kappa$. In other words, any additive terms in the expression of the entropy that are independent of $P^\kappa$ can be discarded.
\newline\newline
This determinant can be evaluated immediately. Defining for each mode the real positive number $N_\eff$ through
\beq \label{noise}
\frac 1 {N_\eff} \equiv \vecv^\dagger N^{-1}\vecv,
\enq
which can be seen as an effective noise term, a simple\footnote{We have namely for any invertible matrix $A$ and vectors $\vecu,\vecv$ the matrix determinant lemma,
\beq
\det \left(A + \vecu \vecv^T\right) = \det\lp A\rp \lp 1 + \vecv^TA^{-1}\vecu \rp. 
\enq}  calculation shows that the joint entropy (\ref{EntropyField}) is equivalent to the following, where the $n$ dimensional determinant has disappeared,
\beq \label{Entropy2}
S = \frac12\sum_{\veck}\ln \left(P^\kappa(\veck)  +  N_\eff(\veck)\right).
\enq
Comparison with equation (\ref{EntropyField}) shows that we have with this equation (\ref{Entropy2}) the entropy of the field $\kappa$ itself, where all the effects of the joint observation of this $n$ fields have been absorbed into the effective noise term $N_\eff$, that contaminates its spectrum. It means that the full combined information in the $n$ probes of the field $\kappa$ is equivalent to the information in $\kappa$, observed with spectral noise $N_\eff$.\newline\newline
Our result (\ref{Infcentr}) applied to (\ref{Entropy2}) puts  bounds on reconstruction of the field $\kappa$ out of the observed samples, which can be at best reconstructed with a contaminating noise term of $N_\eff$ in its spectrum, whose best unbiased reconstruction is given by
\beq
 2\lp P^\kappa(\veck) +N_\eff(\veck)\rp^2.
\enq
\newline
Since the effect of combining these probes at a single mode is only to change the model independent noise term, the parameter correlations and degenaracies as approximated by the Fisher information matrix stay unchanged, whatever the number of such probes is. We have namely from (\ref{Entropy2}) that at a given mode $\veck$, the Fisher information matrix reads
\beq \label{fmatrixpk}
\Fab^X = \frac 12 \frac{\partial \ln \tilde P^\kappa(\veck) }{\partial\alpha}\frac{\partial \ln \tilde P^\kappa(\veck)}{\partial\beta},
\enq
with
\beq
 \tilde P^\kappa(\veck) =  P^\kappa(\veck)  + N_\eff(\veck).
\enq
From the point of view of the Fisher information, it makes formally no difference to extract the full set of $n(n-1)/2$ independent elements of each spectral matrices, or reconstruct the field $\kappa$ and extract its spectrum. They carry indeed the same amount of Fisher information.
\newline\newline
These results still hold when other fields are present in the analysis, which are correlated 
with the field $\kappa$. To make this statement rigorous, consider in the analysis on top of our $n$ samples of the form (\ref{example}) of $\kappa$, another homogeneous field $\theta$, with spectrum $P^\theta(\veck)$, and cross spectrum to $\kappa$ given by $P^{\theta\kappa}(\veck)$ The full spectral matrices are in this case
\beq
P(\veck) = \bem P^\kappa(\veck) \vecv\vecv^T + N  & P(\veck)^{\kappa\theta}\vecv 
\\ P^{\theta\kappa}\vecv^T & P^{\theta}(\veck) \enm.
\enq
Again, the determinant of this matrix can be reduced to a determinant of lower 
dimension, leading to the equivalent entropy
\beq \label{multis}
S = \cst + \frac 12 \ln \det \bem P^\psi(\veck) + N_\eff & P^{\kappa\theta}
(\veck) \\ P^{\theta\kappa}(\veck) & P^\theta(\veck) \enm.
\enq
It shows that the the full set of $n + 1$ 
fields can be reduced without loss to two fields, $\kappa$ and 
$\theta$, with the effective noise $N_\eff$ contaminating the spectrum of $\kappa$.
\newline
\newline
Note that the derivation of our results do not refer to any hypothetical estimators, but 
came naturally out of the expression of the entropy.
\subsection{Weak lensing probes}
We now seek a quantitative evaluation of the full joint information content of the weak lensing probes in galaxy surveys, up to second order in the image distortions of galaxies. The data $X$ consists of a set of fields, which are discrete point fields, which take values where galaxies sits. We work in the two-dimensional flat sky limit, using the more standard notation $\mathbf l$ for the wave vector, and decompose it in modulus and polar coordinate as
\beq
\vecl = l\bem \cos \varphi_l \\ \sin \varphi_l \enm
\enq
We will throughout assume that the intrinsic values of each probe are pairwise uncorrelated, as commonly done. Also, we will assume that the set of points on which the relevant quantities are measured show low enough clustering so that corrections to the spectra due to intrinsic clustering can be neglected. This is however not a limitation of our approach, since corrections to the above assumptions, such as the introduction of some level of intrinsic alignment, can be accommodated for by introducing appropriate terms in the noise matrices $N(\veck)$ in (\ref{noise}). 
As a central field to which all our point fields relates, we take for convenience the isotropic convergence field $\kappa$, with spectrum
\beq
C^\kappa(\vecl) = C^\kappa(l).
\enq
In the case of pairwise uncorrelated intrinsic values that we are following, we see easily from (\ref{noise}) that by combining any number of such probes the effective noise is reduced at a given mode according to
\beq \label{combination}
\frac 1 {N^{\textrm{tot}}_\eff} = \sum_i \frac{1}{N_\eff^i}.
\enq
We therefore only need to evaluate the effective noise for each probe separately, while their combination follows (\ref{combination}). To this aim, the evaluation of the spectral matrices (\ref{specmatrix}), giving us $N_\eff$, is necessary. The calculations for this are presented in appendix \ref{pf} and we use the final results in this section.
\subsubsection{First order, distortion matrix}
To first order, the distortion induced by weak lensing on a galaxy image is described by the distortion matrix that contains the shear, $\gamma$, and convergence, $\kappa$, which come from the second derivatives of the lensing potential field $\psi$, \citep[e.g. ][]{2006glsw.book.....S}
\beq \label{distortion}
\bem \kappa + \gamma_1 & \gamma_2 \\ \gamma_2 & \kappa - \gamma_1 \enm = \psi_{,ij}.
\enq
The shear components read
\beq
\gamma_1 = \frac 12 \left(\psi_{,11} - \psi_{,22}\right), \quad \gamma_2 = \psi_{,12}
\enq
and we assume they are measured from the apparent ellipticities of the galaxies, with identical intrinsic dispersion $\sigma^2_\gamma$. Denoting with $\bar n_\gamma$ the number density of galaxies for which ellipticity measurements are available,
the effective noise is 
\beq \label{neffgamma}
N^\gamma_\eff = \frac{\sigma^2_\gamma}{\bar n_\gamma}.
\enq
The information content of the two observed ellipticity fields is thus exactly the same as the one of the convergence field, with a mode independent noise term as above. 
\newline\newline
\newcommand{\intr}{\textrm{int}}
To reach for the $\kappa$ component of the distortion matrix, we imagine we have measurements of their angular size $s_\obs$, with intrinsic dispersion $\sigma^2_s$. The intrinsic sizes of the galaxies $s_\intr$ gets transformed through weak lensing according to
\beq
s_\obs = s_{\textrm{int}}( 1 +\alpha_s\kappa).
\enq 
The coefficient $\alpha_s$, is equal to unity in pure weak lensing theory, but we allow it to take other values, since in a realistic situation, other effects such as magnification bias effectively enter this coefficient (see e.g. \cite{2010arXiv1009.5590V}). Under our assumption that the correlation of the fluctuations in intrinsic sizes can themselves be neglected, the effective noise reduces to
\beq\label{neffsize}
 N^s_\eff =  \frac 1 {\alpha_s^2}\lp \frac{\sigma_s} { \bar s_\textrm{int}} \rp ^2\frac{1}{\bar n_s}.
\enq
This combination of $\alpha_s$ with the dispersion parameters $\bar s$ and $\sigma_s$ becomes the only relevant parameter in our case, and not the value of each of them.
\subsubsection{Second order, flexion}
To second order, the distortions caused by lensing on the galaxies images are given by third order derivatives of the lensing potential. These are conveniently described by the spin 1 and spin 3 flexion components $\mathcal F$ and $\mathcal G$, which in the notation of  \citep{2008A&A...485..363S} read
\beq \begin{split}
\mathcal F &= \frac 12 \bem \psi_{,111} + \psi_{,122} \\ \psi_{,112} + \psi_{,222} \enm 
\\
\mathcal G &= \frac 12 \bem \psi_{,111} - 3\psi_{,122} \\ 3\psi_{,112} - \psi_{,222} \enm,
\end{split}
\enq
and are extracted from measurements with intrinsic dispersion $\sigma^2_\mathcal F$ and $\sigma^2_\mathcal G$.
The effective noise is this time mode-dependent,
\beq \label{neffflex}
\frac 1 { N^{\mathcal F \mathcal G}_\eff} = l^2\left(\frac{\bar n_F}{\sigma_F^2} + \frac{\bar n_G}{\sigma_G^2}\right).
\enq
\section{Results} \label{results}
\begin{figure}[htbp]
\begin{center}
\centering
  \includegraphics[width = 0.8\textwidth]{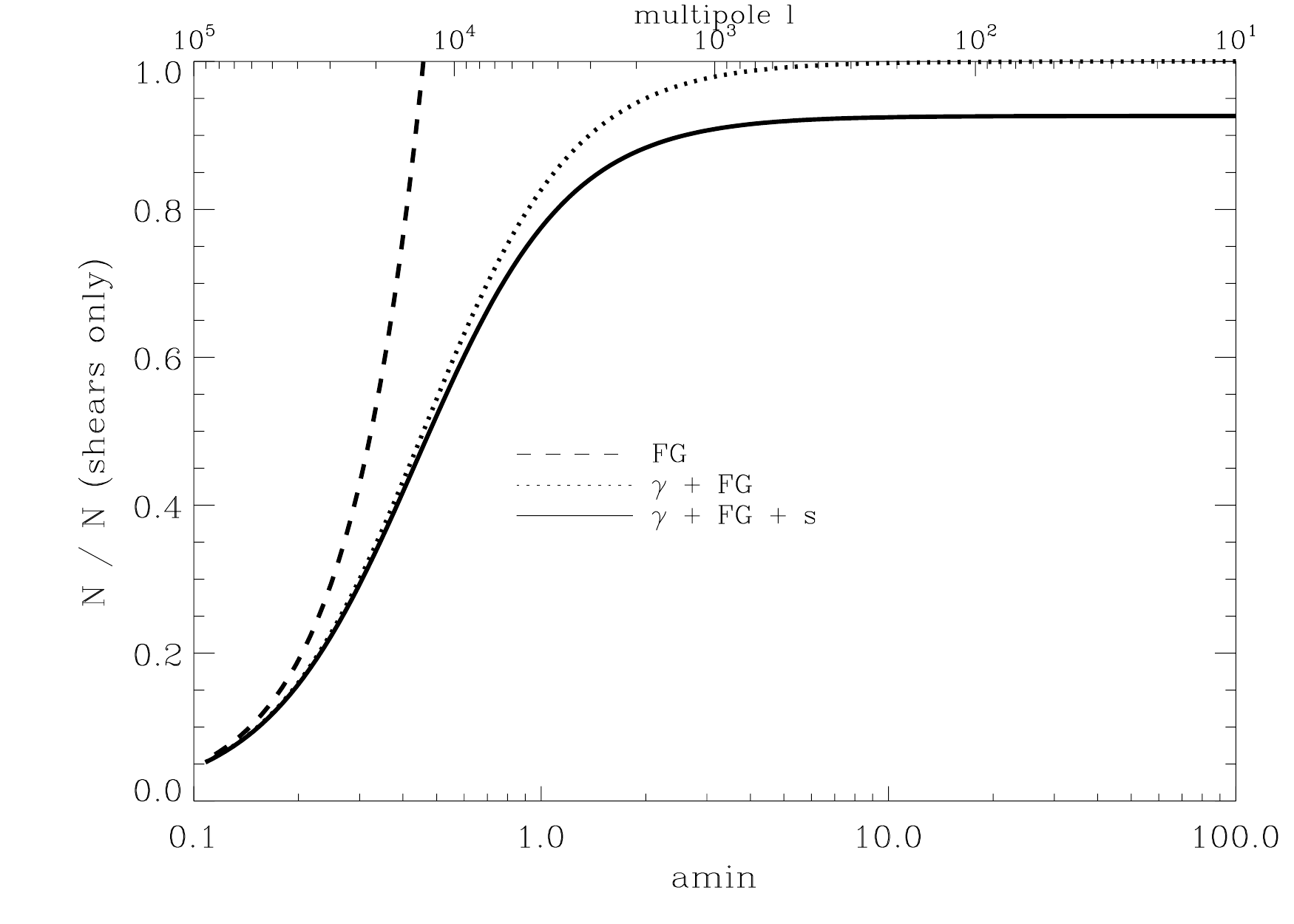}
  \caption[The ratio of the effective noise when including flexion and magnification fields to the level of noise considering the shears only, as a function of the angular scale.]{The ratio of the effective noise to the level of noise considering the shears only, as function of angular scale. The dashed line considers the flexion fields alone. The dotted line shows the combination of the flexion fields with the shear fields,  and the solid line all these weak lensing probes combined. No correlations between the intrinsic values for each pair of probes have been considered.}
  \label{noiseratio}
  \end{center}
\end{figure}

\begin{center}
\begin{figure}[htbp]
 \centering
  \includegraphics[width = 0.8\textwidth]{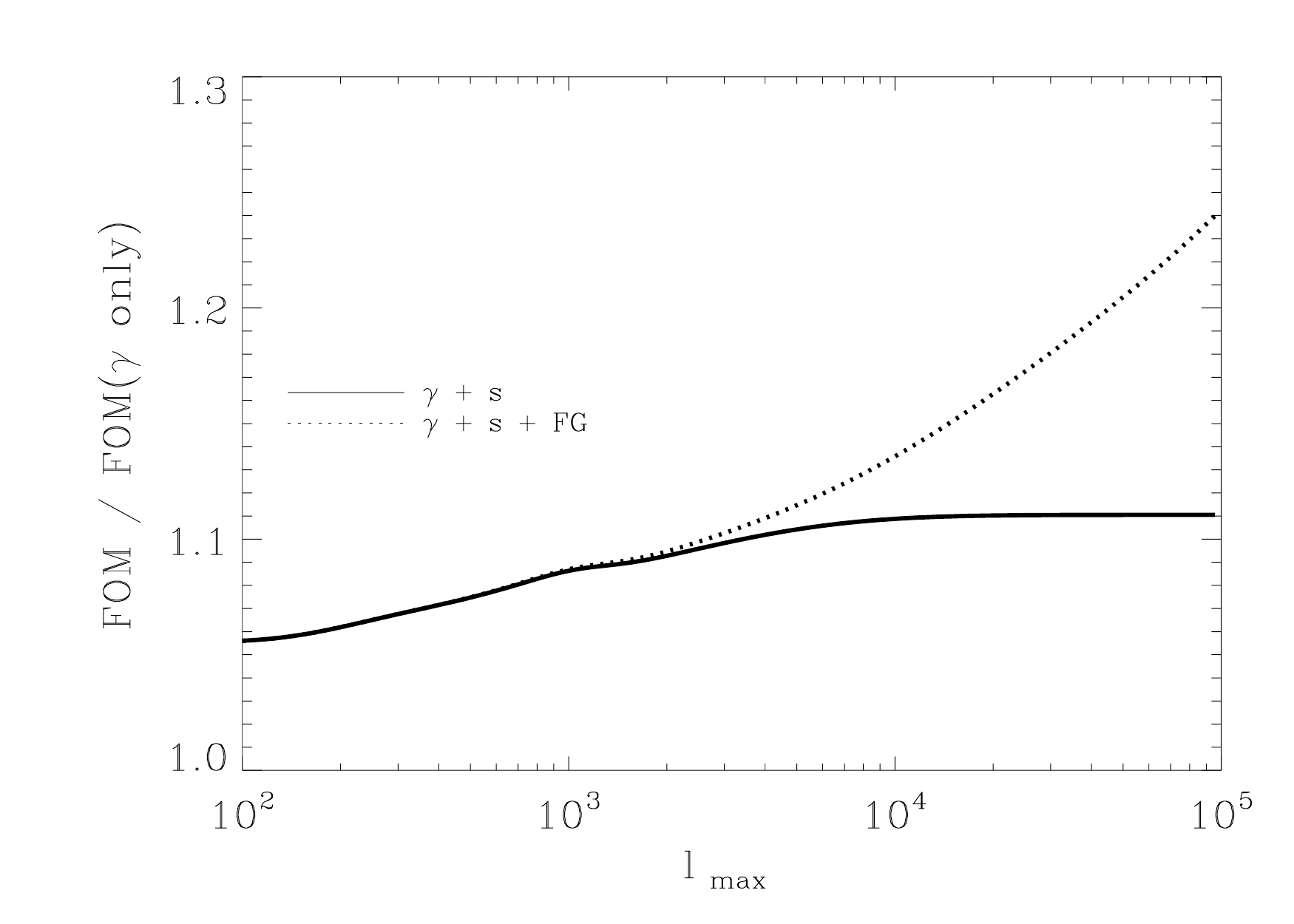}  \caption[The increase of the dark energy FOM including size information as well as flexion information, over the shear only analysis, as a function of the maximal angular multipole.]{The improvement of the dark energy FOM including size information (solid), as well as flexion $\mathcal F$ and $\mathcal G$ information (dotted), over the shear only analysis, as function of the maximal angular multipole included in the analysis.}
  \label{figFOM}
\end{figure}
\end{center}

Figure \ref{noiseratio} shows the ratio of the effective noise to the noise present considering the shear fields only, assuming the same number densities of galaxies for each probe, and the values for the intrinsic dispersion stated in table \ref{paramtable}. The conversion multipole $l$ (upper x-axis) to angular scale $\theta$ (lower x-axis) follows $\theta = \pi / (l + 1/2)$.  We have adopted for the size dispersion parameters the numbers from \citep{2010arXiv1009.5590V}, who evaluated this number for the DES survey conditions \citep{2005astro.ph.10346T}. 
We refer to the discussion in \citep{2010ApJ...723.1507P} for our choice of flexion dispersion parameters.
The curves on this figure are ratios independent of the galaxy number density. They are redshift independent as well, only to the extent that the dispersion in intrinsic values can be treated as such. We can draw two main conclusions from figure \ref{noiseratio}. First, flexion information beings to play role only at the smallest scales, on the arcsecond scales, where it takes over and becomes the most interesting probe. On the scale of $1$ amin, it can bring substantial improvement over shear only analysis, but only in combination with the shears, and not on its own. This is in good agreement with the comparative analysis of the power of the flexion $\mathcal F$ field and shears fields for mass reconstruction done in \cite{2010ApJ...723.1507P}, restricted to direct inversion methods. Second, the inclusion of size of galaxies into the analysis provides a density independent, scale independent, improvement factor of
\beq \label{improve}
\frac{N_\eff^\gamma}{N_\eff^{\gamma+ s}} = 1 + \lp \frac{\sigma_\gamma \bar s}{\sigma_s \alpha_s}\rp^2,
\enq
which is close to a $10\%$ improvement for the quoted numbers. Of course, the precise value depends on the dispersion parameters of the population considered.
\begin{center}
\begin{table} \caption[Dispersion parameters for the shear, flexion and magnification fields used in our numerical analysis.]{Dispersion parameters used in figure \ref{noiseratio}.}\label{paramtable}
\begin{tabular*}{0.3\textwidth}{cccc} 
 $\sigma_\gamma$ & $ \sigma_\mathcal F$ asec$^{-1}$ &$ \sigma_\mathcal G $ asec$^{-1}$ &  $\frac 1 {\alpha_s} \frac{ \sigma_s}{ \bar s}$ \\ \hline
 0.25 & 0.04 & 0.04 & 0.9 
 \end{tabular*}
\end{table}
\end{center}
For the purpose of measuring cosmological parameters rather than mass reconstruction, more interesting are the actual values of the Fisher information matrices. Since with any combination of such probes, these matrices are proportional to each other at a single mode, it makes sense to define the efficiency parameter of the probe $i$ through
\beq
\epsilon_i(l) := \frac{C^\kappa(l)}{C^\kappa(l) + N^i_\eff(l)},
\enq
which is a measure of what fraction of the information contained in the convergence field is effectively captured by that probe. The information in the convergence field is, at a given mode $l$, counting the multiplicity of the mode,
\beq
F^\kappa_{\alpha\beta} = \frac 12 (2l + 1) \frac{\partial \ln C^\kappa(l)}{\partial\alpha}\frac{\partial \ln C^\kappa(l)}{\partial\beta},
\enq
and we have indeed that the total Fisher information in the observed fields is
\beq
\Fab^X = \sum_l  F^\kappa_{\alpha\beta}(l) \epsilon^2_i(l).
\enq
Therefore, according to the interpretation of the Fisher matrix approximating the expected constraints on the model parameters, the factor $\epsilon(l)$ is precisely equal to the factor of degradation in the constraints one would be able to put on any a parameter, with respect to the case of perfect knowledge of the convergence field at this mode.
It is not the purpose of this work to perform a very detailed study on the behavior of the efficiency parameter for some specific survey and the subsequent statistical gain, but its qualitative behavior is easy to see. This parameter is essentially unity in the high signal to noise regime, while it is the inverse effective noise whenever the intrinsic dispersion dominates the observed spectrum. Since information on cosmological parameters is beaten down by cosmic variance in the former case, the latter dominates the constraints. We can therefore expect from our above discussion the size information to tighten by a few percent constraints on any  cosmological parameter. On the other hand, while flexion becomes ideal for mass reconstruction purposes on small scales, it will be able to help inference on cosmological parameters only if the challenge of  very accurate theoretical predictions on the convergence power spectrum for multipoles substantially larger than $1000$ will be met.\newline\newline
To make these expectations more concrete, we evaluated the improvement in information on cosmological parameters performing a lensing Fisher matrix calculation for a wide, EUCLID-like survey, in a tomographic setting. For a data vector consisting of $n$ probes of the convergence field $\kappa_i$ in each redshift bin $i,\: i = 1,\cdots N$, it is simple to see following our previous argument, that the Fisher information reduces to
\beq
F_{\alpha\beta} = \frac 12  \sum_{l}\lp 2l + 1\rp \Tr \:C^{-1}\frac{\partial C}{\partial\alpha} C^{-1}\frac{\partial C}{\partial\beta},
\enq
where the $C$ matrix is given by
\beq \label{cmatrix}
C_{ij} = C^{\kappa_i \kappa_j}(l)  + \delta_{ij}N^i_\eff(l),\quad i,j = 1,N
\enq
with $N_\eff^i$ given by (\ref{noise}). The only difference between standard implementations of Fisher matrices for lensing, such as the lensing part of \cite{2004PhRvD..70d3009H}, being thus the form of the noise component.
we evaluated these matrices respectively for
\beq \label{shearonly}
N_\eff^i = \frac{\sigma^2_\gamma}{\bar n^i} = N^{\gamma,i}_\eff ,
\enq
which is the precise form of the Fisher matrix for shear analysis, for
\beq\label{shearsize}
\frac{1}{N_\eff^i} =\frac 1 {N_\eff^{\gamma,i}} + \frac 1 {N_\eff^{s,i}}
\enq
which account for size information, and
\beq\label{shearsizefg}
\frac{1}{N^i_\eff(l)} = \frac 1 {N_\eff^{\gamma,i}} + \frac 1 {N_\eff^{s,i}} + \frac 1 {N_\eff^{\mathcal F \mathcal G,i}(l)} ,
\enq
which accounts for the flexion fields as well.  We note that in terms of observables, these small modifications incorporate in its entirety the full set of all possible correlations between the fields considered. The values of the dispersion parameters involved in these formulae are the same as in table (\ref{paramtable}). Our fiducial model is a flat $\Lambda$CDM universe, with parameters $\Omega_\Lambda = 0.7$, $\Omega_b = 0.045$, $\Omega_m = 0.3, h = 0.7$, power spectrum parameters $\sigma_8 = 0.8, n = 1$, and Chevallier- Polarski-Linder parametrisation \citep{2001IJMPD..10..213C,2003PhRvL..90i1301L} of the dark energy equation of state implemented as $\omega_0 = -1,w_a = 0$. The distribution of galaxies as function of redshift needed both for the calculation of the spectra and to obtain the galaxy densities in each bin was generated using the cosmological package iCosmo \citep{2008arXiv0810.1285R}, in a way described in \citep{2007MNRAS.381.1018A}. We adopted EUCLID-like parameters of $10$ redshift bins, a median redshift of $1$, a galaxy angular density of $40 / \textrm{amin}^2$, and photometric redshift errors of $0.03(1 + z)$.
\newline\newline
In figure \ref{figFOM}, we show the improvement in the dark energy Figure of Merit (FOM), defined as the square root of the determinant of the submatrix $\lp \omega_0,\omega_a\rp$ of the Fisher matrix inverse $F^{-1}_{\alpha\beta}$ ($\alpha$ and $\beta$ running over the set of eight parameters as described above), as function of the maximal angular mode $l_\max$ considered, while $l_\min$ being always taken to be 10. In perfect agreement with our discussion above, including size information (solid line) increases the FOM steadily until it saturates at a $10\%$ improvement when constraints on the dark energy parameters are dominated by the low signal to noise regime. Also, flexion becomes only useful in the deep non-linear regime, where however theoretical understanding of the shape of the spectra still leaves a lot to be desired.
\begin{table} \caption[Ratio of the marginalised constraints considering size information as well as flexion to that of shear only analysis.]{Ratio of the marginalised constraints $\sigma^2 / \sigma^2_{\textrm{shear only}} $ , for $l_\max = 10^4$. This first line considers the inclusion of the size information in the analysis, while the second the size as well as the flexion fields $\mathcal F$ and $\mathcal G$.\label{ratiotable}}.
\begin{tabular*}{0.59\textwidth}{|cccccccc|}
 $\Omega_{\Lambda}$ & $\Omega_b$ & $\Omega_m$  & $h$  & $n$ &  $\sigma_8$ & $w_0$ & $w_a$ \\  \hline
 0.90 & 0.96 & 0.90 & 0.95 & 0.95 & 0.90 & 0.90 & 0.90 \\
  0.88 & 0.96 & 0.89 & 0.95 & 0.93 & 0.88 & 0.88 & 0.88
 \end{tabular*}
\end{table}
\newline
\newline
These results are found to be very insensitive to the survey parameters, for a fixed $\alpha_s$.  There are also only weakly model parameter independent, as illustrated in table \ref{ratiotable}, which shows the corresponding improvement in Fisher constraints,
\beq
\frac{\sigma^2}{\sigma^2_{\textrm{shear only}}} = \frac{F^{-1}_{\alpha\alpha}}{F^{-1}_{\alpha\alpha, \textrm{shear only}}},
\enq
at the saturation scale $l_\max = 10^4$.
These results are also essentially unchanged using either standard implementations of the halo model \citep[for a review]{2002PhR...372....1C} or the the HALOFIT \citep{2003MNRAS.341.1311S} non linear power spectrum.



\section{Appendix}\label{pf}
The data consists in a set of numbers, at each position where a galaxy sit and a measurement was done. We use the handy notation in terms of Dirac delta function,
\beq
\phi(\vecx) = \sum_i \epsilon_i\delta^D(\vecx - \vecx_i),
\enq
where the sum runs over the positions $\vecx_i$ for which $\epsilon$ is measured. To obtain the spectral matrices, we need the Fourier transform of the field, which reads in our case
\beq
\tilde \phi(\vecl) = \sum_i \epsilon_i \exp\lp - i \vecl\cdot \vecx_i\rp. 
\enq
In this work, we assume that the set of points shows negligible clustering, so that the probability density function for the joint occurrence of a particular set of galaxy positions is uniform.\newline
We decompose in the following the wave vector $\veck$ on the flat sky in terms of its modulus and polar angle as
\beq
\vecl = l\bem \cos\phi_l \\ \sin \phi_l \enm.
\enq
\paragraph{Ellipticities}
When the two ellipticity components are measured, we have two such fields $\phi_1,\phi_2$ at our disposal. For instance, the field describing the first component becomes
\beq \label{ff}
\tilde \phi_1(\vecl) = \sum_{i} \epsilon^1_i \exp\lp - i \vecl\cdot \vecx_i\rp.
\enq
We assume that the measured ellipticities trace the shear fields, in the sense that the measured components are built out of the shear at that position plus some value unrelated to it,
\beq \begin{split} \label{rel2}
\epsilon^1_i &= \gamma_1(\vecx_i) + \epsilon^1_{\textrm{int},\:i}
\\
\epsilon^2_i &= \gamma_2(\vecx_i) + \epsilon^2_{\textrm{int},\:i}.
\end{split}
\enq
The vector $\vecv$ relating the spectral matrices of the ellipticities and the convergence
is then obtained by plugging (\ref{ff}) with the above relations (\ref{rel2}) in its definition (\ref{specmatrix}), and using the relation between shears and convergence in equation (\ref{distortion}). The result is 
\beq
\vecv =\bar n_\gamma  \bem \cos 2\phi_l \\ \sin 2\phi_l\enm.
\enq
where $\bar n_\gamma$ is the number density of galaxies for which ellipticity measurements are available.
Under our assumptions of uncorrelated intrinsic ellipticities,  with dispersions of equal magnitude $\sigma_\gamma^2$ for the two components, the noise matrix $N$ becomes 
\beq
N = \bar n_\gamma \bem \sigma_\gamma^2 & 0 \\ 0 & \sigma_\gamma^2\enm.
\enq
The effective noise, given in equation (\ref{noise}) is readily computed
\beq
N^\gamma_\eff = \frac{\sigma^2_\gamma}{\bar n_\gamma}.
\enq
\paragraph{Sizes}
As noted in the main text, the apparent sizes of galaxies are modified by lensing, in the following way,
\beq
s^i_{\obs} = s^i_{\textrm{int}}\lp 1 + \alpha_s \kappa\rp,
\enq for some coefficient $\alpha_s$ which is unity in pure weak lensing theory. Denoting the number of galaxies for which sizes measurements are available by $n_s$, and the mean intrinsic size of the sample by $\bar s_{\textrm{int}}$,
the spectrum of the size field reduces, under the assumption of uncorrelated intrinsic sizes, to
\beq
C^s(l) = \bar n_s^2 \bar s_{\textrm{int}}^2 \alpha_s^2 C^{\kappa}(l) + \bar n_s \sigma_s^2.
\enq
The vector $\vecv$ and matrix $N$ are now numbers,  that are read out from the above equation, to be
\beq \begin{split}
v &= \bar n_s\bar s_{\textrm{int}} \alpha_s, \\
N &= \bar n_s \sigma_s^2.
\end{split}
\enq
leading to the effective noise
\beq
N_\eff^s = \frac{1}{\alpha_s^2}\lp \frac{\sigma_s}{\bar s_\textrm{int}} \rp^2\frac 1 {\bar n_s}
\enq
\paragraph{Second order, flexion}
Denoting with $\bar n_{\mathcal F}$ and $\bar n_{\mathcal G}$ the number of galaxies for which $\mathcal F$ and $\mathcal G$ are measured, the vectors linking the flexion to convergence are
\beq
\vecv^{\mathcal F} = -il\bar n_{\mathcal F}\bem \cos \phi_l \\ \sin \phi_l\enm\enq
and
\beq
\vecv^{\mathcal G} = -il\bar n_{\mathcal G}\bem \cos 3\phi_l \\ \sin 3\phi_l\enm.
\enq
Using again the assumption of uncorrelated intrinsic components, we have the four dimensional diagonal noise matrix 
\beq
N = \bem \bar n_\mathcal F \sigma^2_\mathcal F \cdot 1_{2x2}& 0 \\0  & \bar n_\mathcal G \sigma^2_\mathcal G\cdot 1_{2x2}\enm,
\enq
leading to the effective noise, this time mode-dependent,
\beq 
\frac 1 { N^{\mathcal F \mathcal G}_\eff} = l^2\left(\frac{\bar n_F}{\sigma_F^2} + \frac{\bar n_G}{\sigma_G^2}\right).
\enq
\clearpage
\newpage

\clearpage
\newpage

\chapter{On the use of Gaussian likelihoods for power spectra estimators \label{chNote}}

The text of this chapter follows very closely that of \cite{2012arXiv1204.4724C}.
\newline
\newline
In this note we revisit the Fisher information content of cosmological power spectra of Gaussian fields, when based on the assumption of a multivariate Gaussian likelihood for estimators, in order to comment on that assumption. We discuss that despite the fact that the assumption of a Gaussian likelihood is motivated by the central limit theorem, it leads if used consistently to a Fisher information content that violates the Cram\'er-Rao inequality, due to the presence of independent information from the parameter dependent covariance matrix. At any fixed multipole, this artificial term is shown to become dominant in the limit of a large number of correlated fields. While the distribution of the estimators does indeed tend to a Gaussian with a large number of modes, it is shown, however, that its Fisher information content does not, in the sense that their covariance matrix never carries independent information content. The reason why the information content of the spectra is correctly described by the usual formula (i.e. without the covariance term) in this estimator perspective is precisely the fact the the estimators have a chi-squared like distribution, and not a Gaussian distribution.  The assumption of a Gaussian estimators likelihood is thus from the point of view of the information neither necessary nor really adequate, and we warn against the use of Gaussian likelihoods with parameter dependent covariance matrices for parameter inference from such spectra.

\section{Introduction}
Starting from the second half of the nineties \citep{1996PhRvD..54.1332J,1996PhRvL..76.1007J,Tegmark97b,1997ApJ...480...22T}, the calculation of Fisher information matrices in order to understand quantitatively the constraining power of an experiment has become ubiquitous in cosmology, with its fundamental aspects now covered in cosmological textbooks \citep[section 11 and 6 respectively, e.g.]{2003moco.book.....D,2008cmb..book.....D}, or \citep{2009LNP...665...51H}. This is especially true for experiments aimed at measuring power spectra of close to Gaussian fields, since in this case very handy analytical expressions can be obtained that can be applied in a variety of major cosmological subfields, such as for instance the CMB, galaxy clustering, weak lensing as well as their combination.
\newline
Nevertheless, even applied to Gaussian variables, Fisher information matrices are not totally exempt of subtleties. In this note, we revisit the two different possible perspectives on the Fisher information content of such spectra. One starting point is often the assumption of Gaussian errors. We point out that this assumption of a multivariate Gaussian likelihood for power spectra estimators is not fully consistent for the purpose of understanding their information content, due to a term violating the Cram\'er-Rao inequality, that we show is not necessarily small. Too much information is therefore assigned to the spectra under this assumption. We show that we can understand why this term is artificial precisely from the non Gaussian properties of the estimators, and discuss the reasons why the usual formula, i.e. without this term, or setting the covariance matrix to be parameter independent, still gives the correct amount of information.
\newline
\newline
The note is built as follows : In section \ref{sectionperspective} the two common approaches to the information content of spectra are discussed in details in the case of a single field.  We clarify to what extent and why one is actually flawed, which is the source our comments on the use of Gaussian likelihoods for spectra. In section \ref{severalfields} we then turn to a correlated family of fields, where the violation of the Cram\'er-Rao inequality is shown to become substantial. We summarize and conclude in section \ref{conclusions}.
\newline
\newline We recall first the specific form of the Fisher information matrix, defined for a probability density function $p$ as $\Fab = \av{\partial_\alpha \ln p\: \partial_\beta \ln p}, \alpha, \beta$ model parameters of interest, in the particular case of a multivariate Gaussian distribution with mean vector $\mu$ and covariance matrix $\Sigma$, \citep{1996ApJ...465...34V,1997ApJ...480...22T}
\beq \label{Fab}
\Fab = \sum_{i,j}\frac{\partial \mu_i}{\partial \alpha}\Sigma^{-1}_{ij}\frac{\partial \mu_j}{\partial \beta} + \frac 12 \Tr \lb \Sigma^{-1}\frac{\partial \Sigma}{\partial \alpha}\Sigma^{-1}\frac{\partial \Sigma}{\partial \beta} \rb.
\enq
Remember that the Fisher information matrix has all the properties a meaningful measure of information on parameters must have, most importantly for us here the fact that any transformation of the data can only decrease its Fisher information matrix.
\section{One field, gamma distribution}\label{sectionperspective}

Consider a zero mean isotropic homogeneous Gaussian random field, in euclidean space or on the sphere. It is well known that the Gaussianity of the field is equivalent to the fact that the Fourier or spherical harmonic coefficients are independent complex Gaussian variables, only constrained by the reality condition. Equivalently, the real and imaginary parts of those coefficients form independent real Gaussian variables. Such fields are entirely described by their spectrum, and so the extraction of the spectrum from the data with the help of an estimator is a fairly natural way to proceed for inference on parameters of interest. We place ourselves on the sphere, adopting the spherical harmonic notation for convenience. With the set of $a_{lm}$ the harmonic coefficients, the model parameter dependent spectrum $C_l$ is defined as
\beq
\av{a_{lm}a^*_{l'm'}} = \delta_{ll'}\delta_{mm'}C_l.
\enq
Standard, unbiased quadratic estimators can be written as a sum over the number of Gaussian modes available, as
\newline
\beq \label{estimator}
\hat C_l = \frac{1}{2l +1} \sum_{m = -l}^l |a_{lm}|^2.
\enq
We do not consider any source of observational noise, incomplete coverage or any other such issue, which are irrelevant for the points of our discussion. At this point, there are two ways to approach the problem of evaluating its information content in the cosmological literature. The first, - let us call this approach the 'field perspective' -, first calculates the information content of the field itself (equal to that of the set of $a_{lm}$'s), and then interprets this information as being the one of the spectrum. In this case, the information in the field is given by formula \eqref{Fab}, with zero mean vector and diagonal covariance matrix $C_l$.
\beq \label{Ffield}
F_{\alpha \beta} = \frac 12 \sum_{l = 0}^\infty (2l +1) \frac 1{C_l}\frac{\partial C_l}{\partial \alpha} \frac 1{C_l} \frac{\partial C_l}{\partial \beta},
\enq
where the factor $2l +1$ accounts for the number of independent Gaussian variables at a given multipole $l$.
The sum is in practice restricted to the multipole range that will actually be measured to obtain the information in the spectrum to be extracted. A very small sample of works in this approach are \cite{1997ApJ...480...22T,2004PhRvD..70d3009H,2009ApJ...695..652B}. This approach is arguably conceptually appealing, as it deals with the information content of the field itself, and does not require the definition of estimators nor the calculation of their covariance. However, for the same reasons, it is only indirectly connected to data analysis as it is not yet specified precisely how this information content is to be extracted.
\newline
In the second approach - that we call the 'estimator perspective' - is defined first an estimator $\hat C_l$ for each $C_l$ to be extracted, within some $\lmin$ and $\lmax$ (maybe with some bandwidth that we ignore here), and its covariance matrix $\Sigma_{ll'} = \av{\hat C_l\hat C_{l'}} - \av{\hat C_l}\av{\hat C_{l'}} $ is calculated. Then it is argued that due to the central limit theorem, the distribution of the estimator will be approximately Gaussian. In the case of spectra of Gaussian fields, this is very well founded, at least for small scales modes, since \eqref{estimator} is a large sum of well behaved identically distributed independent variables. Then, under this assumption of Gaussianity, their information content is given by equation \eqref{Fab} with mean vector this time the set of $C_l$'s itself and (model parameter dependent) covariance matrix $\Sigma_{ll'}$,
\beq
\Fab = \sum_{l,l' = \lmin}^{\lmax} \frac{\partial C_l}{\partial \alpha}\Sigma^{-1}_{ll'}\frac{\partial  C_{l'}}{\partial \beta} + \frac 12 \Tr \lb \Sigma^{-1} \frac{\partial{\Sigma}}{\partial \alpha}\Sigma^{-1}\frac{\partial\Sigma}{\partial \beta}\rb.
\enq
It is well known that for the estimator \eqref{estimator} we have $\Sigma_{ll'} = \delta_{ll'}2C_l^2/(2l +1)$. The Fisher information matrix, in the estimator perspective, reduces thus to
\beq \label{Festimator}
\begin{split}
 \Fab &=  \frac12 \sum_{l = \lmin}^\lmax (2l +1)\frac 1{C_l}\frac{\partial C_l}{\partial \alpha} \frac 1{C_l} \frac{\partial C_l}{\partial \beta} \\
 &\:+ \frac 12 \sum_{l = \lmin}^\lmax 4 \frac 1 {C_l} \frac{\partial C_l}{\partial \alpha} \frac 1{C_l} \frac{\partial C_l}{\partial \beta}.
 \end{split}
\enq
Clearly, the first term in the estimator perspective corresponds to that of the field perspective. However, the second term, coming from the derivative of the covariance matrix, is new. That term is not enhanced by a $(2l +1)$ factor, and is therefore very subdominant at high $l$. It is either usually neglected, or the covariance matrix of the estimators is inconsistently taken to be parameter independent, and in these cases the two approaches give the same results. Some expositions using explicitly this perspective include \citep{Tegmark97b,2003ApJ...598..720S,2007ApJ...665...14S}, where the additional term is neglected, or the approach in \citep[section 11.4.3]{2003moco.book.....D}, where the covariance matrix is treated as parameter independent. A work where this term plays a direct role is \citep{2009A&A...502..721E}, where the authors specifically study the impact of parameter dependent covariance matrices for parameter estimation using such Gaussian likelihoods.
\newline
\newline
Beyond the question of the quantitative relevance of this additional term, its very appearance is however very disturbing. Under this arguably reasonable Gaussian assumption, our estimator \eqref{estimator} is found to carry more information than the full field, even on the smallest scales.  This obviously violates the most fundamental property of Fisher information, i.e. that information can only be at best conserved when transforming the data (in this case reducing the field to its spectrum), a fact essentially equivalent to the celebrated Cram\'er-Rao inequality \citep{1997ApJ...480...22T}. Something must clearly have gone wrong in the assumption of a Gaussian likelihood for our spectra.
\newline
\newline
To understand what has happened, it is worth tracking the exact distribution and information content of the estimator \eqref{estimator}. Since they are independent at different $l$, we can work at a fixed $l$, and the total information content of these estimators will simply be the sum over $l$ of the information of the estimator at fixed $l$. Under our assumptions, the estimator is a sum of squares of $2l +1$ independent Gaussian variables, and its probability density function can be obtained with no difficulty. The exact distribution is the gamma probability density function with shape parameter $k$ and location parameter $\theta$ as follows
\beq \label{pgamma}
p(\hat C_l|\alpha,\beta) = \exp\lp -\hat C_l / \theta\rp\frac{ \hat C_l^{k-1}}{\theta^k \Gamma(k)},
\enq
with
\beq
\quad k := \frac 12 (2l +1),\quad \theta(\alpha,\beta) := \frac{2C_l}{2l +1},
\enq
and where $\Gamma$ is the gamma function. It is well known that the gamma distribution does indeed tend towards the Gaussian distribution for large $k$, with mean $\mu = k\theta = C_l$ and variance $\sigma^2 = k\theta^2 = 2C_l^2/(2l +1)$, as expected. However, its Fisher information content does not tend to that of the Gaussian. In our case, since only $\theta$ is parameter dependent, we have that the Fisher information in the estimator density function \eqref{pgamma} is
\beq
F_{\alpha\beta}^l = \frac{\partial \theta}{\partial \alpha}\frac{\partial  \theta}{\partial \beta}\av{\lp \frac{\partial \ln p (\hat C_l)}{\partial \theta}\rp^2}.
\enq
Since $\partial_\theta \ln p = (\hat C_l - k\theta)/\theta^2$, and $\partial_\alpha \theta = 2\theta \partial_\alpha C_l/C_l$, we obtain with straightforward algebra
\beq
\Fab^l = \frac 12(2l +1) \frac1 {C_l}\frac{\partial C_l}{\partial \alpha}\frac{1}{C_l}\frac{\partial C_l}{\partial \beta}.
\enq
Summing over $l$, we recover the first term of \eqref{Festimator}, but not the second. We have recovered the field perspective result \eqref{Ffield} at any $l$ without the Gaussian assumption but with the exact distribution. It turns out that even though the variance of the gamma distribution is parameter dependent, it does not in fact contribute to the information. This can be seen as the following. Consider the information in the mean only of the estimator. From the Cram\'er-Rao inequality this must be less than the total information,
\beq
\frac 1 {\sigma^2}\frac{\partial \mu}{\partial \alpha}\frac{\partial \mu}{\partial \beta} \le \Fab^l.
\enq
Plugging in the values for the mean and variance leads in fact to the result that the inequality is an equality,
so that the mean of the estimator captures all of its information.
\newline
\newline
In summary, the Gaussian approximation assumes  the mean and the variance of the estimator  are uncorrelated, such that both contributes to the information, while for the exact gamma, they are degenerate in such a way that the variance does not carry independent information.  Another way to see this, that we will use below when the exact form of the distribution will be less convenient, comes from the fact that $\partial_\theta \ln p(\hat C_l)$ is a first order polynomial in $\hat C_l$. It can be shown namely that the first $n$ moments capture all the information precisely when $\partial_\alpha \ln p$ is a polynomial of order $n$ \citep{2011ApJ...738...86C}.
\section{Several fields, Wishart distribution}\label{severalfields}
It is instructive to see how these considerations generalize to a situation of a family of $n$ jointly zero mean Gaussian correlated fields, where the analysis proceeds through the extraction of the spectra and cross spectra.  In this case, the $C_l$ of the above discussion becomes a $n \times n$ (possibly complex) Hermitian matrix
\beq \label{multi}
\av{a_{lm}^ia^{j*}_{lm}} = \delta_{ll'}\delta_{mm'}C_l^{ij}, \quad C^\dagger = C.
\enq
From the hermiticity property there are only $n(n+1)/2$ non redundant spectra. Adequate estimators are defined by a straightforward generalization of equation \eqref{estimator},
\beq
\hat C^{ij}_l = \frac{1}{2l +1}\sum_{m = -l}^l   a^i_{lm}a^{j*}_{lm}.
 \enq
While the estimators are still independent for different $l$'s, the different components at a given $l$ are not. 
The information content of the set of $a_{lm}^i$ in the field perspective is still given by formula \eqref{Fab} for zero mean Gaussian variables. Explicitly, at a given $l$,
\beq \label{Fmean}
\Fab^l = \frac 12 (2l +1) \Tr \lb C_l^{-1} \frac{\partial C_l}{\partial\alpha}C_l^{-1}\frac{\partial C_l}{\partial\beta} \rb.
\enq
In the estimator perspective, assuming the estimators $\hat C_l^{ij}, i \le j$ are jointly Gaussian, we have instead
\beq \label{Fmean2}
\sum_{i<j,k<l = 1}^n\frac{\partial C_l^{ij}}{\partial \alpha}\Sigma^{-1}_{ij,kl}\frac{\partial C_l^{kl}}{\partial \beta}
+
\frac 12 \Tr \lb \Sigma^{-1} \frac{\partial{\Sigma}}{\partial \alpha}\Sigma^{-1}\frac{\partial\Sigma}{\partial \beta}\rb,
\enq
where the covariance matrix is
\beq
\begin{split}
\Sigma_{ij,kl}&= \av{\hat C_l^{ij}\hat C_l^{kl}} - C_l^{ij}C_l^{kl} \\
&=  \frac 1{2l +1}\lp C_l^{ik}C_l^{jl} + C_l^{il}C_l^{jk} \rp.
\end{split}
\enq
While it may not be immediately obvious this time, it has been noted \cite[e.g.]{2004PhRvD..70d3009H} that the first term in \eqref{Fmean2} is rigorously equivalent to the expression from the field perspective \eqref{Fmean}. The estimator perspective under the assumption of a multivariate Gaussian distribution for $\hat C_l$ thus still violates the Cram\'er-Rao inequality due the presence of the second term. Since since this term is not enhanced by a factor of $2l +1$ we expect it to be subdominant again. However, it is less true this time than in the one dimensional setting : using the explicit form of the inverse covariance matrix,
\beq
\begin{split}
\Sigma^{-1}_{ij,kl} &=   \lp 2l +1 \rp\lp  C^{-1,ik}_l  C^{-1,jl}_l + C^{-1,il}_lC^{-1,jk}_l\rp\\
&\quad\cdot \lp 1 - \frac12 \lp \delta_{ij} + \delta_{kl} \rp+ \frac 14 \delta_{ij}\delta_{kl} \rp,
\end{split}
\enq
one can derive with some lengthy but straightforward algebra the following expression for the violating term,
\beq\label{violating}
\begin{split}
& \frac 12 \Tr \lb \Sigma^{-1} \frac{\partial{\Sigma}}{\partial \alpha}\Sigma^{-1}\frac{\partial\Sigma}{\partial \beta}\rb  \\
& = \frac 12 (n + 2) \Tr \lb C_l^{-1} \frac{\partial C_l}{\partial \alpha} C_l^{-1} \frac{\partial C_l}{\partial \beta}\rb \\
&\quad+ \frac12 \Tr \lb C^{-1}_l \frac{\partial C_l}{\partial \alpha}\rb\Tr \lb C^{-1}_l \frac{\partial C_l}{\partial \beta}\rb,
\end{split}
\enq
for any number $n$ of fields. If $n = 1$, we recover indeed \eqref{Festimator}. While the term is still subdominant at high $l$, the situation is yet a bit less comfortable.  The number of fields is not necessary very small in cosmologically relevant situations, such as tomographic joint shear and galaxy densities analysis in redshift slices, to which one may also add magnification, flexion fields, etc. Writing schematically $n = N_fN_{\textrm{bin}}$, where $N_{\textrm{bin}}$ is the number of bins and $N_f$ the number of fields per bin, we have e.g. $N_f = 3$ for the galaxy density and the two shear fields, $N_f = 4$ including magnification, $N_f = 8$ adding hypothetically the four flexion fields, and so on.  Comparing \eqref{Fmean} and \eqref{violating}, and neglecting the second term in \eqref{violating}, we have that at
\beq \label{lim}
l \sim \frac 12 N_fN_{\textrm{bin}},
\enq
the Cram\'er-Rao violating term is actually still the dominant one. Note that this is still optimistic. Due to the product of two traces in the second term in \eqref{violating},  one can expect  roughly the same scaling with $n$ as the first term. Thus, the correct $l$ in \eqref{lim} may generically be closer to
\beq
l \sim  N_fN_{\textrm{bin}}.
\enq
\newline\newline
From the discussion in section \ref{sectionperspective}, we can easily guess what went wrong. Consider the information content of the means of the estimators exclusively. This is given for any probability density function by weighting the derivatives of the means with the inverse covariance matrix, and is thus equal to the correct, first term in \eqref{Fmean2}. Since already the means of the estimators do exhaust the information in the field, we can therefore already conclude that the total information content of the estimators must be equal to that of their means, and in particular that the covariance does not contribute to the information. As before, the second term in the estimator perspective is an artifact of the Gaussian assumption. It is interesting though to derive as above more explicitly why only the means carry information, from the shape of the joint probability density of the estimators. The remainder of this section sketches how this can be simply performed, leading to equation \eqref{form}.
\newline
\newline
We restrict ourselves now for the sake of notation to the case of two fields, $n = 2$ , but the following argumentation holds for any $n$.
The exact joint distribution for the three estimators $\hat C_l = (\hat C_l^{11},\hat C_l^{12},\hat C_l^{22})$, is given from the rules of probability theory as
\beq \label{pcl}
\begin{split}
&p(\hat C_l |\alpha,\beta) = \av{ \prod_{i\le j = 1}^2\delta^D \lp \hat C^{ij}_l - \frac 1 {2l +1}\sum_{ m = -l}^l a^{i}_{lm}a^{j*}_{lm} \rp }
\end{split}
\enq
where $\delta^D$ is the Dirac delta function. The average is over the joint probability density for the two sets of harmonic coefficients $a^1_{lm}$ and $a^2_{lm}$. Define the vector
\beq
\veca_l = (a^1_{l-l},\cdots,a^1_{ll},a^2_{l-l},\cdots,a^2_{ll}).
\enq
Since the $a_{lm}$ are zero mean Gaussian variables with correlations as given in \eqref{multi}, this probability density function is given by
\beq \label{palm}
\frac{1}{Z(\alpha,\beta)}\exp \lp-\frac 12 \veca_l^\dagger\cdot \mathbf C_l^{-1}\veca_l \rp,
\enq
with
\beq
\mathbf C_l = \bem C_l^{11}\cdot 1_{2l+1} &  C_l^{12}\cdot 1_{2l+1} \\  C_l^{21}\cdot 1_{2l+1} &  C_l^{11}\cdot 1_{2l+1}\enm,
\enq
where $1_{2l +1}$ is the unit matrix of size $2l +1$. $Z(\alpha,\beta)$ is the normalization of the density for $\veca$, that does depend on the model parameters through the determinant of the $\bf C_l$ matrix. The inverse matrix $\bf C_l^{-1}$ has the same block structure, with entries being those of $C^{-1}_l$. In the following we are not really interested in keeping track of the exact value of the components of this matrix, but only that they are dependent on the model parameters. With the understanding that $ C_l^{-1} =: D_l$, we have thus, due to the sparse structure of the $\mathbf C^{-1}_l$ matrix and the Dirac delta functions in \eqref{pcl},
\beq
\begin{split}
-\frac 12\veca_l^\dagger\cdot \mathbf C^{-1}_l\veca_l 
=-\frac12(2l +1)  \sum_{i,j = 1,2}D_l^{ij}\hat C_l^{ij} 
\end{split}
\enq
Due to the presence of the Dirac delta functions, we can thus take the exponential \eqref{palm} out of the integral in \eqref{pcl}.  Writing explicitly the dependency of the different terms on $\hat C_l$ and the model parameters, we obtain the following form 
\beq \label{form}
\begin{split}
p(\hat C_l|\alpha,\beta) &  \\
=  \frac {f(\hat C_l)} {Z(\alpha,\beta)} &\exp\lp -\frac12 (2l + 1) \sum_{i,j = 1,2}D_l^{ij}(\alpha,\beta)\hat C_l^{ij} \rp
\end{split}
\enq
which generalizes the gamma distribution, equation \eqref{pgamma}, in this multidimensional case. The factor $f(\hat C_l)$ is what is left from the integral \eqref{pcl} when the density for the set of $a_{lm}$ is taken out, i.e. the volume of the space spanned by the $a_{lm}$'s that satisfies the constraints set by the Dirac delta function. It is thus a factor that depends on $\hat C_l$ but importantly for us not on the model parameters \footnote{The prefactors in \eqref{form} can be obtained in closed form, leading to the Wishart density function. See \citep[e.g.]{2008PhRvD..77j3013H}}. The point of the representation \eqref{form} is that it is immediate that $\partial_\alpha \ln p(\hat C_l)$ is a polynomial first order in the components of $\hat C_l$. Second order terms, corresponding to information within the covariance matrix never appear, however close to a Gaussian the exact density function might be.
It follows that the total Fisher information matrix is always equal to that of the mean, even if we did not derive the exact shape of the distribution.
\section{Summary and conclusions} \label{conclusions}
We discussed two common perspectives (the 'field' and 'estimator' perspectives) on the Fisher information content of cosmological power spectra, and why in the estimator perspective the assumption of a Gaussian likelihood of the spectra estimators violates the Cram\'er-Rao inequality, assigning the estimators more information than there is in the full underlying fields. Under the assumption of Gaussianity of the estimators, their means and covariance matrix are artificially rendered uncorrelated, creating an additional piece of information in their covariance, that we showed was inexistent by calculating the exact information content of the estimators true probability density function. We showed that this violating term can become dominant in the limit of a large number of fields. Using Gaussian likelihoods consistently, i.e. with parameter dependent covariance matrices, as argued for example in \citep{2009A&A...502..721E}, assigns therefore far too much information to the spectra in this regime, and should thus be avoided.
\newline
\newline
In the estimator perspective of the derivation of the Fisher information matrix, this term is usually neglected. This note clarifies why it should not be present in the very first place, and how the agreement between the field and estimator perspective can thus arguably be seen as an happy cancellation of two inconsistencies. It is interesting to note 
that the reason why we still find the exact result in the estimator perspective without this wrong piece is that this expression is also the exact Fisher information content of the exact, for low $l$ strongly non Gaussian, distribution of the estimators, the central limit theorem playing actually no role.
\newline
\newline
The other lesson we can take from this work is that in general, when in doubt about the joint distribution of a set of estimators, a safe choice of information content is always that of their means exclusively, which requires only the knowledge of their covariance. Provided the covariance matrix is correctly chosen, one is indeed sure for any probability density function from the properties of Fisher information to make  a conservative evaluation, that does not rely on any further assumptions on its shape. Thus, leaving apart the question of the very accuracy of the approximation itself, using a Gaussian likelihood with parameter independent covariance matrix, having the entire information in the means, while not entirely consistent remains a safe prescription in the sense that a conservative information content is always assigned to the estimators. 

%

\clearpage
\newpage

\chapter{$N$-point functions in lognormal density fields \label{ch4}}
\newcommand{\muln}{\bar A}
\newcommand{\covln}{\xi_{A}}
\newcommand{\sigln}{\sigma_{\ln\rho}}
\newcommand{\murho}{\bar \rho}
\newcommand{\sigd}{\sigma_\delta}
\newcommand{\sigrho}{\sigma_{\rho}}
\newcommand{\snlnmu}[2]{s_{#1}^{\muln}(#2)}
\newcommand{\snsig}[2]{s_{#1}^{\sigln^2}(#2)}
\renewcommand{\veck}{\boldsymbol \omega}
\newcommand{\idk}{\frac{d^3k}{(2\pi)^3}}
\newcommand{\Vcell}{V_{\textrm{cell}}}
In this chapter we discuss extensively the information content of the lognormal field as a model for the matter fluctuation field in cosmology.
 It is built out of published as well as yet unpublished elements. The text in sections \ref{intro} and \ref{onevariable} follows closely that of \cite{2011ApJ...738...86C}, and that of sections \ref{conventions}, \ref{notationlog}, \ref{otherfields} and \ref{connection} that of \cite{2012ApJ...750...28C}. On the other hand, sections \ref{limitations}, \ref{uncorrfid} and \ref{generalcase} present unpublished material.
 
\section{Introduction \label{intro}}
The cosmological matter density field is becoming more and more directly accessible to observations with the help of weak lensing  \citep{1992grle.book.....S,2001PhR...340..291B,2003ARA&A..41..645R,2006astro.ph.12667M}. Its statistical properties are the key element in trying to optimize future large galaxy surveys aimed at answering actual fundamental cosmological questions, such as the nature of the dark components of the universe \citep{2009Natur.458..587C,2008ARA&A..46..385F}. To this aim, Fisher's measure of information on parameters \citep{fisher25,Rao,vandenbos07} has naturally become of standard use in cosmology. It provides indeed an handy framework, in which it is possible to evaluate in a quantitative manner the statistical power of some experiment configuration aimed at some observable \citep[e.g.]{1997ApJ...480...22T,Tegmark97b,1999ApJ...514L..65H,2004PhRvD..70d3009H,2007MNRAS.381.1018A,2007MNRAS.377..185P,2006astro.ph..9591A,2009ApJ...695..652B}. Such studies are in the vast majority of cases limited to Gaussian probability density functions, or perturbations therefrom, and deal mostly with the prominent members of the correlation function hierarchy \citep{1980lssu.book.....P}, or equivalently their Fourier transforms the polyspectra, such as the matter power spectrum.
\newline
\newline
The approach via the correlation function hierarchy is very sensible in the nearly linear regime for at least two reasons. First, in principle, the correlations are the very elements that cosmological perturbation theory is able to predict in a systematic manner (see \cite{2002PhR...367....1B} for a review, or the more recent \citep{2011PhRvD..83h3518M} and the numerous references in it). Second, primordial cosmological fluctuations fields are believed to be accurately described by the use of Gaussian statistics. It is well known that the correlations at the two-point level provide a complete description of Gaussian fields. It is therefore natural to expect this approach to be adequate throughout the linear and the mildly non linear regime, when departures from Gaussianity are small.
\newline
\newline
Deeper in the non linear regime, fluctuations grow substantially in size, and tails in the matter probability density function do form. A standard prescription for the statistics of the matter field in these conditions is the lognormal distribution, various properties of which are discussed in details in an astrophysical context in \citep{1991MNRAS.248....1C}. It was later shown to be reproduced accurately, both from the observational point of view as well as in comparison to standard perturbation theory and N-body simulations \citep{1995ApJ...443..479B,1994A&A...291..697B,2001ApJ...561...22K,2000MNRAS.314...92T,2005MNRAS.356..247W}, in low dimensional settings. The lognormal assumption is also very much compatible with numerical works \citep{2009ApJ...698L..90N,2011ApJ...731..116N} showing that the spectrum of logarithm of the field $\ln 1 + \delta$ carries much more  information than the spectrum of $\delta$ itself. The first evaluation of the former within the framework of perturbation theory appeared recently \citep{2011ApJ...735...32W}.
\newline
\newline
Lognormal statistics \citep[for a textbook presentation]{Aitchison57} are not innocuous. More specifically, the lognormal distribution is only one among many distributions that leads to the very same series of moments. This fact indicates that, going from the distribution to the moments, one may be losing information in some way or another. A fundamental limitation of the correlation function hierarchy in extracting the information content of the field in the non linear regime could therefore exist, if its statistics are indeed similar to the lognormal. This important fact was already mentioned qualitatively in \citep{1991MNRAS.248....1C}, but it seems no quantitative analysis is available at present.
\newline
\newline
 In this chapter we provide first answers to these issues, looking at the details of the structure of the information within the lognormal field and its $N$-point moments. This chapter is built as follows :
 \newline
 \newline
 We start by exposing briefly two fundamental limitations of the lognormal assumption in section \ref{limitations}. We continue in section \ref{otherfields} defining explicit families of fields that have the very same hierarchy of $N$-point moments than the lognormal field. This is making obvious that the hierarchy does not provide a complete description of the field, and therefore does not contain the entire information. In section \ref{onevariable}, we quantify the importance of this aspect in terms of Fisher information : we solve in that section for the statistical power of the moments of the lognormal distribution exactly at all orders, and  for their efficiency in capturing information. We also compare these predictions to standard perturbation theory. In section \ref{connection} we then use these results derived in section \ref{onevariable} to make a successful connection to the $N$-body simulation results mentioned above, that showed that the statistical power of the spectrum of the logarithm of the field is larger than that of the original fluctuation field. In section \ref{uncorrfid}, we extend the results of \ref{onevariable} by allowing parameters to create correlations between the variables, and discuss in this light the fact that the improvement seen in simulations are seen to be mostly parameter independent.
 We conclude in section \ref{generalcase} with a remark on the derivation of the statistical power of the $N$-point moments for an arbitrary lognormal field.

 \subsection{Notation and conventions \label{conventions}}
We will be dealing throughout this chapter with random vectors $\rho = (\rho_1,\cdots,\rho_d)$, being the sample of a density field $\rho$,
\beq
\rho_i = \rho(x_i) > 0.
\enq
We place ourselves in 3-dimensional cartesian space for convenience.
For a vector $\vecn = \lp n_1,\cdots,n_d \rp$ of non negative integers (multiindex), we write as in chapter \ref{ch2} $\rho^\vecn$ the monomial in $d$ variables,
\beq
\rho^\vecn  = \rho(x_1)^{n_1}\cdots \rho(x_d)^{n_d}.
\enq
Throughout this chapter, we reserve bold letters for vectors of integers exclusively.
\newline
\newline
Let  $p_\rho(\rho)$ be a $d$-dimensional probability density function such that all correlations of the form $\av{\rho^\vecn}$ are finite.  We write the moment $\av{\rho^\vecn}$ with $m_\vecn$. Explicitly,
\beq
m_\vecn = \av{\rho^{n_1}(x_1)\cdots \rho^{n_d}(x_d)}.
\enq
Correlations of order $n$ are given by moments such that the order $|\vecn|$ of the multiindex, defined as
\beq
|\vecn| :=  \sum_{i = 1}^{d}n_i
\enq
is equal to $n$.
These moments coincide with the values of a continuous $n$-point correlation function on the grid sampled by $(x_1,\cdots,x_d)$. We write $\delta$ for the dimensionless fluctuation field, and $A$ for the field defined by $\ln \rho$: 
\beq 
A := \ln \rho,\quad \delta := \frac{\rho -\bar\rho}{\bar \rho}.
\enq
Such assignments involving ratios or logarithms of $d$-dimensional quantities should always be understood component per component.
\subsection{Definition and basic properties of correlated lognormal variables \label{notationlog}}
We say the $d$-dimensional vector $\rho := (\rho(x_1),\cdots,\rho(x_d))$ is lognormal if the $d$-dimensional probability density function for $A$ is Gaussian. Explicitly,
\beq \label{A}
p_A(A) = \frac1{\lp 2\pi |\xi_A| \rp^{d/2}}\exp\lp  -\frac 12 (A - \bar A) \cdot \xi_A^{-1} (A -\bar A) \rp,
\enq
where $\bar A$ is the mean vector of $A$, and $\xi_A$ its covariance matrix, 
\beq
\lb \xi_A \rb_{ij} = \av{\lp A(x_i) - \bar A(x_i) \rp \lp A(x_j) - \bar A(x_j) \rp}.
\enq
The probability density for the vector $\rho$ itself is then by construction a $d$-dimensional lognormal distribution. We name it for further reference as $p^{LN}_\rho$. From the rules of probability theory holds
\beq \label{pLN}
\begin{split}
p^{LN}_\rho(\rho) = \frac{ p_A(\ln \rho) }{\prod_{i = 1}^d \rho(x_i)}.
\end{split}
\enq
The means and two point correlations of $A$ and $\delta$ are in one to one correspondence. We have
\beq \label{meanA}
\bar A = \ln \bar \rho - \frac 12  \sigma^2_A
\enq
where $\sigma^2_A$ is the diagonal of $\xi_A$, i.e. the variances of the individual $d$ points. Also,
\beq \label{reldA}
\lb \xi_A \rb_{ij} = \ln \lp 1 + \lb  \xi_\delta \rb_{ij} \rp,\quad \lb \xi_\delta \rb_{ij}  := \av{\delta(x_i)\delta(x_j)} .
\enq
Especially, the variances are related through
\beq
\sigma^2_A  = \ln \lp 1 + \sigma^2_\delta\rp.
\enq
In the multiidindex notation, the $N$-point moments of $\rho$ take the following simple form,
\beq\label{Correlationrho} 
m_\vecn = \exp \lp  \bar A \cdot \vecn + \frac 12 \vecn \cdot \xi_A \vecn\rp.
\enq
The mapping from $A$ to $\rho$ is invertible, and $A$ is a multivariate Gaussian distribution. Therefore the total Fisher information content of $\rho$ is given by
\beq \label{FILN}
F_{\alpha \beta} = \frac{\partial \bar A^T }{\partial \alpha} \xi^{-1}_A \frac{\partial \bar A }{\partial \beta} + \frac 12 \Tr \lb \frac{\partial \xi_A}{\partial \alpha} \xi_A^{-1}\frac{\partial \xi_A}{\partial \beta}\xi_A^{-1}\rb.
\enq
\section{Fundamental limitations of the lognormal assumption \label{limitations}}
To open this chapter, we discuss two fundamental limitations of the lognormal assumption for the matter density field.
We are interested in the validity of the approximation, i.e. the mere possibility to describe at all the $\Lambda$CDM matter density field with an homogeneous isotropic lognormal field. It is indeed not guaranteed a priori that quantities derived from this assumption will be well defined. For instance, any sensible statistical model requires covariance and information matrices to be positive, and it is not clear yet whether the lognormal assumption on the matter density field satisfies these conditions.
\newline
\newline
It is important to note that this is very different from testing the accuracy of the approximation. For instance, given a two-point correlation function (or equivalently a positive Fourier transform, the power spectrum), it is always possible to use the Gaussian field as a prescription, even though this approximation may be extremely inaccurate. This is because the two-point function is the only relevant ingredient, and the Gaussian field with that two-point correlation function is well defined statistically speaking in all cases.
\newline
\newline
With the lognormal field this is not the case anymore. Namely, given the two-point correlation function $\xi_\delta$ with positive Fourier transform, we see from \eqref{reldA}, in a continuous notation, that there is the additional constraint that
\beq
\xi_A(r) = \ln \lp 1 + \xi_\delta(r)\rp
\enq
must be a valid two-point correlation function as well, since it is that of the $A$ field. Since $A$ is by definition a Gaussian field,  we see that formally for the lognormal assumption to be valid, the following Fourier transform must be positive for all $k$,
\beq \label{PAK}
\int d^3r \ln (1 + \xi_\delta(r)) e^{- i k \cdot r } = P_A(k) \stackrel{!}{\ge} 0. 
\enq
We discuss in the following that these non trivial constraints are not satisfied both on the largest and on the smallest scales in our current understanding of the $\Lambda$CDM power spectrum.
\subsection{Largest scales}
An immediate issue is the power of the $A$ field on the very largest scales. Setting $k = 0$ in \eqref{PAK} gives us
\beq
P_A(0)  = \int d^3 r \: \ln \lp 1+ \xi_\delta(r) \rp.
\enq
On the other hand, we have that for any argument $x$ holds $\ln (1 + x) \le x$, with equality if and only if $x = 0$. For this reason, we have
\beq
P_A(0)  < \int d^3r \: \xi_{\delta}(r)= P_\delta(0).
\enq
Now, the spectrum of the fluctuation field is believed to have the scale free shape of a power law $P_\delta \propto k^n$, with $n$ very close to unity \citep{2011ApJS..192...18K}. Extrapolated to the smallest wavenumbers, this functional form obviously assigns zero power at zero. We find therefore the contradiction
\beq
P_A(0) < 0.
\enq
On the largest scales, the power of the $A$ field is thus found to be negative under this lognormal assumption, rendering it formally untenable.
\newline
\newline
Nevertheless, this issue cannot be considered a real shortcoming of the model. Namely, these large scales are in no way observable and are irrelevant to any realistic situation. In particular the power at zero, formally the variance of the mean of the field from realisations to realisations of the Universe within the statistical model carries no meaning for observations.
\subsection{Smallest scales}
\begin{figure}
  \includegraphics[width =0.4 \textwidth]{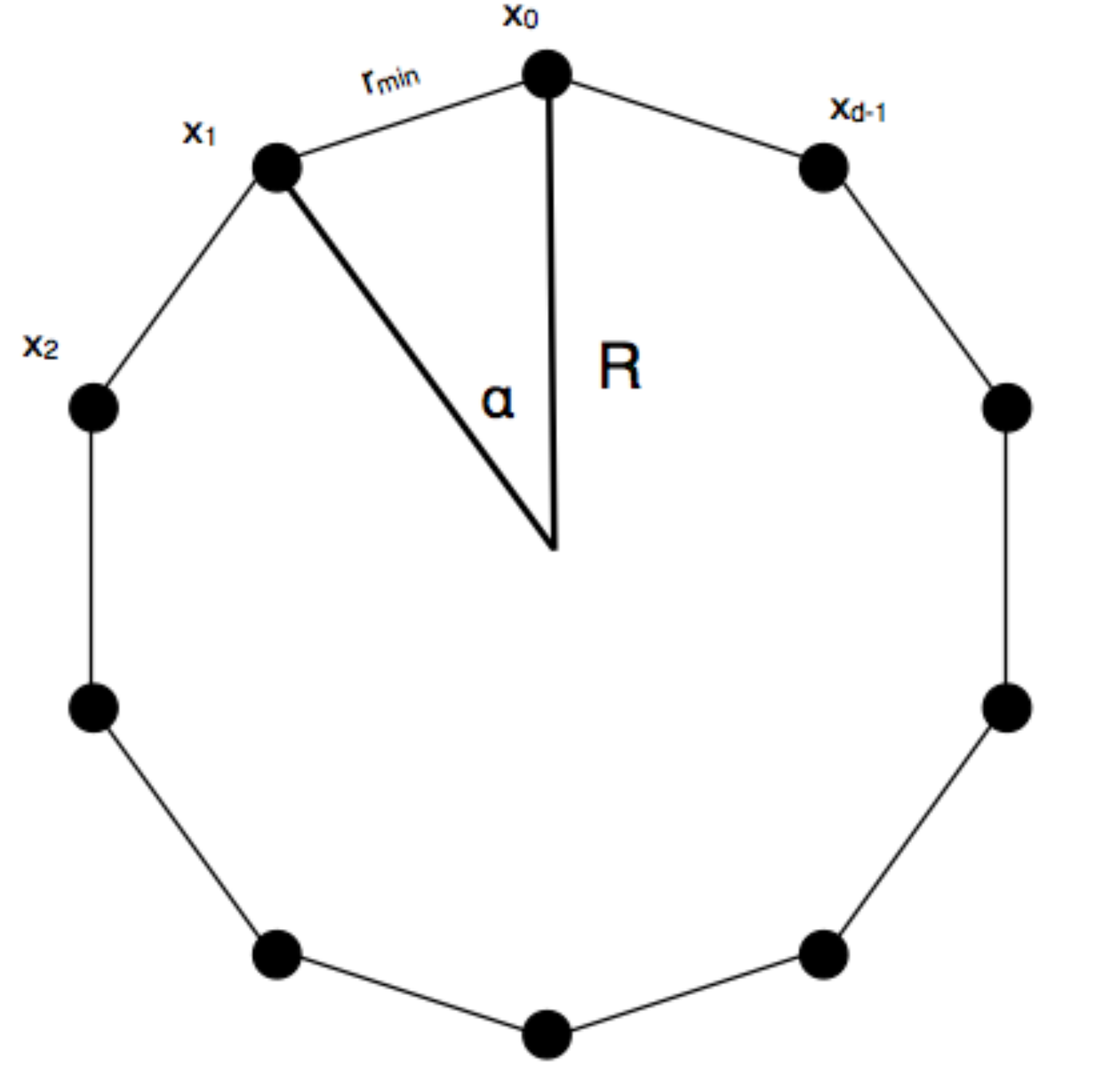}
   \includegraphics[width = 0.5\textwidth]{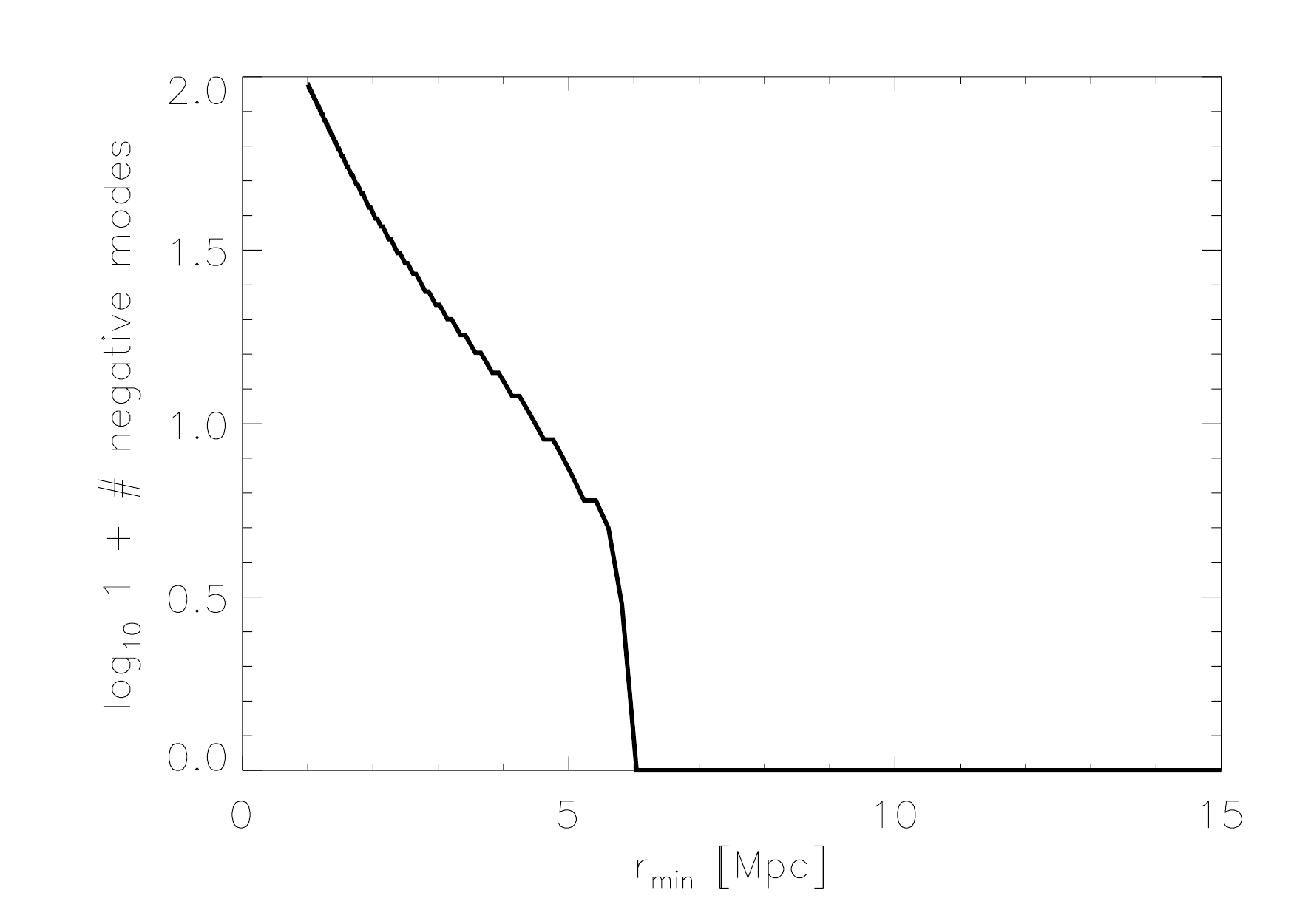}

 \caption[The one dimensional chain of points used to test the positivity of the spectrum of the $A$ field, and the apparition modes with negative spectrum on small scales, indicating that the lognormal field is no longer a well defined assumption for a $\Lambda$CDM cosmology.]{\label{LNchain}Left panel : The maximally symmetric configuration of points, including a finite range of distances, chosen to test the positivity of the spectrum of the $A$ field, for the $\Lambda$CDM  two-point function. Right panel : On a logarithmic scale, the number of wavenumbers such that the spectrum of the $A$ field is found to be negative. Such mode appear once scales below $5-6$ Mpc are included, indicating that a lognormal description is no longer well defined below these scales.}
 \end{figure}

More interesting are the small scales, since these are the ones that are accessible to observations. Performing the angular integration in \eqref{PAK}, the conditions that the two-point function must obey become
\beq \label{cstrfield}
\int_0^\infty dr \:r^2 \ln \lp 1 + \xi_\delta(r) \rp j_0(kr)  \ge 0, \textrm{   for all } k,
\enq
where $j_0(x) = \sin(x)/x$ is the first spherical Bessel function.
These relations are neither convenient to test nor the most meaningful however. On one hand, due to the $r^2$ factor it requires the correlation function at very large argument. On the other hand, the inequalities \eqref{cstrfield} represent the constraints puts on a continuous lognormal field in an infinite volume. They are thus rather formal and do not correspond to a situation that can occur in practice.
We therefore build another set of similar identities that are both simpler to test and more importantly where we can investigate a range of specific scales without ambiguity.
\newline
\newline
For this we use a finite but maximally symmetric configuration of points $x_i$ where the corresponding constraints become the positivity of the two-point correlation matrix 
\beq
\lb \xi_A \rb_{ij}  = \av{ \lp A(x_i) - \bar A \rp \lp A(x_j) - \bar A \rp}
\enq
The ideal configuration for this purpose is the classic chain with $d$ points described in figure \ref{LNchain}.

 \noindent
 In this configuration, periodic boundary conditions are built in by construction, such that we are still able to write the two-point correlation matrix as the transform of a positive spectrum. The two parameters $R$ and the number $d$ of points (or equivalently $r_\min $) allow us to keep complete control over the range of scales that are being tested. From rotational invariance, it is seen that the distances involved are $r_j , j =0 ,\cdots d-1$, given by
\beq
r_j = \left | x_j - x_0 \right| =    2R\sin \lp  j \frac \pi {2d} \rp, \quad j =0,\cdots,d-1.
\enq
The correlation matrix becomes
\beq
\lb \xi_A \rb_{ij} = \xi_A ( r_{|j-i|}).
\enq
 This matrix has the structure of a so-called circulant matrix. Circulant matrices are always diagonalised by the discrete Fourier transform : the eigenvalues are
\beq
P_k = \sum_{ j= 0}^{d-1}  \xi_A(r_j) \exp \lp  - 2\pi i k \frac j N \rp, \quad k = 0,\cdots, d-1,
\enq
and the matrix can be written as
\beq
 \xi_A(r_j) = \frac 1d \sum_{k = 0}^{d-1} P_k \exp \lp 2\pi i k \frac j N \rp.
\enq
The eigenvalues $P_k$ must be positive for $\xi_A$ to be a valid two-point correlation function.
\newline
\newline
We have thus found a new set of constraints :
\beq \label{FFTspec}
\begin{split}
&0 \le \sum_{ j= 0}^{d-1}  \ln \lp 1 + \xi_\delta(r_j)\rp \exp \lp  - 2\pi i k \frac j d \rp, \\& \quad k = 0,\cdots, d-1,\textrm{  and  for any $R$ and $d$}.
\end{split}
 \enq
These constraints can be trivially tested using an FFT algorithm, and involve distances between $r_\min \sim 2\pi R / d$ and $2R$ exclusively. A two-point correlation function $\xi_\delta$ must satisfy these constraints in order for the field to be possibly lognormal.
 \newline
 \newline
In the limit of large number of points, the sum goes over to an integral, and it is easy to show that these constraints become
\beq
0 \le \frac 1 \pi \int_0^\pi d\alpha \: \ \ln \lb 1 + \xi_\delta\lp 2R\sin \lp \frac \alpha 2 \rp \rp \rb  \: \cos \lp \alpha k \rp,
\enq
again for $k= 0,1,\cdots$ and any $R$.
For very large $R$ , $P_k$ becomes the spectrum of the one-dimensional field. 

 \noindent We tested these inequalities for the $\Lambda$CDM correlation function. We generated the power spectrum with the help of the iCosmo package \citep{2008arXiv0810.1285R} smoothed it with a top-hat filter of size $r_\min$, Fourier transformed it to obtain $\xi_\delta$, and then tested straightforwardly equations \eqref{FFTspec} with an FFT algorithm.
 \newline
 \newline
 While we found to the radius $R$ to be of very little relevance, these inequalities break down on scales $r_\min$ below $5-6$ Mpc (corresponding to a variance $\sigma^2_\delta$ of roughly 3) as shown on the right panel of figure \ref{LNchain}, where the spectrum on these scales gets negative. It follows that the lognormal assumption is statistically speaking not well defined on these small scales for a $\Lambda$CDM cosmology.   
 \section{Explicit fields with the same hierarchy of $N$-point functions}\label{otherfields}
 In this section, we present explicit families of density fields with the same $N$-point moments of any order that the lognormal field. This allows us to demonstrate in clearest possible way the fact that the hierarchy of $N$-point functions never provide a complete description of a lognormal field, and discuss in the light of these families some implications. We first begin in \ref{problem} with a discussion on the multidimensional moment problem and its link to some concepts already present in the cosmological literature. The families are then defined in \ref{continuousfields} and \ref{discretefields}. The appendix \ref{proofs} contains the proofs of the key statements. 
\subsection{The problem with tailed fields}\label{problem}
As already mentioned in chapter \ref{ch2}, in one dimension the fact that the hierarchy does not always specify fully the distribution is a well known and still active topic of research in the theory of moments in mathematics \citep[for classical references]{Shohat63,Akhiezer65,Simon97theclassical}. The moment problem is to find a distribution corresponding to a given moment series. When a unique solution exists, it is called a  determinate moment problem. When several exist (in this case always infinitely many), it is called an indeterminate moment problem. We can refer at this point to \citet{1991MNRAS.248....1C} for a discussion in a cosmological context. The theory of the moment problem in several dimensions is less developed, but typical criteria that guarantee determinacy, or indeterminacy, linked to the decay rate of the distribution, stay basically unchanged. Guiding us throughout the discussion in this section will be the following instance: for any dimension $d$, if
\beq \label{criterium}
\av{e^{ c |\rho | }} < \infty, \quad |\rho| = \lp \rho_1^2+ \cdots + \rho_d^2\rp^{1/2}
\enq
for some $c > 0$, then the moment problem corresponding to the moments of that distribution is determinate \citep[theorem 3.1.17]{Dunkl01}.
By a  'tailed' distribution, we have in mind in this work a decay at infinity which is less than exponential, and thus for which this criterion fails. In this regime, there may thus be several distributions with the same hierarchy of correlations.
\newline
\newline
It should be clear why this can have in general a dramatic impact for parameter inference from correlations. Imagine a series of distributions with identical correlations at all orders, one of these distributions being the one that actually describes the observations. Since the distributions are different, they will make in general different predictions for observables other than the correlations. Pick for definiteness an observable $\av{f(\rho)}(\alpha)$ with different predictions among this family of distribution, $\alpha$ any model parameter. The knowledge of the entire hierarchy is unable to distinguish from these different predictions for $\av{f}$, since they result from equally valid distributions.
If $\alpha$ enters the true distribution in such a way that it makes a sharp prediction on the value of $\av{f}$, this is highly valuable information definitely lost to an analyst extracting correlations exclusively.
On the other hand this argument allows us also to see that this effect can become relevant only when perturbation theory breaks down. If the fluctuation field $\delta$ is small, $f$ can be expanded in powers of $\delta$, and thus $\av{f}$ can be obtained in an unique way from the correlation hierarchy of $\delta$.
\newline
\newline
 The formalism developed in chapter \ref{ch2} provides us with a remarkable way to understand what is happening there in terms of Fisher information, familiar to cosmologists.  In particular, we have seen that the distributions for which the Fisher information matrix is within the entire hierarchy are precisely those for which the functions $\partial_\alpha \ln p$ can be written as a power series over the range of $p$. If not, the mean squared residual to the best series expansion is the amount of Fisher information absent from the hierarchy.  It is simple to show that criterion \eqref{criterium}, that guarantees that the distribution is uniquely set by its correlations, implies as well that the entire amount of Fisher information is within the hierarchy : this follows from the very next theorem of the same reference \citep[theorem 3.1.18]{Dunkl01}, that states that the polynomials in the $d$ variables form a dense set of functions with respect to the least mean squared residual criterion, if \eqref{criterium} is met. In particular the functions $\partial_\alpha \ln p$ can be arbitrarily well approximated by polynomials with respect to that criterion,  and therefore the correlations contain all of the Fisher information. 
\newline
\newline
It is important to note that if criterion \eqref{criterium} happens to be met due to a cutoff at a large value $\rho_\textrm{cut}$, on a otherwise tailed distribution, the correlations still are poor probes for any practical purposes.
For instance, if a variable is lognormal over a very long range, but decay quickly at infinity starting from $\rho_\textrm{cut}$. Indeed, if $\rho_\textrm{cut}$ is large enough, the correlations of order up to, say, $2N$, will be identical to that of the lognormal. Since the information content of the first $N$ correlations depends on the first $2N$ only, they will be equally poor probes as for the lognormal (this will be quantified in the next two sections). They will contain the exact same amount of Fisher information as the ones of the lognormal. It is the correlations of order $>N$, that are able to feel the cutoff,  that will make up for the difference between the total information content of the lognormal distribution and its correlation hierarchy (if the cutoff is at a large enough value,  from \eqref{Fab} the two distributions have the same total amount of information). The hierarchy is thus still not well suited for the analysis of data in this regime.
\newline
\newline
For the same reason, even though any lognormal field is indeterminate, this effect plays no role for parameter inference in the linear regime, when the actual range of the variables is still small, and the tail at infinity is not yet felt. This is because in this regime on one hand the lognormal is still very close to a Gaussian over the range where it takes substantial values, and thus the lowest order correlations will still contain most of the Fisher information, and on the other hand a few higher order terms are able to reproduce deviations of the functions $\partial_\alpha \ln p$ from the Gaussian very accurately over this small range. This is consistent with the findings in section \ref{continuousfields} and \ref{discretefields} showing that the families presented there are indistinguishable from the lognormal for any practical purposes in the linear regime.
\newline
\newline
Let us comment in light of the criterion \eqref{criterium} on typical perturbative approaches in cosmology to parametrize (weakly) non Gaussian distributions. These involves moments, such as Gram-Charlier, Edgeworth expansions, or the relation between the moment generating function and the distribution \citep[e.g.]{1985ApJ...289...10F,1994A&A...291..697B,1994ApJ...435..536C,1995ApJ...442...39J,1995ApJ...443..479B,1998A&AS..130..193B}, in one or several dimensions. It is therefore interesting to see to what extent they fit into this picture.  Typically, when applied to the $\delta$ field, to first order these parametrize the non-Gaussianity through a polynomial with coefficients involving the cumulants, or equivalently the moments of the variable. Schematically,
 \beq \label{Edge}
 p_\nu(\nu) \propto e^{-\nu^2/2}\lp1 + \alpha_3H_3(\nu) + \alpha_4H_4(\nu) +\cdots\rp,
 \enq
 with $\nu = \delta/\sigma_\delta$. The coefficient $\alpha_i$ depends on the first $i$ moments. The correction is given in terms of Hermite polynomials $H_{n}$, which are the orthogonal polynomials associated to the Gaussian distribution. Such expansions never produce a tailed distribution, in the sense that \eqref{criterium} is always met. The decay of the distribution namely still is Gaussian. Now, to first order and over the range of $p$, equation  \eqref{Edge} is equivalent to
 \beq
 \ln p_\nu(\nu) \approx \cst - \nu^2/2 + \alpha_3H_3(\nu) + \alpha_4H_4(\nu) + \cdots 
 \enq
 Therefore, the functions $\partial_\alpha \ln p$ will have close to polynomial form.
 This is perfectly consistent with that decomposition of the Fisher information. Indeed, this expansion creates  a probability density for which its Fisher information content is within the moments that were used to build it. This is another way to see that moment-indeterminate distributions cannot be produced by perturbative expansions
 \newline
 \newline
We now present  both continuous as well as discrete families of probability density functions that have the same correlations as the lognormal at all orders, for any dimensionality $d$. In fact, it turns out that a stronger statement is true :  for these families, all observables of the form
\beq
\av{\rho(x_1)^{n_1}\cdots\rho(x_d)^{n_d}},\quad n_i = \cdots -1,0,1 \cdots
\enq
are identical to those of the lognormal field, i.e. any power $n_i$ can also be negative as well. Including the hierarchy of inverse powers and 'mixed' powers to the usual hierarchy thus still does not provide a complete description.
\newline
\newline
These families are generalizations to any number of dimension, means and two-point correlations of known one dimensional examples that can be found in the statistical literature \citep{Heyde63,Stojanov87}.
\newline
\newline
Requirements such as homogeneity and isotropy are actually not needed in the following section. In particular, unless otherwise specified, $\bar A$ is a $d$-dimensional mean vector $(\bar A(x_1),\cdots,\bar A(x_d))$, whose components can differ in principle. Nevertheless, the picture we have in mind is that of statistically homogeneous isotropic fields in a box of volume $V$, where some set of Fourier modes $k_\min$ to $k_\max$ can be probed. The corresponding Fourier representation of the two point correlations, in a continuous notation, is
\beq
\lb \xi_{A,\delta}\rb_{ij} = \int \idk P_{A,\delta}(k)e^{ik\cdot (x_i - x_j)} = \xi_{A,\delta}(x_i-x_j),
\enq
where the integral runs over these modes, and $\xi_{A,\delta}(r)$ is the ordinary two-point correlation function of  $A$ or $\delta$. The matrix inverse is given by
\beq \label{contrepr}
\lb \xi^{-1}_{A,\delta}\rb_{ij} = \int \idk \frac{1}{P_{A,\delta}(k)}e^{ik\cdot (x_i-x_j)}.
\enq
This representation allow us to define a bit more rigorously what we mean by linear and non-linear lognormal field, or linear and non-linear regime, in the following discussion : if needed, it can be formally set as $P_{A}(k) \rightarrow 0 $ or $P_{A}(k) \rightarrow \infty $ respectively, for all $k$.
\subsection{Continuous family \label{continuousfields}}
\begin{figure}
 \includegraphics[width = 0.5\textwidth]{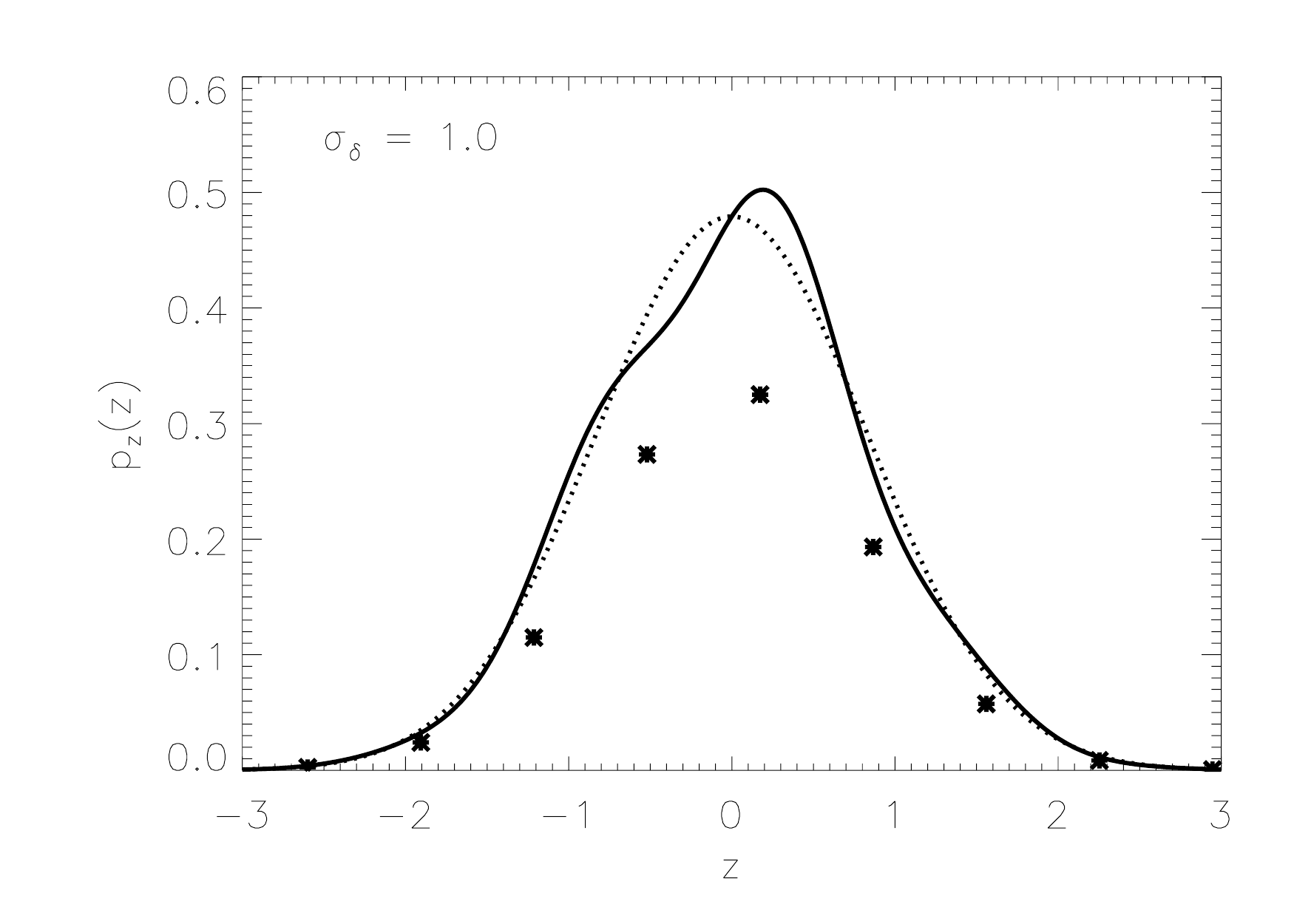}
  \includegraphics[width = 0.5\textwidth]{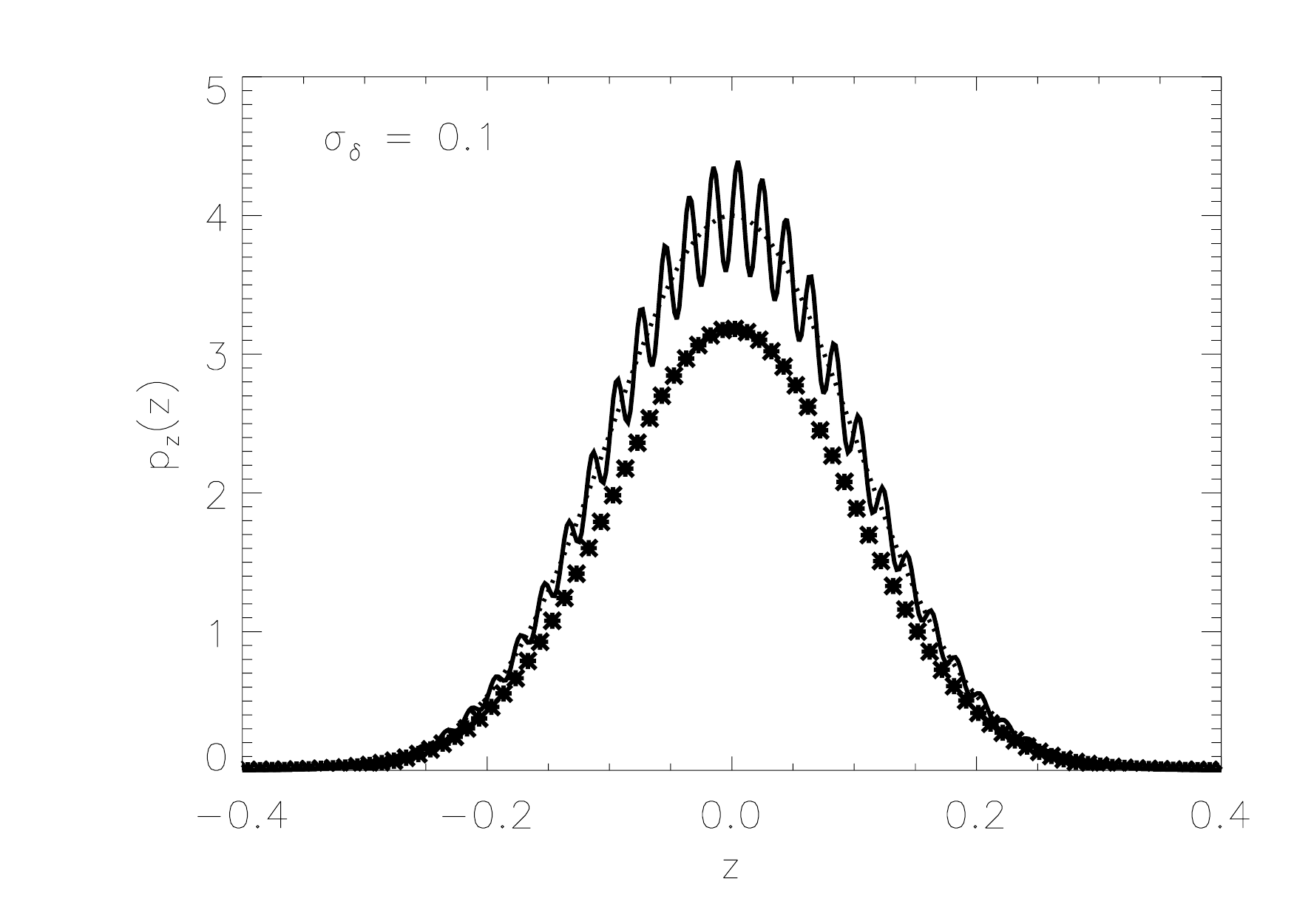}
  \caption[Three density functions with the same moments as those of the lognormal distribution at all orders.]{\label{fig:pdfs} Left panel : Three different one dimensional distributions for $z := A-\bar A$, with identical moments $\av{\rho^n}, \: \rho = e^A$, for all integer $n$, positive or negative. The dashed line is the zero mean Gaussian distribution, so that $\rho$ is lognormal. The solid the member of the family in \eqref{Heydeextended} with the lowest possible frequency, and amplitude $\epsilon = 0.1$. The discrete one is \eqref{discreteP} with shift parameter $\alpha = 0.25$. They are shown at the scale of non linearity $\sigma_\delta = 1$, where this indeterminacy starts to become very relevant for inference. The families in any dimension are qualitatively identical to these. Right panel : Same as the left panel with $\sigma_\delta = 0.1$, when the indeterminacy if far less relevant for inference, for the reasons given in the text. The discrete distribution has been scaled by a constant factor for convenience. } 
 \end{figure}

Define the statistics of $\rho = (\rho(x_1),\cdots,\rho(x_d))$ through the following. Pick a real number $\epsilon$ with $|\epsilon| \le 1$. Pick further a set of angular frequencies $\veck = (\omega_1,\cdots,\omega_d)$. Each of these must be an integer. Fix $p^{LN}_\rho(\rho)$ the $d$-dimensional lognormal distribution with mean $\bar A$ and covariance matrix $\xi_A$ defined earlier. Then set
\beq \label{Heydeextended}
p_\rho(\rho) := p^{LN}_\rho(\rho)\lb 1 + \epsilon \sin \lp  \pi \veck \cdot \xi_A^{-1} \lp A - \bar A \rp \rp \rb
\enq
Since $|\epsilon| \le 1$ this is positive and seen to be a well defined probability density function\footnote{For $d = 1$, there are very slight differences with Heyde original family. Heyde unnecessarily writes $2\pi$ instead of $\pi$, and restricts $\epsilon$ and $\omega$ to be positive.}. The claim that $p_\rho(\rho)$ defined in this way has the same moments $m_\vecn$ as the lognormal for any multiindex $\vecn$ is proved in the appendix. Note that in the above definition, $\bar A$ is the quantity that enters the definition of lognormal variables in equation \eqref{A}. It is however not the mean of $A = \ln \rho$ anymore, when $\rho$ is defined through \eqref{Heydeextended}.
\newline
\newline
The functional form of $p_\rho(\rho)$ consists of the lognormal envelope modulated by sinusoidal oscillations in $A$.  The smaller the two-point function the higher frequency the oscillations. This may sound curious at first, since it seems to imply that the more linear the field, the more different the distributions within this family will thus appear. However, this is precisely when the oscillations are the strongest that this effect is less relevant. This can be seen as the following. Taking the average of any function $f$ with respect to $p_\rho$ leads trivially to
\beq
\av{f} = \av{f}_{LN} + \epsilon \av{f  \sin \lp  \pi \veck \cdot \xi_A^{-1} \lp A - \bar A \rp \rp }_{LN},
\enq
where the subscript $_{LN}$ denotes the average with respect to the lognormal distribution.
In the limit of the very linear regime, other terms fixed, the second term will average out to zero for any reasonable $f$, since it is the integral of an highly oscillating function weighted by a smooth integrand. In the non-linear regime this in general ceases to be the case. 
This is illustrated as the solid lines in the left ($\sigma_\delta  = 1$) and right ($\sigma_\delta  = 0.1$) panels of figure \ref{fig:pdfs}, showing the member of that family in one dimension with minimal frequency $\omega = 1$, and $\epsilon = 0.1$. The dotted lines on these figures are the usual Gaussian for $A- \bar A = z$.  
The probability density function for $\ln \rho$ is not purely Gaussian anymore. It is therefore of interest to see how the correlations of $A$ deviate from those of Gaussian variables. For instance the means $\av{A-\bar A}$ do not vanish anymore as for the lognormal. A straightforward calculation leads to
\beq
\av{(A - \bar A)(x_i)} = -\epsilon \:\pi \omega_i \exp\lp-\frac{\pi^2}2 \veck\cdot \xi_A^{-1} \veck\rp.
\enq
Picking $\veck$ as having a single non zero entry, $\omega$, at $x_i$ we get that they can be as large as
\beq \label{meanz}
\begin{split}
\av{(A - \bar A)(x_i)} &= -\epsilon \:\pi \omega \exp\lp-\frac{\pi^2}2 w^2 \lb \xi_A^{-1}\rb_{ii}\rp \\
&:=  -\epsilon \:\pi \omega\exp\lp-\frac{\pi^2}2 \frac {\omega^2}{\sigma^2_{A,\textrm{eff}}(x_i)} \rp
\end{split}
\enq
Observables as simple as the means of $A$ are therefore not constrained by the knowledge of the entire correlation hierarchy of the lognormal field. While the effect is irrelevant in the linear regime (for say $\sigma_{A,\textrm{eff}} = 0.1$, the maximal value of the mean in equation \eqref{meanz} is only $\approx 10^{-215})$, deep in the non-linear regime this is not the case anymore. It is easy to show from the above expression that the range available to $ \av{\lp A - \bar A\rp(x_i)}$, choosing $\omega$ appropriately, scales to infinity with $\propto \sigma_{A,\textrm{eff}}$. The means are thus left totally unconstrained in that regime. This and the very sharp behavior is of course a generic effect, not limited to that particular observable. It is obvious that the relevance of this effect for parameter inference is very  sensitive to the degree of linearity of the field, and that large amounts of information are lost to the hierarchy in the high variance regime\footnote{Among this family, it turns out that some observables such as the variances $\av{(A-\bar A)^2(x_i)}$ are always identical to $\sigma^2_A(x_i)$ for any choice of $\epsilon$ and $\veck$. We do not attach any significance to this, since this is not the case for the discrete family, though closed analytical expressions cannot be obtained in this case.}.
\subsection{Discrete family \label{discretefields}}
Fix again the dimensionality $d$, the vector $\bar A$ and the matrix $\covln$. For all integer valued $d$-dimensional multiindex $\vecn$ define a realization $A_\vecn$ of $A$ as the following. Pick $\alpha = (\alpha_1,\cdots,\alpha_d)$ any point, and set
\beq \label{An}
A_\vecn := \bar A + \xi_A \cdot \lp \vecn - \alpha \rp.
\enq
While $\alpha$ can in principle be anything,  only components $\alpha_i \in [0,1)$ will actually define different grids. As usual, $\rho$ is given by exponentiation,
\beq \label{rhon}
\rho_\vecn := \exp\lp A_\vecn \rp
\enq
Assign then to these realizations parametrized by $\vecn$ a probability
\beq \label{discreteP}
P_\vecn = \frac 1 Z \exp \lp-\frac 12 \lp A_\vecn - \bar A\rp \cdot \xi_A^{-1}\lp A_\vecn - \bar A\rp\rp.
\enq
These are usual Gaussian probabilities for $A_\vecn$, except that we have only a discrete set of field realizations.
Note that it can be written, maybe more conveniently, as
\beq
P_\vecn = \frac 1 Z \exp \lp -\frac 12  \lp \vecn -  \alpha\rp \cdot \xi_A \lp \vecn - \alpha\rp\rp.
\enq
Since $\xi_A$ is positive definite, the normalization factor $Z$ is seen to be well defined, as for more usual Gaussian integrals, and so are the probabilities. This discrete probability distribution has the same moments of $\rho_\vecn$ than the $d$-dimensional lognormal distribution with associated $\bar A$ and $\xi_A$, as proven in the appendix \ref{proofs}. Again, negative entries in $\vecn$ are allowed.
\newline
\newline
This family is clearly different from the previous, continuous one. Rather than modulating the lognormal distribution with an oscillating factor, it is a series of Dirac delta functions sampling the lognormal on the grid given by \eqref{An}. The role of $\alpha$ is to shift the sample by a small amount. If $\alpha$ is set to zero, then $A = \bar A$ is part of the sample, while it is not if not. The fact that  this indeterminacy is irrelevant in the linear regime comes this time from realizing that for any nice enough function $f$, the average of $f$ will converge to $\av{f}_{LN}$ due to the trapezoidal rule of quadrature. The grid spacing at which $A$ is sampled in this way in \eqref{rhon} becomes namely thinner and thinner. In the non-linear regime, the spacing is however very large, leading again to large deviations. This is also illustrated in figure \ref{fig:pdfs} for the one dimensional version of it, with shift parameter $\alpha = 0.25$.
\subsection{Appendix}\label{proofs}
We prove the claim that the distributions we defined have the same correlations than the lognormal at all orders.  As we will see this is also true including 'negative orders' and 'mixed orders', i.e. when negative powers of the variables are allowed in the correlations. 
\newline
Recall that for lognormal variables $\rho = (\rho_1,\cdots,\rho_d)$ with means and covariance matrix of their logarithms $\bar A = (\bar A _1,\cdots \bar A_d)$ and $\covln$ we have
\beq \label{mln}
m_\vecn := \av{\rho^\vecm}=  \av{\rho_1^{n_1}\cdots \rho_d^{n_d}} =  \exp \lp \vecn \cdot \bar A + \frac 12 \vecn\cdot \covln \vecn \rp,\quad \vecn = (n_1,\cdots,n_d).
\enq
A simple proof of this fact is to make use of the standard formulae for Gaussian integrals, valid for any positive matrix $\xi_A$, mean vector $A$ and vector $z$, that can be complex valued.
\beq \label{Gintegral}
\frac{1}{\lp 2\pi  \rp^{d/2}}\frac{1}{\sqrt {\det \xi_A}} \int d^d A \exp\lp-\frac 12\lp A - \bar A\rp \cdot \xi_A^{-1}\lp A - \bar A\rp + \lp A - \bar A \rp \cdot z \rp = \exp\lp \frac 12 z \cdot \xi_A z\rp.
\enq
Essentially all calculations in this work follow from this formula. Even the proof for the discrete family can be considered a discrete version of that relation.
\paragraph{Continuous family}
\noindent
To prove our claim it is enough to show that
\beq \label{toprove1}
\av{\rho^\vecn \sin \lp \pi  \:\veck \cdot \covln^{-1}\lp A -  \bar A\rp\rp }_{LN} = 0.
\enq
This must hold for any  $d$-dimensional multiindices $\veck $ and $\vecn$ (we allow entries to be negative), where the average is taken with respect to the lognormal density function, equation \eqref{pLN}. We proceed as the following : we evaluate the following integral
\beq \label{A4}
I(\vecn,\veck) := \av{\rho^\vecn \exp \lp i\pi  \:\veck \cdot \covln^{-1}\lp A -  \bar A\rp\rp }_{LN},
\enq
and show that its imaginary part vanishes for $ \veck$ and $\vecn$ as specified.
\newline
Writing equation \eqref{A4} using
\beq
\rho^\vecn = \exp(\vecn \cdot A) = \exp\lb \vecn \cdot \lp A - \bar A \rp + \vecn \cdot \bar A \rb
\enq
leads immediately to the Gaussian integral given in \eqref{Gintegral}, with $z = \vecn +  i \pi \xi_A^{-1}\veck$. It follows from that equation
\beq
I(\vecn,\veck) = \exp \lb \vecn\cdot \bar A + \frac 12 \lp \vecn + i\pi \covln^{-1}\veck \rp \cdot \covln  \lp \vecn +i \pi \covln^{-1}\veck \rp  \rb.
\enq
Separating real from imaginary argument, this expression reduces to
\beq
\begin{split}
I(\vecn,\veck) =& \exp \lp  \vecn\cdot \bar A + \frac 12 \vecn \cdot \covln \vecn - \frac{ \pi^2}{2}\veck\cdot \covln^{-1}\veck   \rp \cdot \exp \lp i\pi  \: \veck \cdot \vecn \rp.
\end{split}
\enq
The imaginary part of that expression is thus proportional to $\sin \pi\: \veck \cdot \vecn$. Whenever $\veck$ and $\vecn$ are integer valued, so is their scalar product $\veck \cdot \vecn = \sum_i \omega_in_i$. Therefore, the sine vanishes and \eqref{toprove1} is proved.
\paragraph{Discrete family}
\noindent
From equation \eqref{An} and \eqref{rhon}, we have
\beq
\rho_\vecn^\vecm = \exp \lp \vecm \cdot \bar A + \vecm \cdot \xi_A\lp \vecn - \alpha \rp \rp.
\enq
It follows that the moments of $\rho$ are given by
\beq
\av{\rho^\vecm} = \frac {e^{\vecm\cdot \bar A}} Z \sum_{\vecn \in \mathbb Z^d} \exp \lb -\frac 12 \lp \vecn -\alpha \rp\xi_A \lp \vecn-\alpha\rp + \vecm\cdot \xi_A\lp \vecn - \alpha \rp\rb.
\enq
The proof is based on completing the square in the exponent, in perfect analogy of standard proofs of the Gaussian integral in \eqref{Gintegral}. Write
\beq
 -\frac 12  \lp \vecn -\alpha \rp \cdot \covln  \lp \vecn -\alpha \rp + \vecm \cdot \covln  \lp \vecn -\alpha \rp = -\frac 12 \lp \vecn - \vecm -\alpha \rp \covln \lp \vecn - \vecm - \alpha\rp + \frac 12 \vecm\cdot \covln \vecm,
\enq
and then perform the shift of summing index $\vecn \rightarrow \vecn + \vecm$, obtaining
\beq
\av{\rho^\vecm} = \exp \lp\vecm \cdot \bar A + \frac 12 \vecm \cdot \covln \vecm\rp \frac 1 Z \sum_{\vecn \in \mathbb Z^d}\exp\lp-\frac 12 \lp \vecn -\alpha \rp \cdot \covln \lp \vecn -\alpha \rp \rp.
\enq
Since the sum ranges over all the multiindices, the shift does not create boundary terms.
This last sum is nothing else than $Z$, so that we recover
\beq
\av{\rho^\vecm} = \exp \lp\vecm \cdot \bar A + \frac 12 \vecm \cdot \covln \vecm\rp,
\enq
which are indeed the same as the lognormal in \eqref{mln}. Again, this is also true if negative entries in $\vecm$ are permitted.
\section{Information at all orders for the one dimensional distribution \label{onevariable}}
We now obtain the information coefficients $s_n$, and thus the matrices $F_{\le N}$, for the lognormal distribution in one dimension $d = 1$ at all orders. By additivity of the information order by order this is equivalent to obtain that of the uncorrelated lognormal field in any number of dimensions.
\newline
\newline
There are two free parameters, the mean $\bar A$ and the variance $\sigma^2_A$. The dependency on cosmological parameters $\btheta = (\alpha,\beta,\cdots)$ can enter one or both of these parameters. We have from section \ref{notationlog}
\beq \label{rel}
\begin{split}
\sigma^2_A = \ln \lp 1 + \sigma^2_\delta \rp \quad \bar A = \ln \bar\rho - \frac 12 \ln \lp 1 + \sigma^2_\delta \rp,
\end{split}
\enq
where $\sigma^2_\delta$ is the variance of the fluctuations.
\newline
\newline
Note that the total Fisher information content of $\rho$ becomes
\begin{equation} \label{exactlnnormal}
F_{\alpha \beta} = \frac{1}{\sigma^2_A} \frac{\partial \bar A}{\partial \alpha} \frac{\partial \bar A}{\partial \beta}  + \frac 1 {2\sigma^4_A}  \frac{\partial \sigma^2_A}{\partial \alpha}  \frac{\partial \sigma^2_A}{\partial \beta}.
\end{equation}
The key parameter throughout this section will be the quantity $q$, defined as
\beq
q := e^{-\sigma^2_A} = \frac{1}{1 + \sigma^2_\delta}.
\enq
Note that $q$ is strictly positive and smaller than unity.  The regime of small fluctuations, where the lognormal distribution is very close to the Gaussian distribution is described by values of $q$ close to unity. Deep in the non linear regime, it tends to zero. These two regimes are conveniently separated at $q = 1/2$, corresponding to fluctuations of unit variance.
\newline
\newline
The moments of the distribution are given by
\beq \label{moments}
m_n =  \bar \rho^n q^{-\frac 12 n(n-1)}.
\enq  
We note the following extremely convenient property of these moments,
\beq \label{separability}
m_{i + j} = m_im_j \:q^{-ij}.
\enq
From chapter \ref{ch2}, equation \eqref{def} we see that the $n$th information coefficient $s_n$ is given by
\beq \label{infcoeff} \begin{split}
s_n(\alpha) = \frac{\partial \ln \bar \rho}{\partial \alpha} \sum_{k = 0}^n C_{nk}\:m_k \:k + \frac{\partial \sigma^2_\delta}{\partial \alpha}\frac q{2} \sum_{k = 0}^n C_{nk}\:m_k \:k(k-1),
\end{split}
\enq
where the $n$th orthornormal polynomial $P_n$ is given by $P_n(\rho) = \sum_{k\le n}C_{nk}\rho^k$, with yet unknown matrix elements $C_{nk}$ to be found.
\newline
\newline
Evaluation of the above sums can proceed in different ways. Notably, it is possible to get an explicit formula for the orthonormal polynomials, and therefore of the matrix $C$, for the lognormal distribution. These are essentially the Stieltjes-Wigert polynomials \citep{Wigert23,Szego39}. We will namely use their specific form later in this section, though they are not needed for the purpose of evaluating (\ref{infcoeff}).
\newline
\newline
We proceed with the following trick : we introduce the $q$-shifted factorial , also called $q$-Pochammer symbol \citep[ section 10]{Kac02,Andrews99}, as
\beq
\pochh t q n := \prod_{k = 0}^{n-1}\lp 1 - tq^k\rp,\quad \pochh t q 0  := 1
\enq
$t$ a real number, and prove (see appendix \ref{proofq} for the proof) that the following curious identity holds,
\beq \label{lemma}
\av{P_n(t\rho)} =(-1)^n\frac{q^{n/2}}{\sqrt{ \pochh qqn }} \pochh t q n.
\enq
By virtue of
\beq
\av{P_n(t\rho)} = \sum_{k = 0}^nC_{nk}\:m_k\:t^k,
\enq
it follows from our identity (\ref{lemma}) that the sums given in the right hand side of equation (\ref{infcoeff}) are proportional to the first, respectively the second derivative of the $q$-Pochammer symbol evaluated at $t = 1$. Besides, matching the powers of $t$ on both sides of equation (\ref{lemma}) will provide us immediately the explicit expression for the matrix elements $C_{nk}$.
\newline
\newline
We consider now the two cases of the parameter of interest $\alpha$ being $\alpha = \ln \bar \rho$ and $\alpha = \sigma^2_\delta$. The general case of a generic parameter is reconstructed trivially from these two from the chain rule of derivation. The total amount of information in the distribution becomes in the first case from (\ref{exactlnnormal})
\beq \label{Fmuex}
I(\ln \bar \rho) = - \frac 1{\ln q}.
\enq
The second case is the most common in cosmology, for instance for any model parameter entering the matter power spectrum.
The exact amount of information on this parameter is this time
\beq \label{Fsigex}
 I(\sigma^2_\delta)= -\frac {q^2} { \ln q }\lp \frac 14  - \frac 1 {2\ln q} \rp.
\enq
\newline
\newline
In both of these situations, we obtain the information coefficients (\ref{infcoeff}) by differentiating once, respectively twice, our relation (\ref{lemma}) with respect to the parameter $t$, and evaluating these derivatives at $t = 1$. The result is
\begin{equation} \label{anbn}
\begin{split}
 s_n(\ln \bar \rho)  = (-1)^{n-1}\sqrt{ \frac{q^n}{1 - q^n}\pochh q q {n-1} }
 \end{split}
 \end{equation}
 and 
 \begin{equation}\label{Ansigma}
 \begin{split}
 s_n(\sigma^2_\delta)  =  - s_n( \ln \bar \rho) q \lb \sum_{k = 1}^ {n-1} \frac{q^k}{1 - q^k}\rb ,\quad n > 1 
\end{split}
\end{equation}
whereas $s_{n = 1}(\sigma^2_\delta)$ is easily seen to vanish, since the mean is independent of $\sigma^2_\delta$.
\subsection{Lack of information in the moments \label{lack}}
\begin{figure}
  \centering
  \includegraphics[width = 0.8\textwidth]{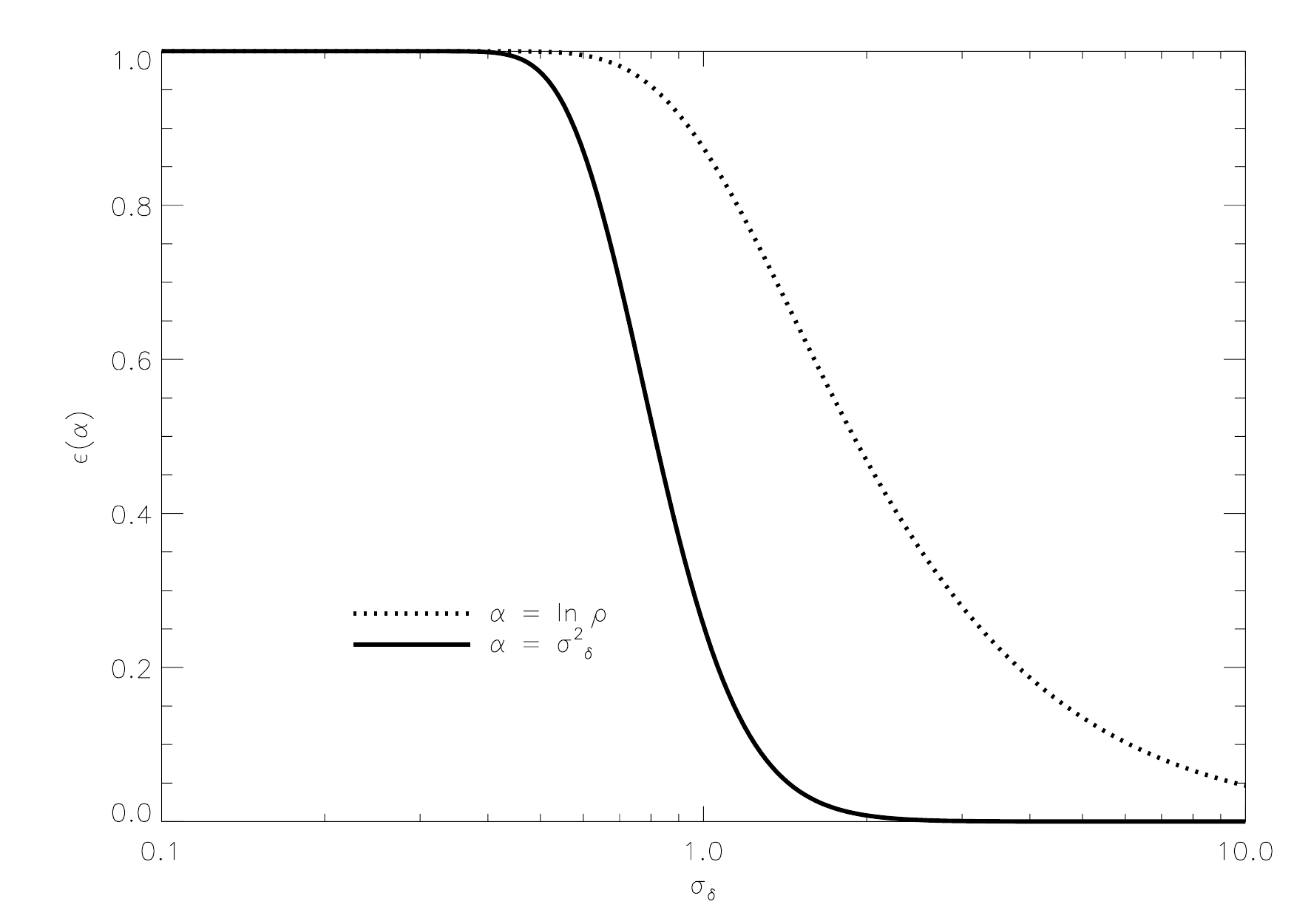}
 \caption[The fraction of the total information content that is accessible through the extraction of the entire series of moments of the lognormal field, as function of the square root of the variance of the fluctuations.]{\label{fig1eps} The fraction of the total information content that is accessible through extraction of the full series of moments of the lognormal field, as function of the square root of the variance of the fluctuations. The solid line for a parameter entering the variance of the field and not its mean, and dotted conversely.}
 \end{figure}

\begin{figure}
  \includegraphics[width = 0.5\textwidth]{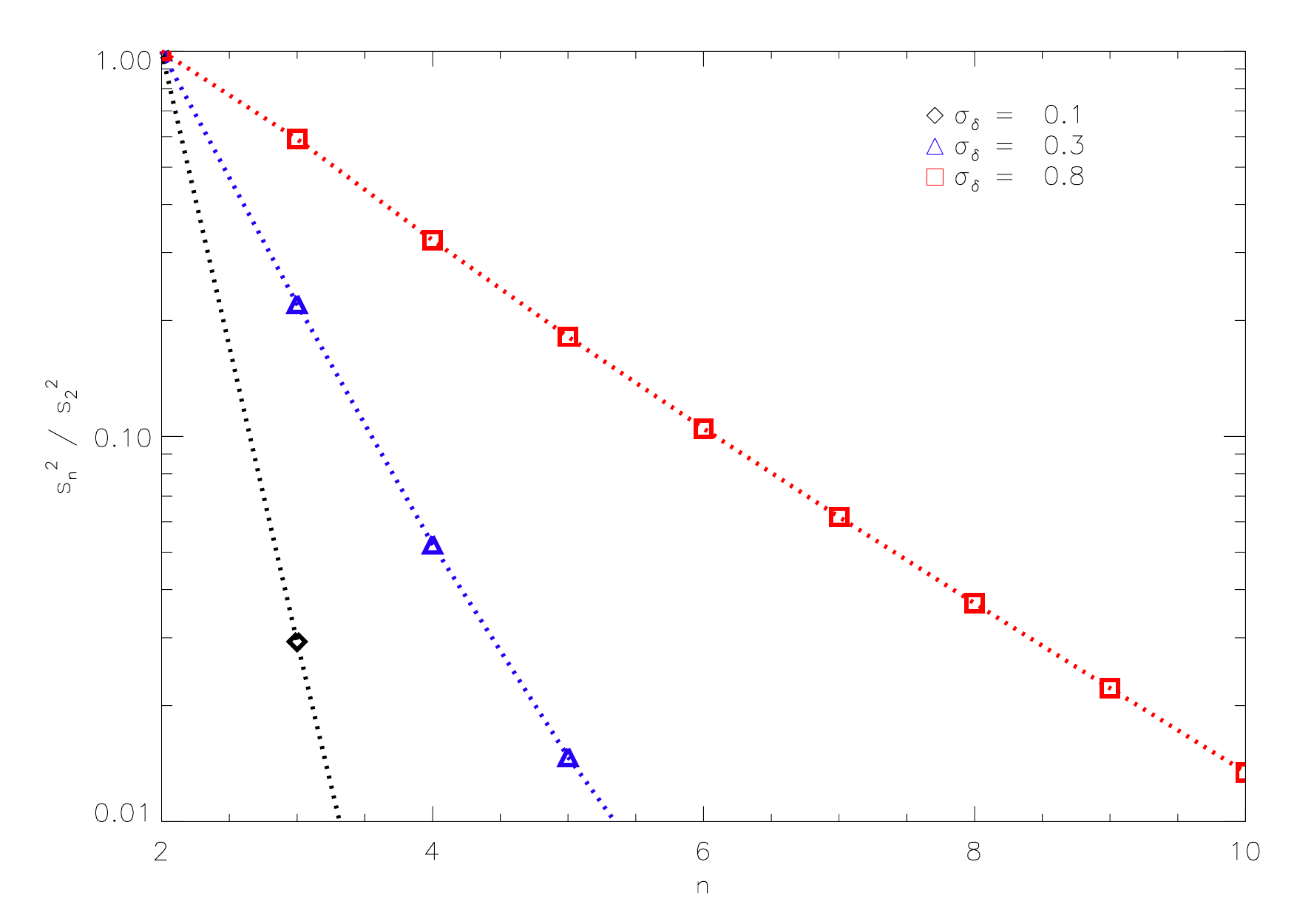}
  \includegraphics[width = 0.5\textwidth]{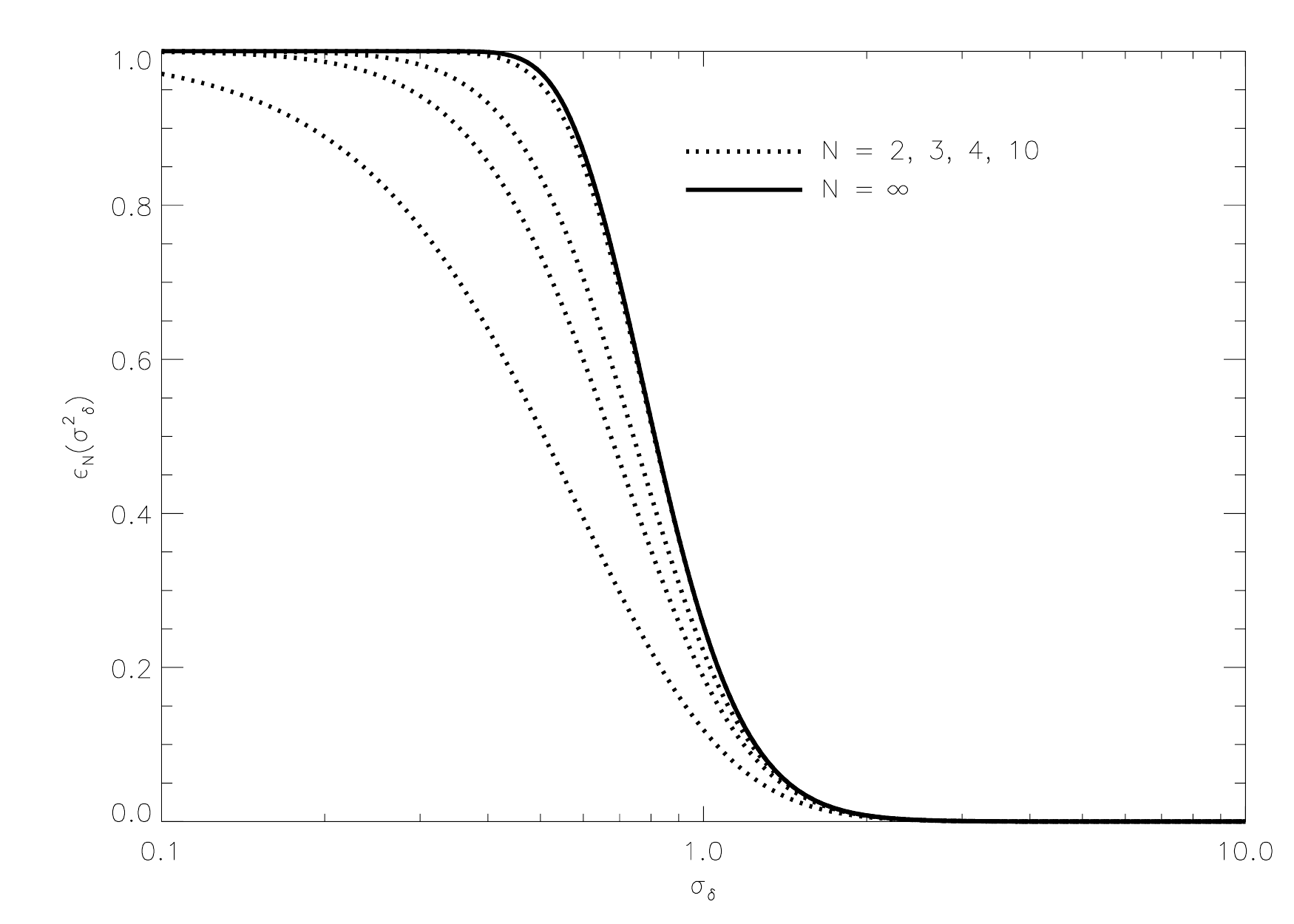}
 \caption[Distribution of the information within the moments of the lognormal distribution]{\label{fig2eps} Left panel : The distribution of the information within the first 10 moments of the lognormal field, given by the coefficients $ s^2_n(\sigma^2_\delta) $, equation (\ref{Ansigma}), normalized to the information content of the second moment, for three different values of $\sigma_\delta$. Note that deeper in the non linear regime, the trend is reversed. Right panel : the cumulative efficiencies of the moments to capture the information on $\sigma^2_\delta$, as function of the root of the fluctuations.}
 \end{figure}

The series
\beq \label{fmoments}
 i(\alpha) :=  \sum_{n =1}^\infty  s^2_n(\alpha)
\enq
are the total amount of information contained in the full series of moments. The efficiencies $\epsilon(\alpha)$ defined as
\beq
\epsilon(\alpha):= \frac{i(\alpha)} {I(\alpha)}.
\enq
are the fraction of the information that can be accessed by extraction of the full set of moments of our lognormal variable. Similarly, we define
\beq
\epsilon_N(\alpha)  := \frac{1}{I(\alpha)}\sum_{n = 1}^N s^2_n(\alpha) 
\enq
for the first $N$ moments.
Note that from the chain rule of derivation, both $\epsilon(\alpha)$ and $\epsilon_N(\alpha)$ are invariant under a reparametrisation of $\alpha$, i.e.
\beq
\epsilon(\beta) = \epsilon(\alpha) \textrm{    whenever   } \alpha = f(\beta)
\enq
for some function $f$, all other parameters fixed. For this reason, these efficiencies refer rather to a parameter type (in our case mean alike or variance alike) rather than  some more specific instance.
\newline
\newline
The two asymptotic regimes of very small and very large fluctuation variance $\sigma_\delta$ can be seen without difficulty. Both for $\alpha = \ln \bar \rho$ and $\alpha = \sigma^2_\delta$, it is seen that in these asymptotic regimes the first non vanishing term of the corresponding series \eqref{anbn} and \eqref{Ansigma} dominates completely its value. For very small variance, or equivalently $q$ very close to unity, both efficiencies $\epsilon$ tend to unity, illustrating the fact the distribution becomes arbitrary close to Gaussian : all the information is contained in the first two moments. The large variance regime is more interesting, and, even tough the information coefficients decays very sharply as well, the series (\ref{fmoments}) are far from converging to the corresponding expressions (\ref{Fmuex}) and (\ref{Fsigex}) showing the total amount of information. Considering only the dominant first term in the relevant series and setting $q \rightarrow 0$, one obtains 
\begin{equation} \label{asymptotics}
\epsilon(\ln \bar \rho)\rightarrow \frac{1}{\sigma_\delta^2} \ln  \lp 1 + \sigma_\delta^2\rp .
\end{equation}
and a much more dramatic decay of $\epsilon(\sigma^2_\delta)$ :
\begin{equation} \label{asymptotics2}
\epsilon(\sigma^2_\delta) \rightarrow \frac{4}{\sigma_\delta^8} \ln \lp 1 + \sigma_\delta^2\rp.
\end{equation}
 Both series given in (\ref{fmoments}) are quickly (exponentially) convergent and well suited for numerical evaluation. Figure \ref{fig1eps} shows the accessible fractions $\epsilon$ of information through extraction of the full series moments. Figure \ref{fig2eps} shows the repartition of this accessible fraction among the first 10 moments. Most relevant from a cosmological point of view in figure \ref{fig1eps} is the solid line, dealing with the case of the  parameters of interest entering the variance only. These figures shows clearly that the moments, as probes of the lognormal matter field, are penalized by two different processes. First, as soon as the field shows non-linear features, following equations (\ref{asymptotics}) and (\ref{asymptotics2}), almost the entirety of the information content cannot be accessed anymore by extracting its successive moments. Within a range of one magnitude in the variance, the moments goes from very efficient probes to highly inefficient. Second, as shown in figure \ref{fig2eps}, in the still close to linear regime, as  the variance of the field approaches unity, this accessible fraction gets quickly transferred from the variance alone to higher order moments.
 \newline
 \newline
 This repartition of the information within the moments is built out of two different regimes. First, for large variance, or  large $n$, we see easily from the above expressions (\ref{anbn}) and (\ref{Ansigma}) that in both cases the information coefficients decays exponentially,
\begin{equation} \label{as1}
s_n^2 \propto q^{n},\quad -n\ln q \gg   1.
\end{equation}
On the other hand, if the variance or $n$ is small enough, we can set $1 - q^n \approx -n\ln q $, and we obtain, very roughly,
\beq\label{as2}
s_n^2 \propto \lb - n \ln q\rb^n,\quad -n\ln q   \ll 1,
\enq
explaining the trend with variance seen in figure \ref{fig2eps}, that puts more importance to higher order moments as the variance grows.
Note that the latter regime can occur only for small enough values of the variance. Deeper in the non linear regime, the trend is therefore reversed, obeying (\ref{as1}) for all values of $n$, with a steeper decay for higher variance. This is clearly shown in the right panel of figure \ref{fig2eps}. In that regime, higher order moments do not carry additional information.

 \subsection{A $q$-analog of the logarithm \label{qanalog}}
 \begin{figure}
  \centering
  \includegraphics[width =0.7 \textwidth]{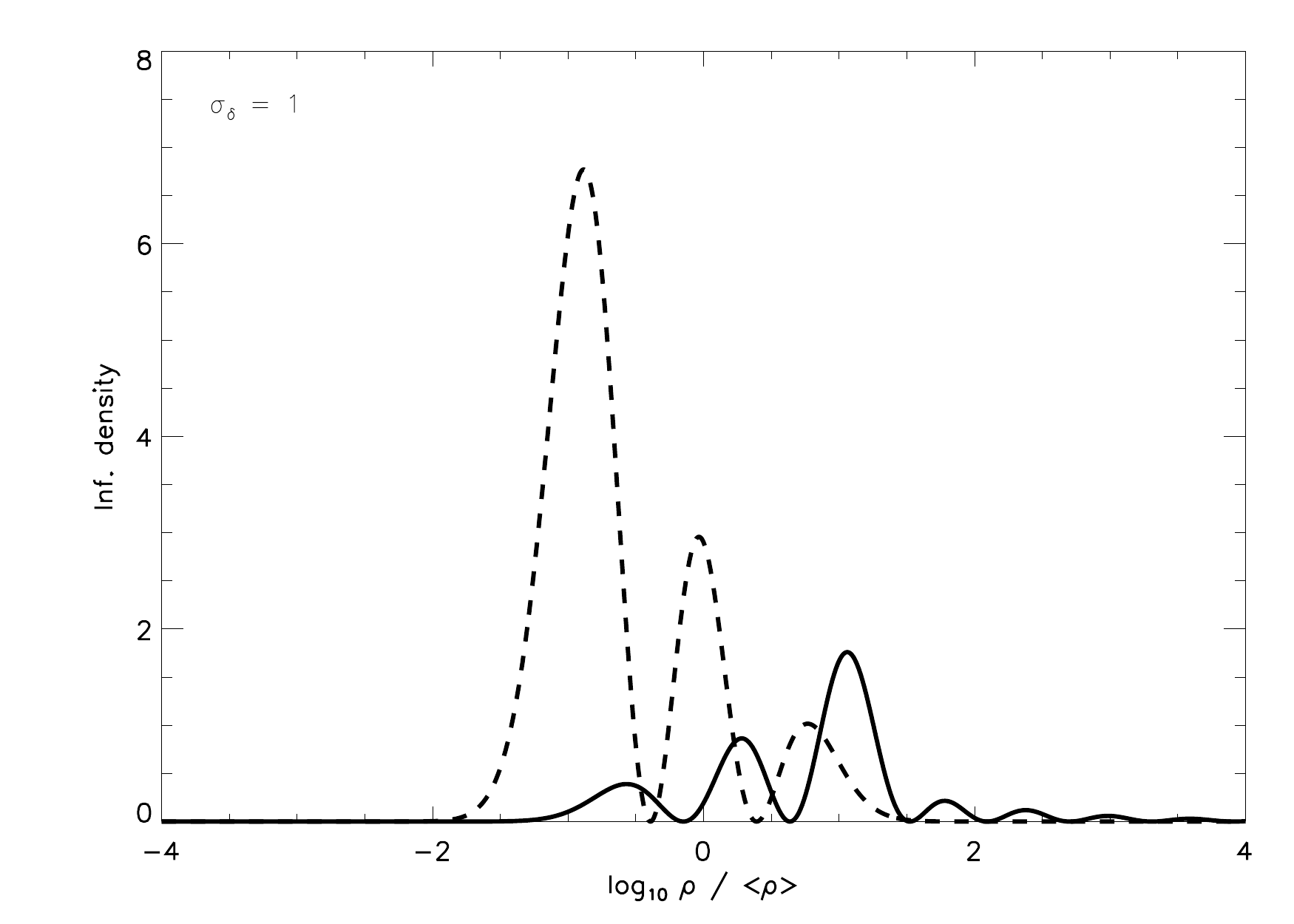}
 \caption[The information density of the lognormal distribution and its approximation through the associated orthonormal polynomials, for fluctuations of unit variance]{\label{fig3eps}The information density of the lognormal distribution, dashed, and, solid, its approximation through the associated orthonormal polynomials, for a variance alike parameter, for fluctuations of unit variance. While most of the information of the lognormal field in this regime is actually contained in the underdense regions, the moments are essentially unable to catch it.}
 \end{figure}
These results show clearly that large parts of the information become invisible to the moments. However, it does not tell us what is responsible for this phenomenon. It is therefore of interest to look into more details of these missing pieces of information. As we have seen, these are due to the inability of the polynomials to reconstruct precisely the score function. 
In the case $\alpha = \ln \bar \rho$, the score function $\partial_\alpha \ln p$ of our lognormal distribution is easily shown to take the form of a logarithm in base $q$,
\beq \label{sfctbarrho}
\begin{split}
s(\rho,\ln \bar \rho)
& = -\frac 12 -  \ln_q \lp \frac  {\rho} { \bar\rho}\rp .
\end{split}
\enq
Therefore the series
\beq
s_{\infty}(\rho,\ln \bar \rho) := \sum_{n = 0}^\infty s_n(\ln \bar \rho) P_n(\rho)
\enq
will represent some function, very close to the logarithm \eqref{sfctbarrho} for $q \rightarrow 1$ over the range of $p(\rho,\btheta)$. It will however fail to reproduce some of its features at lower $q$-values. This is hardly surprising, since it is well known that the logarithm function does not have a Taylor expansion over the full positive axis. For this reason, the approximation
\beq
s_{\infty}(\rho,\ln \bar \rho) := \sum_{n = 0}^\infty s_n(\ln \bar \rho)P_n(\rho) 
\enq
of the score function $s(\rho,\ln \bar \rho)$ through polynomials can indeed only fail when the fluctuation variance becomes large enough. In the appendix, we show that $s_\infty(\rho,\ln \bar \rho)$ takes the form
\beq \label{smu}
s_\infty(\rho,\ln \bar \rho) = - \sum_{k = 1}^\infty \frac{q^k}{1-q^k}\lb 1 +(-1)^k\frac{q^{k(k-1)}}{\pochh  q q k} \lp \frac{\rho}{ \bar \rho}\rp^k\rb.
\enq
 It is interesting to note that this series expansion is almost identical to the one of the $q$-analog of the logarithm $S_q$  defined by E. Koelink and W. Van Assche, with the only difference being the replacement of $q^{k(k-1)/2} $ by $q^{k(k-1)}$ (See \cite{Koelink09}, and also \cite{Gautschi06}). Due to this replacement, $s_\infty$ does not possess several properties $S_q$ has and makes it a real $q$-analog of the logarithm, such as $S_q(q^{-n}) =  n$, for positive integers. The qualitative behavior of $s_\infty$ stays however close to $S_q$. Notably, its behavior in underdense regions, $\rho/ \bar \rho \ll 1$,  where as seen from (\ref{smu}) $s_\infty$ tends to a finite value, is very different from a logarithm.
 \newline\newline
This calculation can be performed as well in the case $\alpha = \sigma^2_\delta$, with similar conclusions. Since it is rather tedious and not very enlightening, we do not reproduce it in these pages. We show in figure \ref{fig3eps} the information density of the lognormal distribution (dashed line), 
 and its approximation by the orthogonal polynomials (solid line),
 \beq
 p(\rho,\btheta)\lp \sum_{n = 0}^\infty s_n(\sigma^2_\delta) P_n(\rho)\rp^2,
 \enq
 when the fluctuation variance $\sigma^2_\delta$ is equal to unity. It is clear from this figure that in this regime, while most of information is located within the underdense regions of the lognormal field, the moments are however unable to catch it.
 \newline
 \newline
 As a non trivial check of the correctness of our numerical and analytical calculations, we compared the total information content as evaluated from integrating the information densities on figure \ref{fig3eps} to the one given by the equation (\ref{Fsigex}), respectively (\ref{fmoments}), with essentially perfect agreement.
\subsection{Comparison to standard perturbation theory \label{compSPT}}
\begin{figure}
    \centering
  \includegraphics[width =0.7\textwidth]{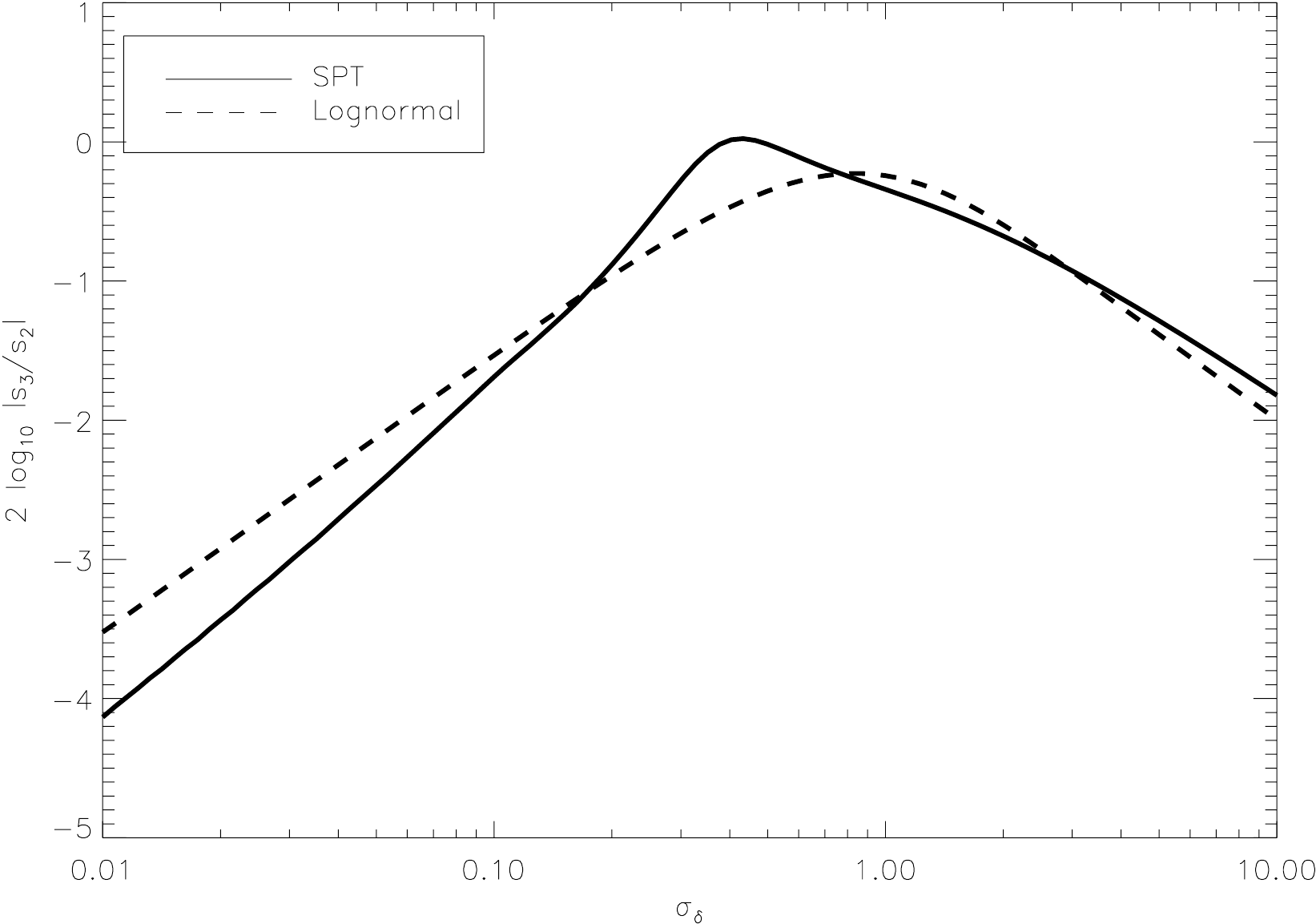}
 \caption[Comparison of the information content of the first moments of the fluctuation field between standard perturbation theory and the lognormal model]{\label{fig4eps}The ratio of the independent information content of the third moment to that of the second moment, for the lognormal field (dashed) and standard perturbation theory (solid), as function of the square root of the variance of the fluctuations. }
\end{figure}

As we have seen in chapter \ref{ch2}, the knowledge of its first $2n$ moments allows  for any distribution the direct evaluation of the independent information content of the first $n$ moments, for instance from equation \eqref{Fn}. This even if the exact shape of the distribution is not known, or too complicated. In particular, we can use the explicit expressions for the first six moments of the density fluctuation field within the framework of standard perturbation theory (SPT) provided by F. Bernardeau in \citep{1994A&A...291..697B}, in order to compare $s_2(\alpha)$ and $s_3(\alpha)$ as given from SPT to their lognormal analogs.\newline\newline
We note that a comparison to \citep{1994A&A...291..697B} can only be very incomplete and, to some extent, it can only fail. It is indeed part of the  approach in \citep{1994A&A...291..697B}, when producing functional forms for the distribution of the fluctuation field, to invert the relation between a moment generating function and its probability density function. For such an inversion to be possible it is of course necessary that the probability density is uniquely determined by its moments. As said, this is not the case for the lognormal distribution. Therefore, that approach can never lead to an exact lognormal distribution, or to any distribution for which the moment hierarchy forms an incomplete set of probes. However, such a comparison can still lead to conclusions relevant for many practical purposes, such as those dealing with the first few moments.
\newline\newline
The variance of the field is explicitly given as an integral over the matter power spectrum,
\beq \label{varbdeau}
\sigma^2_\delta  = \frac{1}{2\pi^2}\int_0^\infty dk \:k^2P(k,\btheta) \left|W(kR)\right|^2,
\enq
where $W(kR)$ is the Fourier transform of the real space top hat filter of size $R$, and any cosmological parameter $\alpha$ entering the power spectrum $P(k)$.
In the notation of \citep{1994A&A...291..697B}, the moments of the fluctuation field $m_n = \av{\delta^n}$ are given by the deconnected, or Gaussian, components, while the connected components $\av{\delta^n}_c ,\quad n \ge 3$ are given in terms of parameters $S_n$,
\beq \label{mbdeau} \begin{split}
\av{\delta^n}_c = \sigma^{2(n-1)}S_n.
\end{split}
\enq
The parameters $S_n$ contain a leading, scale independent coefficient, and deviation from this scale independence are given in terms of the logarithmic derivative of the variance,
\beq
\gamma_i = \frac{d^i \ln \sigma^2_\delta}{d\:\ln R^i},\quad i = 1,\cdots
\enq
Neglecting the very weak dependence of $S_n$ on cosmology, from (\ref{mbdeau}) we can write
\beq
\begin{split}
 \frac{\partial m_n}{\partial \alpha} = \frac{\partial \sigma_\delta^2}{\partial \alpha}\cdot \begin{cases} 0,&\quad n = 1 \\  1,&\quad n = 2
 \\ 2 m_3 \:/ \: \sigma_\delta^2,&\quad n = 3  \end{cases}
\end{split}
\enq
With the coefficients $S_n$ up to $n = 6$ given in  \citep[page 703]{1994A&A...291..697B}, and the above relations, we performed a straightforward evaluation of the information coefficients $s_2^2(\alpha)$ and $s_3^2(\alpha)$, from equation \eqref{Fn}. The variance was obtained from (\ref{varbdeau}) within a flat $\Lambda$CDM universe ($\Omega_\Lambda = 0.7,\Omega_m = 0.3,\Omega_b = 0.045, h = 0.7$),  with power spectrum parameters ($\sigma_8 = 0.8, n = 1$) and we used the transfer function from Eisenstein and Hu \citep{1998ApJ...496..605E}. The needed derivatives $\gamma_i, \: i = 1,\cdots, 4$ were obtained numerically through finite differences.
\newline\newline
In figure \ref{fig4eps}, we show the ratio
\beq
\lp \frac{s_3(\alpha)}{s_2(\alpha)} \rp^2,
\enq
both for the lognormal distribution and the SPT predictions, for $\alpha = \sigma^2_\delta$. 
Note that this ratio is actually independent of the parameter $\alpha$ as long as it does not enter the mean of the field but only the power spectrum. It is the relative importance of the third moment with respect to the second, as function of the variance, This ratio is identically zero for a Gaussian distribution.
The models stands in good agreement over many orders of magnitude. It is striking that both models consistently predict that a the entrance of the non-linear regime,  this ratio takes a maximal value close to unity. Surely, the SPT curve for larger values of the variance is hard to interpret, since out of its domain of validity. 

\subsection{Appendix} \label{proofq}
\paragraph{Derivation of \ref{lemma}}
To prove (\ref{lemma}),  we note that both sides of the equation are polynomials of degree $n$ in $t$, and that the zeroes of the right hand side are given by
\beq
t = q^{-i}, \quad i = 0,\cdots, n-1.
\enq 
We first show that the left hand side evaluated at these points does vanish as well, so that the two polynomials must be proportional. We then find the constant of proportionality by requiring $P_n$ to have the correct normalization.
\newline
\newline
The first step is performed by noting that
\beq
\av{P_n(q^{-i}\rho)} = \frac 1 {m_i}\av{P_n(\rho)\rho^i }, \quad i = 0,1,\cdots
\enq
an identity which is proven by expanding $P_n $ in both sides of the equation  in terms of the transition matrix $C$, and using the relation (\ref{separability}) between the moments.
Since $P_n$ is by construction orthogonal to any polynomial of strictly lower degree, we have indeed
\beq
\av{P_n(q^{-i}\rho)} = 0,\quad i = 0,\cdots,n-1. 
\enq
This implies
\beq \label{sumrule}
\sum_{k = 0}^n C_{nk} m_k\:t^k  =\alpha_n \pochh t q n
\enq
for some constant of proportionality $\alpha_n$. To find it, 
we note that by expanding the normalization condition of $P_n$,
\beq \begin{split}
1 &= \av{P_n^2(\rho)}, 
\end{split}
\enq
using again property (\ref{separability}), it must hold that
\beq
1 = \sum_{i,j = 0}^nC_{ni}m_j\: C_{nj}m_j\: q^{-ij}.
\enq
The sums can be performed using equation (\ref{sumrule}), leading to the following equation for $\alpha_n$,
\beq
1 = (-1)^n\:\alpha_n^2\: q^{n(n-1)/2}\pochh {q^{-n}}qn.
\enq
This expression simplifies to
\beq
\alpha_n^2 = \frac{q^n}{\pochh q q n}
\enq
and the sign of $\alpha_n$ must be $-1^n$ in order to have a positive matrix element $C_{nn}$. This concludes the proof of (\ref{lemma}).
\paragraph{Derivation of  the representation (\ref{smu})}
In order to get the explicit series representation of (\ref{smu}), we first obtain from relation (\ref{lemma}) the exact expression of the transition matrix $C$. The expansion of the $q$-Pochammer symbol on the right hand side of (\ref{lemma}) in powers of $t$ is the Cauchy binomial theorem,
\beq
\pochh t q n = \sum_{k = 0}^n \binq n k q q^{k(k-1)/2}(-t)^k,
\enq   
where
\beq
\binq n k q = \frac{\pochh q q n}{\pochh q q k \pochh q q {n-k}}
\enq is the Gaussian binomial coefficient. Matching powers of $t$ in (\ref{lemma}) we obtain the explicit form
\beq \label{}
C_{nk}=(-1)^{n-k} \frac{q^{n/2}}{\sqrt{\pochh q q n}} \binq n k q q^{k^2}{\bar \rho}^{-k}.
\enq
Therefore, interchanging the $n$ and $k$ sums in (\ref{smu}) , it holds
\beq
s_\infty(\rho,\ln \bar \rho) =-\sum_{n = 1}^\infty \frac{q^n}{1- q^n} + \sum_{k = 1}^\infty q^{k^2} \lp -\frac \rho  {\bar \rho}\rp^k \sum_{n = k}^\infty\frac{q^n}{1-q^n}\binq n k q.
\enq
With the help of some algebra the following identity is not difficult to show
\beq
\sum_{n = k}^\infty\frac{q^n}{1-q^n}\binq n k q = \frac 1 {\pochh q q k}\frac{q^k}{1 - q^k},\quad k \ge 1.
\enq
Consequently, the series expansion of $s_\infty(\rho,\bar\rho)$ is given by
\beq
s_\infty(\rho,\ln \bar\rho) = - \sum_{k = 1}^\infty \frac{q^k}{1-q^k}\lb 1 +(-1)^k\frac{q^{k(k-1)}}{\pochh  q q k} \lp \frac{\rho}{\bar \rho}\rp^k\rb
\enq
 \section{Connection to N-body simulations}\label{connection}
\begin{figure}
\centering
\includegraphics[width = 0.8\textwidth]{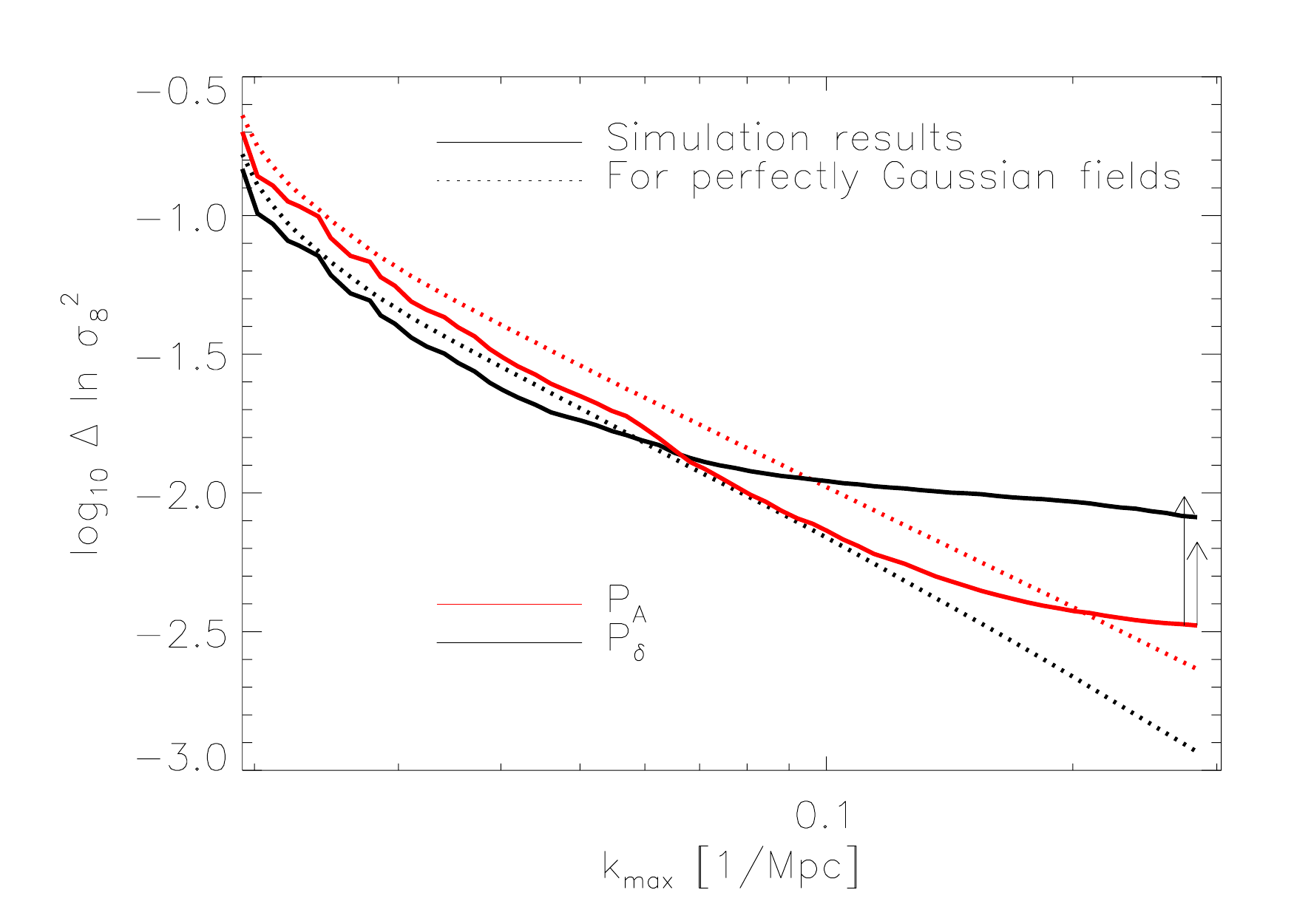}
\caption[Comparison of various estimates of the error bar on the linear power spectrum amplitude, constrained using the power spectra of the overdensity field or of the log-density field in an $N$-body simulation. ]{  \label{fig:cstr} Comparison of various estimates of the error bar on the linear power spectrum amplitude, $\ln\sigma_8^2$, constrained using power spectra of the overdensity $\delta$ (black) and the log-density $A$ (red) in an $N$-body simulation.  Solid curves show how the error bars tighten as the maximum $k$ analyzed increases up to the Nyquist frequency, as in e.g. \cite{2011ApJ...742...91N}, equation \eqref{FGnum}. Dotted curves neglect the non-Gaussian component of the covariance matrices, as well as the discrete nature of the Fourier-space mode lattice, equation \eqref{FG}.  The arrows (one for each choice of $\sigma_A$, 0.7 and 0.9) show the expected degradation of the error bars from analyzing $\delta$ instead of $A$ in our model given by equation \eqref{ratioss}; these factors appear numerically in the first column of Table \ref{table}.}
\end{figure}

Mark C. Neyrinck analysed in \citep{2011ApJ...742...91N} the Coyote Universe N-body simulations suite \citep{2010ApJ...715..104H,2010ApJ...713.1322L} in a box of volume $V = 2.2 $Gpc$^3$, with $256^3$ cells, extracting the spectrum  $P(k)$ of $A$ and $\delta$ over the range $0.02/$Mpc $ \lesssim k \lesssim 0.6 /$Mpc, comparing their statistical power as function of the smallest scale $k_\max$ included in the analysis for several cosmological parameters. It was found that the spectrum of $A$ has more constraining power on cosmological parameters than that of $\delta$, when the non linear scales are included in the analysis. We refer to that paper for more details on the procedures and results. In this framework, $\rho$ is $1 + \delta$, and thus $A   =\ln (1 + \delta)$. The fields are statistically homogeneous and isotropic.
\newline\newline
Given the considerations of the previous sections, and the fact that the density field is known to be somewhat  close to lognormal, these results can hardly be considered surprising. The field $A$ must be indeed closer to a Gaussian field for all values of the cosmological parameters, so that low order $N$ point functions of $A$ must contain a larger fraction of the information than those of $\delta$ (it is useful to remember that the full fields $A$ and $\delta$ carry in all cases the very same total amount of information, since the mapping between them is parameter independent and invertible). In this section we want to go a step further from these qualitative considerations and make a quantitative comparison of these results to simple analytical methods using the results of the previous sections.
\newline
\newline
First, we need to make sure that a Gaussian description of the field $A$ seen in the simulations is reasonable, at least for what concerns the information  content. In particular, this is not the case for the smallest scales of $A$, since the covariance matrix of $P_A$ in the $256^3$ box clearly shows substantial off diagonal elements starting from $k \simeq 0.3$/Mpc. The same analysis was therefore repeated, performing the logarithmic transform on the $\delta$ field only after smoothing $\delta$ on twice the original length scale, by merging the $256^3$ into $128^3$ cells. This allowed us to extract the spectra of $A$ and $\delta$ over the range  $0.02/$Mpc $ \lesssim k \lesssim 0.3 /$Mpc,  with a diagonal covariance matrix over the full range to a very good approximation. It is important to realize that sadly it is not identical to the much simpler approach of considering the original $A$ field only up to the new $k_\max$: since all the scales of $\delta$ have an impact on the large scales of $A$, the operations of smoothing $\delta$ and then log transforming $\delta$ are not identical to log transforming $\delta$ and then smoothing $A$.
\newline
\newline
As we discussed already on several occasions, for a purely Gaussian field with spectrum $P$, the information content on $\alpha$ in the spectrum is given by
\beq \label{FG}
I(\alpha)  = \frac V 2 \int \idk \lp \frac{\partial \ln P(k)}{\partial \alpha} \rp^2,
\enq
where the integral runs over the modes extracted,
and
\beq \label{cstrFG}
\frac 1 {\sqrt{I(\alpha)}} =: \Delta(\alpha)
\enq
can be thought of as approximating the constraints on $\alpha$ achievable with these modes.  We focus for reasons that become clear below primarily on the parameter $\ln \sigma^2_8$, which has a roughly constant impact both on $\ln P_{\delta}$ and $\ln P_{A}$. In figure \ref{fig:cstr}, we compare this for the $\delta$ field and the $A$ field as function of $k_\max$. The solid lines are the simulation results, evaluating the covariance matrix $C_{kk'}$ between the modes $k$ and $k'$ and setting
\beq \label{FGnum}
\Delta(\ln \sigma^2_8) = \sum_{k,k'\le k_\max}\frac{\partial \ln P(k)}{\partial{\ln \sigma^2_8}}  C^{-1}_{kk'}\frac{\partial \ln P(k')}{\partial{\ln \sigma^2_8}},
\enq
while the dashed lines are in both cases equation \eqref{cstrFG} given by \eqref{FG}, with the derivatives being those extracted from the simulations. Since the derivatives are roughly constant, the dashed lines scale like $k^{-3/2}$, i.e. the inverse root of the number of modes. It is clear that the log transform extends the (rough) validity of the Gaussian approximation in terms of Fisher information to the full range of scales we are dealing with. Note however that this is a statement only up to the four point level, since those are the only ones that enter \eqref{FG} and \eqref{FGnum}.
\newline
\newline
\noindent To compare these results to analytical predictions from lognormal statistics, we first note the following. For a parameter, such as $\ln \sigma^2_8$, that obeys roughly
\beq
\frac{\partial \ln P_A(k)}{\partial \alpha} \approx \cst = : c,
\enq
the correlated Gaussian field $A$ is equivalent, from the point of the view of the information on that parameter, to a field with the same variance but with $\xi(r) = 0$ for $r > 0$. This may not sound like an obvious statement so let us show this explicitly : start from equation \eqref{FG} which leads to
\beq
I(\alpha) = c^2\: \frac V 2 \int \idk
\enq
The integration on the right, in a discrete description, is the number of available modes, equal to the number $d$ of grid points, times the spacing of the modes $\Delta k = (2\pi)^3/V$. It follows
\beq \label{Ftot}
I(\alpha) = c^2 \: \frac d 2.\
\enq
On the other hand, the observation of $d$ uncorrelated Gaussian variables with variance $\sigma^2_A$ always carries the information
\beq \label{Fvar}
 d\: \lp \frac{\partial \sigma^2_A}{\partial \alpha}\rp^2 \frac 1{2\sigma^4_A}
\enq
in their variances.
If the derivative of $\ln P$ is the constant $c$, we have
\beq
\begin{split}
\frac{\partial  \sigma^2_A}{\partial \alpha} &=\int \idk  \frac{\partial P_A(k) }{\partial\alpha } \\ &= \int \idk P_A(k) \frac{\partial\ln P_A(k)}{\partial \alpha} \\& =  c
\: \sigma^2_A.
\end{split}
\enq
and thus expressions \eqref{Ftot} and \eqref{Fvar} are identical. In terms of information on such parameters, the correlated, Gaussian $A$ field is thus exactly equivalent to $d$ uncorrelated Gaussian variables with the same variances. These parameters  can be seen as entering therefore predominantly the variance, the two point correlation function at zero lag, that contains most information, and the correlations at non zero lag carrying little independent information. This is also expected to hold for the $\delta$ field, since it is very non-linear and the variance dominates  over the clustering in the two point correlation matrix, i.e. the two point correlation matrix is close to diagonal, so that the variance will dominate in any covariance matrix, as well as in the sensitivity to the parameter.
\newline
\newline
Since information just adds up for any number independent variables, this means that we can try and use directly the exact results we derived in the previous section for the one dimensional lognormal distribution to get a rough but still reasonable estimate of the improvement in the constraints from analyzing the $A$ field. In that section were derived all necessary quantities to obtain the cumulative efficiencies
\beq
\epsilon_N(\alpha) := \frac{1}{I(\alpha)} \sum_{n = 2}^N s^2_n(\alpha )\in (0,1)
\enq
of the first $N$ moments of the $\delta$ field to catch the information in $A$. These coefficients are extremely sensitive functions of $\sigma^2_A$, decaying like $\exp(-4\sigma^2_A) \sim \sigma_\delta^{-8} $ as soon as $\sigma_A$ becomes close to unity.
\newline\newline
There is a slight modification to make to these coefficients so that we can confront them to the simulations results. From the simulations only the spectrum of $A$ were extracted, but not the mean of $A$, which also carries information in principle, even if $\delta$ itself has zero mean. For a one dimensional lognormal variable with unit mean, we have from equation \eqref{meanA} that $\bar A = -\frac 12 \sigma^2_A $. For that lognormal variable the total information is given by the usual formula for the Gaussian $A$,
\beq
I(\alpha) = \frac{1}{\sigma^2_A}\lp \frac{\partial \bar A}{\partial \alpha} \rp^2+ \frac{1}{2\sigma^4_A}\lp\frac{\partial \sigma^2_A}{\partial \alpha}\rp^2.
\enq
It reduces thus to
 \beq
I(\alpha) = \frac{1}{2\sigma^4_A}\lp\frac{\partial \sigma^2_A}{\partial \alpha}\rp^2 \lp 1 + \frac{\sigma^2_A}{2}\rp,
\enq
where the rightmost term contains the part of the information in the mean of $A$.
The efficiencies ratios of the moments of $\delta$ to that of the variance of $A$ only, excluding the mean, becomes thus
\beq
\tilde \epsilon_N(\alpha):= \epsilon_N(\alpha)\lp 1 + \frac {\sigma^2_A} 2\rp.
\enq
Note that in principle theses efficiencies $\tilde \epsilon(\alpha)$ can now be larger than unity, if the moments of $\delta$ would capture not only the information in $\sigma^2_A$, but also that in $\bar A$.
\newline
\newline
The improvement factors, i.e. the ratio of the constraints on $\alpha$ from analyzing the first $N$ correlation functions of $\delta$, to the the constraint from the two-point function of A, are thus in this model
\beq \label{ratioss}
\lb \tilde \epsilon_N(\alpha) \rb^{-1/2} =: \Delta^\delta_N(\alpha)/ \Delta^A_2(\alpha).
\enq
They are independent of the parameter $\alpha$ in this one dimensional picture, since the only relevant parameter is $\sigma^2_A$, or equivalently $\sigma^2_\delta$. Remember that the denominator on the right hand side can actually be calculated for any lognormal field from \eqref{FG}, our additional assumptions can be seen thus as entering only the numerator. We argued that these ratios are expected to be roughly correct for parameters such as $\ln \sigma^2_8$, but they become in all cases exact for a lognormal field whose variance dominates enough the clustering, $\xi_\delta(r) / \sigma_\delta^2 \ll 1$, for all $r$. The effective nearest neighbor distance given the modes we used can be evaluated as $r_\min \approx \int \idk ^{-1/3}$,  and we find $\xi_\delta(r_\min) / \sigma^2_\delta = 0.3$.
\newline\newline
Finally, there is slight ambiguity in evaluating $\tilde \epsilon_n(\alpha)$. A purely lognormal field has $\sigma_A = [\ln (1+\sigma^2_\delta)]^{1/2}$, but this relation is not fulfilled precisely in our simulations. We obtain $\sigma_A = 0.7, \sigma_\delta = 1.1$ and so $, [\ln (1 + \sigma_\delta^2)]^{1/2} = 0.9$ rather than $0.7$. This discrepancy may be due of course to an intrinsic failure of the lognormal assumption, or to the presence of the smallest scales, slightly correlated, as seen from the start of saturation in figure \ref{fig:cstr}. 
\newline\newline
We show in the first two rows of Table \ref{table} the factors of improvement for these two values of $\sigma_A, 0.7 $ and $0.9$, for $N= 2,3$ and $\infty$.  In the third row is shown the improvement found extracting $P_A$ rather than $P_\delta$ in the simulations. Given our assumptions, and the very high sensitivity of $\epsilon_N(\alpha)$ to the variance of the field, they agree remarkably : for the sake of comparison, a variance twice as large of $\sigma_A = 2 \rightarrow \sigma_\delta = 7.3$ would have predicted a factor of $\Delta_2 (\delta)/ \Delta_2(A) = 522$, and for $\sigma_A = 3 \rightarrow \sigma_\delta = 90$ a factor of $\approx 5\cdot10^6$. 
\newline
 \newline
We also performed this analysis for the tilt parameter $n_s$, which from its very definition has a very differentiated impact over different modes, and finding, just as in the original analysis \citep{2011ApJ...742...91N}, that the improvement factor is roughly parameter independent as shown in the fourth row of the table. This is another argument supporting the view that the dynamics of the information are indeed captured by such a simple picture. It may be due to the fact that the smallest scales, containing the largest number of modes, contributes the majority of the information in $A$ for any parameter, and thus that the sensitivity can be effectively treated as constant, equal to its value on small scales, making our argument above valid for basically any parameter. Note that for both values of $\sigma_A$ the spectrum of $A$ still outperforms the entire hierarchy of $\delta$ by a sizeable factor for the lognormal model. Of course, this is much more speculative.
\begin{table} \caption[The factors of improvements in constraining power of the spectrum of the log-density field in the lognormal model and as seen in simulations]{  \label{table} The factors of improvements in constraining power of the spectrum of the log-density field in the lognormal model and as seen in simulations. See text for more details.}
\centering
\begin{center} 
\begin{tabular*}{\textwidth}{cccc} 
&$\Delta^\delta_2 / \Delta^A_2$ & $\Delta^\delta_3 / \Delta^A_2$ &$\Delta^\delta_\infty / \Delta^A_2$ \\ 
LN, $\sigma_A = 0.7$ & 2.0 & 1.6 & 1.3 \\
LN, $\sigma_A = 0.9$ & 2.9 & 2.4 & 2.1 \\
Sim. $\alpha = \ln \sigma^2_8$ & 2.5 & & \\
Sim. $\alpha =  n_s$ & 2.4 & & \\
\end{tabular*}
\end{center}
\end{table}


\noindent As a concluding remark, we did not consider observational noise issues in this section. It remains therefore unclear to what extent these improvements can be achieved with actual galaxy survey data. Generically, it is reasonable to expect that noise will reduce these improvement factors. This section nonetheless makes clear that in this case, improving the specifications of a survey in order to decrease the observational  (e.g. shot) noise will be at the same time actually reducing the efficiency with which cosmological parameters can be extracted with the hierarchy of $\delta$ (i.e. the fraction of information that is contained in the hierarchy with respect to the total). First elements towards noise issues in this context are presented in section \ref{deffects}

 \section{Information at all orders for uncorrelated fiducial \label{uncorrfid}}
 In this section, we extend the results of section \ref{onevariable}, going to the next simplest step. In section \ref{onevariable} we derived the information coefficients at all order for the uncorrelated lognormal field. Now we solve the situation of an uncorrelated fiducial. That is to say we allow parameters to create correlations between the different variables according to the lognormal score functions, while keeping the fiducial model model  uncorrelated. In this case, all correlations contribute in general to the information. This is equivalent to use the exact derivatives of the $N$-point moments, while keeping their covariance matrices as those of the uncorrelated model.
 \newline
 \newline
 Of course, this is still a very much simplified picture. Nevertheless it corrects one of the defect of the model we used earlier to compare to the $N$-body simulations, namely that its predictions are model parameter independent, since the only relevant parameter was the variance. In the following differentiated impact on the two-point function allow to make different predictions for different parameters. Also, it provides us for the first time with an easy to implement model for the statistical power of the point functions at all orders, where $N$-point moments at nonzero lag genuinely contribute to the information. This model can also be extended to deal with the presence of noise, see section \ref{deffects}.
 \newline
 \newline
  We work throughout this section first in $d$ dimensions, and then extend these results to the continuous field limit, in Fourier space notation.  We assume throughout statistical homogeneity and isotropy. Also, since we will be interested to compare this to the simulations we assume again that the mean of the field is independent of the parameter. In this case, we have $\partial_\alpha \bar A = - \partial_\alpha \sigma^2_A /2$.
 \newline
 \newline
 A parameter $\alpha$ can enter the two-point correlation function at any argument. As in previous sections $p^{LN}_\rho(\rho,\btheta)$ is the $d$-dimensional lognormal density function for $\rho = (\rho_1,\cdots,\rho_d)$, and $p^{LN}_{\rho}(\rho_i,\btheta)$ are the one dimensional marginals, identical for all $i$.
 \newline
 \newline
 Remember that for an uncorrelated fiducial, a most convenient choice of orthogonal polynomials, that we adopt in the following, is given by
 \beq
 P_\vecn(\rho) := \prod_{i = 1}^d P_{n_i}(\rho_i),
 \enq
 where in our case $P_n$ are the Stieltjes-Wigert polynomials introduced earlier in \ref{onevariable}.
 We will therefore be able to express our results in terms of those for the one dimensional distribution in that section. 
 \newline
 \newline
 From the definition of lognormal variables, the score function of the $d$ dimensional distribution reads
 \beq
 \frac{\partial \ln p^{LN}_{\rho}(\rho,\btheta)}{\partial \alpha} = -\frac 12 \sum_{i,j = 1}^d\lp A-\bar A \rp_i\frac{\partial \lb \xi_A^{-1} \rb_{ij}}{\partial \alpha}\lp A - \bar A \rp_j  - \frac 12 \frac {\partial \sigma^2_A}{\partial \alpha} \sum_{i,j = 1}^d \lb \xi_A^{-1}\rb_{ij}(A - \bar A)_j + \cst.
 \enq
 Under our assumption of uncorrelated fiducial, we have
 \beq
 \frac{\partial \xi_A^{-1}}{\partial \alpha} = -\xi_A^{-1} \frac{\partial \xi_A}{\partial \alpha}\xi_A^{-1} =  -\frac{1}{\sigma^4_A} \frac{\partial \xi_A}{\partial \alpha}.
 \enq
The easiest way to obtain the information coefficients $s_\vecn(\alpha)$ is to rewrite this last expression in terms of the score function of the one dimensional distribution, for which we have already calculated the coefficients for $\alpha = \ln \bar \rho$ and $\alpha = \sigma^2_\delta$.
 \newline
 \newline
 Separating diagonal and non diagonal elements of $\xi_A$ we obtain after a straightforward calculation
 \beq \label{scoreshape}
 \frac{\partial \ln p^{LN}_{\rho}(\rho,\btheta)}{\partial \alpha} =  \frac{\partial \sigma^2_\delta}{\partial \alpha} \sum_{i = 1}^d \frac{\partial \ln p_\rho^{LN}(\rho_i,\btheta)}{\partial \sigma^2_\delta}+\frac 12 \sum_{i \ne j = 1}^d \frac{\partial \lb  \xi_A\rb_{ij}}{\partial \alpha} \frac{\partial \ln p^{LN}_\rho(\rho_i,\btheta)}{\partial \ln \bar \rho} \frac{\partial \ln p^{LN}_\rho(\rho_j,\btheta)}{\partial \ln \bar \rho}.
 \enq
 We can now evaluate easily the information coefficients. Consider a fixed order $N = |\vecn|$. From the product form of the polynomials and the fact that the score function couples variables at most in pairs, we can conclude that
 \beq
 s_\vecn(\alpha) = 0, \quad \textrm{if  }\vecn \textrm{  has more than two nonzero indices.}
 \enq
 Then, using the fact that $\av{\partial_\alpha \ln p} = 0$ for any parameter (normalisation of the density), it follows that for $\vecn$ with only one non zero index the second term in \eqref{scoreshape} does not contribute to $s_\vecn(\alpha)$. We obtain
 \beq \label{infoln1}
 s_\vecn(\alpha) =  \frac{\partial \sigma^2_\delta}{\partial \alpha} s_N(\sigma^2_\delta),\textrm{  for  } \vecn = (0,\cdots,0,N,0,\cdots,0),
 \enq
 wherever the nonzero entry is. Finally, for $\vecn$ with two nonzero entries we have this time that the first term does not contribute, with the result
 \beq \label{infoln2}
 s_\vecn(\alpha) = \frac{\partial \lb \xi_A \rb_{ij}}{\partial \alpha}s_N(\ln \bar \rho)s_{N-k}(\ln \bar \rho),\textrm{  for  } \vecn = (0,\cdots,0,k,0,\cdots,0,N-k,0,\cdots,0),
 \enq
 and $i,j$ are the indices of the nonzero entries.  Summing up over all such multiindices of order $N$, we obtain
 \beq \label{FNun}
 \lb F_{N}\rb_{\alpha \beta} =d\: \frac{\partial \sigma^2_\delta}{\partial \alpha}\frac{\partial \sigma^2_\delta}{\partial \beta} s_N^2(\sigma^2_\delta) + \lp \frac 12 \sum_{i \ne j = 1}^d\frac{\partial \lb \xi_A\rb_{ij}}{\partial \alpha}\frac{\partial \lb \xi_A\rb_{ij}}{\partial \beta} \rp \sum_{i = 1}^{N-1}s^2_N(\ln \bar \rho)s^2_{N-i}(\ln \bar \rho).
 \enq
\subsection{Cumulative efficiencies}
For our uncorrelated fiducial, we have from \eqref{FILN} that the total information is given by
 \beq \label{totalinfo}
 \begin{split}
 F_{\alpha \beta} =d \frac{\partial \sigma^2_A}{\partial \alpha}\frac{\partial \sigma^2_A}{\partial \beta}\lp \frac{1}{4\sigma^2_A} + \frac{1}{2\sigma^4_A} \rp &+ \frac 1{2\sigma^4_A} \sum_{i \ne j= 1}^d \frac{\partial \lb \xi_A \rb_{ij}}{\partial \alpha}\frac{\partial \lb \xi_A \rb_{ij}}{\partial \beta}
 \\ := F^{\sigma^2}_{\alpha \beta} &+ F^{\xi}_{\alpha \beta}.  
 \end{split}
 \enq
Summing the information coefficient over $N$ and dividing by the total information content, we obtain the corresponding cumulative extraction efficiencies :
 \beq \label{formula}
 \lb F _{\le N}\rb_{\alpha \beta} = F_{\alpha \beta}^{\sigma^2} \:\epsilon_N(\sigma^2_\delta)
  + F_{\alpha \beta}^{\xi}\:\epsilon_N(\xi),
 \enq
 where $\epsilon_N(\sigma^2_\delta)$ was defined in section \ref{onevariable} (and shown in the right panel of figure \ref{fig2eps}), and the second efficiency is
 \beq
 \epsilon_N(\xi) := \sigma^4_A \sum_{n = 2}^N \sum_{i = 1}^{N-1}s_i^2(\ln \bar \rho)s_{n-i}^2(\ln \bar \rho) \in (0,1).
 \enq
  \begin{figure}
  \centering
  \includegraphics[width = 0.8\textwidth]{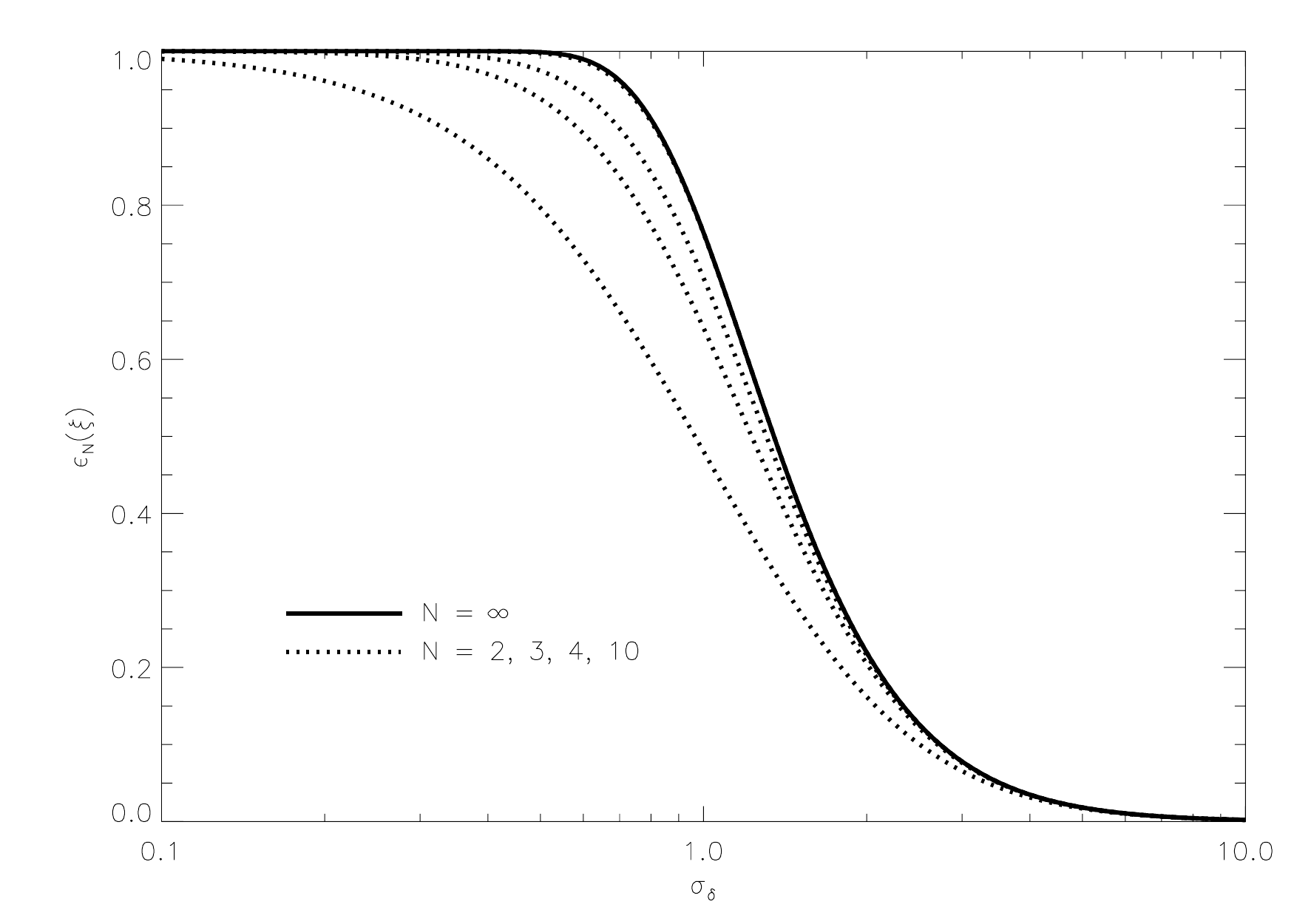}
 \caption{\label{effxi} The fraction of the total information content that is accessible through extraction of the full series of $N$-point moments of the lognormal field for an uncorrelated fiducial model, for a parameter entering the two-point function at non zero lag, but not the variance.}
 \end{figure}
 Figure \ref{effxi} shows the behavior of $\epsilon_N(\xi)$ as function of $\sigma_\delta$. It is obviously qualitatively very similar to figure \ref{fig2eps} for $\epsilon(\sigma^2_\delta)$, showing a slower rate of decay to zero though.
\subsection{Impact of discreteness effects \label{deffects}}
\newcommand{\noise}{\textrm{noise}}
It is a very interesting to note that these exact results for an uncorrelated fiducial can be extended with little additional (numerical) effort to the case of a noisy field.
\newline
\newline
Consider that the observed field $\rho_\obs$ is not exactly the lognormal density field but a noisy tracer : 
\beq
p_{\rho_\obs,\btheta}(\rho_\obs) = \int d^d \rho \: p_\rho^{LN} (\rho,\btheta) \prod_{i = 1}^d p_{\noise}(\rho_{\obs,i} |\rho_i),
\enq
for some one dimensional, parameter independent, density function $p_\noise (\rho_\obs |\rho)$. It can represent for instance additive noise, or discreteness effects if $\rho_\obs$ is for instance a number of dark matter particles, or galaxies, in a cell of a given volume. For an uncorrelated fiducial, direct calculation shows that equation \eqref{scoreshape} stays in fact formally unchanged :
\beq \label{scoreshape2}
\begin{split}
 \frac{\partial \ln p_{\rho_\obs}(\rho,\btheta)}{\partial \alpha} &=  \frac{\partial \sigma^2_\delta}{\partial \alpha} \sum_{i = 1}^d \frac{\partial \ln p_{\rho_\obs}(\rho_{\obs,i},\btheta)}{\partial \sigma^2_\delta} \\
 &\:\:+\frac 12 \sum_{i \ne j = 1}^d \frac{\partial \lb  \xi_A\rb_{ij}}{\partial \alpha} \frac{\partial \ln p_{\rho_\obs}(\rho_{\obs,i},\btheta)}{\partial \ln \bar \rho} \frac{\partial \ln p_{\rho_\obs}(\rho_{\obs,j},\btheta)}{\partial \ln \bar \rho}.
 \end{split}
\enq
It follows that the results \eqref{infoln1} and \eqref{infoln2} for the information coefficents, as well as the information of order $N$, equation \eqref{FNun}, that we derived in the last section hold unchanged, with the understanding that the coefficients $s_n$ present there are not those of the one dimensional lognormal density function but now those of the one dimensional density for $\rho_\obs$. Those coefficients can generically be obtained numerically without much difficulty even for high orders.
\newline
\newline
The total information becomes from \eqref{scoreshape2}
\beq
F_{\alpha \beta} = d\:F^{1D}_{\alpha \beta} + \frac 12 \sum_{i \ne j = 1}^d \frac{\partial \lb \xi_A \rb_{ij}}{\partial \alpha}\frac{\partial \lb \xi_A \rb_{ij}}{\partial \beta} \lb F^{1D}_{\ln \bar \rho \ln \bar \rho}\rb^2,
\enq
where $F^{1D}$ is the information matrix of the $1$ dimensional density for $\rho_\obs$, that can also be evaluated numerically for a prescribed shape of 
$p_\noise$.
\newline
\newline
We evaluated how the efficiencies change from the purely lognormal case, figure \ref{fig2eps} and \ref{effxi}, when including discreteness effects in the form of the Poisson model, introduced already earlier in \ref{poissoneffects}. In this case, the observations are Poisson variables, with intensity in each cell given by the value of the lognormal field in that cell. The one-point density for $ \rho_\obs =:n$ object in a cell is given by
\beq \label{Poissona}
p_n(n,\btheta) = \int d\rho\: p_\rho^{LN}(\rho,\btheta) e^{-\rho n_c} \frac{\lp \rho n_c\rp^n}{n!},\quad n = 0,1 \cdots.
\enq
Here the lognormal distribution is set to have mean unity, so that $n_c$ is the mean number of particles per cell.
We already know from section \ref{poissoneffects} that the matrices $F_{\le N}$ must always be reduced, as well as the total information. Of course, they might be reduced at a different rate, so that the efficiencies need not be reduced, but can increase drastically, if the observed density function is made closer to a Gaussian.
\newline
\newline
  \begin{figure}
   \centering
  \includegraphics[width = 0.8\textwidth]{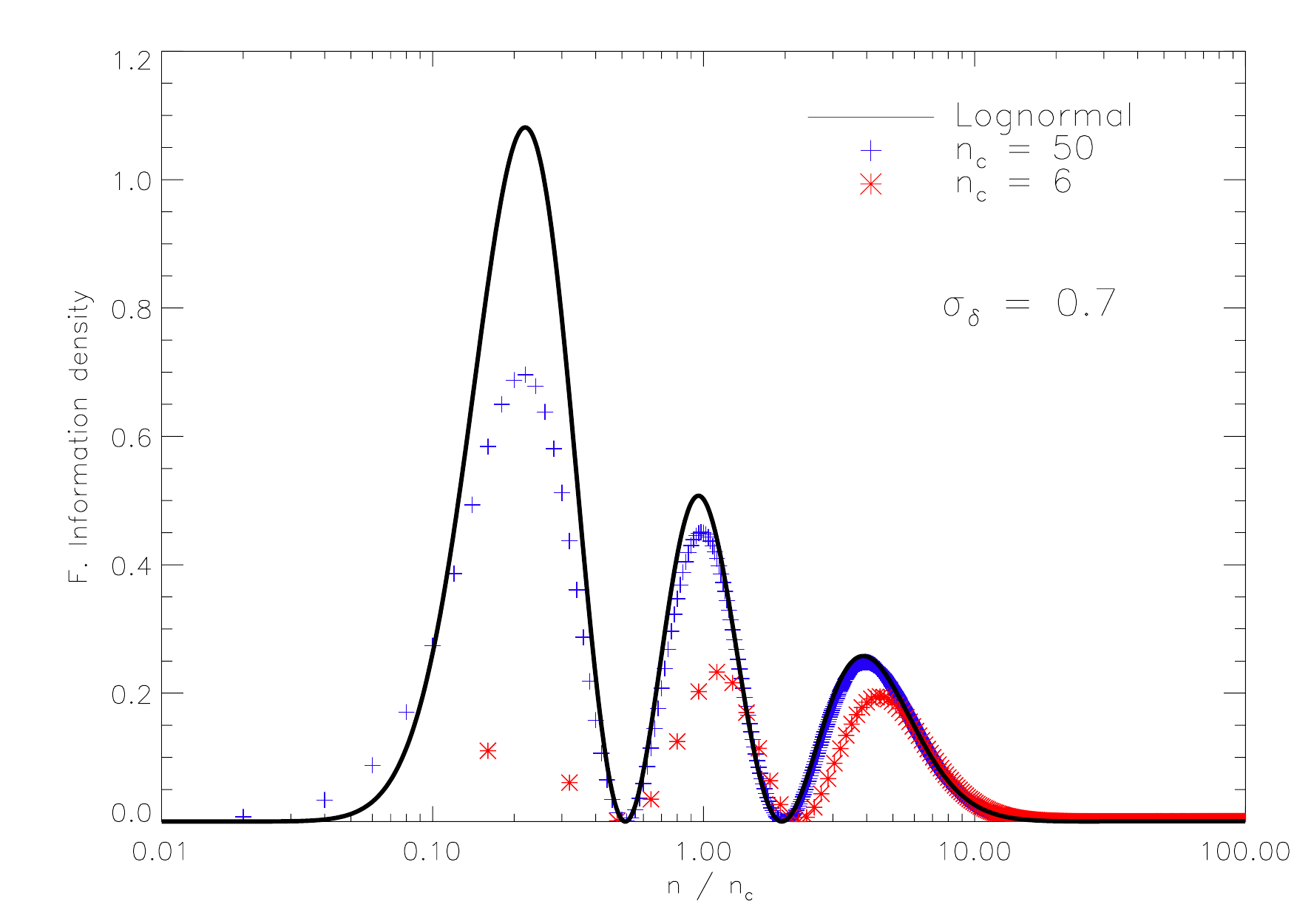}
   \caption[Discreteness effects destroying the information contained in the underdense regions of the lognormal distribution]{\label{densitydiscrete} The Fisher information density of the Poisson sample of the lognormal distribution with $\sigma_\delta = 0.7$, as given in \eqref{Poissona}, for two different values of $n_c$ as indicated. The solid line is that of the (continuous) lognormal distribution. The vast majority of the information  was initially within the underdense regions but is destroyed if the mean number of object per cell is low, making the $N$-point moments efficient again.}
 \end{figure} 
  \begin{figure}[ht]
  \centering
  \includegraphics[width = 0.8\textwidth]{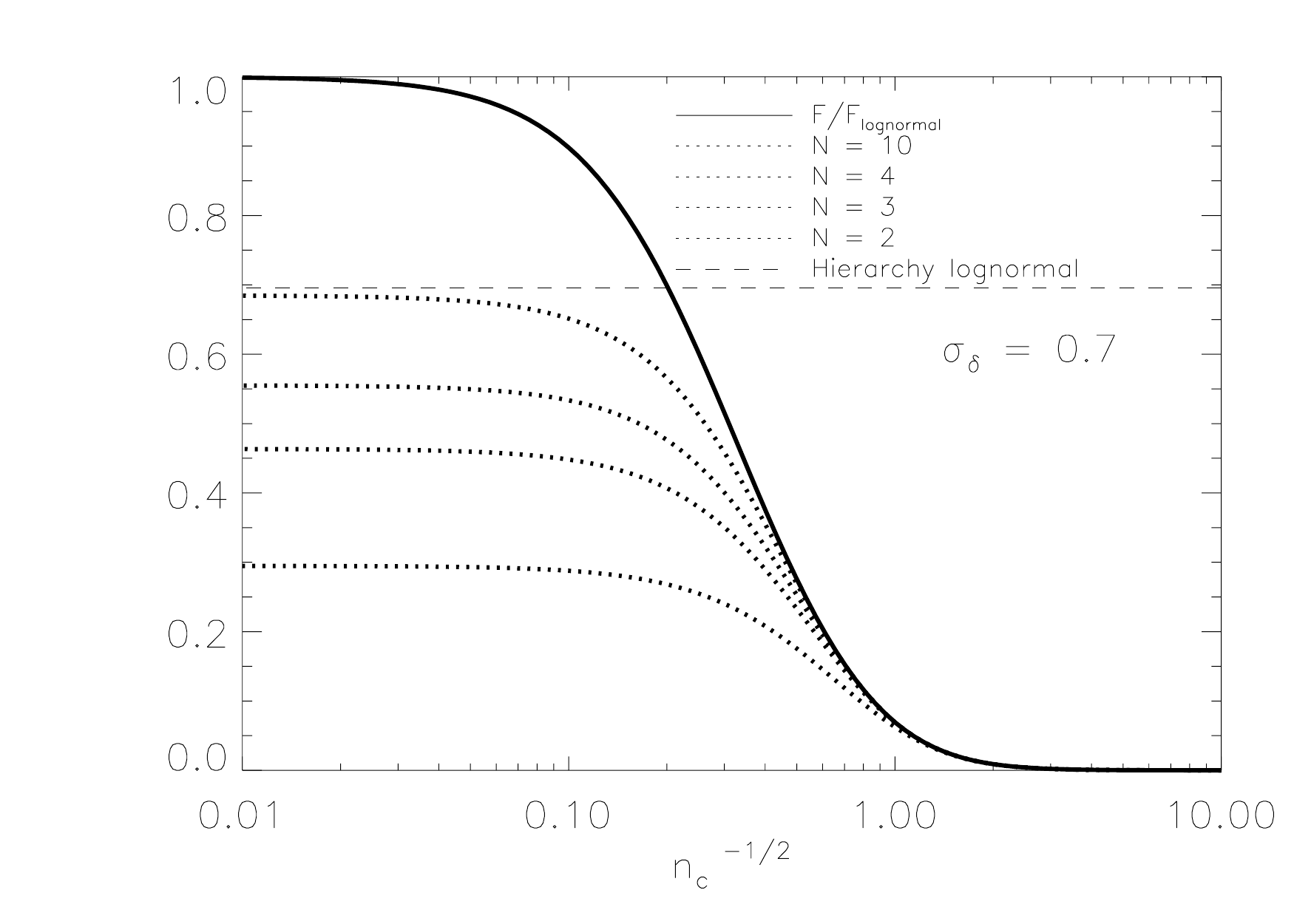}
  \includegraphics[width = 0.8\textwidth]{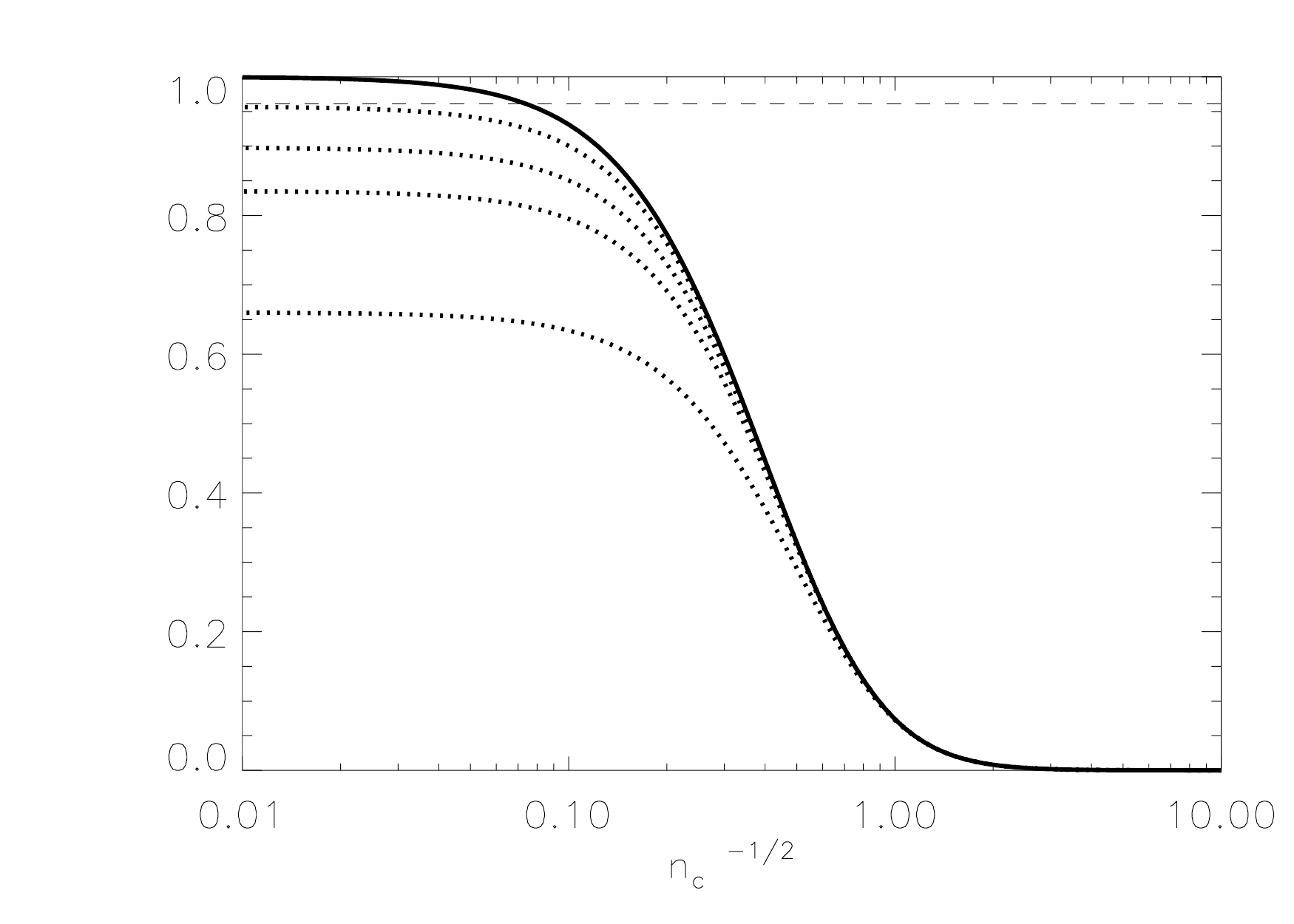}
   \caption[The information content of the first $N$-point moments in a Poisson sample of a lognormal field, as function of the mean number of objects per cell, normalised to the information content of the lognormal field.]{\label{discreteness}Upper panel : the information content of the first $N$-point moments in Poisson sample of a lognormal field with uncorrelated fiducial with $\sigma_\delta  = 0.7$, as function of the inverse root of the mean number of objects per cell, for a parameter entering the variance of the field but not the correlations. The solid line denotes the total information content, and the dotted lines that of the point functions up to order $N$. All lines are normalised to the total information content of the noise-free lognormal field. Dashed is that of the entire hierarchy of the lognormal for comparison. Lower panel : same for a parameter entering correlations but not the variance. $N$-point functions become efficient again in the noisy regime, but only due to the massive decrease in the total information content.}
 \end{figure}
\noindent The results are shown on figure \ref{discreteness}. It is shown, as function of $n_c$, the total information $F$, solid line, as well as $F_{\le{N}}$, dashed lines, all normalised by the total information content of the noise free lognormal distribution, at a fixed variance $\sigma^2_\delta = 0.7$. The upper panel is for a parameter that enter the variance only and not the correlations, and the lower panel for a parameter that enter the correlations exclusively. It is clear that the strongest effect is to reduce the total information content, as soon as the mean number of objects is close to unity. It follows that in the very noisy regime, the $N$-point moments are efficient again, in that the fraction of the information that they capture is close to unity again. 
\newline
\newline

\noindent We have already argued that the reason for which the $N$-point moments of the lognormal field are inefficient is that they are unable to probe the underdense regions, rich in information. This is seen very clearly in figure \ref{densitydiscrete}.  Shown are the Fisher information density for the parameter $\sigma^2_\delta$ of the one dimensional distribution $p_n(n)$, for a mean number of objects $n_c$ of $50$ (crosses) and $6$ (diamonds). The solid line shows that of the noise-free, continuous lognormal. Unsurprisingly, it is the information in the underdense regions ($n/n_c < 1$), precisely the information that the moments were unable to catch, that is mostly destroyed by discreteness effects. On the other hand, the information in the overdense regions ($n /n_c > 1$), accessible to the moment hierarchy, is left untouched.
 \clearpage
\subsection{The reason for a roughly parameter independent factor of improvement \label{Fspace}}
\newcommand{\kmax}{\textrm{kmax}}
\renewcommand{\veck}{\mathbf k}
We now go back to our exact expressions \eqref{formula} for the information in $N$-point moments of a lognormal field with uncorrelated fiducial. We seek to transform these expressions in Fourier space such that it is easier to compare or use $N$-body simulation results within this model.
\newline
\newline
 Within our assumptions, the fiducial power spectrum is constant and directly proportional to the variance, such that we can write \beq
\frac{\partial  \ln \sigma^2_A}{\partial \alpha} = \frac 1 d \sum_{\veck} \frac{\partial \ln P_A}{\partial \alpha}.
\enq
Turning the sum over the $d$ available modes, with spacing $\Delta k = (2\pi)^3/V$ into an integral, it becomes
\beq
\frac{\partial  \ln \sigma_A^2}{\partial \alpha} = V_{\textrm{cell}} \int_0^{k_\textrm{max}}\frac {dk\:k^2} {2\pi^2} \frac{\partial \ln P_A}{\partial \alpha}.
\enq
In this equation, $V_{\textrm{cell}}$ is the volume associated to a real space cell, given  formally by
\beq
V_{\textrm{cell}} = \frac{(2\pi)^3}{d \Delta k} = \frac V d 
\enq
that can be written in the continuous notation as
\beq
\frac 1 {V_{\textrm{cell}}} = \int_0^{k_\textrm{max}}\frac {dk\:k^2} {2\pi^2}.
\enq
Thus, we can write the first term in \eqref{formula} as
\beq
\begin{split}
\frac d 2 \frac{\partial \ln \sigma^2_A}{\partial \alpha}\frac{\partial \ln \sigma^2_A}{\partial \beta}  = V_{\textrm{cell}}\frac V 2 \int_0^{k_\textrm{max}}\frac {dk\:k^2}{2\pi^2} \frac{\partial \ln P_A(k)}{\partial \alpha}\int_0^{k_\textrm{max}}\frac {dk\:k^2}{2\pi^2}\frac{\partial \ln P_A(k)}{\partial \beta}.
\end{split}
\enq
On the other hand, to write the second term of equation \eqref{formula} as an integral over the power spectrum, we first complete the sum over $i \ne j$, and remember that within our assumptions, $\xi_A^{-1}$ is $1/\sigma^2_A$ times the identity matrix. We obtain
\beq 
\begin{split}
 \frac{1}{2\sigma^4_A}\sum_{i \ne j} \frac{\partial \xi_{ij}}{\partial \alpha} \frac{\partial \xi_{ij}}{\partial \beta}
 =\frac 12 \Tr  \lb \frac{\partial \xi_{A}}{\partial \alpha} \xi_A^{-1} \frac{\partial \xi_{A}}{\partial \beta}  \xi_A^{-1} \rb 
  - \frac d 2 \frac{\partial \ln \sigma^2_A}{\partial \alpha}\frac{\partial \ln \sigma^2_A}{\partial \beta}.
 \end{split}
\enq 
The first term on the right is nothing else than the well known expression for the information content of the two point correlation function of the Gaussian field $A$. Fourier transformation of the first term leads to
\beq\label{infoSpec}
 \frac V 2 \int_0^{k_\textrm{max}}\frac {dk\:k^2} {2\pi^2} \frac{\partial \ln P_A(k)}{\partial \alpha}\frac{\partial \ln P_A(k)}{\partial \beta} =: F^{P_A}_{\alpha \beta}. 
\enq
This last expression \eqref{infoSpec}, the information content of the spectrum of the Gaussian field $A$, is exact for any lognormal field, the assumptions we are making above not entering it.
\newline\newline
We are now ready to put all these relations together. In section \ref{connection} (see also \cite{2011ApJ...742...91N}) we just studied the constraints on the parameters given by extracting the spectrum of $A$ or $\delta$. We already know what the information in the spectrum of $A$ is, equation \eqref{infoSpec}. We will therefore find convenient to consider the ratios $F_{\le N}/ F^{P_A}$. Define first the ratios $R_{\alpha\beta}$ as
\beq \label{Rratio}
R_{\alpha\beta} := \frac{V_{\textrm{cell}}}{2\pi^2} \frac{\int_0^{k_\textrm{max}}dk\:k^2  \partial_\alpha \ln P_A\: \int_0^{k_\textrm{max}}dk\:k^2\partial_\beta \ln P_A}{\int_0^{k_\textrm{max}} dk\:k^2 \partial_\alpha \ln P_A \partial_\beta \ln P_A}.
\enq
These ratios are unity for parameters that obeys $\partial_\alpha \ln P_A \approx \cst$, such as $\ln \sigma_8^2$. They can vanish for parameters with differentiated impact, such as the spectral index $n_s$.
\newline
\newline
We obtain from \eqref{formula} and the relations in this section that the loss of information by extracting the first $N$ correlation functions of $\delta$ rather than the spectrum of $A$ is given by
\beq \label{masterf}
\frac{ \lb F_{\le N} \rb_{\alpha\beta}}{F^{P_A}_{\alpha\beta}} = \epsilon_N(\xi)\lb  1- R_{\alpha \beta}\rb + \epsilon_N(\sigma^2_\delta)R_{\alpha \beta}\lp 1 + \frac 12 \sigma^2_A \rp.
\enq
   \begin{figure}
   \centering
  \includegraphics[width = 0.8\textwidth]{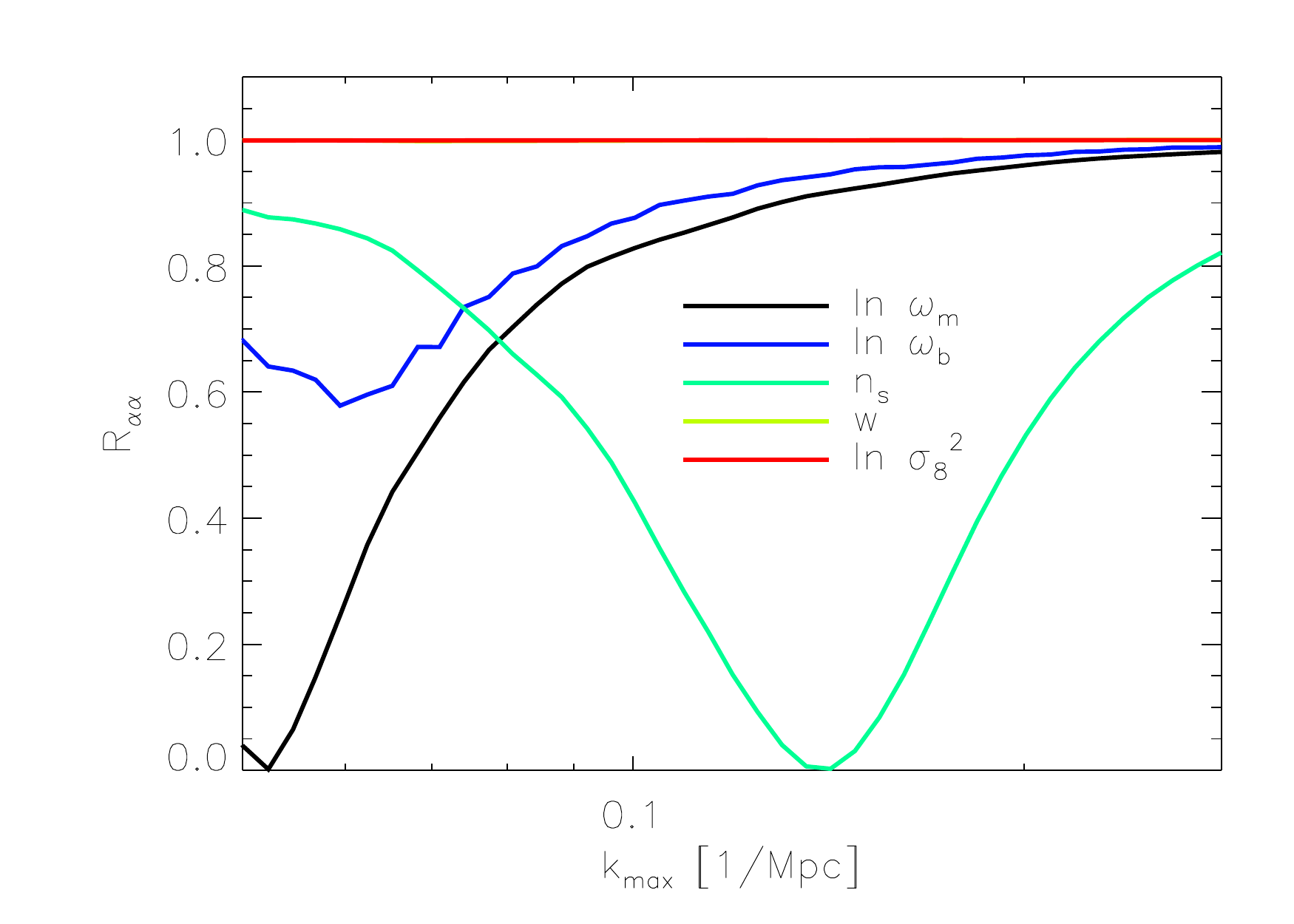}
   \caption[Ratios measuring the sensibility of the spectrum to that of the variance for different cosmological parameters]{\label{figRab} The ratios $R_{\alpha\alpha}$, equation \eqref{Rratio}, for the five cosmological parameters studied in the analysis of \cite{2011ApJ...742...91N}, as function of the smallest scale present in the analysis. The constancy of these factors once all modes are included implies a generic, roughly parameter independent gain from the analysis of the spectrum of $A$ rather than $\delta$. See text for more details.}
 \end{figure} 
Recall that while the denominator in \eqref{masterf} equation \eqref{infoSpec} is exact for any lognormal field, the numerator must be exact only within the assumption of an uncorrelated fiducial as we specified.
All the terms of this equation are strong functions of the variance of the field, or equivalently of $k_\max$. The variance of $A$ being set by
\beq \label{varA}
\sigma^2_A(k_\max) = \int_0^{k_\max} \frac{dk\:k^2}{2\pi^2} P_A(k).
\enq
\newline
\newline
We are now in position to test whether our more sophisticated model changes our conclusions from section \ref{connection}, where we reproduced the correct factor of improvements with our one dimensional model. Of course, it is apparent that the key element is the magnitude of the coefficients $R_{\alpha\alpha}$ in \eqref{masterf}: if they are close to unity then the prediction of the model is essentially unchanged from that of the one dimensional distribution. These factors, obtained from the derivatives of the spectrum measured from the simulations as in section \ref{connection}, are shown in figure \ref{figRab} as function of $k_\max$ for the parameters $\omega_b = h^2 \Omega_b$, $\omega_m = h^2 \Omega_m$, the tilt $n_s$, as well as the dark energy equation of state $w$ and $\ln \omega^2_8$. The last two curves are indistinguishable from unity on this figure.
\newline
\newline
It is obvious that when all the scales are included, these coefficients are all very close unity, leading to nearly identical predictions for the improvement factors for all these parameters. The tilt parameter is the one that differs the most, where the prediction in this improved model is slightly below the one dimensional case, consistently with the findings in simulations that we presented in section \ref{connection}(see table \ref{table}).
The other parameters show essentially identical dynamics, and the predictions do not differ from the one dimensional case. This is also remarkably consistent with the study of \cite{2011ApJ...742...91N}, where $n_s$ is the only parameter to show a slightly different behavior (remember though that that work includes some modes $0.3/$Mpc $<k < 0.6$/Mpc  that cannot be described as lognormal).
\newline
\newline
Maybe surprisingly, our simple, essentially one dimensional treatment thus definitely seems to capture very well the dynamics of the information as seen in $N$-body simulations.

 
\section{The general case : a difficult Lagrange interpolation problem \label{generalcase}}
We end this chapter by a remark on the problem of the derivation the matrices $F_N$ in the case of a lognormal field with arbitrary correlations, a still unsolved problem, except in the situations dealt with in section \ref{onevariable} and \ref{uncorrfid}.
\newline
\newline
Remember that we could solve the one dimensional case making use of a curious identity of the polynomials orthogonal $P_n$ to the lognormal distribution. We showed that the polynomial in $t$ defined as
\beq
\av{P_n(t\rho)}
\enq
has zeroes in $q^{-i}, i = 0,n-1$. Since a polynomial in one variable factorizes in product of roots, we could easily obtain explicit form for the polynomials as well as of the information coefficients.
\newline
\newline
It is very interesting that this property of the set of orthogonal polynomials generalises to any dimensioniality $d$.
In the multiindex notation, recall that the $N$-point moments are given by
\beq
m_\vecn = \av{\rho(x_1)^{n_1}\cdots \rho(x_d)^{n_d}} = \exp \lp \bar A \cdot \vecn + \frac 12 \vecn \cdot \xi_A \vecn \rp.
\enq
The following property is easy to verify :
\beq \label{propcool}
m_{\vecn + \vecm} = m_\vecn m_\vecm Q_{\vecn \vecm} ,
\enq
where we defined
\beq
Q_{\vecn \vecm} := \exp \lp \vecn \cdot \xi_A \vecm \rp. 
\enq
This identity is the generalisation of the property $m_{i +j} = m_im_j q^{-ij}$ which we made good use of in section \ref{onevariable}.
We can write the matrix in an other useful form. Define the points $t_\vecn$ as
\beq
t_\vecn = \exp \lp \xi_A \cdot \vecn \rp,\textrm{   i.e.  }  \lp t_\vecn \rp_i = \exp \lp \sum_{j = 1}^d \lb \xi_A \rb_{ij}n_j\rp. 
\enq
We have the following relation 
\beq
Q_{\vecn\vecm} = t_\vecn^\vecm.
\enq
The matrix Q is thus a Vandermonde matrix in several variables. 
\newline
\newline
We can now prove the following property of the orthogonal polynomials in any dimensionality. For any $\vecn$ and $\vecm$ holds:
\beq
\av{P_\vecn(t_\vecm \rho)} = \frac{1}{m_\vecm} \av{\rho^\vecm P_\vecn(\rho)}.
\enq
To prove that relation expand $P_\vecn(\rho) = \sum_\veck C_{\vecn\veck} \rho^\veck$ on both sides and use \eqref{propcool}.
Remember that the orthogonal polynomials of a given order are defined such that they are orthogonal to all polynomials of lowest order.
If for convenience a further ordering of the multindices of same order $|\vecn|$ is chosen, we can say without ambiguity whether for two multindices $\vecn$ and $\vecm$ hold $\vecn < \vecm$, $\vecn > \vecm$ or $\vecn = \vecm$. We can then write
\beq
\av{P_\vecn(t_\vecm \rho)} = 0, \textrm{  for  } \vecm < \vecn.
\enq
This identity is is the direct analog of the relation we used to solve the one dimensional problem as just discussed at the very beginning of this section. We know enough zeroes of the polynomial 
\beq
\pi_\vecn(t) = \av{P_\vecn(t\rho)}
\enq
in $t = (t_1,\cdots,t_d)$, to specify it uniquely, and with it the orthogonal polynomials. 
\newline
\newline
The problem becomes therefore that to find the polynomials in $d$ variables with the given zeroes as prescribed above, which is nothing else than a familiar Lagrange interpolation problem, with associated Vandermonde matrix $Q$. Unfortunately, this is of course a considerably more involved task in several dimensions than in one, where many properties such as factorization of the polynomials does not hold anymore.

%

\clearpage
\newpage
\chapter{Information escaping the correlation hierarchy of the convergence field\label{chPRL}}
The text of this chapter follows that of the letter we published in \cite{PhysRevLett.108.071301}. There we evaluated with numerical methods the information content of the moment hierarchy of the noise-free weak lensing convergence field. The approach presented in chapter \ref{ch2} is applied this time not to the somewhat ad-hoc model of the lognormal distribution of chapter \ref{ch3}, but to the most realistic case of fits to $N$-body simulations outputs already present in the literature. We find the dynamics of the information within the hierarchy to be qualitatively very similar than for the lognormal, the moments becoming quickly (even quicker than the lognormal ) dramatically inefficient. For these reasons, even though this chapter is comparatively short, we believe it to be an important part of this thesis, as it makes clear that the dynamics explored in chapter \ref{ch3} with the help of the lognormal field generically affect cosmological non linear fields. It if also found that a simple logarithmic mapping makes the moment hierarchy well suited again for parameter extraction.

\section{Introduction}

N-point correlation functions, first introduced in cosmology by Peebles and collaborators to describe the large scale distribution of galaxies \citep{1980lssu.book.....P}, are now ubiquitous in this field. They are at the heart of many cosmological probes like the CMB, galaxy clustering, or notably weak lensing, which was recognized as one of the most promising probe of the dark components of the universe \citep{2001PhR...340..291B,2004PhRvD..70d3009H,2006astro.ph..9591A,2009ApJ...695..652B}, and which traces the cosmological convergence field.
\newline
\newline
 On large scales, or in the linear regime, correlations are a particularly convenient approach to tackle the difficult problem of statistical inference on cosmological parameters. Indeed, primordial cosmological fluctuation fields are believed to obey Gaussian statistics, and the first two members of the hierarchy, the mean and the two-point correlation function, provide a complete description of such fields. However, much less is known about the pertinence of the correlation hierarchy in the non-linear regime, or on small scales, where in principle a lot of information is contained, if only due to the large number of modes available for the analysis. More elaborated statistical models  must be made in this regime. For instance, the statistics of the matter field and its weighted projection the convergence field were shown to be closer to lognormal, at least in low dimensional settings \citep{1991MNRAS.248....1C,2000MNRAS.314...92T,2002ApJ...571..638T,2006ApJ...645....1D}, though with sizeeable deviations still.
 \newline
 \newline
 Two effects relevant for statistical inference can in principle play a role entering the non linear regime, departing from Gaussian initial conditions. First, information may propagate to higher order correlators. Second, the correlation function hierarchy may not provide a complete description of the field anymore 
 , so that information escapes the hierarchy. Even though this second possibility was pointed out qualitatively in an astrophysical context already in \cite{1991MNRAS.248....1C}, it seems it was not given further attention in the literature.
In this \textit{Letter} we show, using accurate fits of the convergence one-point probability density function to numerical simulations \citep{2006ApJ...645....1D} that the second effect very quickly completely dominates the convergence field, and thus that the hierarchy is not well suited for inference on cosmological parameters anymore. 
\paragraph{Fisher information and orthogonal polynomials.}
The approach is based on decomposing the Fisher's matrix valued information measure in components unambiguously associated to the independent information content of the correlations of a given order. It was recently proposed in \cite{2011ApJ...738...86C}, building upon \cite{Jarrett84}. Exact results at all orders were obtained only for the moment hierarchy of a idealized, perfectly lognormal one dimensional variable, where analytical methods could be applied. In cosmology, the Fisher information matrix is widely used for many years now to estimate the accuracy with which cosmological parameters will be extracted from future experiments aimed at some observables \citep[e.g.]{Tegmark97b,2004PhRvD..70d3009H,2009ApJ...695..652B}, assuming Gaussian statistics.

For a general probability density function $p(x,\alpha,\beta)$, $\alpha$, $\beta, \cdots$ any model parameters, its definition is
\beq \label{FI}
F_{\alpha\beta} = \av{\frac{\partial \ln p}{\partial \alpha}\frac{\partial \ln p}{\partial \beta}}.
\enq
Its inverse can be seen through the Cram\'er-Rao bound \citep{1997ApJ...480...22T} to be the best covariance matrix of the relevant parameters achievable with the help of unbiased estimators.
The general procedure to decompose the Fisher information content into uncorrelated pieces, corresponding to an orthogonal system, was presented in a statistical journal in  \cite{Jarrett84}.  When the observables of interest are products of the variables, i.e. moments or more generally correlation functions, the orthogonal system are orthogonal polynomials. It is discussed in detail in an cosmological context in \cite{2011ApJ...738...86C}.
In particular, the variables for which the Fisher information content on $\alpha$ is entirely within the first $N$ pieces, such as the Gaussian variables for $N = 2$, are those for which the function $\partial_\alpha  \ln p$ entering
(\ref{FI}), called the score function, is a polynomial of order $N$ in $x$.
In the case of a single variable, 
the uncorrelated contribution of order $N$ to the Fisher information matrix $\Fab$ is given by
 \beq
s_N(\alpha)s_N(\beta),
 \enq
 where the Fisher information coefficients $s_N$ are the components of the score function with respect to the orthonormal polynomial of order $N$,
 \beq \label{sndef}
 s_N(\alpha) = \av{\frac{\partial \ln p}{\partial \alpha}P_N(x)},
 \enq
 \beq
 \av{P_n(x)P_m(x)} = \delta_{mn},\quad n,m \ge 0
 \enq
 For any $N$, the following relation holds
 \beq \label{sumsn}
 \sum_{n = 1}^Ns_n(\alpha)s_n(\beta) = \sum_{i,j = 1}^{N}\frac{\partial m_i}{\partial \alpha } \lb \Sigma^{-1}\rb_{ij}\frac{\partial m_j}{\partial \beta },
 \enq
 where $m_i = \av{x^i}$ and $\Sigma_{ij} = m_{i+j} - m_im_j$ is the covariance matrix.
 The right hand side being the expression describing the Fisher information content of the moments $m_1$ to $m_N$. Whether one recovers the full matrix $\Fab$ with $N\rightarrow \infty$ or only parts of it depends on the distribution under consideration. A sufficient condition is that the polynomials $P_n$ form a complete basis set, which is then essentially equivalent to the condition that the distribution can uniquely be recovered from its moments hierarchy \cite[and references therein]{1991MNRAS.248....1C,2011ApJ...738...86C}. This and other sufficient criteria for completeness are tightly linked to the decay rate of the probability density function at infinity.
 \newline
 \newline
 We define the cumulative efficiency $\epsilon_N$ of the moments up to order $N$ to capture Fisher information on $\alpha$ as
 \beq
 \label{eff}
 \epsilon_N(\alpha):= \frac{\sum_{n = 1}^Ns_n^2(\alpha)}{F_ {\alpha\alpha}}.
 \enq
 From the Cram\'er-Rao bound, $\sqrt{\epsilon_N}$ is the ratio of the the best constraints achievable on $\alpha$ with any unbiased estimator to the expected constraints on $\alpha$ from the extraction of the first $N$ moments .
 \section{Fisher information coefficients}
\begin{figure}
  \centering
  \includegraphics[width = 0.8\textwidth]{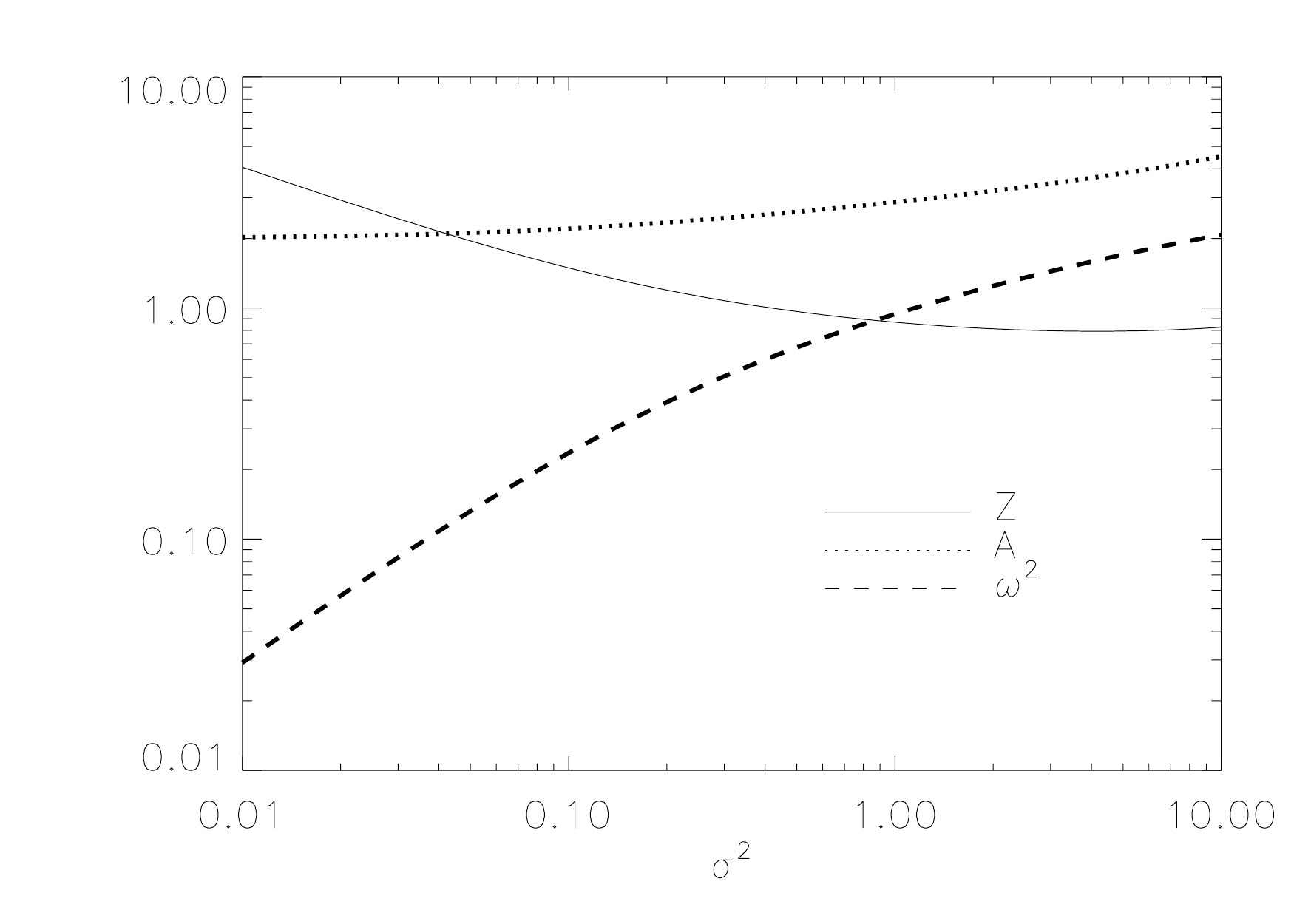}
 \caption[The three parameters entering the generalized lognormal model, as function of the variance of the effective matter density fluctuations]{\label{fig1}The three parameters $Z$,$A$ and $\omega^2$ entering the generalized lognormal model, as function of the variance of $\delta_m^\eff $.} 
 \end{figure}

\begin{figure}
  \centering
  \includegraphics[width = 0.8\textwidth]{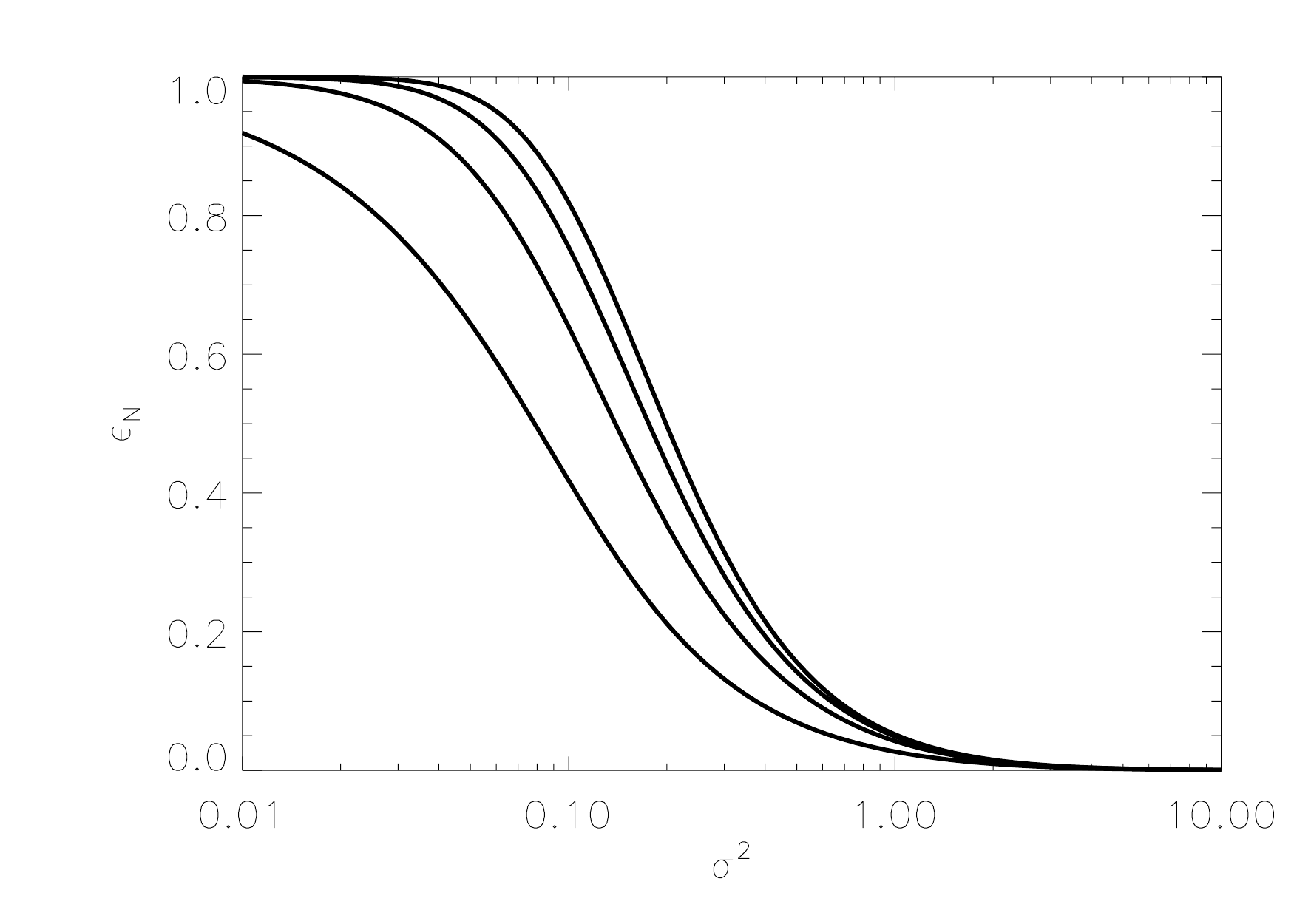}
 \caption[The cumulative efficiencies of the moments of the convergence in capturing Fisher information as function of the variance  of the effective matter density fluctuations.]{\label{fig2} The cumulative efficiencies $\epsilon_N$ of the moments of the convergence in capturing Fisher information, for $N =2$ to $N = 5$, defined in eq. (\ref{eff}), from bottom to top, as function of the variance  of $\delta_m^\eff $.} 
 \end{figure}

 We use the fits to simulations from \cite{2006ApJ...645....1D}, valid down to the arcsecond scales.  Initially built to correct for the failure of the lognormal distribution to reproduce the high and low density tails of the convergence $\kappa$ on a single lens plane, it reproduces accurately the cosmological convergence as well, taking into account the broader lensing kernel \citep{2011ApJ...742...15T}. In terms of the reduced variable $x$,
  \beq
 x = 1 + \frac \kappa{ \left |\kappa_{\textrm{empty}} \right |} =: 1 + \delta_m^\eff,
 \enq
 where $\kappa_\textrm{empty}$ is the minimal value of the convergence, corresponding to a light ray traveling an empty region, it takes the form of a generalized lognormal model for the associated effective matter fluctuations $\delta_m^\eff$,
 \beq \label{model}
 p(x,\sigma) = \frac{Z}{x} \exp \lb -\frac{1}{2\omega^2} \lp \ln x + \frac{\omega^2}{2}   \rp^2\lp 1 +\frac{A}{x} \rp \rb.
 \enq
 In this equation, the three parameters $Z$, $A$ and $\omega$ are such that the mean of $x$ is unity, and its variance $\sigma^2 = \sigma^2_\kappa / \kappa^2_{\textrm{empty}}$ (we are neglecting here a small but non-zero mean of the convergence argued in \cite{2011ApJ...742...15T}). Therefore, the only relevant parameter is the variance of the associated matter fluctuations $\sigma^2$, fixed by the cosmology from $\kappa_{\textrm{empty}}$ and the convergence power spectrum, together with some filter function corresponding to the smoothing scale, determining the level of non linearity of the field \citep[figure 1]{2006ApJ...645....1D}. Linear and non-linear regime being separated at $\sigma^2 \approx 1$. We obtained 
 $Z,A$ and $\omega^2$, shown in figure \ref{fig1}, 
 with the help of a standard implementation of the Newton-Raphson method for non-linear systems of equations.
 \newline
 \newline
Orthogonal polynomials can very conveniently be generated by recursion, as exposed in details in \cite{Gautschi04}, since they satisfy a three terms recurrence formula. We define for convenience
 \beq
 \hat \pi_N(x) := \sqrt{p(x,\sigma)}\pi_N(x), 
 \enq
 where $\pi_N(x)$ is $P_N(x)$ rescaled such that the coefficient of $x^N$ is unity. The recursion relations in \cite {Gautschi04} 
 become
  \beq
 \begin{split}
\hat \pi_{k + 1} &= (x - \alpha_k)\hat\pi_k -\beta_k \hat\pi_{k-1}, \\
 \alpha_k &:= \frac{\int_0^\infty dx\: x \:\hat \pi^2_k(x) }{\int_0^\infty dx\: \hat \pi^2_k(x) } \\
 \beta_k &:= \frac{\int_0^\infty dx\:\hat \pi^2_k(x) }{\int_0^\infty dx\:\hat \pi^2_{k-1}(x) },
 \end{split}
 \enq
 and $\pi_{-1}(x) = 0, \pi_1(x) = 1, \beta_0 = 1$, that we implemented using an appropriate discretization of the $x$-axis.  Proper normalization of the polynomials can be performed afterwards. The Fisher information coefficients were then obtained with the help of equation (\ref{sndef}), using a precise five point finite difference method for the derivatives of $Z,A$ and $\omega^2$ with respect to $\sigma^2$ that are needed to obtain the score function.\newline
 \newline
In figure \ref{fig2}, we show the cumulative efficiency $\epsilon_N(\sigma^2)$, for $N = 2$ to $N = 5$, from bottom to top. (Note that $s_1(\sigma^2)$ vanishes since the mean of $x$ is unity for any value of the variance). The uppermost line contains therefore the variance, the skewness, the kurtosis as well as the 5th moment of the field.
The contribution of each successive moment can be read out from the difference between the corresponding successive curves. For higher $N$ quick convergence of $\epsilon_N$ occurs, presented in figure \ref{fig3} as the solid line, showing $\epsilon_{10}$. For small values of the variance, the field is still close to Gaussian, so that the Fisher information is close to be entirely within the the 2nd moment, and accordingly the ratio $\epsilon$ is close to unity in this regime. It is obvious from these figures that the main effect for larger values of the variance is not that Fisher information is transferred to higher order moments, but rather the dramatic cutoff as soon as the variance crosses $0.1$.  At redshift 1, this corresponds to the scale of $\approx 1'$ \cite[figure 1 ]{2006ApJ...645....1D}, so still within scales probed by weak lensing. For $\sigma \sim 1$, the ratio is close to $0.05$, meaning that all moments completely fails to capture the information. Optimal constraints on any cosmological parameter entering $\sigma$ are thus for this value of the variance a factor $1/\sqrt{0.05} \sim 4.5$ tighter than those achievable with the help of the entire hierarchy. 
 \newline
 \newline\noindent
In figure \ref{fig3} we compare these results to the exact analytical expressions given in \cite{2011ApJ...738...86C} for the lognormal distribution, shown as the dashed line. These are given by, accounting for the different normalization,
\beq
s^2_N(\sigma^2) = q^2 \frac{q^N}{1-q^N} \lp \prod_{n = 1}^{N-1} \lp1 - q^n \rp \rp\lp \sum_{n = 1}^{N-1} \frac{q^n}{1-q^n}\rp^2,
\enq
with $q := 1/ (1 + \sigma^2)$. The total Fisher information content being in this case $(q/\ln q)^2 / 2 -q^2/ (4\ln q)$.
There also the information content of the moments saturates quickly as $N$ grows. It is striking that the incompleteness of the moment hierarchy occurs much earlier in the convergence field than in the lognormal. This can be understood from the following considerations. The main effect of the improved model (\ref{model}) for the convergence is to reproduce accurately the very sharp cutoff of the probability density function at low convergence values \cite[figure 3-6]{2006ApJ...645....1D}. This cutoff is very sensitive to the variance of the field, more sensitive than the cutoff of the lognormal. However, there the contribution to the moment $m_n$, $x^n$, is beaten down by orders of magnitude. To make this point clearer, we show in figure \ref{fig4} the Fisher information density $p \lp \partial_{\sigma^2}\ln p\rp^2$ for the lognormal distribution (dashed) and the model we used (solid), at the scale of $\sigma = 1$ It is obvious in both cases that a large fraction of the information is contained in the underdense regions, describing the cutoff of the distribution, but unaccessible to the moments of $x$. Since this is even more the case for the convergence field, the efficiency is accordingly even worse.
\section{Restoration of the information} 
\begin{figure}  \includegraphics[width = 0.8\textwidth]{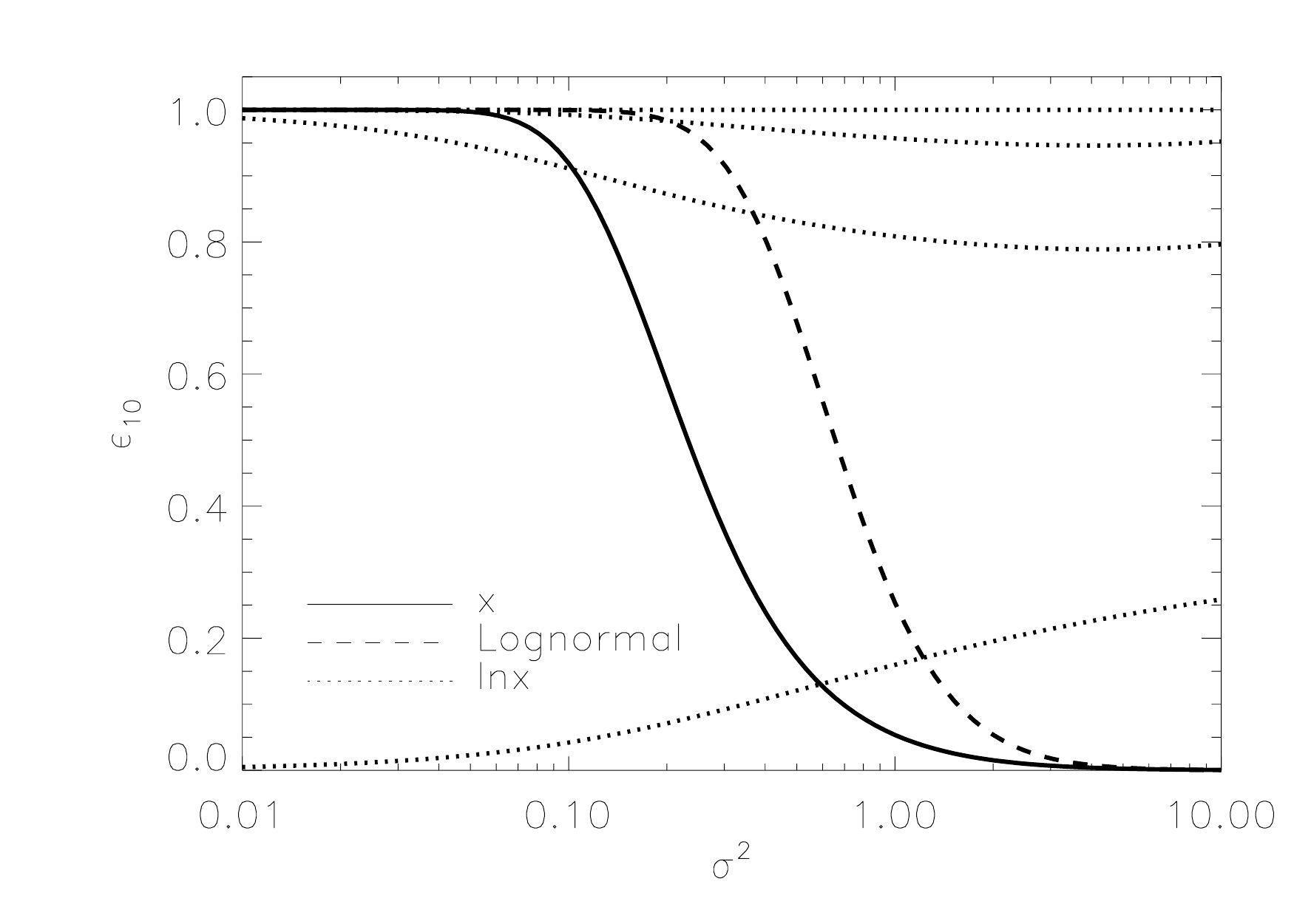}
 \centering
 \caption[The efficiency $\epsilon_{N = 10}$ of the first $10$ moments to capture Fisher information, for the convergence field and lognormal field, as well as for the logarithmic transform of the convergence field.]{\label{fig3}Solid and dashed line : the efficiency $\epsilon_{N = 10}$ of the first $10$ moments to capture Fisher information, for the convergence field (solid) and lognormal field (dashed). The curves do not change anymore with increasing $N$. Dotted : $\epsilon_1$,$\epsilon_2$, $\epsilon_3$ and $\epsilon_{10}$ for the logarithmic transform of the field, from bottom to top.} 
 \end{figure}

\begin{figure}
     \centering
  \includegraphics[width = 0.8\textwidth]{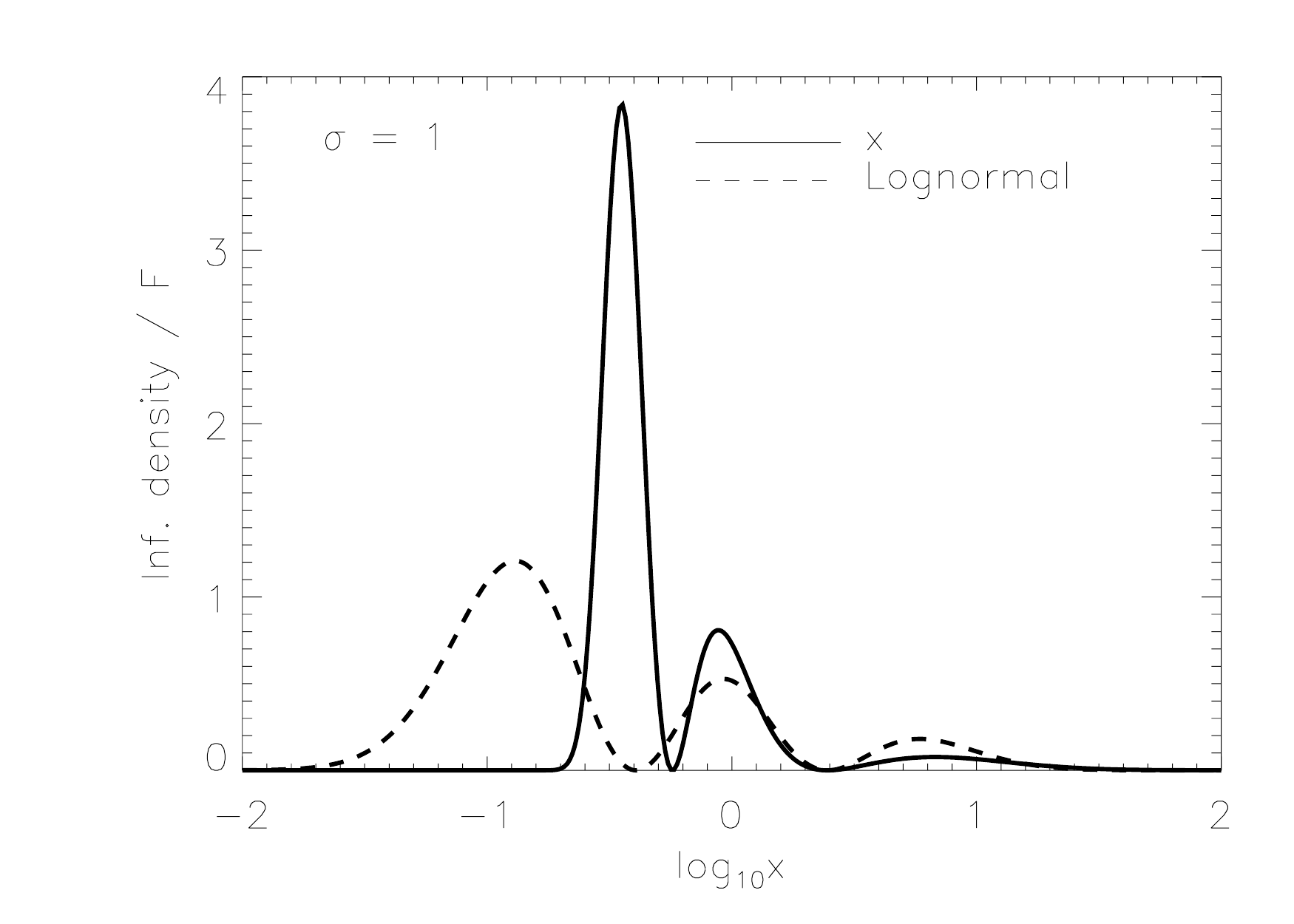}
 \caption[The Fisher information density of the lognormal and the convergence field.]{\label{fig4} The Fisher information density of the lognormal (dashed) and the convergence field (solid), renormalized such that it integrates to unity. Clearly, the Fisher information is mostly contained within the underdense regions. The moments are however sensitive to the tail.}
  \end{figure}

Finally we investigate to what extent the moment hierarchy of $\ln x$ contains more Fisher information than the hierarchy of $x$. Though our method is completely independent, this can be seen as complementary to recent works looking at the statistics of the field after local transforms, and at the statistical power of its power spectrum initiated in \cite{2009ApJ...698L..90N,2011ApJ...731..116N,2011ApJ...729L..11S}, even though in these works the fact that information actually completely escapes the hierarchy is not appreciated. This is done with the very same method used above, by obtaining the polynomials orthogonal to the distribution of $\ln x$, or equivalently decomposing the score function of $x$ in polynomials in $\ln x$ rather than in $x$. This is seen to perform very well, as shown by the dotted lines in figure \ref{fig3}. From bottom to top are plotted $\epsilon_1$,$\epsilon_2$ and $\epsilon_3$. Also shown in the figure is $\epsilon_{10}$ but it is not to be distinguished from unity, meaning that completeness of the hierarchy is restored. We see that over the full range at least $80\%$ of the information is back in the two first moments, and $95\%$ in the first three.

\section{Conclusions}
We have studied the statistical power of the moment hierarchy of the convergence field, when leaving the linear regime. Notably, the hierarchy ceases to provide a complete description of the statistics of the convergence, letting an increasingly large fraction of the Fisher information actually escape the hierarchy, and thus making constraints on cosmological parameters achievable with measurements of the hierarchy suboptimal by increasingly large factors. While our results are exact only for the one point distribution (or equivalently the full correlation function hierarchy of the convergence field in the limit of vanishing correlations), the correlation function hierarchy will also show a similar behavior, though the amplitude of the loss in information and constraining power may vary from parameter to parameter in the details. This is because this defect, for any number of variables, is due to the very slow decay rate at infinity of the field distribution, which cannot be reproduced by the exponential of a polynomial in the relevant variables. Our findings are consistent with previous analytical results on the lognormal distribution \citep{2011ApJ...738...86C}, and numerical work from $N$-body simulations at the power spectrum level \citep{2009ApJ...698L..90N,2011ApJ...731..116N}. Making a tighter connection to such simulation results with the methods presented here is the subject of future work. Of course, the quest for the information in the non linear regime already has problems of its own, such as shot noise issues, or accurate modeling, that we did not consider here. Nonetheless, these results clearly shows that if the correlation function hierarchy is to play a substantial role in getting constraints out of the mildly or non-linear regime, then an approach similar to a Gaussianizing transform \citep{2011ApJ...731..116N,2011ApJ...729L..11S}, in this work a simple logarithmic mapping, can hardly be avoided though the details still needs to be figured out. It is reassuring that this approach seems to work well to first order, and that first steps have recently already been taken in that direction in perturbation theory \citep{2011ApJ...735...32W}, for the matter field. Our work also points toward low convergence regions as carrying large amounts of information, though the importance of noise issues needs to be clarified in this regime. Thus, many promising ways have still to be explored to make profit of mildly and non-linear scales.

%

\clearpage
\newpage
\chapter*{Summary of main results and outlook}
\addcontentsline{toc}{chapter}{Summary of main results and outlook}
\markboth{SUMMARY AND OUTLOOK}{}

\parindent=0.0 cm

\begin{figure}
   \includegraphics[width = 0.49\textwidth]{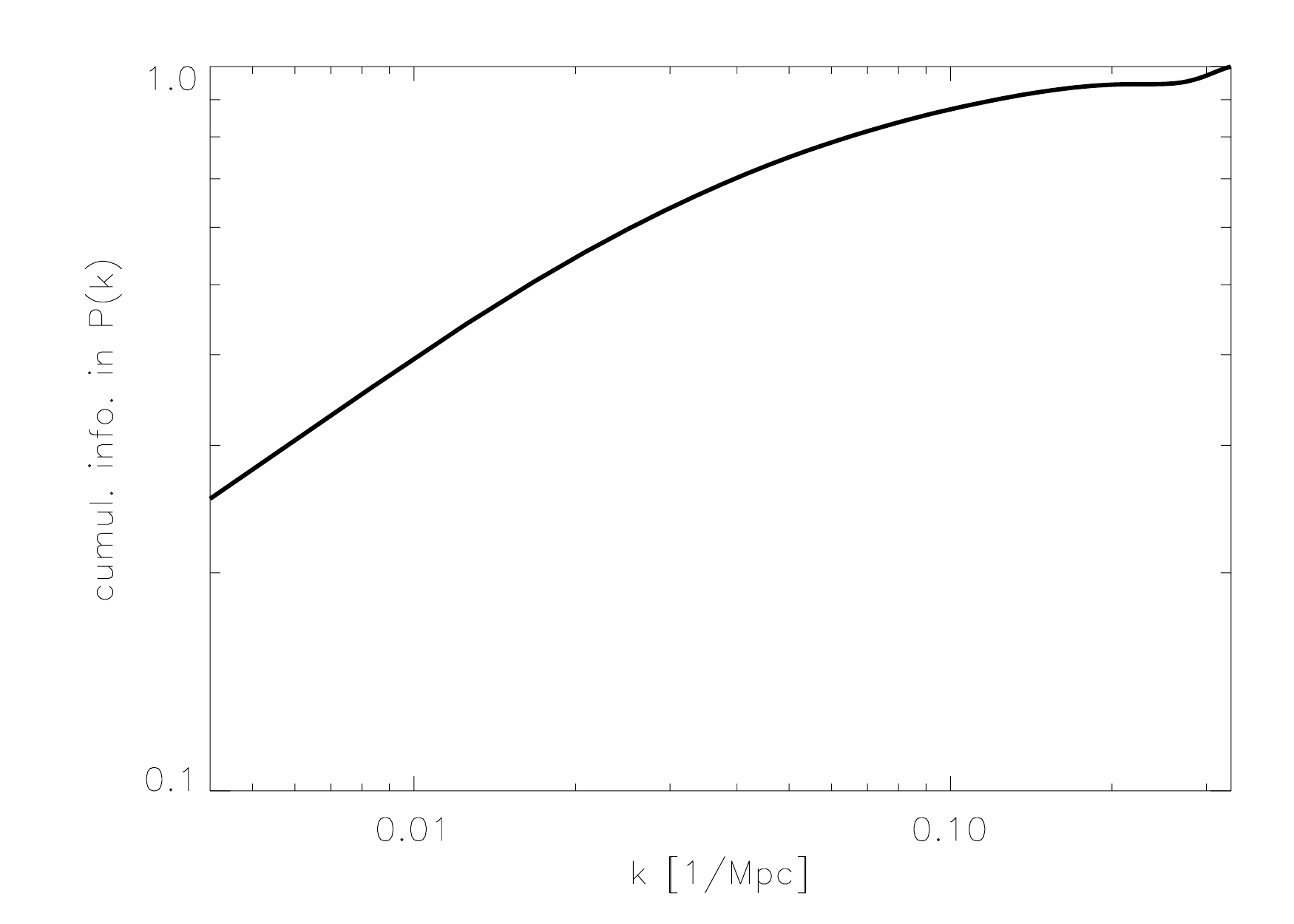}  
   \includegraphics[width = 0.49\textwidth]{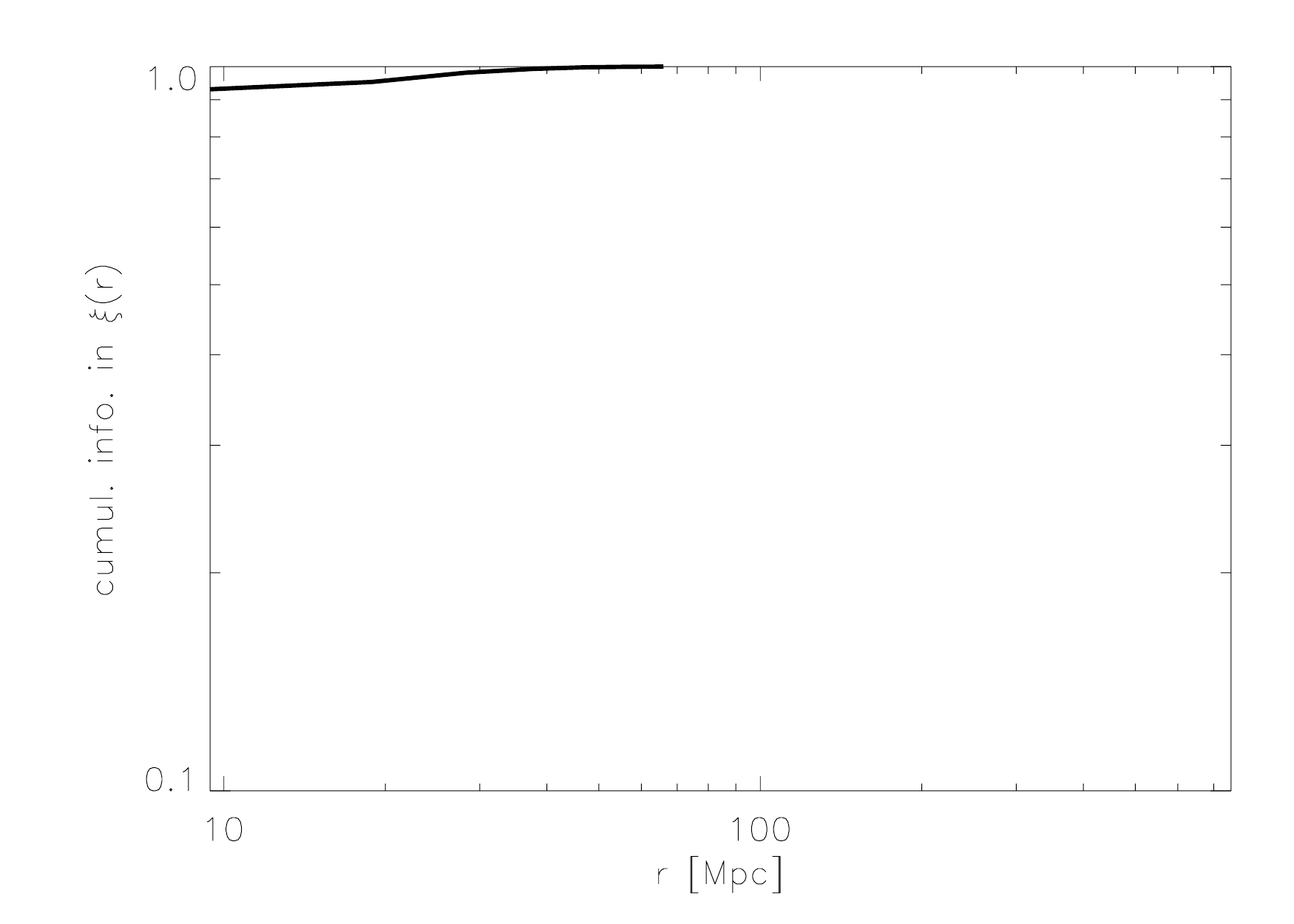}
   \centering
\caption[The cumulative information on the amplitude of the spectrum within either the spectrum or two-point function of the one dimensional lognormal field.]{On the left panel it is shown the exact cumulative information within the spectrum of the one-dimensional lognormal field as function of the maximal wavenumber included in the analysis, for a variance of unity. The model parameter of interest is here the amplitude of the spectrum. For a Gaussian field, this line would be a straight line, but in the lognormal field saturation appears after the linear scales. On the right the same cumulative information but within the different arguments of the two-point function. The variance of the field carries the vast majority of the information. \label{figcumul}}
\end{figure}

We have researched the information content of the cosmological matter density field and of the convergence field in order to gain insights on the statistical power of important cosmological observables and their combination, using tools borrowed from information theory.
\newline
\newline
In a first step, we have introduced distributions of maximum entropy, and discussed with their help the combination of magnification and flexion fields to the more usual shear fields, all tracers of the convergence, for the purpose of extracting cosmological parameters, or to reconstruct the underlying mass field. We found that size information contribute a modest but scale independent amount to the information, and one can expect constraints on any model parameter to increase by some $10\%$ with respect to a shear only analysis.  The flexion field has very different noise properties, that are strongly scale dependent. We found the information from flexion alone to take over that of shear only on the scales of the arcsecond, becoming the most interesting observable for the purpose of mass reconstruction. We believe that these scales are rather small in order for flexion to become a useful cosmological tool due to the fact that the nonlinear matter power spectrum is not extremely well understood on these scales at present, but flexion can complement well the shears on the scale of the arcminute.
\newline
\newline
After a brief discussion of the information content of power spectra estimators in chapter \ref{chNote}, we have then turned to the main part of thesis, which is the study of the statistical power of the hierarchy of $N$-point moments of a given density function, with applications to the noise-free lognormal and convergence field. We found that the hierarchy is clearly inadequate to describe such fields in the nonlinear regime, as the hierarchy fails to capture increasingly large fractions of the information. The reason is that the information that is captured becomes completely correlated with that of the overdense regions, and the information in the underdense regions, increasingly important, becomes inaccessible. A few large, cosmic variance dominated density peaks dominate the correlations and hide most of the information, making parameter inference from the $N$-point moments inefficient by orders of magnitude in the deeply nonlinear regime. The fact that information escapes the hierarchy could also been seen directly from the different fields we defined with the very same hierarchy of $N$-point moments than the lognormal field. We could successfully confront these results to numerical simulations, and show how simple non linear mappings, already present in the literature, are able to correct for these defects. 
\newline
\newline
An aspect of our results that we found most intriguing is the clear success of the model with uncorrelated lognormal fiducial to reproduce correctly the behavior of the Fisher information that is observed in the $N$-body simulations for a wide class of model parameters. For all the standard cosmological parameters that we tested in chapter \ref{ch4}, we found that the correlations played very little role in understanding the dynamics of their information. The only relevant parameter was found to be the variance of the field, and not the details of the correlation structure. This is also illustrated in figure \ref{figcumul}. There we evaluate numerically with the methods we exposed in this thesis the cumulative information on the amplitude of the $\Lambda$CDM power spectrum assuming lognormal statistics, for the one-dimensional configuration of points introduced in chapter \ref{ch4}, figure \ref{LNchain}. The minimal distance we took is $r_\min = 10$ Mpc, corresponding to a variance of unity. On the other hand, the radius $R$ plays no role. On the left panel is shown the contribution of the different wavenumbers, as a function of the maximal wavenumber extracted.  We see clearly that, as expected, when entering the non linear regime the spectrum ceases slowly to contain further independent information. On the other hand, the same analysis with the real space correlation function, shown in the right panel is very different. There, almost the entirety of the information is contained in the variance, the two-point function at non-zero lag carrying very little information.  In other words, for such parameters, there is as much information in the variance of the field than in its entire two-point function. We find this result an interesting starting point for further investigations, as this implies that it may be possible to understand the information content of the non linear regime with much simpler means than previously thought.
\newline
\newline
Certainly, a number of important issues still have to be answered. For instance, we have reached our conclusions using rather simple models such as the lognormal field. These models are good enough so that we can be sure that our claims on the $N$-point moment hierarchy of the noise-free fields do hold, at least for the first members of the hierarchy, which for practical reasons are the one that can be measured. However, when it comes to more stringent tests or measurements of a cosmological model, it is necessary to go beyond these simple prescriptions. Also, we pointed out on several occasions that underdense regions carry the largest part of the information in the nonlinear regime, a part unaccessible to the hierarchy. It is however not clear yet how this information can be extracted. As we discussed, in recent times several works have been looking at non linear transformations  for the purpose of parameter inference, using expensive $N$-body simulations. Our simple analytical results bring strong support towards such methods, and provide as well an unambiguous explanation of their results, which are natural consequences of the statistics of fields with heavy tails. Nevertheless, it still remains to try and apply these methods to actual data. In light of our research, we judge this approach as promising, as it is a very straightforward manner to try and capture these large amount of information within the underdense regions. Clearly, the issue of the noise pervading the measurements will be essential for the success of these nonlinear transforms.  We have tried to give very first elements of answers in this thesis, but a lot of work is still ahead. We hope that the methods introduced in this research will contribute to these efforts.

\cleardoublepage

\renewcommand{\theequation}{A-\arabic{equation}}
\setcounter{equation}{0}
\renewcommand{\thefigure}{A.\arabic{figure}}
\setcounter{figure}{0}
\renewcommand{\thetable}{A.\arabic{table}}
\setcounter{table}{0}
\renewcommand{\thesection}{A.\arabic{section}}
\setcounter{section}{0}
\renewcommand{\thesubsection}{A.\arabic{section}.\arabic{subsection}}
\setcounter{subsection}{0}

\fancyhead[LE]{\leftmark}
\bibliographystyle{plainnat}
\bibliography{astro_references}
\addcontentsline{toc}{chapter}{Bibliography}
\cleardoublepage
\cleardoublepage

 \end{document}